\documentclass[a4paper,11pt]{article}
\usepackage{jcappub}

\usepackage[a4paper, total={6in, 8in}]{geometry}
\usepackage{subfigure}
\usepackage{wrapfig}
\usepackage{graphicx}
\usepackage[utf8]{inputenc}
\usepackage{hyperref}
\usepackage{aas_macros}
\usepackage{amsmath,amssymb}
\usepackage{doi}
\usepackage{array,longtable}
\usepackage{multirow}
\usepackage[table,usenames,dvipsnames]{xcolor}
\usepackage{color}
\usepackage{orcidlink}

\newcommand{\Ms}{{\rm ~M}_\odot}

\newcommand{\citeg}[1]{\citep[e.g.][]{#1}}

\newcommand{\changes}[1]{#1}

\newcommand{\irm}{{\rm i}}
\newcommand{\beq}{\begin{equation}}
\newcommand{\eeq}{\end{equation}}

\newcolumntype{L}[1]{>{\raggedright\let\newline\\\arraybackslash\hspace{0pt}}m{#1}}

\hypersetup{
    colorlinks=true,
    linkcolor=blue,
    citecolor=blue,
    filecolor=magenta,      
    urlcolor=cyan,
    pdftitle={LGWA Science White Paper},
    pdfpagemode=FullScreen,
    }

\def\gtsim {>\kern-1.2em\lower1.1ex\hbox{$\sim$}~}   
\def\ltsim {<\kern-1.2em\lower1.1ex\hbox{$\sim$}~}   

\title{\boldmath The Lunar Gravitational-wave Antenna: Mission Studies and Science Case}

\author[60]{Parameswaran Ajith \orcidlink{0000-0001-7519-2439}}
\author[36,72,73]{Pau Amaro Seoane \orcidlink{0000-0003-3993-3249}}
\author[1,2,15]{Manuel Arca Sedda \orcidlink{0000-0002-3987-0519}}
\author[4]{Riccardo Arcodia \orcidlink{0000-0003-4054-7978}}
\author[13,14]{Francesca Badaracco \orcidlink{0000-0001-8553-7904}}
\author[1,2]{Biswajit Banerjee \orcidlink{0000-0002-8008-2485}}
\author[5,6]{Enis Belgacem \orcidlink{0000-0003-4920-0911}}
\author[71]{Giovanni Benetti}
\author[37]{Stefano Benetti \orcidlink{0000-0002-3256-0016}}
\author[54]{Alexey Bobrick \orcidlink{0000-0002-4674-0704}}
\author[43]{Alessandro Bonforte \orcidlink{0000-0003-0435-7763}}
\author[9,10]{Elisa Bortolas \orcidlink{0000-0001-9458-821X}}
\author[23,55]{Valentina Braito \orcidlink{0000-0002-2629-4989}}
\author[1,2,15]{Marica Branchesi \orcidlink{0000-0003-1643-0526}}
\author[65]{Adam Burrows \orcidlink{0000-0002-3099-5024}}
\author[37]{Enrico Cappellaro \orcidlink{0000-0001-5008-8619}}
\author[23]{Roberto Della Ceca \orcidlink{0000-0001-7551-2252}}
\author[61]{Chandrachur Chakraborty \orcidlink{0000-0003-4380-3033}}
\author[49]{Shreevathsa Chalathadka Subrahmanya \orcidlink{0000-0002-9207-4669}}
\author[3]{Michael W. Coughlin  \orcidlink{0000-0002-8262-2924}}
\author[23]{Stefano Covino \orcidlink{0000-0001-9078-5507}}
\author[47,48]{Andrea Derdzinski\orcidlink{0000-0001-9880-8929}}
\author[56]{Aayushi Doshi \orcidlink{0009-0006-2205-3391}}
\author[29,30]{Maurizio Falanga \orcidlink{0000-0003-3095-6065}}
\author[5,6]{Stefano Foffa \orcidlink{0000-0002-4530-3051}}
\author[40,10]{Alessia Franchini \orcidlink{0000-0002-8400-0969}}
\author[27]{Alessandro Frigeri \orcidlink{0000-0002-9140-3977}}
\author[51]{Yoshifumi Futaana \orcidlink{0000-0002-7056-3517}}
\author[49]{Oliver Gerberding \orcidlink{0000-0001-7740-2698}}
\author[66]{Kiranjyot Gill \orcidlink{0000-0003-4341-9824}}
\author[7,8]{Matteo Di Giovanni \orcidlink{0000-0003-4049-8336}}
\author[67,68]{Ines Francesca Giudice}
\author[12]{Margherita Giustini \orcidlink{0000-0002-1329-658X}}
\author[53]{Philipp Gläser \orcidlink{0000-0002-7552-5800}}
\author[1,2]{Jan Harms\footnote{Corresponding author.} \orcidlink{0000-0002-7332-9806}}
\author[45,46]{Joris van Heijningen \orcidlink{0000-0002-8391-7513}}
\author[5,6]{Francesco Iacovelli \orcidlink{0000-0002-4875-5862}}
\author[24]{Bradley J. Kavanagh \orcidlink{0000-0002-3634-4679}}
\author[50]{Taichi Kawamura \orcidlink{0000-0001-5246-5561}}
\author[22]{Arun Kenath \orcidlink{0000-0002-2183-9425}}
\author[18]{Elisabeth-Adelheid Keppler \orcidlink{0000-0003-3453-2606}}
\author[64]{{Chiaki Kobayashi} \orcidlink{0000-0002-4343-0487}}
\author[44]{{Goro Komatsu} \orcidlink{0000-0003-4155-108x}}
\author[32]{Valeriya Korol \orcidlink{0000-0002-6725-5935}}
\author[60]{N. V. Krishnendu \orcidlink{0000-0002-3483-7517}}
\author[60]{Prayush Kumar \orcidlink{0000-0001-5523-4603}}
\author[18,19]{Francesco Longo \orcidlink{0000-0003-2501-2270}}
\author[5,6]{Michele Maggiore \orcidlink{0000-0001-7348-047X}}
\author[9,10,11]{Michele Mancarella \orcidlink{0000-0002-0675-508X}}
\author[1,2]{Andrea Maselli \orcidlink{0000-0001-8515-8525}}
\author[28,71]{Alessandra Mastrobuono-Battisti \orcidlink{0000-0002-2386-9142}}
\author[39]{{Francesco Mazzarini} \orcidlink{0000-0002-3864-6558}}
\author[16]{Andrea Melandri \orcidlink{0000-0002-2810-2143}}
\author[38]{{Daniele Melini} \orcidlink{0000-0002-5383-2375}}
\author[41]{{Sabrina Menina} \orcidlink{0000-0003-1044-6877}}
\author[12]{Giovanni Miniutti \orcidlink{0000-0003-0707-4531}}
\author[58]{{Deeshani Mitra} \orcidlink{0000-0002-1350-019X}}
\author[33]{Javier Morán-Fraile \orcidlink{0000-0002-8918-5130}}
\author[31]{Suvodip Mukherjee \orcidlink{0000-0002-3373-5236}}
\author[5,6]{Niccol\`o Muttoni \orcidlink{0000-0002-4214-2344}}
\author[17,27]{Marco Olivieri \orcidlink{0000-0002-7333-8809}}
\author[15]{Francesca Onori \orcidlink{0000-0001-6286-1744}}
\author[63]{Maria Alessandra Papa \orcidlink{0000-0002-1007-5298}}
\author[42]{Ferdinando Patat \orcidlink{0000-0002-0537-3573}}
\author[69,70]{Andrea Perali \orcidlink{0000-0002-4914-4975}}
\author[59]{Tsvi Piran \orcidlink{0000-0002-7964-5420}}
\author[16]{Silvia Piranomonte \orcidlink{0000-0002-8875-5453}}
\author[5,6]{Alberto Roper Pol \orcidlink{0000-0003-4979-4430}}
\author[62]{Masroor C. Pookkillath \orcidlink{0000-0002-7199-8037}}
\author[60]{R. Prasad \orcidlink{0000-0002-6602-3913}}
\author[60]{Vaishak Prasad \orcidlink{0000-0001-6712-2457}}
\author[27]{Alessandra De Rosa \orcidlink{0000-0001-5668-6863}}
\author[34,35]{Sourav Roy Chowdhury \orcidlink{0000-0003-2802-4138}}
\author[16]{Roberto Serafinelli \orcidlink{0000-0003-1200-5071}}
\author[9,10,23]{Alberto Sesana \orcidlink{0000-0003-4961-1606}}
\author[23]{Paola Severgnini \orcidlink{0000-0001-5619-5896}}
\author[17]{Angela Stallone \orcidlink{0000-0002-8141-017X}}
\author[1,2]{Jacopo Tissino \orcidlink{0000-0003-2483-6710}}
\author[52]{Hrvoje Tkal\v{c}i\'{c} \orcidlink{0000-0001-7072-490X}}
\author[37]{Lina Tomasella \orcidlink{0000-0002-3697-2616}}
\author[9]{Martina Toscani \orcidlink{0000-0001-5997-7148}}
\author[57]{David Vartanyan \orcidlink{0000-0003-1938-9282}}
\author[25,26]{Cristian Vignali \orcidlink{0000-0002-8853-9611}}
\author[17]{Lucia Zaccarelli \orcidlink{0000-0002-4053-7625}}
\author[20,21]{Morgane Zeoli \orcidlink{0009-0007-1898-4844}}
\author[39]{Luciano Zuccarello \orcidlink{0000-0003-0094-9577}}

\affiliation[1]{Gran Sasso Science Institute (GSSI), Via Michele Iacobucci 2, I-67100 L'Aquila, Italy}
\affiliation[2]{INFN, Laboratori Nazionali del Gran Sasso, Via Giovanni Acitelli 22, I-67100 Assergi, Italy}
\affiliation[3]{School of Physics and Astronomy, University of Minnesota, 116 Church Street SE, Minneapolis, Minnesota 55455, USA}
\affiliation[4]{MIT Kavli Institute for Astrophysics and Space Research, 70 Vassar Street, Cambridge, MA 02139, USA}
\affiliation[5]{D\'epartement de Physique Th\'eorique,
Universit\'e de Gen\`eve, 24 quai Ansermet, CH-1211 Gen\`eve 4, Switzerland}
\affiliation[6]{Gravitational Wave Science Center (GWSC), Universit\'e de Gen\`eve, CH-1211 Geneva, Switzerland}
\affiliation[7]{La Sapienza Università di Roma, I-00185 Roma, Italy}
\affiliation[8]{INFN, Sezione di Roma, I-00185 Roma, Italy}
\affiliation[9]{Dipartimento di Fisica ``G. Occhialini'', Universit\'a degli Studi di Milano-Bicocca, Piazza della Scienza 3, 20126 Milano, Italy}
\affiliation[10]{INFN, Sezione di Milano-Bicocca, Piazza della Scienza 3, 20126 Milano, Italy}
\affiliation[11]{Aix Marseille Univ, Universit\'e de Toulon, CNRS, CPT, Marseille, France}
\affiliation[12]{Centro de Astrobiolog\'ia (CAB), CSIC-INTA, ESAC campus, Camino Bajo del Castillo s/n, 28692 Villanueva de la Ca\~nada, Spain}
\affiliation[13]{Dipartimento di Fisica, Università di Genova, I-16146 Genova, Italy}
\affiliation[14]{INFN Sez. di Genova, I-16146 Genova, Italy}
\affiliation[15]{INAF - Osservatorio Astronomico d'Abruzzo, Via Mentore Maggini, s.n.c., I-64100 Teramo, Italy}
\affiliation[16]{INAF - Osservatorio Astronomico di Roma, via Frascati 33, I-00078 Monte Porzio Catone (Roma), Italy}
\affiliation[17]{Istituto Nazionale di Geofisica e Vulcanologia, Sezione di Bologna, viale Berti Pichat 6-2, I-40127 Bologna, Italy}
\affiliation[18]{Dipartimento di Fisica, Universit\`a degli Studi di Trieste, via Valerio 2, 34127, Trieste, Italy}
\affiliation[19]{INFN, Sezione di Trieste, via Valerio 2, 34127 Trieste, Italy}
\affiliation[20]{Centre for Cosmology, Particle Physics and Phenomenology (CP3), UCLouvain, B-1348 Louvain-la-Neuve, Belgium}
\affiliation[21]{Precision Mechatronics Laboratory, ULiège, A\&M Dept., Allée de la Découverte 9, B52/Quartier Polytec 1, B-4000, Liège, Belgium}
\affiliation[22]{Department of Physics and Electronics, Christ University, Hosur Road, Bangalore - 560029, India}
\affiliation[23]{INAF - Osservatorio Astronomico di Brera, via Brera 28, I-20121 Milano, Italy}
\affiliation[24]{Instituto de F\'isica de Cantabria (IFCA, UC-CSIC), Av.~de Los Castros s/n, 39005 Santander, Spain}
\affiliation[25]{Dipartimento di Fisica e Astronomia ``Augusto Righi", Alma Mater Studiorum, Universit\`a degli Studi di Bologna, Via Gobetti 93/2, I-40129 Bologna, Italy}
\affiliation[26] {INAF -- Osservatorio di Astrofisica e Scienza dello Spazio di Bologna, Via Gobetti 93/3, I-40129 Bologna, Italy}
\affiliation[27]{INAF -- Istituto di Astrofisica e Planetologia Spaziali, Via Fosso del Cavaliere 100, I-00133, Roma, Italy}
\affiliation[28]{GEPI, Observatoire de Paris, PSL Research University, CNRS, Place Jules Janssen, 92190 Meudon, France}
\affiliation[29]{International Space Science Institute (ISSI), Hallerstrasse 6, 3012 Bern, Switzerland}
\affiliation[30]{Physikalisches Institut, University of Bern, Sidlerstrasse 5, 3012 Bern, Switzerland}
\affiliation[31]{Department of Astronomy and Astrophysics, Tata Institute of Fundamental Research, Homi Bhabha Road, Mumbai-400005, India}
\affiliation[32]{Max-Planck-Institut f{\"u}r Astrophysik, Karl-Schwarzschild-Stra{\ss}e 1, 85748 Garching, Germany}
\affiliation[33]{Heidelberger Institut f{\"u}r Theoretische Studien (HITS), Schloss-Wolfsbrunnenweg 35, 69118 Heidelberg, Germany}
\affiliation[34]{Research Institute of Physics, Southern Federal University, 344090 Rostov on Don, Russia.}
\affiliation[35]{Department of Physics, Vidyasagar College, 39, Shankar Ghosh Lane, Kolkata, India.}
\affiliation[36]{Universitat Politècnica de València, C/Vera s/n, València, Spain}
\affiliation[37]{INAF - Osservatorio Astronomico di Padova, Vicolo dell'Osservatorio 5, I-35122 Padova, Italy}
\affiliation[38]{Istituto Nazionale di Geofisica e Vulcanologia, Sezione di Roma 1, Via di Vigna Murata 605, I-00143 Roma, Italy}
\affiliation[39]{Istituto Nazionale di Geofisica e Vulcanologia, Sezione di Pisa, Via C. Battisti  53, I-56125 Pisa, Italy}
\affiliation[40]{Institut für Astrophysik, Universität Zürich, Winterthurerstrasse 190, CH-8057 Zürich, Switzerland}
\affiliation[41]{SYRTE,  Observatoire de Paris - Université PSL, CNRS, Sorbonne Université, LNE, 77 Av. Denfert Rochereau, 75014 Paris, France}
\affiliation[42]{European Southern Observatory, Karl-Schwarzschild-Str. 2, 85748 Garching b. München, Germany}
\affiliation[43]{Istituto Nazionale di Geofisica e Vulcanologia, Sezione di Catania - Osservatorio Etneo  Piazza Roma, 2 - 95125 Catania, Italy}
\affiliation[44]{International Research School of Planetary Sciences, Università d'Annunzio, Viale Pindaro 42, 65127 Pescara, Italy}
\affiliation[45]{Department of Physics and Astronomy, VU Amsterdam; De Boelelaan 1081, 1081, HV, Amsterdam, The Netherlands}
\affiliation[46]{Nikhef; Science Park 105, 1098, XG Amsterdam, The Netherlands}
\affiliation[47]{Department of Life and Physical Sciences, Fisk University, 1000 17th Avenue N., Nashville, TN 37208, USA}
\affiliation[48]{Department of Physics \& Astronomy, Vanderbilt University,
2301 Vanderbilt Place, Nashville, TN 37235, USA}
\affiliation[49]{Institute of Experimental Physics, University of Hamburg, Luruper Chaussee 149, 22761 Hamburg, Germany}
\affiliation[50]{Institut de physique du globe de Paris, CNRS, Université Paris Cité, Paris, France}
\affiliation[51]{Solar System Physics and Space Technology Programme, Swedish Institute of Space Physics, Kiruna, SE98128, Sweden}
\affiliation[52]{Research School of Earth Sciences, The Australian National University, 142 Mills Road, Canberra, Australia}
\affiliation[53]{Technical University Berlin, Department of Geodesy and Geoinformation Science, Str. des 17. Juni 135, 10623 Berlin, Germany}
\affiliation[54]{Technion -- Israel Institute of Technology, Physics department, Haifa Israel 3200002}
\affiliation[55]{Dipartimento di Fisica, Universit\`a di Trento, Via Sommarive 14, I-38123 Trento, Italy}
\affiliation[56]{Department of Physics and Astronomy, Rice University, Houston, TX}
\affiliation[57]{Carnegie Observatories, 813 Santa Barbara St, Pasadena, CA 91101, USA}
\affiliation[58]{Dept. of Physics, St Xavier's College (Autonomous), Kolkata, India}
\affiliation[59]{Racah Institute of Physics, The Hebrew University, Givat Ram,  Jerusalem, 91904 Israel}
\affiliation[60]{International Centre for Theoretical Science, Tata Institute of Fundamental Research, Bangalore - 560089, India}
\affiliation[61]{Manipal Centre for Natural Sciences, Manipal Academy of Higher Education, Manipal 576104, India}
\affiliation[62]{Centre for Theoretical Physics and Natural Philosophy, Mahidol University, Nakhonsawan Campus,  Phayuha Khiri, Nakhonsawan 60130, Thailand}
\affiliation[63]{Max Planck Institute for Gravitational Physics (Albert Einstein Institute), Callinstraße 38, D-30167 Hannover, Germany}
\affiliation[64]{Centre for Astrophysics Research, University of Hertfordshire, College Lane, Hatfield  AL10 9AB, UK}
\affiliation[65]{Department of Astrophysical Sciences, 4 Ivy Lane, Princeton University, Princeton, New Jersey 08544, USA}
\affiliation[66]{Center for Astrophysics \textbar{} Harvard \& Smithsonian, 60 Garden Street, Cambridge, MA 02138-1516, USA}
\affiliation[67]{Department of Physics "Ettore Pancini", Università degli Studi di Napoli Federico II, Via Cinthia, 21, I-80126 Napoli, Italy}
\affiliation[68]{INAF - Osservatorio Astronomico di Capodimonte, Salita Moiariello 16, I-80131, Naples, Italy}
\affiliation[69]{School of Pharmacy, Physics Unit, University of Camerino, Via Madonna delle Carceri 9B, I-62032 Camerino (MC), Italy}
\affiliation[70]{INAF, Section of Camerino, Via Gentile III da Varano 7, I-62032 Camerino (MC), Italy}
\affiliation[71]{Dipartimento di Fisica e Astronomia ``Galileo Galilei'', Università di Padova, Vicolo dell’Osservatorio 3, I-35122 Padova, Italy}
\affiliation[72]{Max-Planck Institut für Extraterrestrische Physik, Garching, Germany}
\affiliation[73]{Higgs Centre for Theoretical Physics, Edinburgh, UK}

\emailAdd{jan.harms@gssi.it}

\abstract{The Lunar Gravitational-wave Antenna (LGWA) is a proposed array of next-generation inertial sensors to monitor the response of the Moon to gravitational waves (GWs). Given the size of the Moon and the expected noise produced by the lunar seismic background, the LGWA would be able to observe GWs from about 1 mHz to 1 Hz. This would make the LGWA the missing link between space-borne detectors like LISA with peak sensitivities around a few millihertz and proposed future terrestrial detectors like Einstein Telescope or Cosmic Explorer. In this article, we provide a first comprehensive analysis of the LGWA science case including its multi-messenger aspects and lunar science with LGWA data. We also describe the scientific analyses of the Moon required to plan the LGWA mission.}

\begin{document}
\maketitle
\flushbottom

\tableofcontents

\section*{List of abbreviations}

\begin{tabular}{ll}
AGN & Active galactic nuclei \\
BH & Black hole \\
BBH & Binary black hole \\
BNS & Binary neutron star \\
CBC & Compact binary coalescence \\
CCSN & Core-collapse supernova \\
CGWB & Cosmological gravitational-wave background \\
DE & Dark Energy \\
DM & Dark Matter \\
DWD & Double white dwarf \\
EM & Electromagnetic \\
EMRI & Extreme mass ratio inspiral \\
EoS & Equation of State\\
GR & General Relativity \\
GW & Gravitational wave \\
IMBH & Intermediate mass black hole \\
IMRI & Intermediate mass ratio inspiral \\
LIGS & Lunar inertial gravitational-wave sensor \\
NS & Neutron star \\
NSWD & Neutron star - white dwarf \\
PBHs & Primordial black hole \\
PNS & Proto-neutron star\\
ppE & Parametrized post-Einsteinian \\ 
PPISN & Pulsational pair-instability supernova \\
PSR & Permanently shadowed region \\
QNM & Quasi normal mode\\
QPE & Quasi-periodic eruptions \\
RMS & Root mean square \\
SBN & Seismic background noise \\
SGWB & Stochastic gravitational wave background \\
SMBH & Supermassive black hole \\
SN & Supernova \\
SNR & Signal-to-noise ratio \\
TDE & Tidal disruption event \\
WD & White dwarf \\

\end{tabular}

\section{Introduction}
The first generation of GW detectors Virgo and LIGO have opened a new observational window to the Universe \cite{AbEA2016a,LIGOScientific:2021djp}. Additionally, the Pulsar Timing Array collaborations have recently reported the first signs of a possible GW background in their data \cite{antoniadis2023epta,agazie2023nano,reardon2023ppta}. Expanding the current observational window can be accomplished with two technological approaches: (1) by developing new instruments with better sensitivity within existing observation bands, as for example the proposed Einstein Telescope \cite{ET2020} and Cosmic Explorer \cite{Evans:2021gyd}, and (2) by opening new frequency bands for GW observations. The LISA mission was recently adopted by ESA and is scheduled for launch in 2035 \cite{colpi2024lisa}. New probes of the cosmic-microwave background aimed at revealing imprints of GWs on its polarization pattern are in the planning \cite{litebird2023}. Despite this well-developed roadmap for GW science, it is important to realize that the exploration of our Universe through GWs is still in its infancy. In addition to the immense impact expected on astrophysics and cosmology, this field holds a high probability for unexpected and fundamental discoveries.

The Lunar Gravitational-wave Antenna (LGWA) is a proposed first-generation lunar GW detector \cite{HaEA2021a}. The LGWA utilizes the Moon as a planetary-scale antenna for space-time fluctuations. The mission concept is to deploy an array of inertial sensors on the surface of the Moon to measure the Moon's vibrations caused by GWs. The LGWA is enabled by the unique geophysical conditions on the Moon: namely, its extreme seismic silence, the cryotemperature environment inside its PSRs, and its tidal lock with respect to Earth. The Moon may indeed be the only suitable planetary body in the whole solar system for LGWA.

The LGWA will serve as the missing link in the decihertz band, filling the gap between the sensitivity range of the LISA detector, which peaks at a few millihertz, and terrestrial detectors that may become sensitive down to a few hertz in the future. Unique and breakthrough contributions to GW science by LGWA are expected, which would also enable a new class of multi-messenger astronomy. In the following, we list some highlights of its science case:
\begin{itemize}
    \item {\bf Studying astrophysical explosions}. Only LGWA can observe astrophysical events that involve WDs like tidal disruption events and SNe Ia. Only LGWA can provide early warnings weeks to months before the mergers of solar-mass compact binaries that include NSs together with excellent sky localizations.
    \item {\bf Exploring black-hole populations and their role for structure formation in our Universe}. Only LGWA can observe lighter IMBH binaries in the early Universe and understand their role in the formation of today's SMBHs.
    \item {\bf Hubble constant measurement}. Only LGWA can detect GW signals from DWDs beyond our galaxy \changes{and nearby dwarf galaxies}. Only DWDs produce strong enough signals and are so ubiquitous in the local Universe that a large enough number can be detected with identified host galaxies to measure the Hubble constant accurately with the help of GW observations.
    \item {\bf Enabling the next level of high-precision waveform measurements}. Especially LGWA's unique role as a partner for multiband observations of GW sources enables a new level of precision in waveform measurements and searches for signs of a new fundamental physics.
    \item {\bf Imaging the lunar interior and shedding light on the formation history of the Moon.} LGWA might be the first mission to be able to observe lunar normal modes complementing seismic body waves, which give direct insight into the Moon's deep internal structure and hold information about its formation history.
    \item {\bf Revealing the Moon's meteoritic hum}. LGWA's inertial sensors will be more than two orders of magnitude more sensitive than any other lunar seismometer planned for geophysical studies, which enables the observation of the weakest ever resolvable lunar seismic events and of the incessant meteoritic hum.
\end{itemize}
Through multiband observations, LGWA emerges as a formidable partner of future laser-interferometric detectors both on Earth and in space. Moreover, LGWA marks the first step of a long-lasting utilization of the Moon for lunar GW detection \cite{Branchesi2023}. 

The SciSpacE white papers of the European Space Agency (ESA) on planetary science, astrophysics, and fundamental physics \footnote{\href{https://www.esa.int/Science_Exploration/Human_and_Robotic_Exploration/Research/The_SciSpacE_White_Papers}{ESA SciSpacE white papers}}, the Decadal Survey in Astrophysics \footnote{\href{https://www.nationalacademies.org/our-work/decadal-survey-on-astronomy-and-astrophysics-2020-astro2020}{Astro Decadal 2020 report}}, in Biological and Physical Sciences Research in Space\footnote{\href{https://www.nationalacademies.org/our-work/decadal-survey-on-life-and-physical-sciences-research-in-space-2023-2032}{BPS Decadal 2023 report}}, and in Planetary Science and Astrobiology \footnote{\href{https://www.nationalacademies.org/our-work/planetary-science-and-astrobiology-decadal-survey-2023-2032}{Planetary Decadal 2023 report}} all recognize the importance of GW detection and lunar science. The LGWA pathfinder mission \emph{Soundcheck} was selected by ESA in 2023 into the Reserve Pool of Science Activities for the Moon.

The purpose of this paper is to provide a description of the LGWA science case and an outline of the scientific studies needed for mission planning. Instead, we do not enter into the details of the payload design and technological requirements, and of the possible deployment scenarios, which will be provided with a dedicated document. \changes{There are also important limitations to the analyses presented in this paper. First and foremost, except for selected cases, there are no accurate simulations available yet to estimate the sky-localization capabilities of LGWA alone or as part of a network with LISA-type detectors or terrestrial detectors. This problem has analytical as well as numerical-precision aspects that must be solved. Furthermore, we lack accurate waveform models of some signals where matter effects are important, and population models are generally less developed compared to most models used in science-case studies of terrestrial detectors and LISA.}

\subsection{Science traceability matrices}
\label{sec:stm}

\subsubsection{Gravitational-wave science and multi-messenger astronomy}

\begin{longtable}{|L{2.8cm}|L{3cm}|L{2.5cm}|L{2.4cm}|L{2.4cm}|}
\hline
\cellcolor{Dandelion}\bf Science goals & \cellcolor{Dandelion}\bf Science questions & \cellcolor{Dandelion}\bf Observational requirements & \cellcolor{Dandelion}\bf Instrument requirements & \cellcolor{Dandelion}\bf Top-level mission requirements \\
\hline\endhead
\multirow{2}{2.8cm}{Understanding the origin and evolution of massive black holes} & What were the seeds of today's SMBHs? & Observe BBHs beyond redshift $z=10$ & Sensitivity better than $10^{-20}$\,Hz$^{-1/2}$ at 0.1\,Hz & 3\,yr mission lifetime \\
\cline{2-5}
& What is the merger rate of MBHs as function of z and what is the distribution of physical and orbital parameters regulating MBHs in pairs? 
& Observe binary MBH and IMBH mergers over a wide range of redshifts.  & Sensitivity better than $10^{-18}$\,Hz$^{-1/2}$ at 0.01\,Hz  & 5\,yr mission lifetime \\
\cline{2-5}
& What physics governs the correlations between the mass of a galaxy's central MBH and the velocity dispersion $\sigma$?  & Observe binary MBH and IMBH mergers over a wide range of redshifts and combine GW and EM observations & Sensitivity better than $10^{-18}$\,Hz$^{-1/2}$ at 0.01\,Hz &   5\,yr mission lifetime \\
\hline
\multirow{2}{2.8cm}{Understanding astrophysical explosions and their progenitors} & What processes govern the final years of the lifetime of DWD and NSWD binaries? & Observe short-period ($<100\,$s) binaries containing WDs. & Sensitivity better than $10^{-20}$\,Hz$^{-1/2}$ at 0.1\,Hz & 3\,yr mission lifetime \\ 
\cline{2-5}
& What processes drive the low-frequency GW signals of CCSN explosions? & GW memory signal detection coincident with an optical CCSN &  Sensitivity better than $10^{-21}$\,Hz$^{-1/2}$ around 0.3\,Hz & 10\,yr mission lifetime \\
\cline{2-5}
& What are the jet-launching conditions in gamma-ray bursts?& Observation of a GW memory signal coincident with a gamma-ray burst & Sensitivity better than $10^{-22}$\,Hz$^{-1/2}$ around 0.3\,Hz & 10\,yr mission lifetime (remains improbable detection) \\
\cline{2-5}
& What are the progenitors of SN type Ia? & Observe the last years of the life of a few DWDs & Sensitivity better than $10^{-21}$\,Hz$^{-1/2}$ around 0.3\,Hz & 10\,yr mission lifetime (remaining an improbable detection) \\
\cline{2-5}
& How often do tidal disruptions of WDs near BH horizons occur? & Observe the (unmodeled) GW transient produced by a WD tidal disruption in the dHz band. &  Sensitivity better than a few times $10^{-22}$\,Hz$^{-1/2}$ around 0.3\,Hz & 10\,yr mission lifetime \\
\hline
\multirow{2}{2.8cm}{Observations of rare stellar mergers or transients} & What is the asymmetry of the helium flash event? & Observation of a Galactic helium flash event & TBD & 10\,yr mission lifetime \\
\cline{2-5}
& What is the rate and mass distribution of stellar core mergers? & Observations coincident with luminous red nova events & TBD & 10\,yr mission lifetime \\
\hline
\multirow{2}{2.8cm}{Cosmology with dHz GW signals} & Are there systematic errors in the EM estimations of the Hubble parameter? & Detection of GWs from identified host galaxies with known redshift & Sensitivity better than $10^{-21}$\,Hz$^{-1/2}$ around 0.3\,Hz & 10\,yr mission lifetime \\
\cline{2-5}
 & Are there dark-matter particles around BBHs? & High-precision BBH waveform measurements possibly as a multiband observation. & TBD  & 3\,yr mission lifetime \\
\cline{2-5}
 & What processes create a CGWB in the dHz band? & Detection of a stochastic GW background via correlation measurements (or setting important upper limits on the CGWB). & Sensitivity better than a few times $10^{-22}$\,Hz$^{-1/2}$ around 0.3\,Hz & 10\,yr mission lifetime; presence of a second dHz detector \\
\cline{2-5}
& Is there a primordial population of BHs? & Detection of BBHs out to and beyond $z=30$. & Sensitivity better than $2\times 10^{-21}$\,Hz$^{-1/2}$ between 0.1\,Hz and 0.5\,Hz & 3\,yr mission lifetime \\
\hline
\end{longtable}

\subsubsection{Lunar science}

\begin{longtable}{|L{2.8cm}|L{3cm}|L{2.5cm}|L{2.4cm}|L{2.4cm}|}
\hline
\cellcolor{ProcessBlue}\bf Science goals & \cellcolor{ProcessBlue}\bf Science questions & \cellcolor{ProcessBlue}\bf Observational requirements & \cellcolor{ProcessBlue}\bf Instrument requirements  & \cellcolor{ProcessBlue}\bf Top-level mission requirements \\
\hline\endhead
 \multirow{3}{2.8cm}{Understanding of the internal structure of the Moon} & What are the resonance frequencies and Q-values of the lunar normal modes? & Observe ringdown of normal modes produced by moonquakes. & Sensitivity better than 0.1\,$\mu$m/Hz$^{1/2}$ at 1\,mHz and better than 0.1\,nm/Hz$^{1/2}$ at 0.1\,Hz  & 1\,yr mission lifetime\\
 \cline{2-5}
& What is the level of fracturing in the megaregolith? & Observe the effect of seismic scattering on the wavefield generated by moonquakes and GW signals. & Sensitivity better than 0.1\,nm/Hz$^{1/2}$ at 0.1\,Hz and better than 0.1\,pm/Hz$^{1/2}$ at 1\,Hz & 1\,yr mission lifetime \\
\cline{2-5}
& What are the rheological and thermal properties and composition of the Moon's core? & Discriminate and model P- and S-waves travelling through the Moon's core and interacting with its boundary.  & Sensitivity better than 1\,pm/Hz$^{1/2}$ from 0.1\,Hz to 1\,Hz & 1 yr mission lifetime; possibly as part of a lunar seismic network \\
 \hline
 \multirow{2}{2.8cm}{Understanding the nature of lunar seismicity} & What is the origin of moonquakes (e.g.,  tectonic, thermal or tidal) and how to determine their locations (latitude, longitude, and depth)?  & Perform array analysis of at least a few 100 moonquake waveforms.  & Sensitivity better than 0.1\,nm/Hz$^{1/2}$ at 0.1\,Hz and better than 0.1\,pm/Hz$^{1/2}$ at 1\,Hz & 3\,yr mission lifetime; possibly as part of a lunar seismic network \\ 
\cline{2-5}
&Is the Moon a competitive platform for GW detection beyond LGWA?  & Determine the seismic noise level in PSRs and its daily and seasonal variation.  & Sensitivity better than 1\,pm/Hz$^{1/2}$ at 0.1\,Hz and better than 1\,fm/Hz$^{1/2}$ at 1\,Hz & 1\,yr mission lifetime \\
 \hline
 Observing geological processes through their seismic signatures & How to characterize the Moon's surface and crustal structure from tectonic and meteoroid impact events? How does it change in time? & Ambient noise and event coda imaging and interferometry through autocorrelations and cross-correlations. & Continuous recordings in the frequency range 0.1--5\,Hz & $> 1$\,yr mission \\
 \hline
 Understanding the Moon's formation history and evolution & What can the deep internal structure of the Moon tell us about its formation history? & Observe the lowest-frequency normal modes. &  Sensitivity better than 0.1\,$\mu$m/Hz$^{1/2}$ at 1\,mHz and better than 3\,nm/Hz$^{1/2}$ at 10\,mHz & 1\,yr mission lifetime\\
 \hline
\end{longtable}

\section{The Lunar Gravitational-wave Antenna}
\label{sec:lgwa}

The planning of a lunar GW detector is a complex task, which needs to consider many environmental factors and address operational challenges. This section provides an overview of the scientific studies that need to be carried out in preparation of the LGWA. Most of the information required for the scientific preparation of the mission will come from LGWA's pathfinder mission Soundcheck, which is a geophysical station to be deployed inside a PSR equipped with accelerometers, a thermometer and a magnetometer. However, a wealth of scientific lunar data is already available from past lunar missions, whose analysis from the perspective of lunar GW detection still needs to be completed. New important data will also be collected with upcoming lunar missions that will explore the lunar polar regions and include geophysical payloads. 

The LGWA mission concept is summarized in section \ref{sec:concept}. The LGWA sensitivity model is used in section \ref{sec:observational} to highlight some of LGWA's observational capabilities. While these first sections mostly serve as an introduction to the white paper, they also point out some open problems like the development of a calibration procedure for LGWA or the accurate modeling of waveforms given the complex motion of a lunar GW detector. In section \ref{sec:environment}, we provide an overview of the geophysical variables and site characteristics that need to be studied to be able to plan the LGWA mission. We then conclude this part with an overview of planned lunar missions that will provide information about the Moon relevant to LGWA and its pathfinder mission Soundcheck \ref{sec:lunarmissions}.

\subsection{Mission Concept}
\label{sec:concept}

The LGWA was originally proposed in 2020 in response to a call for ideas by the European Space Agency for lunar science payloads to measure vibrations of the Moon caused by GWs \cite{HaEA2020b}. The mission concept was refined in the following year \cite{HaEA2021a}. The targeted observation band is 1\,mHz to 1\,Hz, where the lower bound is given by the frequency of the lowest order quadrupole normal mode of the Moon. Above about 1\,Hz, the lunar GW response is expected to be so weak that instrumental and SBN start to dominate the LGWA signal \cite{Har2022a, Cozzumbo2023}. The SBN is unknown, i.e., it is so weak that it was not possible to observe it with the Apollo seismometers, but it is predicted to be several orders of magnitude quieter in the 0.1\,Hz -- 1\,Hz band than on Earth \cite{LoEA2009}.

The LGWA is proposed as an array of four stations deployed in a PSR at one of the lunar poles. The PSRs are formed by craters and are defined as regions where sunlight can never directly hit the ground. Only sunlight reflected from Earth or scattered from upper parts of the crater walls can still reach the ground. The PSRs are possible only because the Moon's rotation axis is almost perpendicular to the vector pointing from the Moon towards the Sun. Some PSRs have surface temperatures continuously below 40\,K, and are thermally very stable \cite{PaEA2010}. These temperature conditions together with the extremely low SBN enable lunar GW detection at frequencies well below what most people think is possible on Earth with the proposed Einstein Telescope or Cosmic Explorer \cite{ET2020,Evans:2021gyd}.

Each LGWA station will be equipped with two horizontal LIGS measuring surface displacements along two orthogonal directions. The horizontal measurement is preferred in terms of achievable instrument noise and current models predict stronger GW response in horizontal direction. A disadvantage is that one must deal with noise produced by ground tilt. 

A more detailed payload description is given in section \ref{sec:payload}. Four stations are required for effective SBN reduction (see section \ref{sec:background}). An important advantage of LGWA over space detectors like LISA is that there is no known hard limit to its mission lifetime. At the poles, the rate of meteoroid impacts is smaller compared to the equatorial regions \cite{SzEA2019}, and power sources like a radioisotope thermoelectric generator can operate for decades \footnote{\href{https://rps.nasa.gov/system/downloadable_items/30_eMMRTG_onepager_LPSC20140317.pdf}{Radioisotope
thermoelectric generator concept}}. Such long mission lifetimes, besides increasing the chances to detect rare GW signals, would enable future developments of the lunar GW detector network. For example, deploying stations at both poles makes it possible to carry out very sensitive searches for primordial GW backgrounds \cite{CoHa2014}. By burying stations or deploying them in lava tubes, other regions on the lunar surface might become suitable for GW detection, which makes it possible to carry out interesting additional tests of general relativity \cite{WaPa1976}.

\subsubsection{Payload description}
\label{sec:payload}
{\it Main contributors:} Joris van Heijningen, Oliver Gerberding, Shreevathsa Chalathadka Subrahmanya, Morgane Zeoli\\

\begin{figure}[ht!]
\centering
\includegraphics[width=0.75\textwidth]{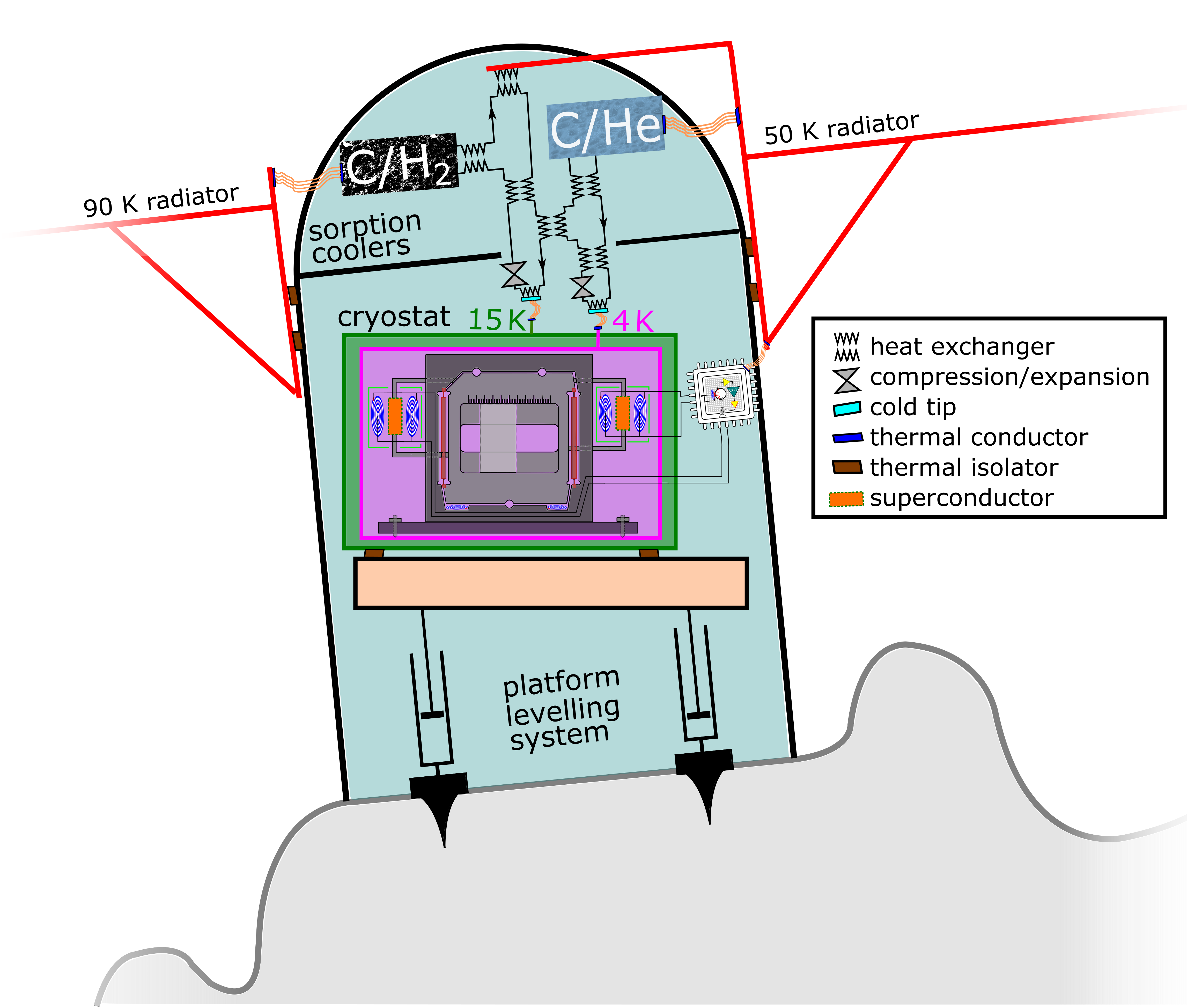}
\caption{Conceptual overview of an LGWA seismic station on a tilted surface on the lunar regolith. The roughness and tilt of the lunar surface are exaggerated for illustrative purposes. Several subsystems vital to successful operation are depicted and further detailed in the text. Subsystems are not shown to scale. Figure adjusted from \cite{vHeEA2023}.}
\label{fig:lgwapay}
\end{figure}

The LGWA seismic stations each have a platform with a leveling system on which the two horizontal LIGS stand. An onboard cryocooler to achieve liquid helium temperatures allows for the use of superconducting elements. Current plans for the payload for LGWA are shown in Fig.~\ref{fig:lgwapay}. The aim is to achieve a displacement sensitivity of 10$^{-12}$\,m/$\surd$Hz at 0.1\,Hz and 10$^{-15}$\,m/$\surd$Hz at 1\,Hz surpassing commercial state-of-the-art by about 3 orders of magnitude. Such sensitivities can be demonstrated on Earth only with the help of specialized vibration-isolation facilities to emulate the low-noise conditions on the Moon as well as possible. 

The LGWA payload aims to achieve its sensitivity by high quality mechanics, cryocooling, superconducting actuation and interferometric or superconducting sensing. LGWA will employ niobium or silicon as proof mass and suspension material, thereby reaching mechanical quality factors of 10$^4$ and 10$^6$, respectively. Together with additional cooling using a sorption cooler\,\cite{Burger2002}, this leads to a thermal noise limited sensitivity below 0.3\,Hz. The sorption coolers also bring the inertial sensor temperature down below the critical temperature of niobium, a superconductor traditionally used for sensing\,\cite{Paik1976}. Similar coil technologies can also be used for ultra low-noise actuation of the inertial sensor proof mass, which can be used to keep the suspended proof mass in operating range.

As detailed in \cite{vHeEA2023}, the LGWA payload currently has two baseline designs: a niobium Watt's linkage with interferometric readout and a silicon Watt's linkage with superconducting readout. Fig.~\ref{fig:lgwapay} presented the latter option. The development of a 1-kg niobium and superconducting actuators is ongoing; the rest of this concept has been developed and used before\,\cite{Heijningen2018}. The hybrid fabrication of a silicon Watt's linkage\,\cite{Ferreira2021} and also the superconducting sensing has only been conceptualized. The second concept therefore requires more R\&D, at the benefit of an expected 10-fold sensitivity increase in inertial displacement sensitivity directly translating to a 10-fold sensitivity to GWs. A more sensitive laser-interferometric readout with sub-fm/$\surd$Hz sensitivity is being investigated as well, but requires an ultra-stable laser frequency reference, see \cite{vHeEA2023}.

The seismic stations are constructed and shielded such that effects due to varying temperatures in the local terrain are minimized. A leveling system will be needed for initial compensation of the ground slope, and later to maintain the horizontal alignment of the LIGS platform accurate to within 10\,$\mu$rad to prevent the proof mass and components attached to it from touching the sensor frame. Regular or maybe even continuous adjustment of the alignment might be needed due to time-varying ground tilts (see section \ref{sec:tilt}).
    
\subsubsection{Main instrument-noise contributions} 
{\it Main contributors:} Joris van Heijningen, Shreevathsa Chalathadka Subrahmanya, Morgane Zeoli\\

\begin{figure}[ht!]
\centering
	\subfigure[~]{
	\includegraphics[width=0.47\textwidth]{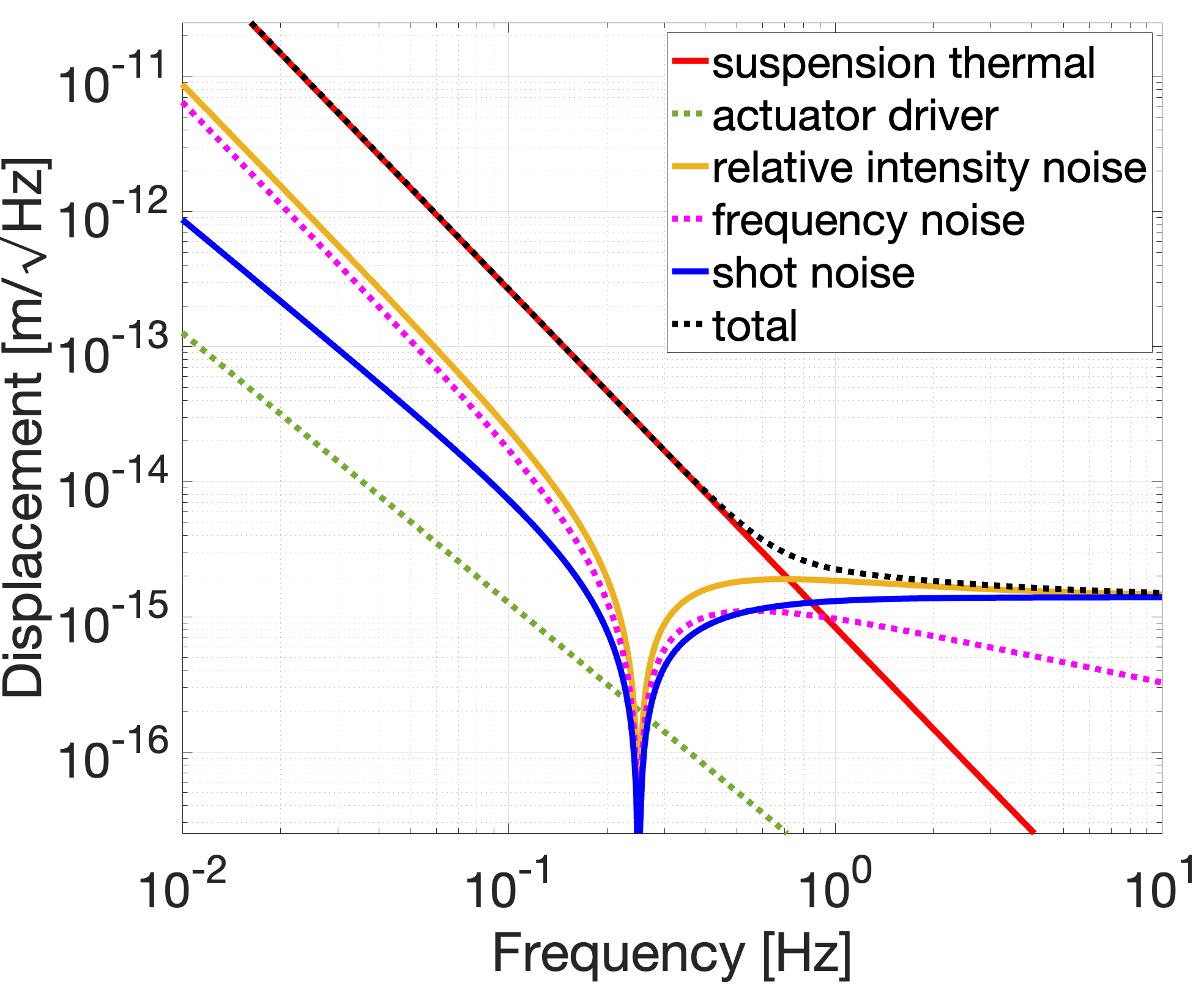}
    \label{sfig:NbIFO}}
	\subfigure[~]{
	\includegraphics[width=0.47\textwidth]{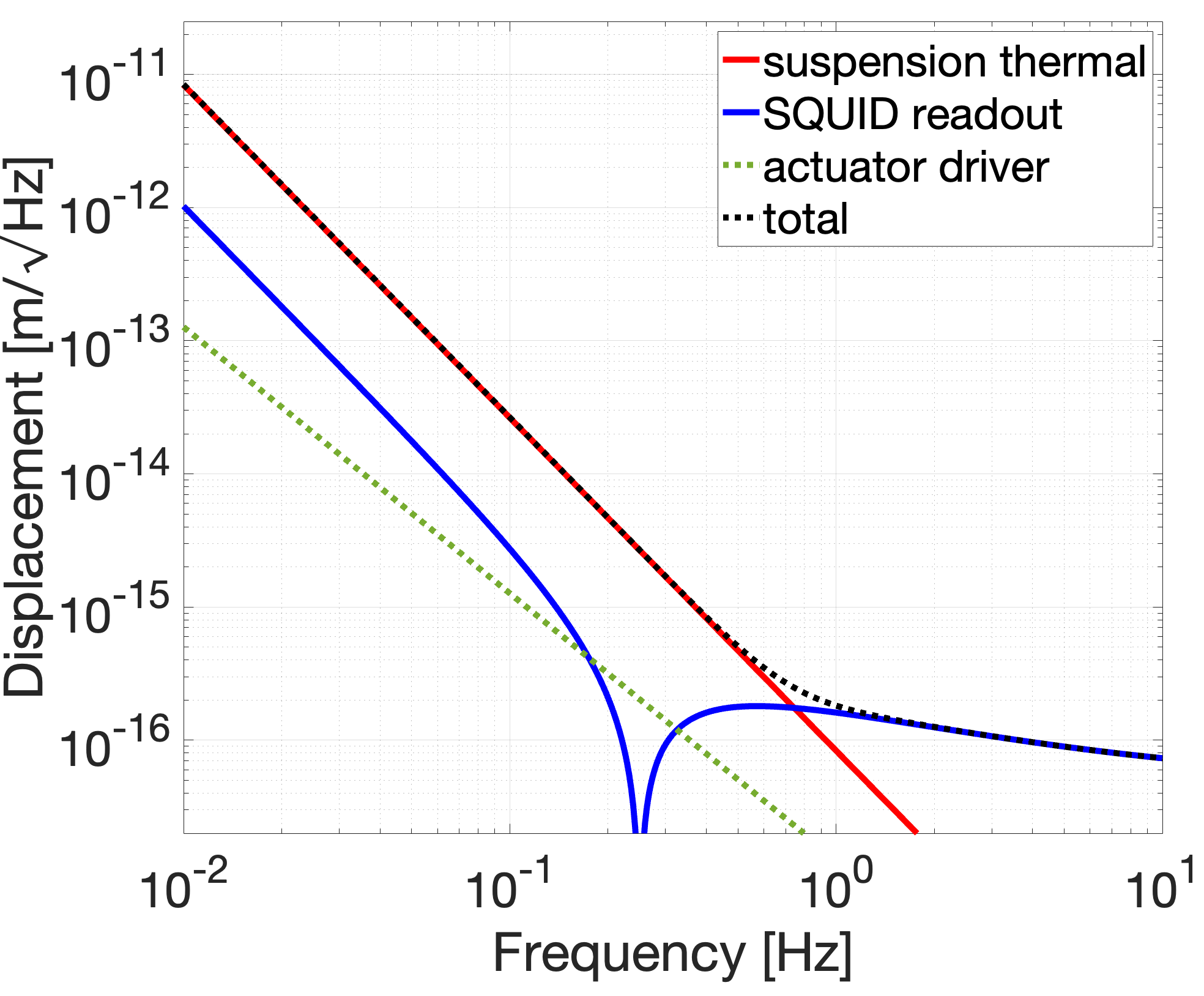}
    \label{sfig:SiSQUID}}
\caption{Minimum detectable inertial displacement for the two LGWA concepts: \subref{sfig:NbIFO} a niobium Watt's linkage with laser-interferometric readout and \subref{sfig:SiSQUID} a silicon Watt's linkage with superconductive readout. All readout noises are corrected by the mechanical transfer function of the proof mass suspension.}
\label{fig:LGWAsens}
\end{figure}

The general LGWA sensitivity curves are obtained by dividing the inertial sensor displacement sensitivity curve by the lunar response to a passing GW. The latter is discussed further in section \ref{sec:LunarResp}. The sensitivity curves of the 2 concepts -- LGWA niobium interferometric and LGWA silicon superconducting -- are presented in Fig.\,\ref{fig:LGWAsens}. 

At low frequency, say below 0.5\,Hz, thermal noise is dominant for both LGWA concepts and depends on the temperature $T$, the mechanical resonance frequency $\omega_0$ and mechanical quality factor $Q$ as \,\cite{CaWe1951,Saulson2017}  
\begin{equation}
x_{\rm{th}}^2 = \frac{4k_{\rm{B}} T}{m \omega \big[\big(\omega_0^2-\omega^2\big)^2+\big(\omega_0^2/Q\big)^2\big]} \frac{\omega_0^2}{Q}
\end{equation}
where $k_{\rm{B}}$ denotes the Boltzmann constant and $\omega$ the angular frequency. The quantity $\omega_0^2/Q$ is constant for structurally damped mechanical oscillators\,\cite{HaML2018} and is relatively low for high-quality materials. We model for a resonance frequency of 0.25\,Hz, a $Q$ of $10^4$ for niobium and $10^6$ for silicon. 

At higher frequencies, the readout noise is dominant. The readout for two studied LGWA concepts are interferometric and superconductive. For the interferometric readout, the shot noise limit can be calculated as amplitude spectral density
\begin{equation}
	i_{\rm{sn}} = \sqrt{2 e I_{\rm{PD}}} = \sqrt{2e\rho P_{\rm{PD}}}, 
\end{equation} 
where $e$ denotes the elementary charge, $\rho$ the responsivity in A/W of the photodiodes and $I_{\rm{PD}}$ and $P_{\rm{PD}}$ the photocurrent and power on the photodiode, respectively. For solid state lasers the relative intensity noise (RIN) spectrum can be roughly expressed as
\begin{equation}
i_{\rm{RIN}} = i_{\rm{sn}} \sqrt{\frac{\omega_{\rm{c}}}{\omega}+1},
\end{equation}
where $\omega_{\rm{c}}$ represents the corner frequency above which the light source intensity fluctuations converge to shot noise limit. Thanks to the differential configuration of the interferometer, $\omega_{\rm{c}}$ can be pushed to low frequency. The effective value of $\omega_{\rm{c}}$ can be determined experimentally. Laser frequency noise can also impact the total noise budget since a frequency noise $\nu_{\rm{L}}$ (in Hz/$\sqrt{\rm{Hz}}$) translates into a readout displacement noise 
\begin{equation}\label{eq:FreqNoise}
x_{\rm{f}} = \frac{\nu_{\rm{L}}}{\nu_{\rm{0}} } \Delta L_0,
\end{equation} 
where $\nu_{\rm{L}}$ represents the laser-frequency noise, $\nu_{\rm{0}} = c / \lambda$ the central laser frequency and $\Delta L_0$ the static arm length difference. 

A sub-fm/$\surd$Hz sensitivity can also be achieved with a superconductive readout. Referred to ground displacement, this readout noise can be calculated as follows \cite{PaVe2009}
\begin{equation}
    x_{\rm{squid}}^2 = \frac{2 E_{\rm{A}}(1+f_{\rm{c}}/f)}{m \omega_0 \eta\beta}, 
\end{equation}
where $m$ is the suspended proof mass, $\eta\beta$ denotes the coupling efficiency from mechanical motion to the readout and $E_{\rm{A}}$ the energy resolution. At 4.5\,K, the latter is estimated to be $E_{\rm{A}}$ = 50\,$\hbar$\,\cite{Falferi2008}. The superconductive readout has a 1/$\surd$f characteristic below $f_{\rm{c}}$.

Superconducting coil actuators can be used to exert forces on a superconducting surface by means of the Meissner effect. They channel the digitization noise $n_{\rm{DAC}}$ from the digital-to-analog conversion (DAC) to an effective displacement noise of the proof mass motion as
\begin{equation}
  x_{\rm{act}} = \frac{\alpha n_{\rm{DAC}}}{R_{\rm{s}} m \omega^2}, 
\end{equation}
where $\alpha$ denotes the actuator strength in N/A and $R_{\rm{s}}$ the shunt resistance. It will depend on the lunar SBN if actuation on the proof mass of the inertial sensor is necessary. 

\subsubsection{Reduction of the seismic background}
\label{sec:background}
{\it Main contributors:} Jan Harms, Francesca Badaracco, Lucia Zaccarelli\\

The superposition of seismic waves produced by a large number of small events like meteoroid impacts (see section \ref{sec:seismicbackground} for details) produces a continuous SBN in LGWA measurements \cite{HaEA2021a,Har2022a}. It is not yet known how strong the SBN is, but simulations indicate that it might limit LGWA sensitivity above 0.1\,Hz \cite{LoEA2009}. For this reason, LGWA was proposed as an array of four stations with distances small enough so that the seismic signal observed in one LIGS can be modeled accurately using data from the other LIGSs, and large enough so that there is an observable phase delay or difference in amplitude between stations. If these conditions are fulfilled --- how well might depend on frequency --- then it is possible to distinguish between a GW signal, which is approximately the same at all LGWA stations, and the SBN.

Methods of array optimization have been developed for noise cancellation in terrestrial GW detectors like Virgo \cite{BaEA2020, KoEA2023}. Compared to background reduction in Virgo, the problem is different in LGWA since all LIGS available to LGWA also contain the GW signal, while the GW signal in terrestrial seismometers can be neglected. This means that standard optimization methods for background reduction need to be modified for LGWA. This will generally lead to a nonlinear optimization problem, considering properties of the GW signal as well. First studies of the efficiency of the LGWA SBN reduction are under way. The sensitivity models used in this white paper assume that the impact of the SBN is fully removed from the LGWA data.

\subsubsection{Lunar response to GWs and calibration}
\label{sec:LunarResp}
{\it Main contributors:} Jan Harms, Oliver Gerberding\\

The response of an elastic halfspace to GWs was first modeled by Dyson \cite{Dys1969} and later analyzed for a spherically symmetric body in the normal-mode formalism \cite{Ben1983}. The equivalence of the standard Newtonian gravitational coupling in an elastic body to mass density, and the general relativistic coupling to shear-modulus gradients were shown in \cite{Harms2019}. The Dyson model was used in \cite{CoHa2014,CoHa2014c} and the normal-mode formalism was used in \cite{CoHa2014b,HaEA2021a} to estimate the sensitivity of GW searches using the Earth and the Moon as antennas. In 2024, three studies on the lunar GW response emerged with the aim to provide a deeper physical understanding of the lunar GW response in the Dyson field-theoretical formalism and in the normal-mode formalism \cite{belgacem2024lunarresp,yan2024lunarresp}, and to extend the Dyson model to a horizontally layered half space \cite{bi2024response}.
\begin{figure}[ht!]
\centering
\includegraphics[width=0.9\textwidth]{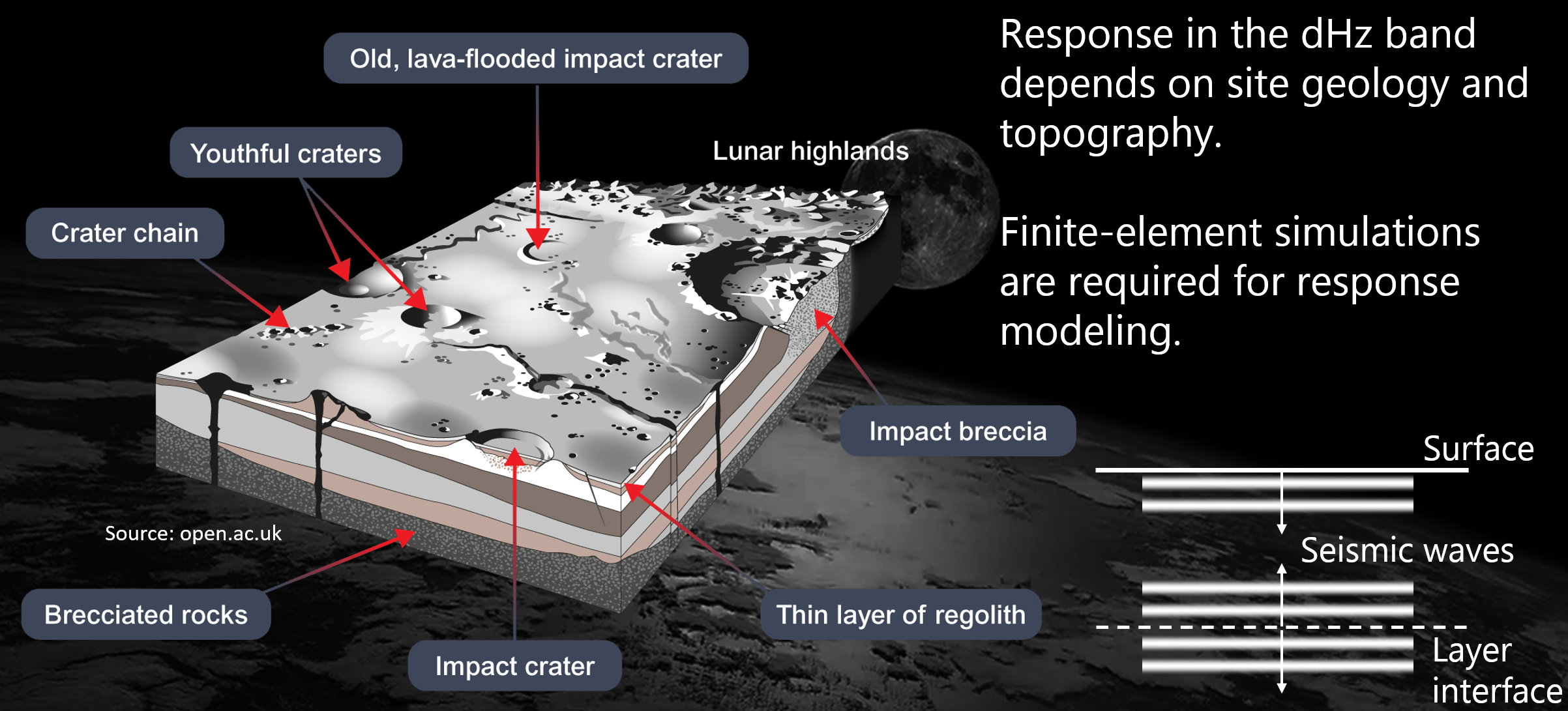}
\caption{Modeling of the lunar GW response in the decihertz band requires analytical studies and numerical simulations based on a geological and topographic model of the deployment site. Gravitational waves generate seismic waves propagating in normal direction to the surface and interfaces.}
\label{fig:lunarfem}
\end{figure}

Especially the GW response modeling in the 0.1\,Hz -- 1\,Hz band is important to LGWA, where most of its breakthrough science can be achieved. In lack of a well-founded numerical model, the lunar GW response remains an important uncertainty in LGWA sensitivity models at frequencies $>0.1\,$Hz. At these frequencies, topography and regional geology as shown in figure \ref{fig:lunarfem} are expected to play an important role. In this band, the approach for current LGWA sensitivity curves is to start with a simple Dyson model, and then to modify it so that it becomes consistent with lunar seismic observations. Most importantly, we include an amplification of the ground response. The quality factor inferred from moonquake observations (wave coda, wave envelopes) with Apollo seismometers is on the order of a few thousand \cite{LaEA1970,Garcia2019}. However, seismic scattering is an important effect on the Moon, and it is unclear today how much of the Q-factor can be exploited in GW signal analyses and how scattering effects fields locally at the LGWA site. This would ultimately be a problem of response calibration (see below), and assistance might come from multiband observations with space-borne or terrestrial GW detectors. For the LGWA sensitivity, we assume on-peak amplification factors of a few 100 in the decihertz band with respect to a simple Dyson response model for a homogeneous half space.

\paragraph{Calibration of LGWA data} Calibration requires models of the (1) laser-interferometric/coil response to proof-mass displacement, (2) mechanical response to ground displacement, and (3) lunar GW response. The calibration of the mechanical and laser-interferometric response is well understood \cite{Gerberding_2015} and any errors will not be a dominating factor. Calibration procedures have also been developed for a superconducting coil + SQUID readout \cite{CMP1987,SHIRRON1996805}, and calibration errors are expected to be minor also in this case compared to the uncertainty of the lunar GW response.

The main calibration error will come from the model of the lunar GW response, which depends on our understanding of the lunar internal structure ($f\lesssim 0.1$\,Hz) and the geology and topography at the deployment site ($f\gtrsim 0.1$\,Hz). The lunar GW response model will be obtained from numerical simulations including geological and topographic information. It can be refined with time by analyzing moonquakes observed with LGWA and using single-station inference methods of local subsurface structure \cite{Hobiger2021,Compaire2021,Carrasco2022}. We note that ground tilt produced by GWs is very small since GWs only excite quadrupolar modes of the Moon. The dominant gradients along the lunar surface of fields excited by GWs are likely connected to geological heterogeneity and topography. The tilt contribution to the horizontal displacement signal of a GW is likely negligible, but a careful analysis is required. It is impossible today to say what calibration errors can be achieved in this way. While the calibration task seems more complicated for LGWA than for terrestrial GW detectors, one advantage is that the lunar GW response does not change with time. 

A highly accurate modeling of the lunar GW response might be possible with multiband observations; see section \ref{sec:multiband_gw_obs}. For example, solar-mass or intermediate mass BBH could first be observed with LGWA and later with a network of terrestrial GW detectors. In this case, parameters of the GW signal can be inferred based on an accurate calibration of the data of the terrestrial detectors, and be used to calculate the waveform during the period when the signal was in the LGWA observation band. The calibration is then obtained as the function that relates the (known) waveform with the GW signal observed by LGWA. A practical challenge of this procedure is that the lunar GW response depends on the propagation direction of the GW due to geological and topographic inhomogeneities. Consequently, it might be that many multiband observations are required before an accurate model of the lunar GW response can be reached.

\subsubsection{Sensitivity models of LGWA}
\label{sec:sensModels}
{\it Main contributors:} Jan Harms\\

The power-spectral densities (PSDs) of the LGWA sensitivity models can be obtained from the GWFish repository\footnote{\href{https://github.com/janosch314/GWFish/}{GWFish git repository}}. As mentioned earlier, we developed two models, one being more conservative, i.e., higher readout noise and lower quality factor of the proof-mass suspension (referred to as the niobium model), and another being a more ambitious design with low readout noise and high-quality factor (referred to as silicon model). 

\begin{figure}[ht!]
\centering
\includegraphics[width=0.8\textwidth]{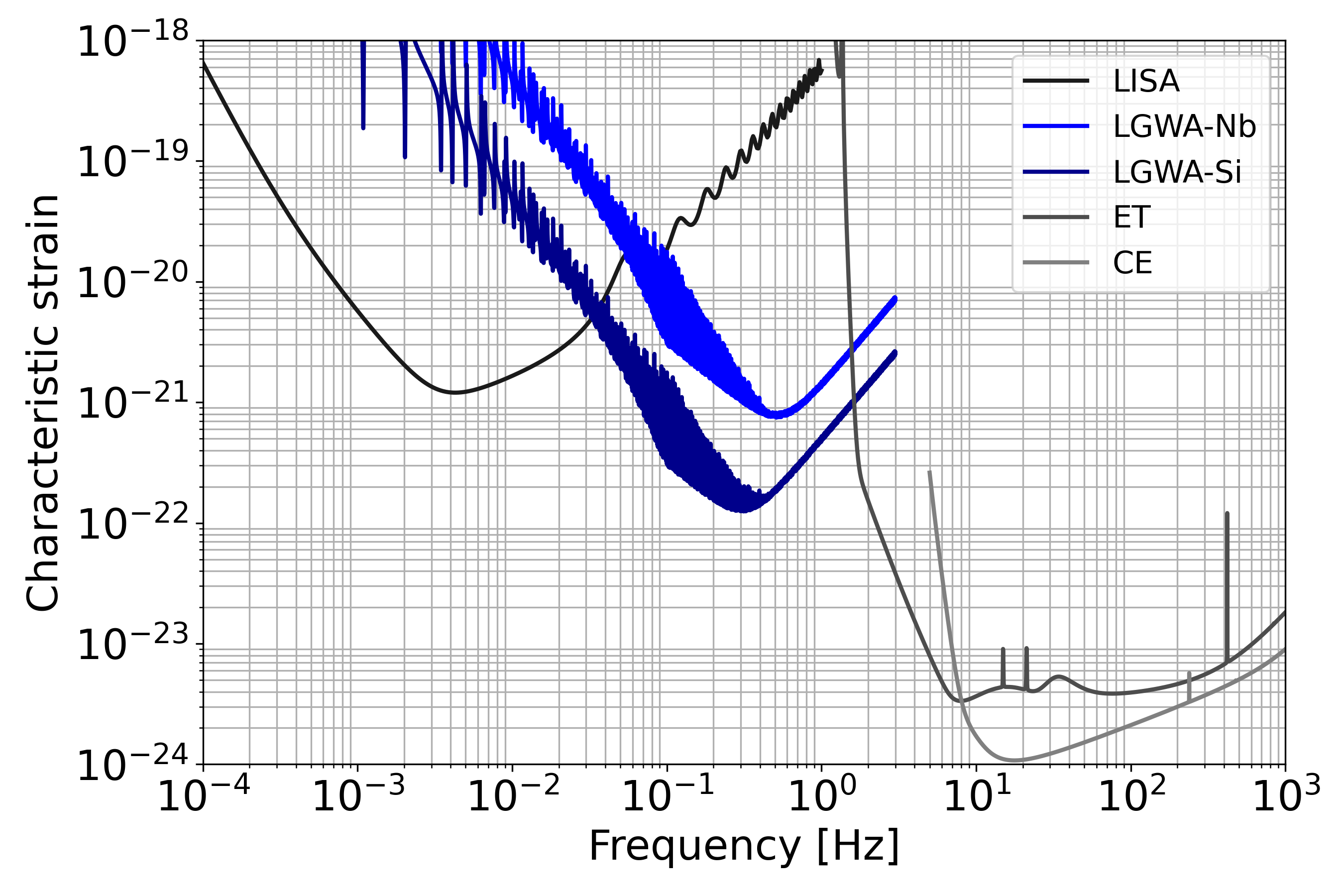}
\caption{Models of characteristic strain sensitivities.}
\label{fig:strainnoise}
\end{figure}

The LGWA sensor array has a total of 8 horizontal seismic channels: each station monitors surface displacement along two orthogonal horizontal directions. The observation of GW signals is accurately modeled in GWFish taking contributions of the two GW polarizations into account, but for some of the analyses presented in this white paper, separate simulation software was used and a simplified approach is taken: the instrument-noise PSD of a single LGWA channel is divided by 4 and only one polarization is observed. The LGWA sensitivity models (LGWA-Nb and LGWA-Si) including the division of noise by a factor 2 (in amplitude) are shown in figure \ref{fig:strainnoise} as characteristic strain $(fS(f))^{1/2}$, where $S(f)$ is the instrument-noise PSD.

\subsubsection{Outlook: LGWA upgrades}
{\it Main contributors:} Jan Harms\\

There are possibilities to improve the LGWA LIGS network beyond its baseline configuration and increase its science reach. The three approaches are
\begin{itemize} 
\item Improving LIGS instrument sensitivity 
\item Increasing the number of array stations for improved SBN reduction
\item Extending the station network across the lunar surface for enhanced GW analyses, e.g., to enable correlation measurements between stations, or GW polarization measurements.
\end{itemize}
Improvements of the LIGS sensitivity are conceivable over the entire frequency band from 1\,mHz to 1\,Hz, i.e., no known fundamental sensitivity limitations prevent improvements beyond the LGWA-Si model. Especially, innovative technologies like superconducting levitation as proof-mass suspension could have an enormous impact on LGWA sensitivity below 0.1\,Hz \cite{MPC2002,PaVe2009,PaEA2016}. 

Improving LIGS technologies, it is increasingly likely that LGWA will become limited by SBN. This is especially true for improvements of the LIGS sensitivity in the frequency range 0.1\,Hz -- 1\,Hz. In this case, techniques to reduce the SBN must be improved as well. Adding stations to the LGWA array would provide a better analysis of the ambient seismic field and better noise reduction. Below 0.1\,Hz, LGWA sensitivity is not expected to be limited by the SBN, and improvements of the LIGS sensitivity would directly lead to improvements of LGWA sensitivity to GWs. 

It is hard to imagine LIGS and SBN technologies today that would make GW observations above 1\,Hz with LGWA possible, since the lunar GW response is very weak at these frequencies. We believe that observations above 1\,Hz must be done with long-baseline laser interferometry (see \cite{JaLo2020,ASEA2021,Branchesi2023} for proposed lunar detector concepts). 

The station network can be extended across the lunar surface. For example, deployment at both lunar poles would make it possible to carry out correlation measurements between stations at the two poles to search for a stochastic GW background \cite{CoHa2014}. It was also pointed out that a certain distribution of LIGS over the lunar surface would make it possible to precisely measure the polarization of a GW including polarization states not predicted by general relativity \cite{WaPa1976}. This however would require deployments outside the PSRs. In order to avoid excess noise from ground tilts due to temperature variations at stations outside PSRs, it would be necessary to bury the LIGS. It is not known how deep such a vault would have to be, but temperature stability is predicted to be very high already a few 10\,cm below surface \cite{MALLA2015196}. To further reduce the noise from ground tilt, a sensor could be deployed with the ability to measure ground tilts in addition to ground displacement as recently proposed by Li et al \cite{LiEA2023}. In this case, the tilt measurement could be used to subtract the tilt-induced noise in the displacement measurement. 

\subsection{Observational capabilities}
\label{sec:observational}

\subsubsection{Detection horizons}
\label{sec:horizons}
{\it Main contributors:} Jacopo Tissino, Jan Harms, Martina Toscani, Manuel Arca Sedda, Alberto Sesana\\

Detection horizons are an important figure-of-merit of the observational capabilities of GW detectors. The detection horizon is the maximum distance of a source observed with a certain signal-to-noise ratio (SNR). Such a source emits GWs with the optimal polarization, e.g., for a compact binary, this means that its orbital plane is perpendicular to the line of sight to the source. For short GW signals, there is also an optimal sky location with respect to the detector orientation. However, for GW sources observed for months and years, the concept of optimal sky location must be generalized to take into account the change of the orientation of LGWA sensors due to the rotation of the Moon. 

An often used alternative figure of merit similar to the detection horizon is the detection range, which, given some SNR threshold, quantifies the maximum distance at which a source can be seen after averaging over its orientation and sky-location angles. The horizon is about a factor 2.24 larger in luminosity distance than the detection range, but this is true only at smaller distances $z\ll 1$ where redshift effects on the waveform can be neglected.

\paragraph{Binary black holes --}
\begin{figure*}[ht!]
    \centering
    \includegraphics[width=.59\textwidth]{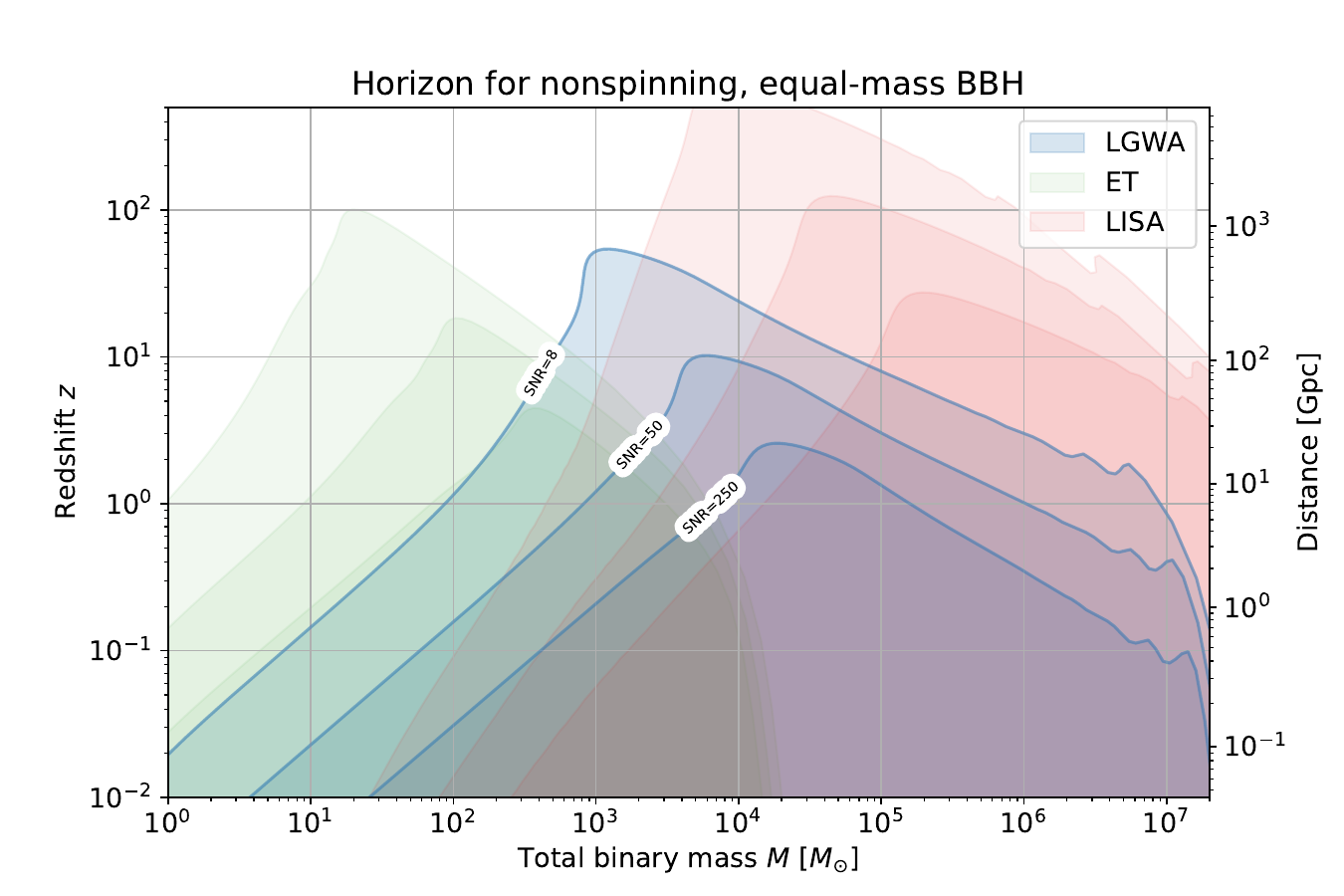}
    \includegraphics[width=0.39\textwidth]{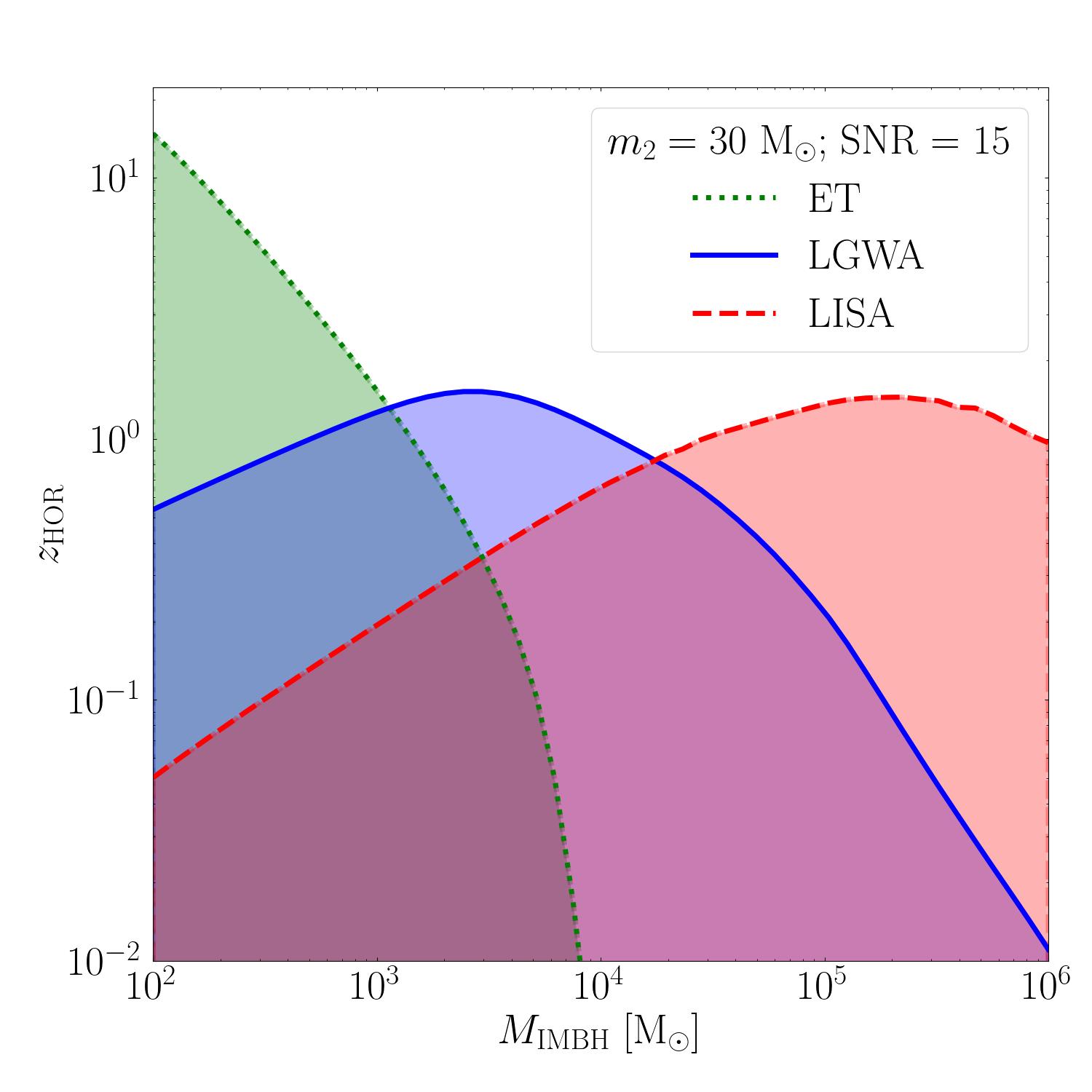}
    \caption{Left: LGWA horizon for equal-mass black hole binaries, compared to Einstein Telescope and LISA. The mass value is given in the source frame. Right: Redshift threshold for detection (SNR$\> 15)$ for an IMRI with primary mass $M_{\rm IMBH}$ and secondary mass of $30 M_{\odot}$. The central blue region corresponds to LGWA, which can be compared with the horizons of the Einstein Telescope (left region, green) and LISA (right region, red).}
    \label{fig:horizon-et-lgwa-lisa}
\end{figure*}
Here, we present three analyses to characterize the detection capabilities of LGWA. The first is the so-called waterfall plot shown in figure \ref{fig:horizon-et-lgwa-lisa}. The plot shows the horizon for SNR values 8, 50, 250, and 1000. The x-axis shows the total mass $M$ of the observed binary in the detector frame, which is connected to the mass in the source frame by $M = (1+z)M_{\rm src}$ ($z$ is the redshift). At 1000 $M_\odot$, the LGWA detection horizon is about a factor 8 larger in redshift compared to ET and LISA, which means that LGWA would complement the study of BBH populations possible with ET and LISA. Instead, the detection horizon of solar-mass BBH and BNS is comparable to the detection horizon of the LIGO detectors during their third observation run. The most distant detection was GW200308 at a redshift of $z=1.04$ (albeit the analysis of this event might be biased by unphysical modes of the posterior)  \cite{LIGOScientific:2021djp}. In the case of more extreme mass ratio events, the ideal detections for LGWA consist of light IMRIs, for which the primary IMBH is of order $10^{2-4}$ masses, up to redshift $z\sim1$. The horizon distance limits for IMRIs consisting of a $30 M_{\odot}$ secondary are shown in the right plot of figure~\ref{fig:horizon-et-lgwa-lisa}, with comparisons to the Einstein Telescope and LISA. 

\begin{figure*}[ht!]
    \centering
    \includegraphics[width=.8\textwidth]{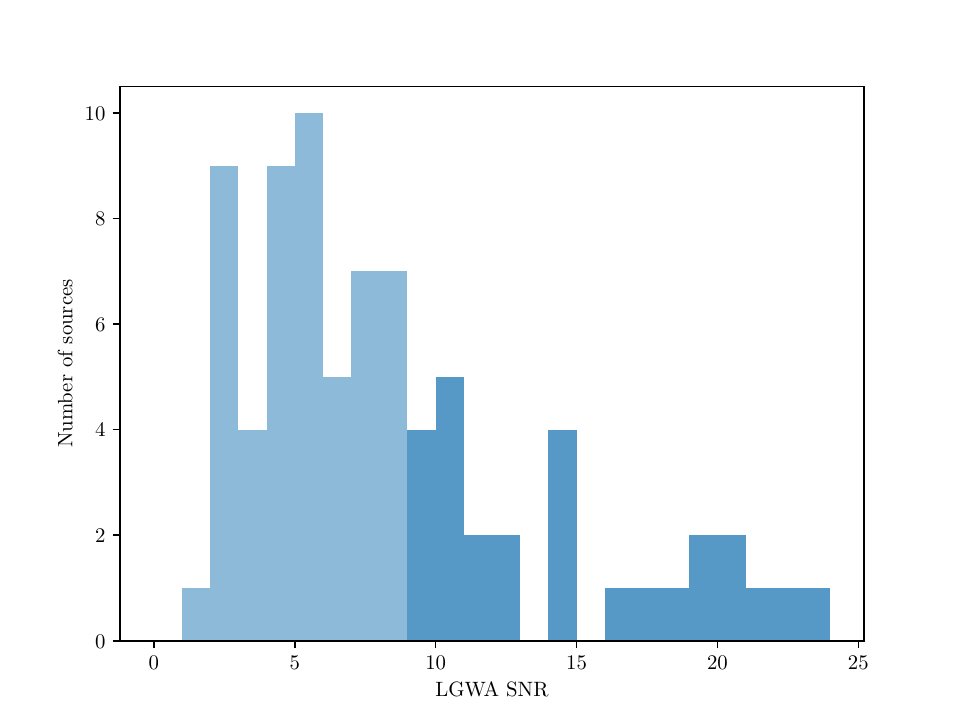}
    \caption{Signal to noise ratios of GWTC-3 signals simulated in LGWA. Even when assuming the same volume-time product that underlies the GWTC-3 detections (about 0.004\,Gpc$^3$\,yr referred to BNS \cite{LIGOScientific:2021djp}, which corresponds to a year of LGWA observation), the total number of BNS and solar-mass BBH detected by LGWA will be higher than indicated in this plot since GWTC-3 is merely a subsample of all such systems within the detection horizon. The LGWA detection range for 1.4+1.4 M$_\odot$ BNS is about 100\,Mpc, and for 30+30 M$_\odot$ BBH about 1.9\,Gpc ($z=0.4$).}
    \label{fig:gwtc3-lgwa}
\end{figure*}

\paragraph{Detection of the GWTC signals --}
In order to have a better idea of the detection capability of LGWA with respect to solar-mass binaries, we simulate the observation of the signals in the Gravitational-wave Transient Catalogues GWTC-1 -- GWTC-3 with LGWA \cite{abbott2021gwtc21,LIGOScientific:2021djp}. As shown in figure \ref{fig:gwtc3-lgwa}, of the 80 signals detected with Virgo and LIGO, 28 would have been seen with LGWA, and the same volume-time product for 1.4 + 1.4 $M_\odot$ would be achieved with LGWA within one year. These 28 detected GWTC signals are not the total number of expected solar-mass sources that LGWA would see per year since GWTC signals are only a subset of all solar-mass binaries within the Virgo/LIGO or LGWA horizons (see also section \ref{sec:multiband_gw_obs}). Also the exact number of detections depends on the absolute time of the simulation, and might vary from simulation to simulation under changes of absolute time, i.e., changing the orientation of the Moon.

\begin{figure*}[ht!]
    \centering
    \includegraphics[width=.8\textwidth]{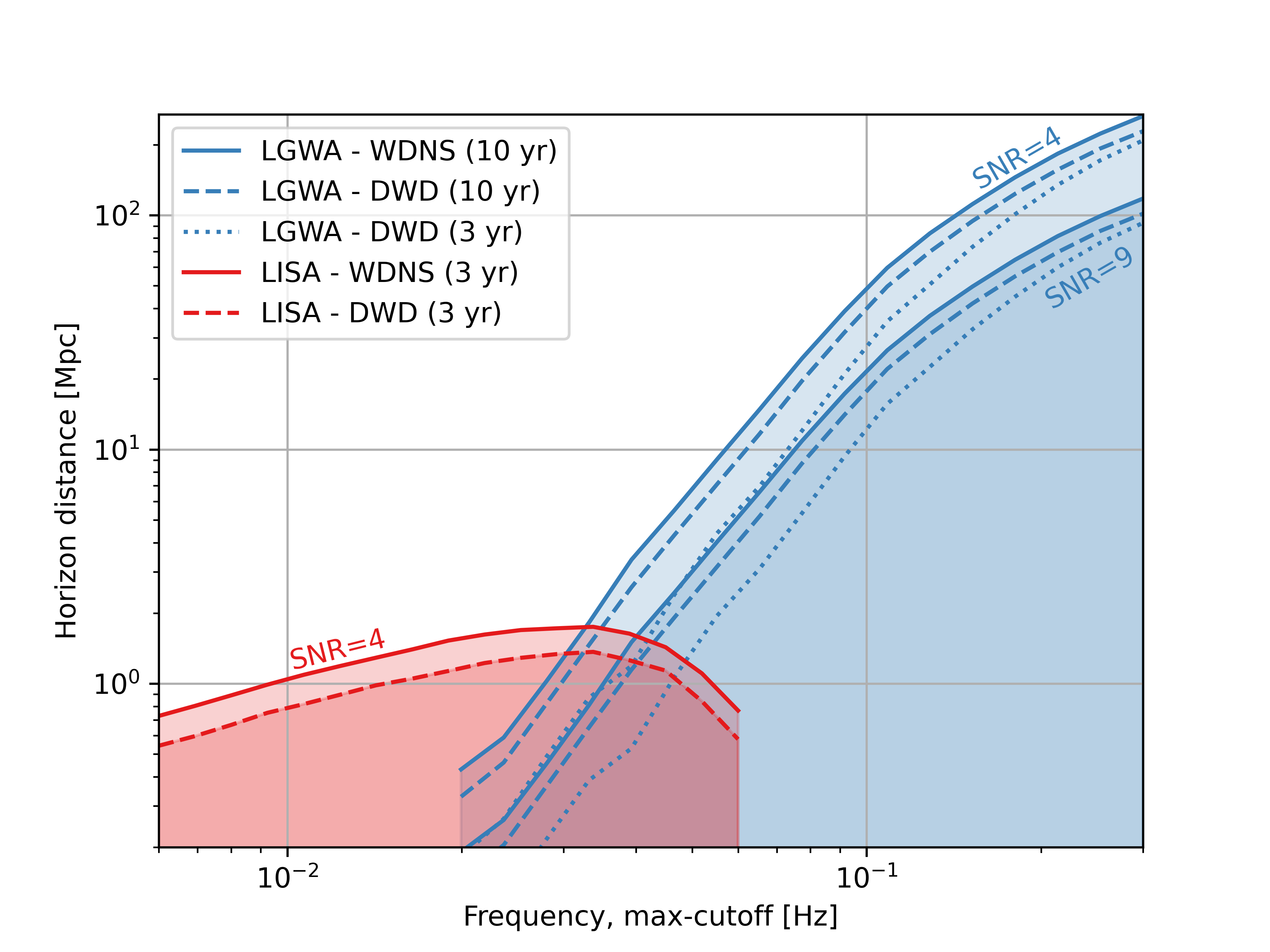}
    \caption{LGWA (LISA) detection horizons for DWD and WD/NS inspiral signals observed over 10 years or 3 years and with SNR threshold as indicated in the plot. The horizon is drawn as a function of the GW frequency reached by the DWD or WD/NS at the end of the observation time. The masses are $1M_\odot+1M_\odot$ for DWDs and $1M_\odot+1.4M_\odot$ for WD/NS.}
    \label{fig:horizon_dwd}
\end{figure*}

\paragraph{Binaries containing white dwarfs --}
Another important horizon calculation concerns DWDs. They can either be observed as signals whose GW frequency increases very slowly never reaching the point of contact during the observation, or the signal inspirals and eventually comes to an end when the two WDs disrupt each other or merge. We calculate the horizon for $1M_\odot+1M_\odot$ DWDs and $1M_\odot+1.4M_\odot$ WD/NS, all observed over 10 years (3 years) corresponding to the designated LGWA (LISA) mission lifetime. The plot in figure \ref{fig:horizon_dwd} shows the horizons as a function of the GW frequency reached by the binary at the end of the observation time. Signals from DWDs and even WD/NS are not expected to reach 1\,Hz as the merger should happen at lower frequencies depending on the masses of the white dwarfs; see equation (\ref{eq:kepler}). So, LGWA's maximum horizon for DWDs is around 100\,Mpc. 
This horizon can be confronted with the LISA's DWD horizon, which only extends over the volume of the Galaxy. The reason why the LISA horizon decreases below 30\,mHz even though its sensitivity keeps increasing down to a few mHz is because at such low frequencies, the 10 years inspiral time is so short that the SNR is accumulated over a very narrow band. These signals are different in nature to the broadband spectra produced by compact binaries closer to their merger.

\begin{figure}[ht!]
    \centering
    \includegraphics[width=0.7\textwidth]{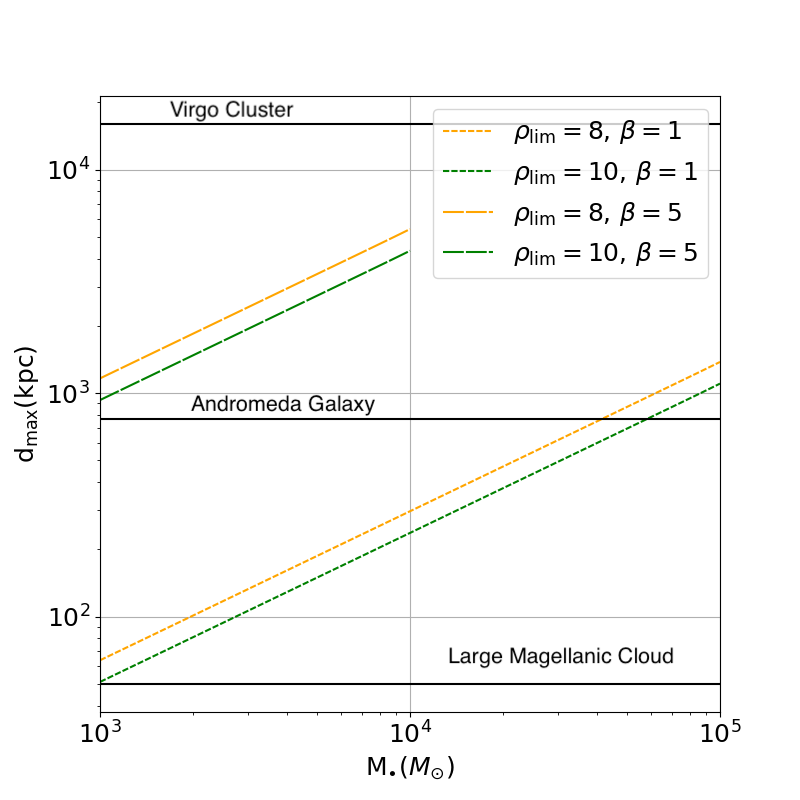}
    \caption{Maximum distance $d_{\max}$, at which LGWA can see a WD-TDE ($M=0.5\text{M}_{\odot}$, $R_*=0.01\text{R}_{\odot}$) as a function of the BH mass, for two fixed values of the penetration factor (tidal ratio to stellar pericenter): 1 (short-dashed curves) and 5 (long-dashed curves). We consider two possible thresholds for detection: 8 (orange curves) and 10 (green curves). For comparison, we also display the distances at which the Large Magellanic Cloud, the Andromeda Galaxy and the Virgo Cluster are located (horizontal black lines).}
    \label{fig:TDE_WD_beta1_5_sicurve}
\end{figure}

\paragraph{Tidal disruption events --} In figure \ref{fig:TDE_WD_beta1_5_sicurve}, we show the maximum distance at which LGWA can observe a WD-TDE as a function of the BH mass, for two fixed values of $\beta$: 1 and 5. We assume two signal-to-noise thresholds for detectability: 8 and 10, calculated in a similar way as outlined in \cite{2022MNRAS.510.2025P}. The plot indicates that LGWA is capable of detecting these events in the satellites of the Milky Way, extending up to the Andromeda galaxy and almost reaching the Virgo cluster. This finding is indeed encouraging, despite the challenge of determining the expected rate of occurrence for these events (further details on the rates are discussed in section \ref{sec:TDE}).

\subsubsection{Localization of GW sources in space and time}
{\it Main contributors:} Jan Harms, Jacopo Tissino, Michele Mancarella, Francesco Iacovelli\\
\label{sec222}

The localization of GW sources is an important aspect of GW observations. Unless one considers the multiband scenario described in section \ref{sec:multiband_gw_obs}, LGWA is likely going to be an isolated detector whose localization capabilities are limited compared to detector networks. However, there are important exceptions. Solar-mass compact binaries consisting of white dwarfs, neutron stars and lighter black holes will be observed by LGWA for several weeks to years. The modulation of the signal's amplitude and phase due to the Moon's rotation and orbital motion provide information about the source's location in the sky. We also point out that contrary to laser-interferometric detectors, LGWA carries out independent surface displacement measurements along two orthogonal horizontal directions, which also helps with the sky localization. Vice versa, being able to measure sky location accurately improves estimates of other parameters like distance and polarization. Such a scenario is specifically attractive for BNS observations, where the detection and localization of the source would become an early warning of a BNS merger in the band of terrestrial detectors, and electromagnetic observatories can prepare for the event \cite{liu2022realistic,kang2022electromagnetic}. Also the disruption or merger of white dwarfs observable in the decihertz band would be accompanied by EM counterparts (see sections \ref{sec:GWAstro} and \ref{sec:SNe}), and sky localization might be crucial for the detection of an EM signal.

However, we do not have a simulation yet that is able to estimate the sky-localization errors of generic sources for LGWA. The problem is outlined in the following, and we leave it to a future task to improve the current numerical simulations.

The calculation of sky-localization and merger-time estimation errors has non-trivial technical challenges especially for LGWA connected to the Fourier-domain phase term appearing in the signal at a detector: $\exp(\irm (\vec k\cdot \vec r(t(f))-2\pi f t_{\rm c}))$, where $t_{\rm c}$ is the merger time, $f$ the frequency of the GW, $\vec k$ is its wavevector, and $\vec r(t)$ is the position of the detector. The function $t(f)$ provides the time, at which the GW signal emits at frequency $f$, and it is used to simulate the changing detector position and orientation while the GW signal is being observed. This works only for certain types of GW signals and under certain conditions with respect to the frequency evolution of the signal and how quickly the detector is moving \cite{marsat2018fourierdomain}. The principal problem is that with this phase term the likelihood (or Fisher matrix) is dependent on the choice of the origin of the coordinate system. There is a method to obtain coordinate independent parameter-estimation errors \cite{WeCh2010}, but it has not been integrated in general simulation software yet. 

\begin{figure}[ht!]
    \centering
    \includegraphics[width=0.6\linewidth]{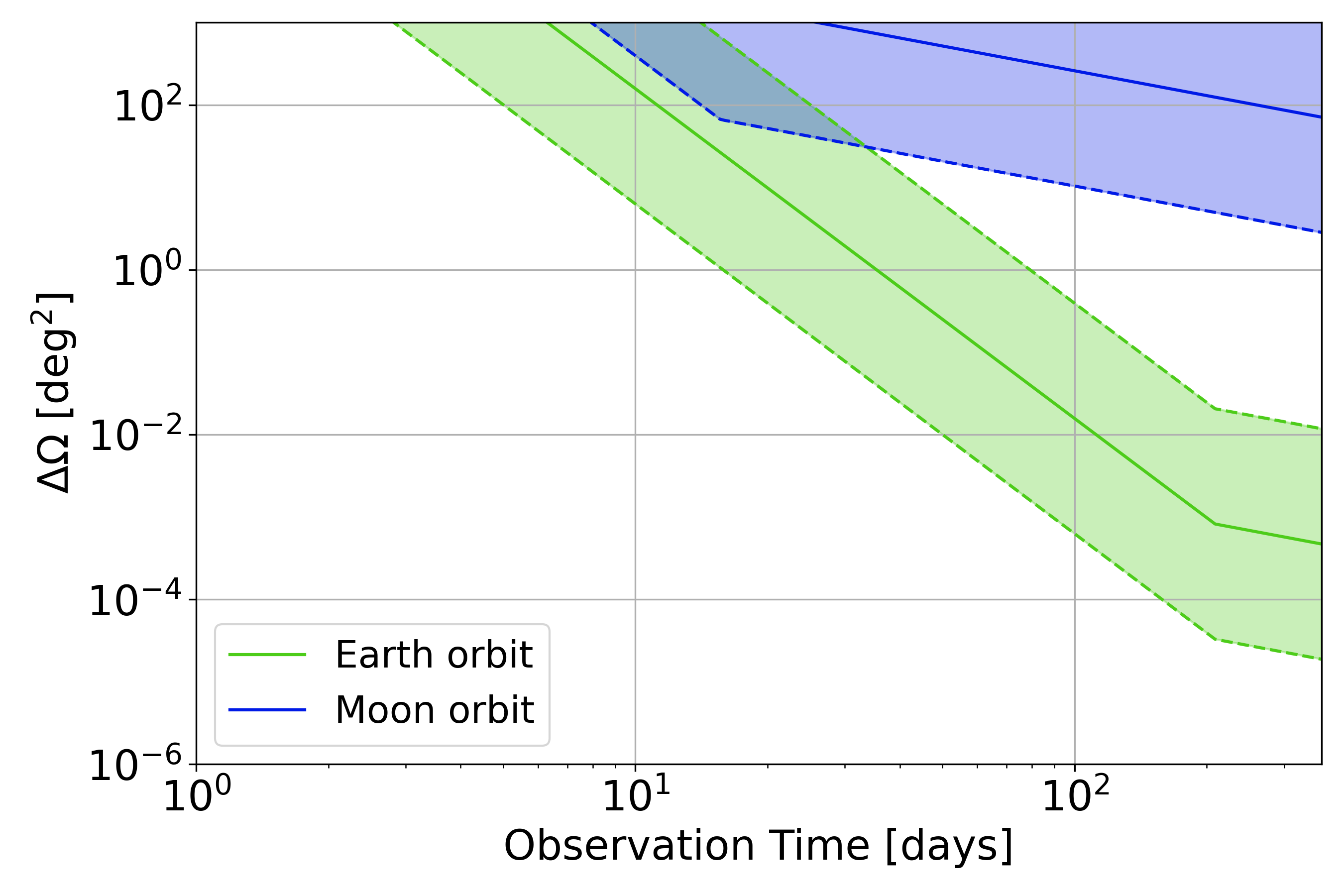}
    \caption{\changes{Sky-localization estimates based on \cite[eqs.~(51),(52)]{WeCh2010} coming from the orbital motion of the Moon (around the Earth) and the Earth (around the Sun). The bands represent a range of frequencies from $f=0.02$\,Hz (worse localization) to 0.5\,Hz (better localization) with the solid lines representing 0.1\,Hz. The signal SNR increases with the square-root of the observation time reaching a value of 30 after one year.}}
    \label{fig:skyloc}
\end{figure}
A correction can however be applied for selected signals by expressing the detector positions with respect to a reference point that minimizes coordinate artifacts. This was done to analyze the localization of GW170817 with LGWA using the Fisher-matrix approximation \cite{Cozzumbo2023}, and it was shown that few-arcmin$^2$ sky-localization errors can be obtained. The order of magnitude of this result can also be estimated analytically under simplified assumptions, e.g., neglecting changing antenna patterns and assuming a fixed signal frequency \cite[eqs.~(51),(52)]{WeCh2010}. \changes{Results from the analytical expressions are shown in figure \ref{fig:skyloc}.

Accordingly, the signal modulation caused by the orbital motion around the Sun leads to the dominant sky-localization capabilities of LGWA compared to the contribution of the Moon's orbital motion. Sky localization improves quadratically with the SNR and signal frequency.}

\subsubsection{Estimates of spin and mass parameters}
\label{sec:intrestimates}
{\it Main contributors:} Jan Harms, Jacopo Tissino \\

Estimation of mass and spin parameters can be important for various reasons. Component masses together with the distance of compact binaries are the fundamental quantity to start a population study. Furthermore, spins can help to distinguish formation channels of compact binaries. The LGWA science case on population studies and the role of the various parameters are described in section \ref{sec:populations_and_formation_channels}. In this section, we provide information about the capability of LGWA to measure these intrinsic parameters. The method is based on the Fisher-matrix approximation. The waveforms were simulated with the approximant IMRPhenomXPHM \cite{pratten2021imrphenomxphm}, which can simulate precession and higher-order modes. \changes{The Fisher-matrix method leads to accurate error estimates whenever the SNR is high enough and there is no degeneracy of the waveform model involving the analyzed parameters. These conditions can be assessed post factum when finding that the parameter-estimation errors are small. Large parameter errors point to potential issues typically with model degeneracies, which can be cured by using prior distributions as well \cite{dupletsa2024}, or by switching to a full Bayesian analysis tool without assuming a Gaussian likelihood \cite{AsEA2019}. This means that the errors presented in this section for mass parameters are a good indication of what a full Bayesian analysis would obtain, while errors of the spin parameters might deviate significantly from full-Bayesian calculations when the error approaches unity. Error estimates from a full Bayesian analysis are typically significantly smaller in this case due to the help of (even uniform) prior distributions \cite{dupletsa2024}.}

\begin{figure*}[ht!]
    \centering
    \includegraphics[width=.49\textwidth]{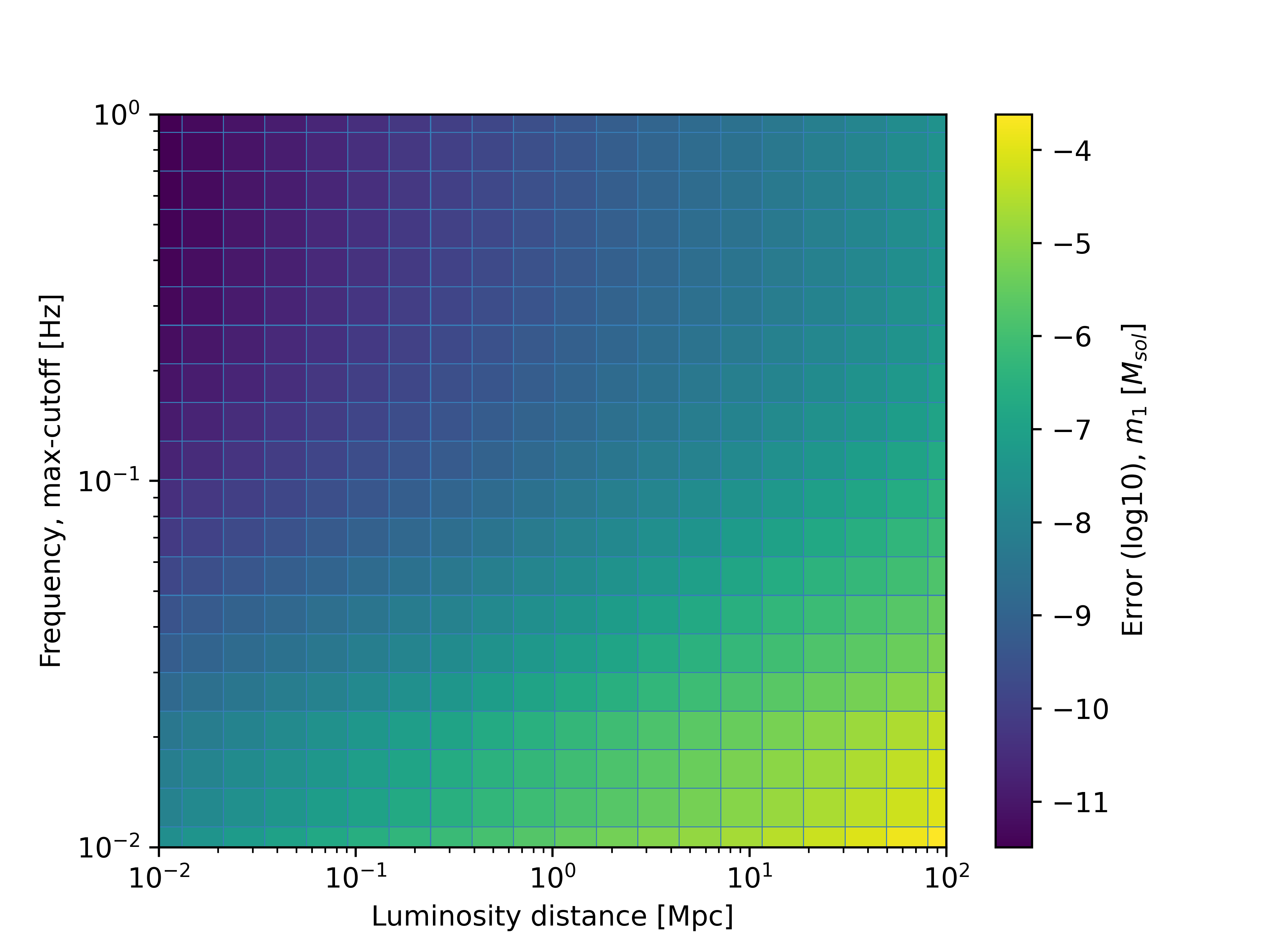}
    \includegraphics[width=.49\textwidth]{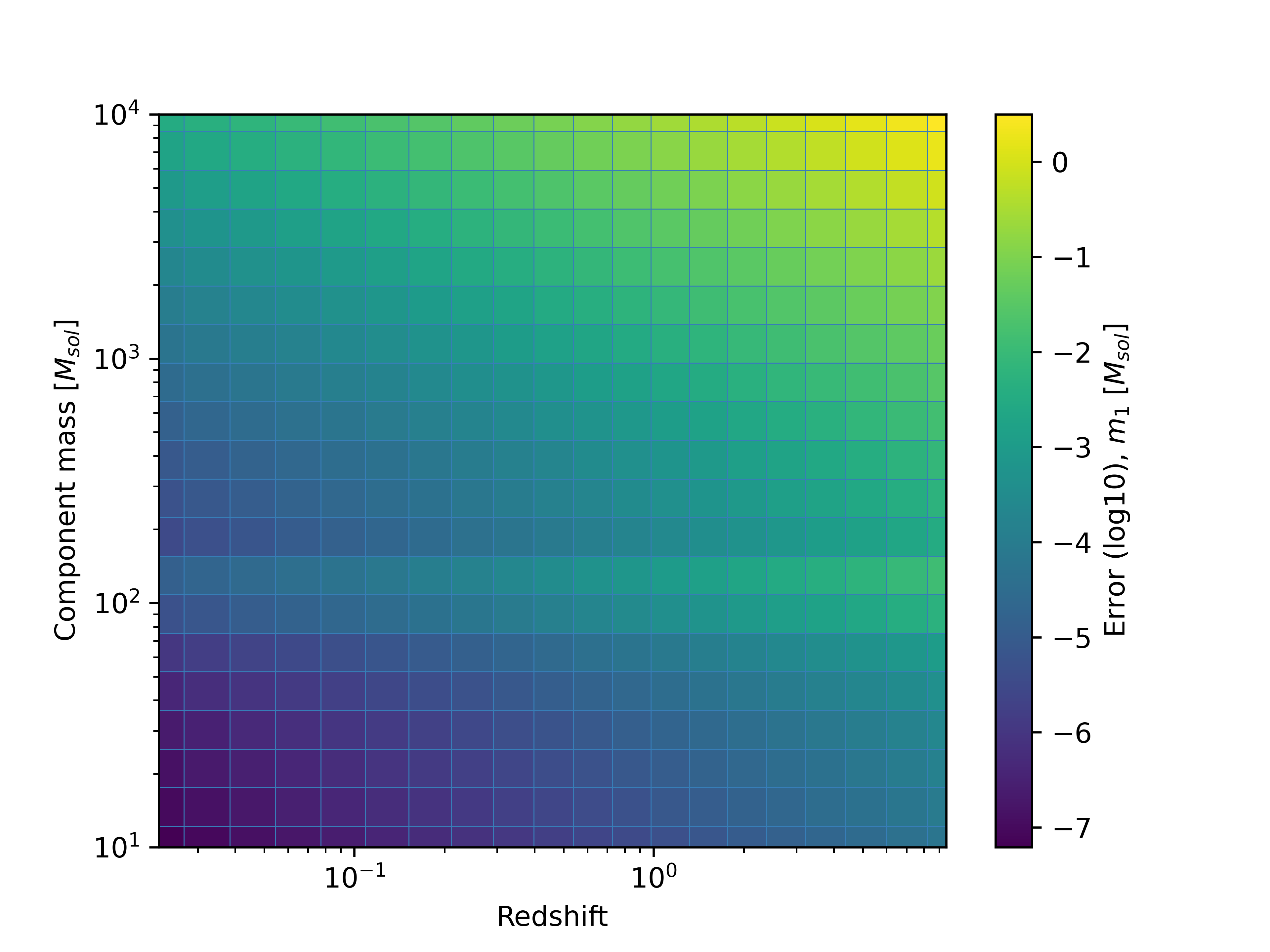}
    \caption{Plots of the estimation errors for the component mass $m_1$. Left: For a $1M_\odot+1M_\odot$ DWD as a function of the distance to the source and the maximum frequency reached by the binary after a 10-year observation time. Right: For equal-mass BBHs expressed in the observer frame.}
    \label{fig:dwd_bbh_m1_errors}
\end{figure*}

The results in figure \ref{fig:dwd_bbh_m1_errors} concern the estimation of the component mass $m_1$. The left plot shows the parameter-estimation error in the case of $1M_\odot+1M_\odot$ DWDs as a function of distance and maximum frequency reached after a 10-year observation time with LGWA. The errors all lie orders of magnitude below the component mass even for large distances of the source close to the LGWA detection horizon. The right plot shows the mass errors for an equal-mass BBH with varying component masses expressed in the observer frame and over a range of redshifts. Also in this case, mass-estimation errors are orders of magnitude lower than the component masses. LGWA's ability to provide such precise mass estimates either comes from long observation times (DWDs) or high SNR (BBH). 

\begin{figure*}[ht!]
    \centering
    \includegraphics[width=.49\textwidth]{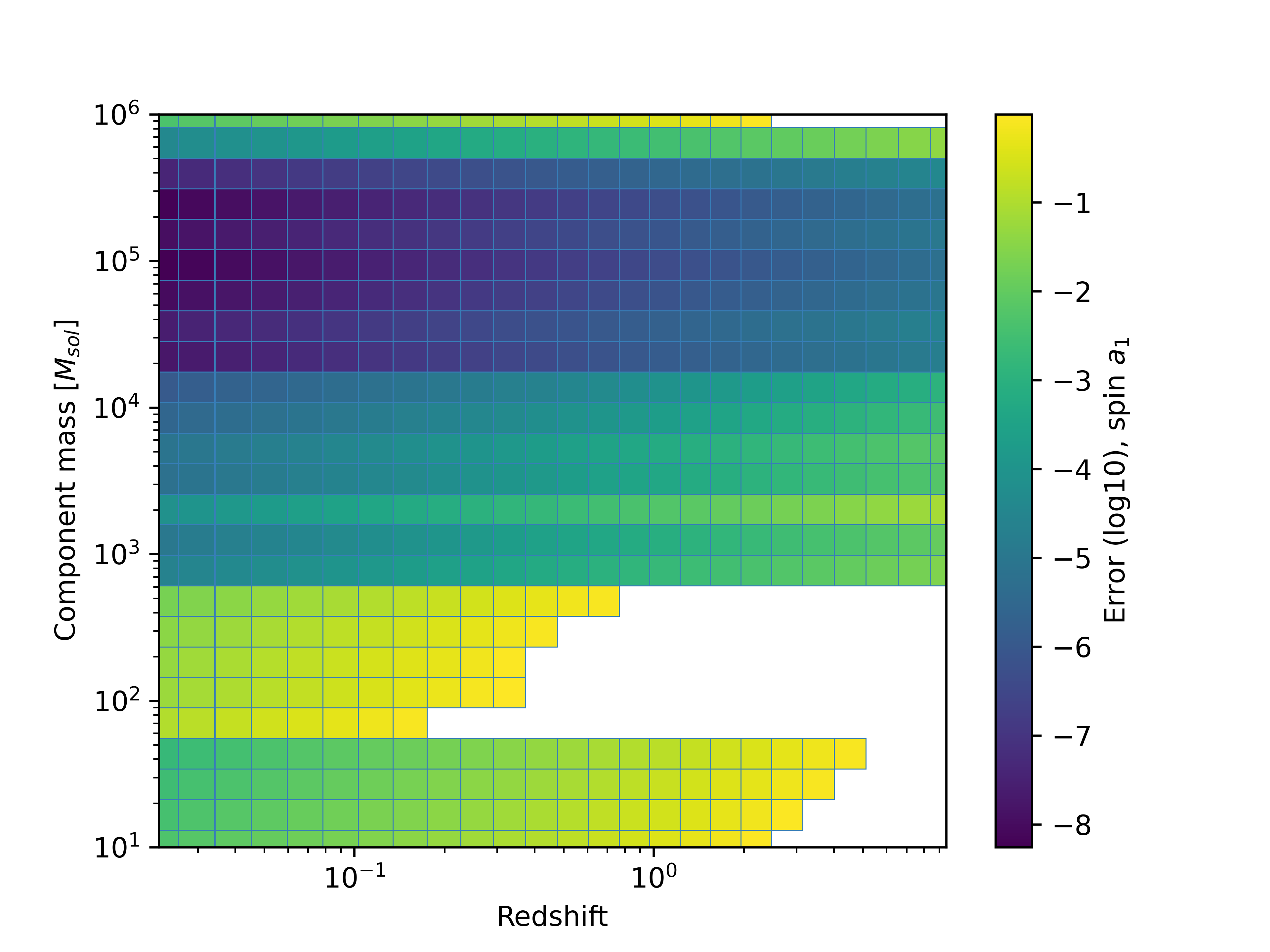}
    \includegraphics[width=.49\textwidth]{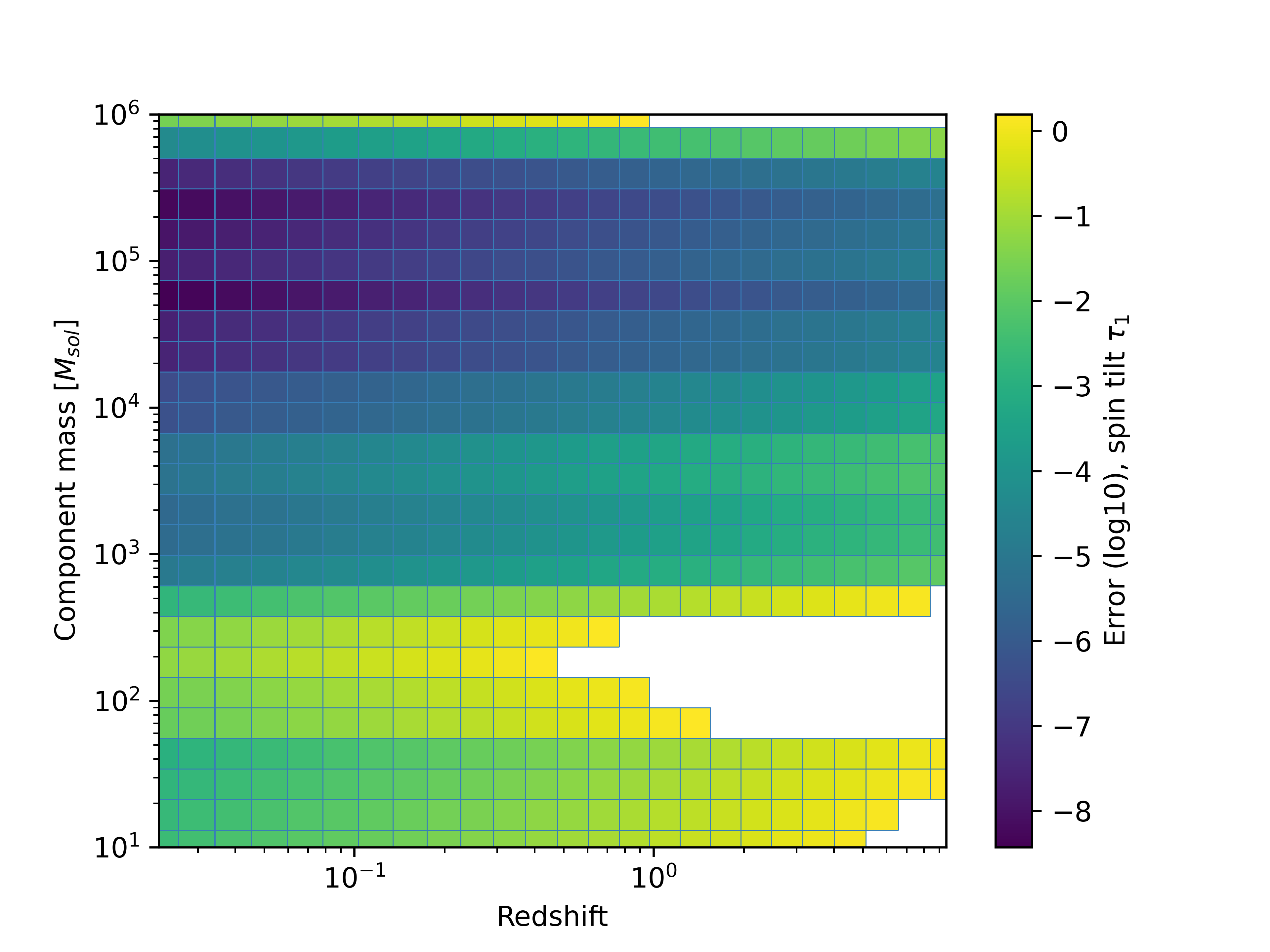}
    \caption{Plots of the estimation errors of spin amplitude $a_1$ (left) and spin tilt angle $\tau_1$ (right) as a function of the distance to the source and for equal-mass BBHs expressed in the observer frame. White areas in the plots mean that the parameter-estimation error exceeds the range of possible values of the parameter.}
    \label{fig:bbh_spin_errors}
\end{figure*}

The next result concerns spin measurements. We expect that LGWA can only provide decent spin estimates for BBHs; especially those merging within the LGWA observation band. This is confirmed by the results in figure \ref{fig:bbh_spin_errors}. The left plot shows the estimation error for the spin amplitude $a_1=0.8$ ($a_2=0.3$) of the first black hole, and the right plot of the tilt of the spin direction $\tau_1=\pi/3$ ($\tau_2=-\pi/5$). As expected, the best spin estimation is achieved for BBHs with component masses around $10^5M_\odot$ consistent with the maximal LGWA detection horizon at $z=50$ and total masses around 1000$M_\odot$ (intrinsic mass) in the horizon plot in section \ref{sec:horizons}. The spin estimates are very poor for BBHs with components masses under 600$M_\odot$ and above $8\cdot10^5M_\odot$.

\subsubsection{Multiband GW observations}
\label{sec:multiband_gw_obs}
{\it Main contributors:} Michele Mancarella, Francesco Iacovelli, Pau Amaro Seoane, Niccolò Muttoni, Alberto Sesana \\

The next decade will witness a flourishing of GW experiments covering different frequency bands. Ground-based interferometers of "third generation" (3G) are under design and entering in advanced stages of planning, both in Europe with the Einstein Telescope (ET) \cite{PuEA2010,MaEA2020,Kalogera:2021bya,Branchesi:2023mws} and in the USA with the Cosmic Explorer (CE) \cite{Reitze:2019iox,Kalogera:2021bya,Evans:2021gyd,Gupta:2023lga}. They may be further complemented by observatories in the southern hemisphere \cite{Gardner:2023znk}. These experiments are mostly sensitive in the frequency range between a few Hz and a few kHz.
The Laser Interferometer Space Antenna (LISA), recently adopted by ESA and scheduled for launch in 2035, will instead target signals in the band 0.1\,mHz to 0.1\,Hz.
The presence of multiple experiments covering such a broad range of frequencies opens the tantalising possibility of multi-band observations \cite{Ses2016} that would have tremendous scientific impact. In this context, LGWA would provide a crucial bridge between LISA and ET/CE with several potential multi-band targets covering the intermediate frequency band around the dHz \cite{IsEA2018,GrHa2020}.

A number of benefits of multi-band detections are shared between different sources and observatories. Detections and determination of the merger time in the LGWA and/or LISA bands can be used to pre-alert ground-based GW detectors and EM observatories, adjusting their time schedules or even tuning them for specific science goals \cite{Tso:2018pdv,NiCa2021,Banerjee2023}; parameter estimation can either be performed jointly, or sequentially, using results from detectors at lower frequencies as priors for data analysis at higher frequencies \cite{Vit2016}; detections in ground-based interferometers can also be used to look back in LGWA/LISA data to find events at lower SNR \cite{Wong:2018uwb}.
\begin{figure*}[ht!]
    \centering
    \includegraphics[width=1.\textwidth]{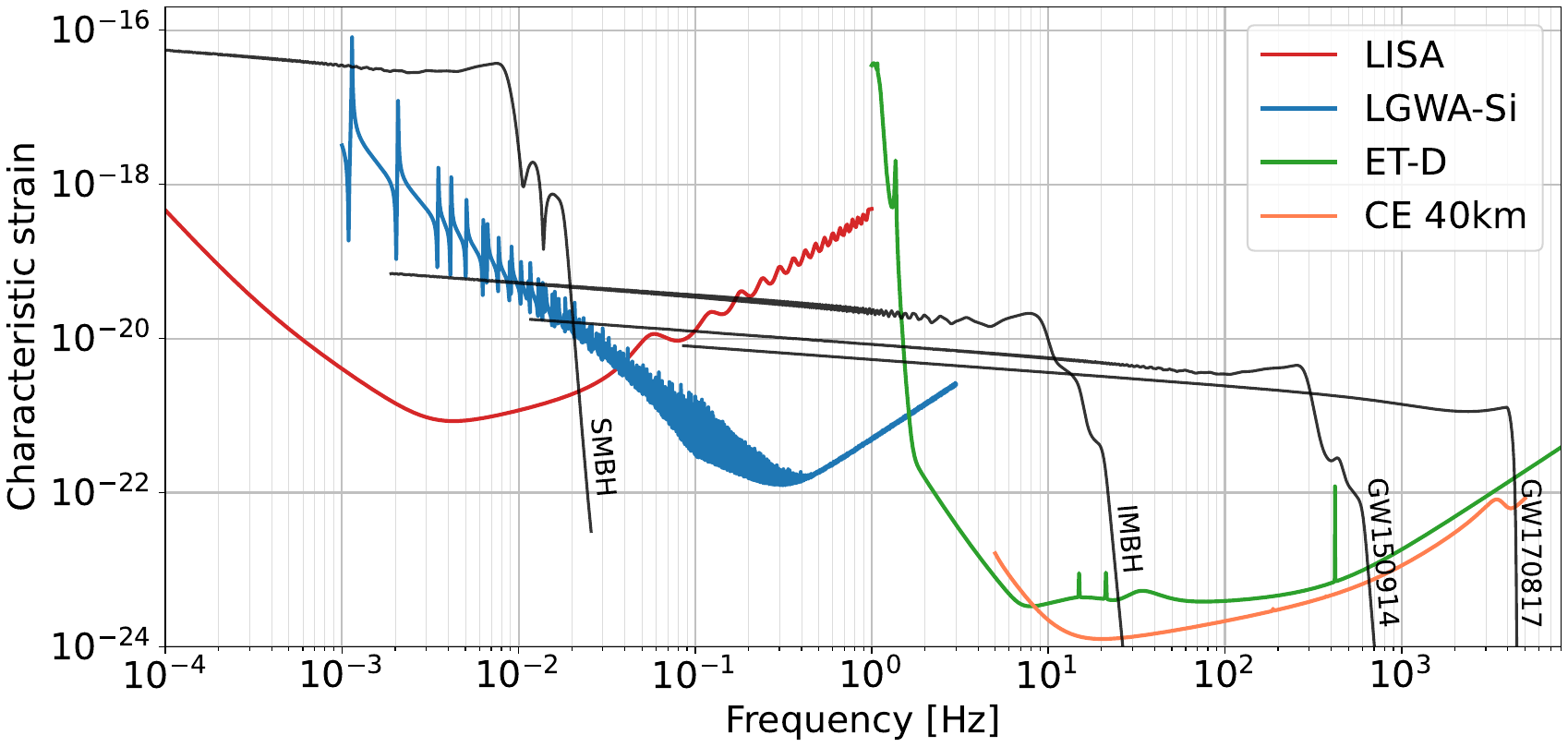}
    \caption{Conceptual illustration of the multi-band approach across the GW frequency spectrum. Black solid lines mark few representative GW signals of different astrophysical origin sweeping through their inspiral, merger and ringdown phases over the frequency bands probed by space-borne and ground-based detectors, colors as in legend. BBH (BNS) signals - which are not projected in the interferometers for illustrative purpose - are modelled by the \texttt{IMRPhenomXHM} (\texttt{IMRPhenomD\_NRTidalv2}) waveform template. For reference, the SMBH binary has source masses $M_1 = 10^{6}$ M$_{\odot}$, $M_2 = 8 \times 10^{5}$ M$_{\odot}$ and $d_L = 1.7$\,Gpc, while the IMBH binary has source masses $M_1 = 10^{3}$ M$_{\odot}$, $M_2 = 4 \times 10^{2}$ M$_{\odot}$ and $d_L = 2$\,Gpc.}
    \label{fig:multiband_concept}
\end{figure*}
An illustration of multi-band GW sources is provided in Fig.~\ref{fig:multiband_concept}, where we show the LISA, LGWA, ET and CE sensitivity curves together with representative GW waveforms from different classes of compact objects. We discuss them in turn.
\begin{itemize}
\item \emph{Stellar Origin Binary Black Holes.} These sources are the primary target of the LIGO and Virgo detectors, with a total of $\sim90$ sources already observed~\cite{LIGOScientific:2021djp}. Their multi-band potential is actively studied in the context of LISA and ground-based interferometers~\cite{Ses2016,Breivik:2016ddj,2016PhRvD..94f4020N,2017MNRAS.465.4375N,Kremer:2018cir,Gerosa:2019dbe,MoEA2019,Seto:2022xmh,Klein:2022rbf,Toubiana:2022vpp}. For example, a binary Black Hole like GW150914 on a circular orbit, with source-frame masses of $36$ and $29$ solar masses \cite{AbEA2016a}, was emitting at $\sim0.016\, \rm Hz$ $5 \, \rm yrs$ before merger, well inside the LISA band. The same system $1$ day before merger emits at $\sim0.26\, \rm Hz$, which is around the peak sensitivity of LGWA. This shows that  Stellar-origin BBHs are natural candidates for joint detections even in all the three bands accessible to LISA, LGWA, and ground-based detectors. A particularly interesting possibility would be the measurement in LGWA of binary parameters hardly accessible to ground-based detectors, such as eccentricity. \changes{Detecting signals that cross a wide band from low to high frequencies would furthermore strengthen the constraints on GW speed propagation~\cite{Baker:2022eiz}, tests of GR~\cite{Barausse:2016eii,Chamberlain:2017fjl,Carson_2020,Gnocchi:2019jzp} and the whole set of source parameters~\cite{Ses2016}. In particular, the accurate sky localization can lead to the unique possibility of a pre--alert of telescopes for the detection of potential counterparts, which are possible in some astrophysical environments; knowledge of the eccentricity can be combined with spin measurements by ground--based detectors to constrain BBH formation channels; precise knowledge of the inspiral part of the signal will enable unprecedented accuracy in constraining the merger physics.}

It is possible to use the constraints provided by the LVK collaboration on the population of stellar-origin BBHs to obtain preliminary estimates of the joint detection rates. We consider a population calibrated on the latest LVK results~\cite{KAGRA:2021duu}, where the source-frame mass distribution is described by the ``PowerLaw+Peak'' (PLP) model~\cite{LIGOScientific:2020kqk}, the spin distribution by the ``Default'' spin model of ~\cite{LIGOScientific:2020kqk}, and the evolution of the merger rate with redshift by a Madau-Dickinson profile~\cite{Madau:2014bja}. We refer to Appendix A of~\cite{Iacovelli:2022bbs} for the specific numerical values adopted. The value of the local rate is taken to be the central value inferred from the GWTC-3 catalog, i.e. $\mathcal{R}_0 = 17 \, \rm Gpc^{-3} \, yr^{-1}$. With these assumptions, we expect $\sim6600$ sources per year up to redshift $\sim1$, which is approximately the LGWA horizon at $100\,M_{\odot}$~\footnote{We remind that the maximum BBH mass in the PLP mass distribution is $m_{\rm max} = 87\,M_{\odot}$ }. We simulate 10 years of observations of LGWA, and associate to each source a time to coalescence extracted from a flat distribution in the interval $[10^{-5}, 10 ]\,\rm yrs$. The time to coalescence $\tau$ and the chirp mass $\mathcal{M}_c$ of the system determine the initial frequency of each GW source according to the GR prediction, $f_{\rm in} = 134 \times (1.21 \rm M_{\odot}/\mathcal{M}_c)^{5/8} \times ( 1 \rm s /\tau)^{3/8} \, \rm Hz$. We compute the SNR of each source assuming a maximum observing time of $10\, \rm yrs$ in ET in its triangular configuration, LGWA and LISA. We use the public package \texttt{gwfish}~\cite{DuEA2022}.
With a single realization of the population according to the above prescriptions, we obtain $\sim 960$ sources observed by both ET and LGWA with $\rm SNR>8$, among which one source is observed also by LISA. These numbers are subject to the uncertainty on the population model, but they show that joint detections of LGWA and ground-based facilities should be expected in large numbers. The SNRs in LGWA span from the minimun of 8 to a maximum of $\sim20$, while those in ET are always $>100$, with the best events reaching $\rm SNR\sim\mathcal{O}(10^3)$. As for joint detections with LISA, a few of them are possible, even if not guaranteed. A detection in three separate observatories would however represent an extraordinary science achievement. This result is in line with the findings in~\cite{MoEA2019}.

\item  \emph{Binary Neutron Stars.} A multiband detection of these systems would be a unique possibility opened by LGWA, since those are not accessible to LISA.
For example, a Binary Neutron Star system with source-frame chirp mass $\mathcal{M}_c = 1.188\, M_{\odot}$, corresponding to the value measured for GW170817 \cite{LIGOScientific:2017vwq} was emitting at $\sim0.05 \rm Hz$, at the edge of the LISA band, more than 45 yrs before the merger. This timescale is clearly too long for the possibility of a joint detection. On the contrary, such a source emits at $\sim0.27\, \rm Hz$, around the peak sensitivity of LGWA, 6 months before merger, being a perfect candidate for joint observations of LGWA and ground-based detectors, which can observe BNS systems at high rates \cite{Iacovelli:2022bbs}. Considering 10 years as a limit timescale for the possibility of a joint detection, a GW170817-like source could be observed from a starting frequency of  $\sim 0.09\, \rm Hz$.
We estimate the joint detection rate in LGWA and ET using the same procedure as for stellar-origin BBHs. For BNS systems, the uncertainty on the population model is much larger; in particular, the value of the local merger rate inferred from the GWTC–3 catalog, assuming a flat mass distribution, is $\mathcal{R}_0 = 105^{+190}_{-83} \, \rm Gpc^{-3} \, yr^{-1}$. Using the central value of $105 \, \rm Gpc^{-3} \, yr^{-1}$, a flat mass distribution between $[1.1, 2.5]\, \rm M_{\odot}$, and values for the Madau-Dickinson profile obtained as described in Appendix A of~\cite{Iacovelli:2022bbs}, we obtain $\sim36$ sources per year within redshift $z\sim 0.1$, which corresponds to the LGWA horizon at $\sim \rm few \, M_{\odot}$. Over a period of $10\, \rm yrs$, we obtain 5 joint detections between LGWA and ET with $\rm SNR>8$. Given the large uncertainty on the local merger rate, we also compare with the results obtained using the upper and lower rates values of $295 \, \rm Gpc^{-3} \, yr^{-1}$ and $22 \, \rm Gpc^{-3} \, yr^{-1}$, which give 24 and 1 joint LGWA-ET detections respectively. We conclude that a joint detection is a possibility that should be expected in LGWA. The particularly long time spent by BNS systems in the LGWA band would enable their precise localization prior to merger. \changes{Combined with the early detection by ground-based 3G detectors, the sky patch to scan for EM signals would further shrink making use of the long baseline between Earth and Moon. There are prediction of radio and X-ray counterparts produced well before the merger of two neutron stars \cite{10.1046/j.1365-8711.2001.04103.x}. Early warnings from LGWA would make it possible to prepare for the observation of these faints signals.} Also, long-duration multi-band studies using LGWA and ground-based GW detectors \cite{GrHa2020,liu2022neutron} can improve the estimation of parameters, in particular the tidal deformability~\cite{IsEA2018}. This will lead to better constraints, ruling out many more theoretical EOS models \cite{liu2022neutron}.

\item \emph{Intermediate Mass Black Holes.} 
While the rates of these systems are still largely uncertain, their potential is high since they could accumulate a high enough SNR in LISA, LGWA and ET during the inspiral, merger and ringdown. A combined measurement could then allow a joint analysis, possibly breaking degeneracies among waveform parameters that are best constrained in different phases of the evolution. A system such as the one in Fig.~\ref{fig:multiband_concept} emits at $\sim0.002\, \rm Hz$ 5 years before the merger, and at $\sim0.27\, \rm Hz$ around 10 minutes before the merger. This makes imaginable an early warning by LISA to both LGWA and ground-based detectors, with the subsequent detection in all the three bands.

\item \emph{Supermassive Black Holes.} 
These systems represent a target for joint LISA-LGWA detections. In particular, there is ample overlap in the mass and redshift range of the two observatories between $10^3 - 10^6 \, M_{\odot}$ up to redshift $\sim10$. For example, Ref.~\cite{Bonetti:2018tpf} finds between $\approx25$ and $\approx75$ events per year depending on the seed model. Most or all of them will be detected also by LGWA, which can in particular accumulate enough late inspiral-merger-ringdown SNR allowing a better localization of the sources, especially at high redshift. This is of paramount importance since a combined LISA-LGWA analysis might help pinpointing the source parameters with higher accuracy, in particular the source distance. In fact, although LISA can see MBHBs at $z>20$, for faint sources the distance estimate is rather poor and determining the actual high redshift nature of the source might be challenging~\cite{Sesana:2012nf,Mancarella:2023ehn}. Moreover, LGWA will be sensitive to the merger and ringdown part of the signal, which for systems of $M<10^4$ at $z>5$ is not accessible to LISA.

\end{itemize}

\begin{figure*}
    \centering
    \includegraphics[width=1.\textwidth]{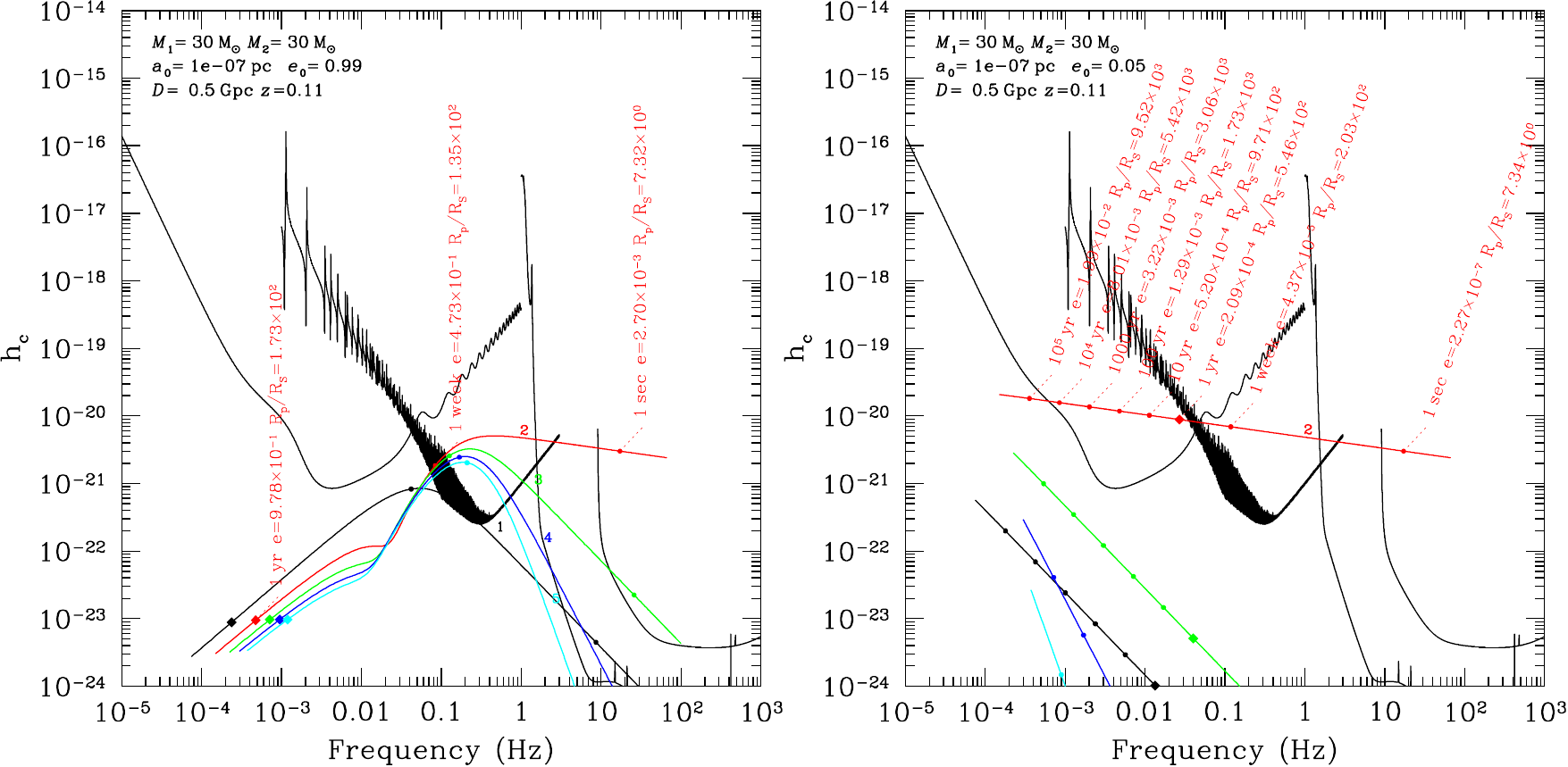}
    \caption{Characteristic amplitude $h_c$ of the first four harmonics (indicated with numbers) emitted by a binary of two black holes with masses $M_1=M_2=30\,M_{\odot}$ and at a luminosity distance of $D=0.5$\,Gpc.  The amplitude and the orbital evolution correspond to the quadrupole approximation. We display a binary starting at a semi-major axis of $a_0=10^{-7}$\,pc and with initially two different eccentricities: (i) $e_0=0.99$ (left panel), and (ii) $e_0=0.05$ (right panel).  Along the harmonics we mark several particular moments with dots, where the labels show the time before the coalescence of the binary and the corresponding orbital eccentricities. Additionally to the LGWA, we depict the noise curves for LISA and LIGO.}
    \label{fig:LGWA_multi_ecc_circ}
\end{figure*}

For some of the systems discussed above, it is important to mention the role of eccentricity. This can be a signature of formation channels~\cite{ChenAmaroSeoane:2017}, though its detection might be challenging. Indeed, eccentricity dampens the characteristic amplitude of each harmonic compared to a circular one. Figure~\ref{fig:LGWA_multi_ecc_circ} shows the signal of two GW150914--like sources with different initial eccentricities. Eventually, about an hour before the merger, the signal becomes indistinguishable from the circular case.
On the other hand, increasing the eccentricity shifts the peak of the relative power of the GW harmonics towards higher frequencies. Therefore, more eccentric orbits will emit their maximum power at frequencies close to the sweet spot of the LGWA. More precisely, when the eccentricity $e=0$, all the GW power is radiated by the $n=2$ harmonic, so that GWs have a single frequency of $2/P$, where $P=2\pi]G(M_1+M_2)/a^3]^{-1/2}$ is the orbital period.  On the other hand, at $e\simeq1$, the $n=2.16(1-e)^{-3/2}$ harmonic becomes dominant \cite{FaPh2003}, so most of the GW power is radiated at the peak frequency of $f_{\rm peak}=2.16(1-e)^{-3/2}P^{-1}$.

\begin{figure*}
    \centering
    \includegraphics[width=\textwidth]{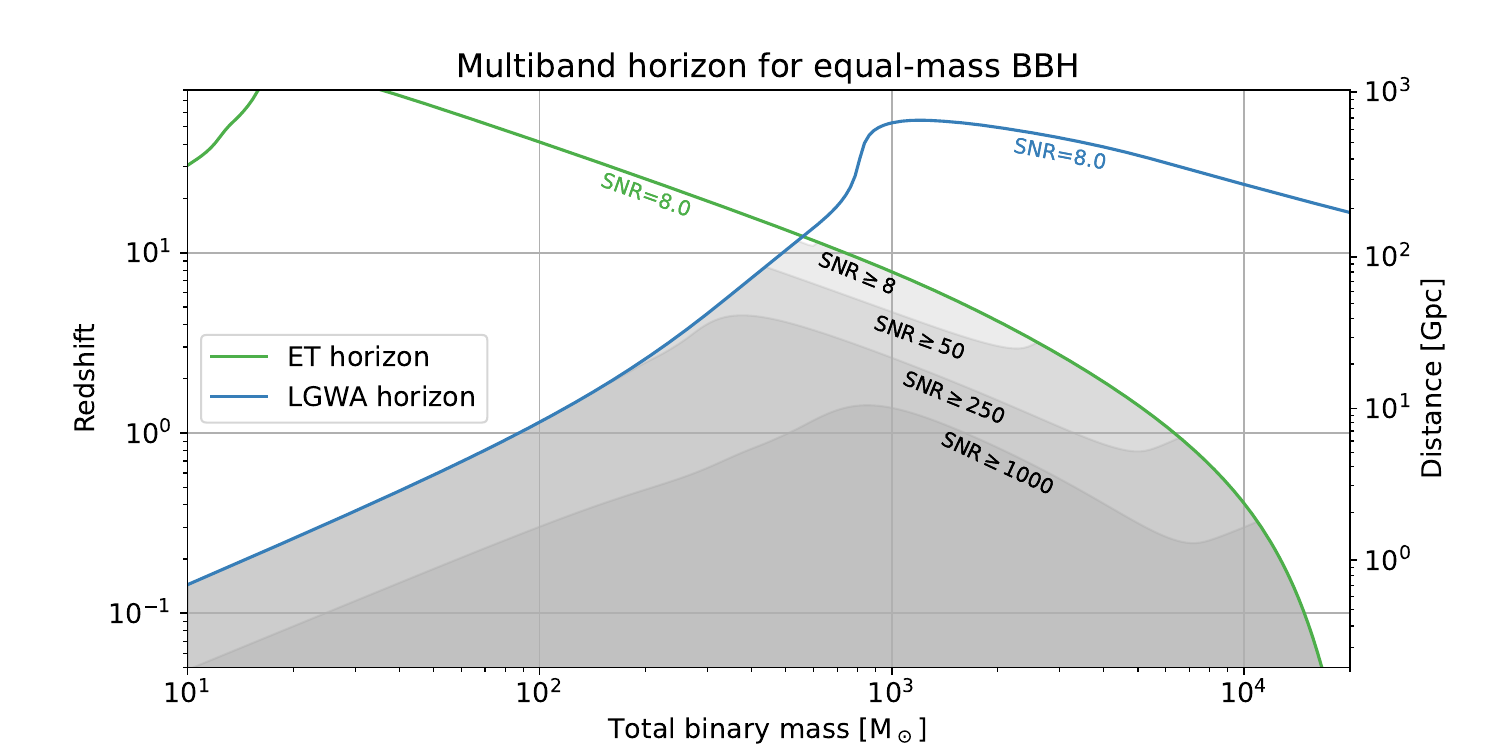}
    \caption{Multiband horizon for a network of Einstein Telescope and LGWA (with the Silicon sensitivity curve). As discussed in section \ref{sec:horizons}, this is showing the maximum distance and redshift for which an optimally-oriented source can be detected with a given signal to noise ratio (SNR). In the grey shaded areas, the source has at least an SNR of 8 in both detectors, allowing for multi-band parameter estimation. The optimal sky position is computed for every mass. The SNRs in the shaded region are referred to the whole network, including the contribution from both detectors. The masses reported on the horizontal axis are in the source frame, i.e.~as they would be measured by an observer in the same reference as the binary.}
    \label{fig:muliband-horizon-et-lgwa}
\end{figure*}

\begin{figure*}
    \centering
    \includegraphics[width=\textwidth]{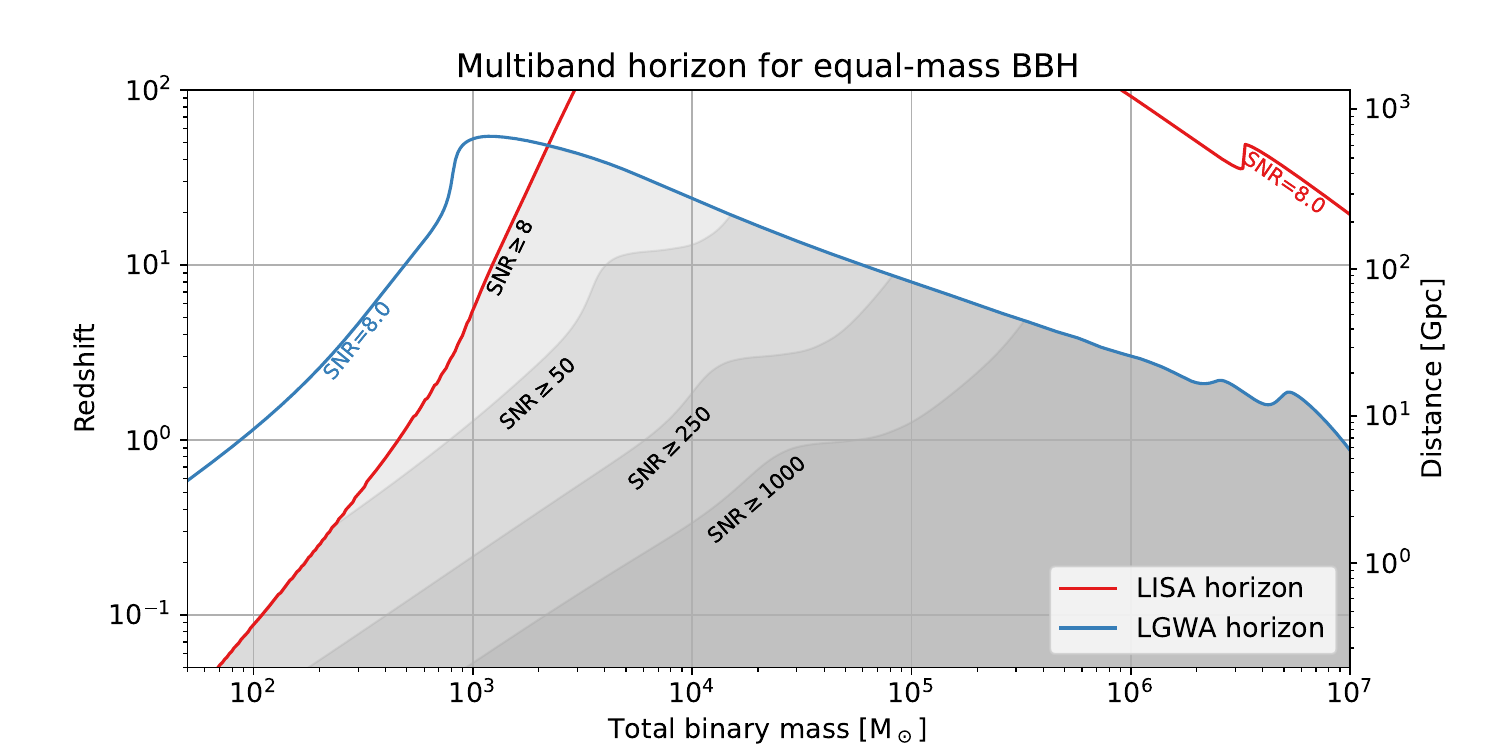}
    \caption{Multiband horizon for a network of LISA and LGWA, computed as in figure \ref{fig:muliband-horizon-et-lgwa}.}
    \label{fig:muliband-horizon-lisa-lgwa}
\end{figure*}

A useful figure-of-merit to assess multi-band capabilities of LGWA is the detection horizon. In Figures~\ref{fig:muliband-horizon-et-lgwa} and~\ref{fig:muliband-horizon-lisa-lgwa} we show the joint detection horizons for LGWA-ET and LGWA-LISA respectively, for binary systems between 10--10$^{7}\, M_{\odot}$.
In general, the horizon is dominated by the least sensitive detector. For stellar-origin BBHs ($\sim5-100 \, \rm M_{\odot}$), the LISA sensitivity will be the limiting factor for LISA+LGWA+ET multi-band sources, limiting the potential targets to $z\lesssim0.3$. On the contrary, a joint LGWA+ET detection would enlarge the observable window to $z\lesssim1$. Furthermore, as already pointed out, those sources would be "golden events" in ET with SNR $\geq \mathcal{O}(100)$. As mentioned earlier, a detection in ET could also be used for an \emph{a-posteriori} targeted search in LGWA data, allowing to further reduce the SNR threshold and increasing the horizon. Similarly, at the opposite edge of the mass range considered here, stellar-origin BBHs could be observed in both LISA and LGWA up to redshifts $>1$ with very high SNR  $\sim \mathcal{O}(10^3)$ in LISA.
Finally, we note that an interesting "sweet spot" exists for IMBHs around $10^3 \rm M_{\odot}$ where both LISA and ET can reach horizons of $z\sim7$--8 while LGWA is at its peak, $z\sim40$. As a consequence the multi-band horizon is quite large even for a LISA+LGWA+ET detection. 
The highest potential for observing high-redshift sources is for systems between $ [2-5]\times 10^3 \, \rm M_{\odot}$ where the joint LISA-LGWA horizon exceeds redshift $30$.

Finally, we stress that a crucial aspect of the multi-band science case is the time overlap between different experiments. A particularly strong point in favour of multi-band perspectives for LGWA is the absence of a strong time limit for the mission duration. The overlap with the LISA mission can be limited by the time schedule of the latter (even if an extension up to a maximum of 10 yrs may be possible), while the schedule of 3G ground-based detectors is still much more flexible. 

\subsection{Studies of the Lunar Surface Environment and of the Deployment Site}
\label{sec:environment}

This section summarized the studies of the Moon's surface and subsurface required to prepare and model LGWA deployment scenarios. Theoretical and numerical modeling of what we expect from the Moon will help in minimizing approximation in instrument design and in selecting most appropriate deployment scenarios for the sensors. In detail, we require modeling of the geology (section \ref{modeling:geology}),  the surface temperature field (section \ref{modeling:temp}), the ground tilt (section \ref{sec:tilt}), background seismic noise (section \ref{sec:background}) and magnetic fluctuations (section \ref{modeling:magnetic}).

\subsubsection{Geologic models}
\label{modeling:geology}
{\it Main contributors:} Alessandro Frigeri, Angela Stallone \\

The geology at any site of a planetary body is the result of its unique evolution. Over time, a sequence of geological processes, deposits or erodes materials, revealing the environmental conditions of specific periods. A geological model serves as a comprehensive framework for conceptualizing the subsurface geological structures and processes representing our current state of knowledge of the subsurface of a solid planetary body. 

\begin{figure}[htp!]
    \centering
  \includegraphics[width=.72\textwidth]{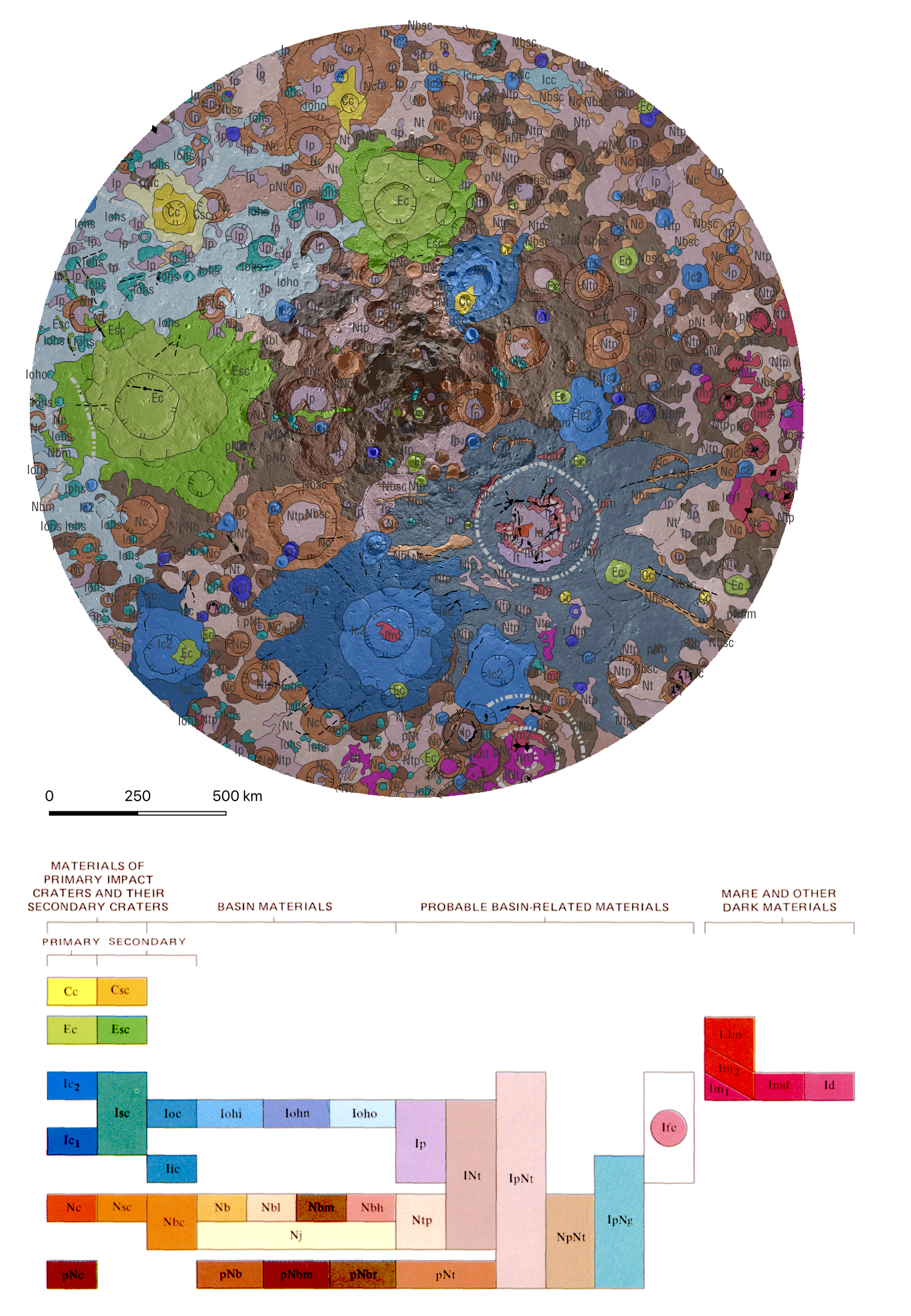}
      \caption{Top: Global-scale geological map of the Moon's southern polar region, displaying how the geological units are spatially distributed \cite{Wilhelms1979, Fortezzo2020}. The map shows a stereographic projection spanning from 55 degrees of latitude south to the South Pole. The graphical scale is referred to the lunar South Pole. Colors indicate the areal extent of different geologic units. Bottom: the vertical and lateral geometric relationship of the units gives the information to read the map in the third dimension of space and the relative timing of the emplacement of the units. The correlation of map units is not fully represented in the map and units in the map have been updated for a global stratigraphy.}
      \label{fig:geologic_map}
\end{figure}

Figure~\ref{fig:geologic_map} shows the geological map for the Moon's southern polar region. Initially compiled after the Apollo program by Wilhelms et al \cite{Wilhelms1979}, it has recently been updated by Fortezzo et al \cite{Fortezzo2020}. The map's fundamental components are surface portions classified by similar morphological and/or compositional characteristics, referred as \emph{geological units} represented by different colors. Besides the map, the lateral and vertical geometric relationship of the \emph{geological units} enables ordering the units in space and in time of emplacement.

The geological map is the foundation for constructing any \emph{geological model}, providing both geometric and physical parameters for creating a related \emph{geophysical model}.  The latter is used as input for numerical modeling techniques to simulate or predict physical processes occurring on Earth or any other planetary body.  In the context of LGWA, our aim is to develop geophysical models that can simulate seismic background response across different deployment scenarios. This begins with one-dimensional modeling \cite{garcia2011very} and progresses towards multi-dimensional and multi-resolution models (See Section \ref{sec:seismicbackground}). 

As our understanding evolves with new instrumental data, the geologic map and model must be continuously updated to incorporate existing and new insights about the Moon's subsurface.  This implies that modeling is an iterative process, where geological evidence and geophysical modeling dynamically interplay with each other.

\subsubsection{Solar illumination and surface temperature}
\label{modeling:temp}
{\it Main contributors:} Philipp Gläser\\

Due to the small angle of $1.5^\circ$ between the ecliptic plane and the lunar equatorial plane, extreme illumination conditions occur near the polar regions of the Moon. Here we find both, extensive, almost continuous illumination right next to PSRs. PSRs are found on the floors of near-polar craters and represent the darkest and coldest places on the lunar surface. Extensively illuminated regions are typically found on elevated topography such as ridges and crater rims. The PSRs are ideal deployment sites for LGWA, and in fact, it is currently believed that deployment inside a PSR is necessary since temperature changes and gradients elsewhere on the lunar surface are strong and would cause excess noise in LGWA data (see section \ref{sec:tilt}). Study of surface temperature data and surface temperature modeling are necessary to identify suitable LGWA deployment site candidates.

\begin{figure}[ht!]
    \centering     \includegraphics[width=1\textwidth]{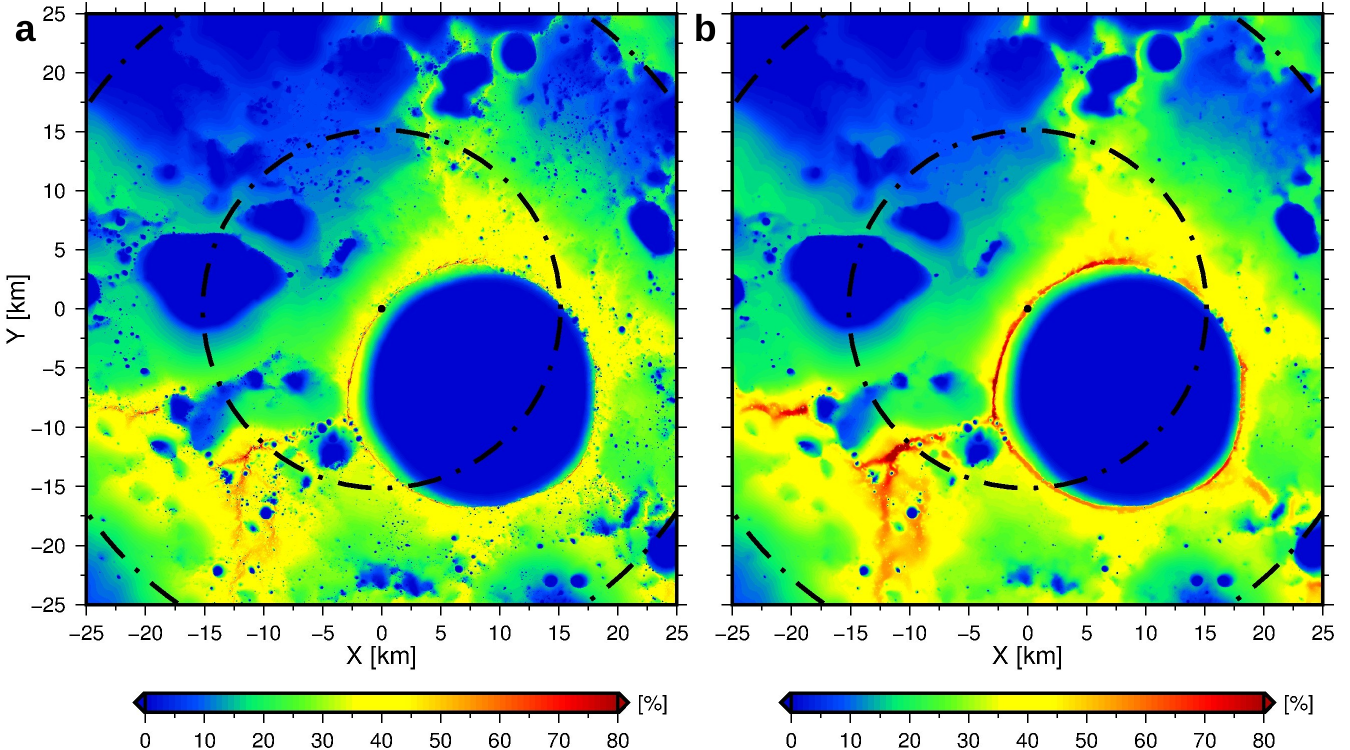}
      \caption{Accumulated illumination at the lunar south pole calculated over a twenty-year period and one-hour intervals. Illumination is color-coded where dark blue refers to PSRs and dark red refers to extensively illuminated spots. \textbf{a:} Accumulated illumination at the lunar surface. \textbf{b:} Accumulated illumination for an observer at two meters above the ground. Figure adapted from \cite{2018P&SS..162..170G}.}
      \label{fig:SP_ill}
\end{figure}

Illumination conditions are calculated based on a high-resolution digital terrain model (DTM) derived from Lunar Orbiter Laser Altimeter (LOLA) data. Here, the LOLA DTM can be synthetically illuminated and it provides illumination conditions for any chosen time period and location \cite{2018P&SS..162..170G,Mazarico2011}. Based on the resulting illumination maps, PSRs and extensively illuminated spots can be identified (compare blue and red areas in Fig. \ref{fig:SP_ill}). Further, illumination maps can be calculated for observers at various heights above the ground, i.e. to model illumination at elevated solar panels. It was found that there is a significant increase in illumination in the first $\approx 10$\,m above ground. The illumination at the surface and 2\,m above ground of one of the PSRs is shown in figure \ref{fig:SP_ill}a+b. 

\begin{figure}[ht!]
    \centering
      \includegraphics[width=1\textwidth]{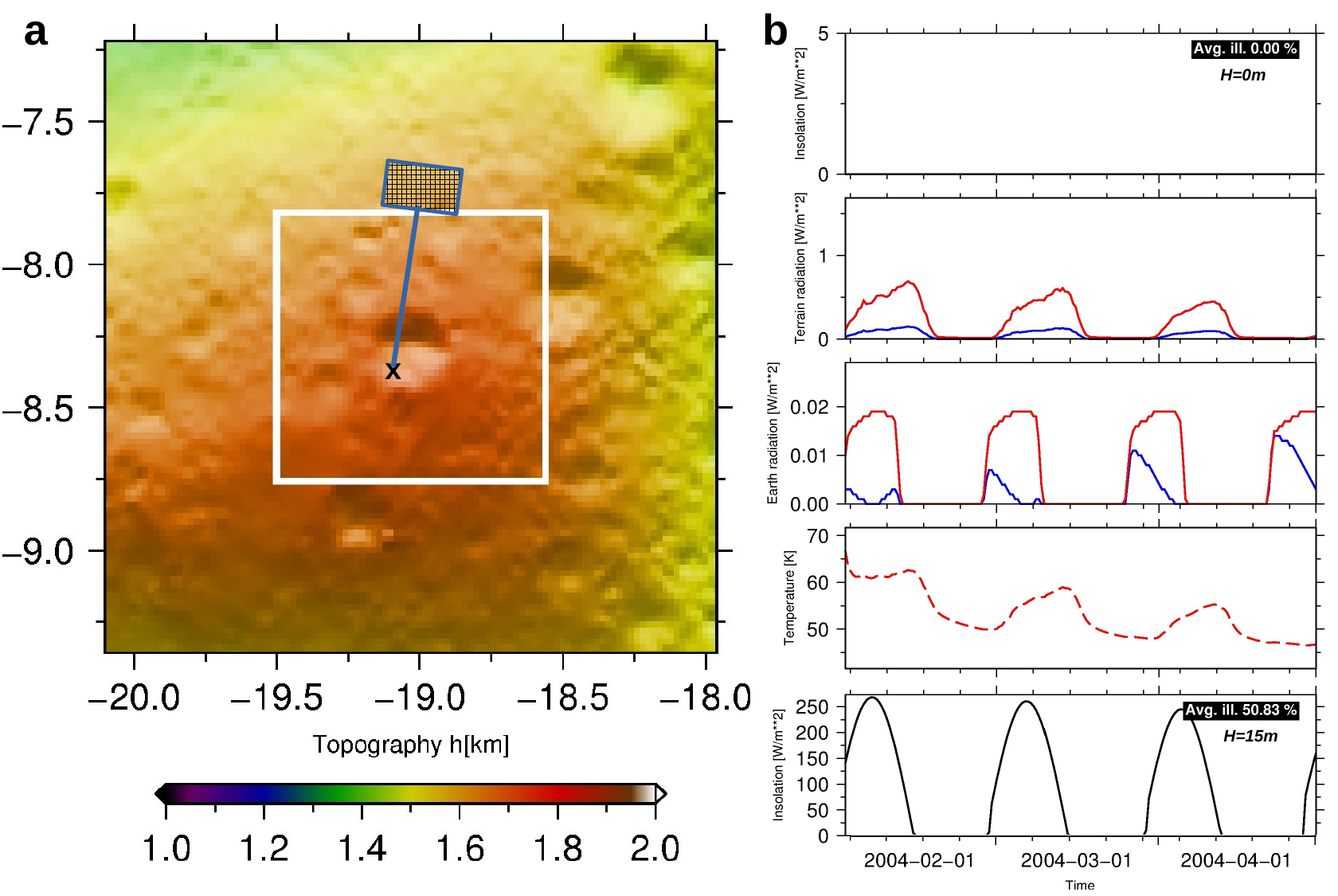}
      \caption{A PSR with significant illumination on a solar panel mounted on a beam at 15\,m above ground. \textbf{a:} The small crater showing the location of the possible landing site (black x). A sketch of a solar panel mounted directly and 15\,m above the landing site is shown in blue. \textbf{b:} The five panels from top to bottom show the direct illumination at the surface, radiation coming from the surrounding terrain, radiation coming from Earth, surface temperature, and the direct illumination at 15\,m above ground (solar panel). Note that the radiation coming from terrain and Earth is divided into thermal radiation (red profile) and scattered sunlight (blue profile). Panels 1--3, 5 show radiation in W/m$^2$ over time and panel 4 shows temperature in K over time.}
      \label{fig:SP_lsite}
\end{figure}

Illumination maps also serve as an input for a thermal model to derive (sub-)surface temperature estimates for the uppermost 2\,m of regolith \cite{2019A&A...627A.129G}. In order to derive meaningful temperature values inside PSRs it is crucial to consider internal heating as well as multiple scattered sunlight and thermal radiation from the surrounding terrain as these are the only energy sources. For LGWA a landing site that offers low temperatures with little to no variations would be beneficial. On the contrary, to generate power using solar panels a location offering extensive illumination is needed. In an initial survey we searched for small craters that offer PSRs on their crater floors and significant illumination for a solar panel which would be mounted on a beam at several meters above ground. A potential candidate landing site was found in a small crater located between Shackleton and de Gerlache craters for which these constraints were met, see Fig. \ref{fig:SP_lsite}. 

Since the thermal characteristics of a PSR are crucial to the performance of LGWA, it is important to carry out a systematic analysis of all PSRs at both lunar poles. Other aspects might be considered in these analyses such as the vertical distance to sunlight with the idea to realize a solar-energy powered experiment inside the PSR. 

\subsubsection{Regolith composition}
\label{sec:regolith}
{\it Main contributors:} Francesco Mazzarini, Goro Komatsu \\

On the lunar surface, regolith forms the boundary between the Moon's interiors and the space environment, preserving key information about the geological evolution of the satellite but also  and also providing some important clues for the history of the inner Solar System impact flux, the solar wind, and galactic cosmic rays \cite{mckay1991lunar}. Lunar regolith is a fine-grained layer of fragmental debris overlying the coherent substrate consisting largely of the upper heavily reworked dust layer, fragmented rocks, impact ejecta, breccia lenses, glass fragments, and agglutinates \cite{wilcox2005,zhang2023}. The regolith has been estimated to be 5\,m to 20\,m thick globally, with a shear velocity of less than 100\,m/s \cite{cooper1974,lognonne2005}. 

\begin{figure}[ht!]
    \centering
      \includegraphics[width=1\textwidth]{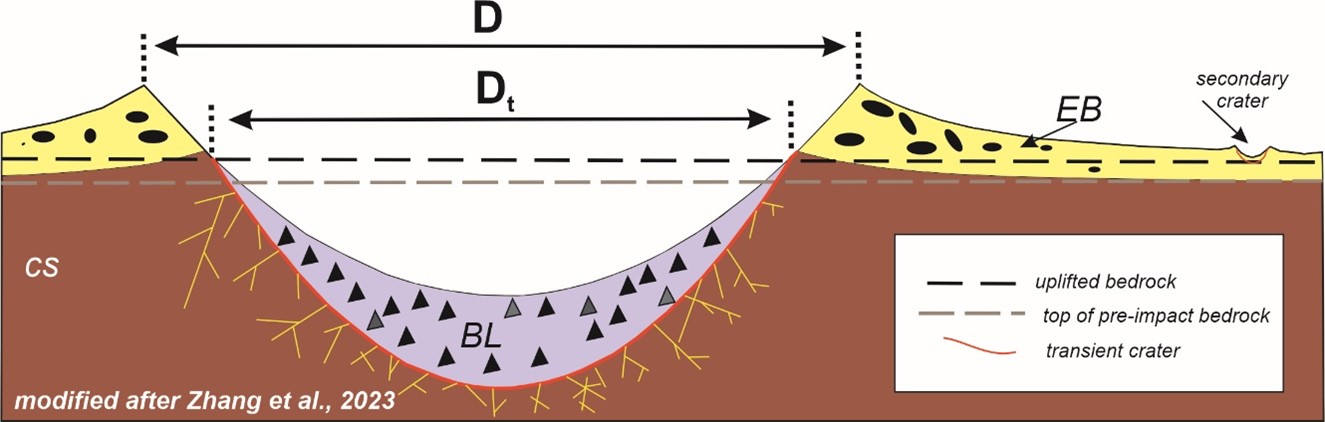}
      \caption{Conceptual model of the regolith formation by impact modified after \cite{zhang2023}. Yellow thin lines are the fractured bedrock at the impact site; BL: breccia lens; CS: coherent substrate (bedrock); EB: Ejecta blanket. D is the rim to rim crater diameter; Dt is the transient crater diameter measured with respect to the pre-impact surface.}
      \label{fig:regolith}
\end{figure}

Regolith in the Moon forms by high‐energy physical weathering processes (e.g., \cite{head2020}) such as: i) micrometeorite comminution, ii) agglutination (quenched impact glass and welded particles up to 25--30 vol\% of the regolith), and iii) crater impacts (\cite{zhang2023} and references therein). In general, the continuous impacts of large and small asteroids led to the formation of the lunar regolith \cite{shoemaker1969,zhang2023}. The impact on the pristine bedrock surface generates breccia lens within the crater and a blanketing by ejecta around the crater (Figure \ref{fig:regolith}).

The initial formation of the regolith starts with impact bombardment onto the pristine crust, a stochastic process deforming pulverizing, and melting the pristine crust (the protolith). Recent studies suggest the occurrence of H$_2$O and other volatile species linked to magma (\cite{head2020} and references therein).

The strength of bedrock significantly affects the crater size and hence the volume of regolith produced especially for sub-decameter impactors. The regolith volume produced by an individual impact crater is quantitatively characterized as a function of crater diameter and pre-impact regolith thickness \cite{zhang2023}.

For the lunar maria, the surface structure can be modelled as a fine-grained regolith layer atop the underlying coherent bedrock (Figure \ref{fig:regolith}). Depending on the ratio of crater rim-to-rim diameter (D) to pre-impact regolith thickness (T) different morphologies may form (\cite{zhang2023} and references therein). The thickness of lava flows in lunar maria are poorly constrained,
although observation within pit craters suggested that the bedrock of the regolith consists of several flows up to 10--14\,m thick \cite{ROBINSON201218}.

Being the impact the most likely forming process, the composition of the impacted surface will determine the composition of the regolith. The old anorthositic crust formed by fractional crystallizations of the pristine Lunar Magma Ocean \cite{Rossi2018,borg2023}. After the formation of an early anorthositic crust (the present-day highlands) the partial melting of the mantle produced a secondary crust (basalts) within impact craters. Large impact basins floored by flood basalts formed the lunar maria and oceanus, which are mostly located at equatorial and mid latitude regions \cite{Rossi2018}. In the south pole of the Moon are present both pristine crust as well as basalts filling craters \cite{zhang2020}.

The basalts in the maria and the anorthosites in the highland are low-silica magmatic effusive and intrusive rocks, respectively. The main compositional difference between them is their mineralogical composition \cite{JAUMANN201215}. Basalts contains MgO, FeO, TiO$_2$ -rich mafic minerals (olivine, pyroxene, ilmenite). Anorthosites are composed of mostly plagioclase feldspar (90–-100\%), with a minimal mafic component (0–-10\%).

The lunar surface mineralogy-petrology can thus be simplified, and it has been characterized by four main mineral phases: plagioclase, pyroxenes, olivine, and ilmenite, with other oxides as accessory minerals. Beneath Mare Imbrium (mare basalts), hundreds of meter beneath the surface, SAR processed radar recordings highlighted the occurrence of well-defined layering formed by levels having the dielectric constant ($\epsilon=3.3$) typical of that lunar regolith \cite{BANDO2015144}.

The presence of a sequence of regolith, lava and ejecta layers to a depth of 360\,m has been confirmed also for the Von Karman crater on the far side of the Moon in the South-Pole Aitken basin crater by using dual-channel lunar penetrating radar (LPR), suggesting that the site underwent remodelling for multiple impacts \cite{zhang2020}. Similar multiple regolith layers are interpreted to exist also from Chang’e-3 Yutu rover data \cite{Ding2020}.

\begin{figure}[ht!]
    \centering
      \includegraphics[width=0.5\textwidth]{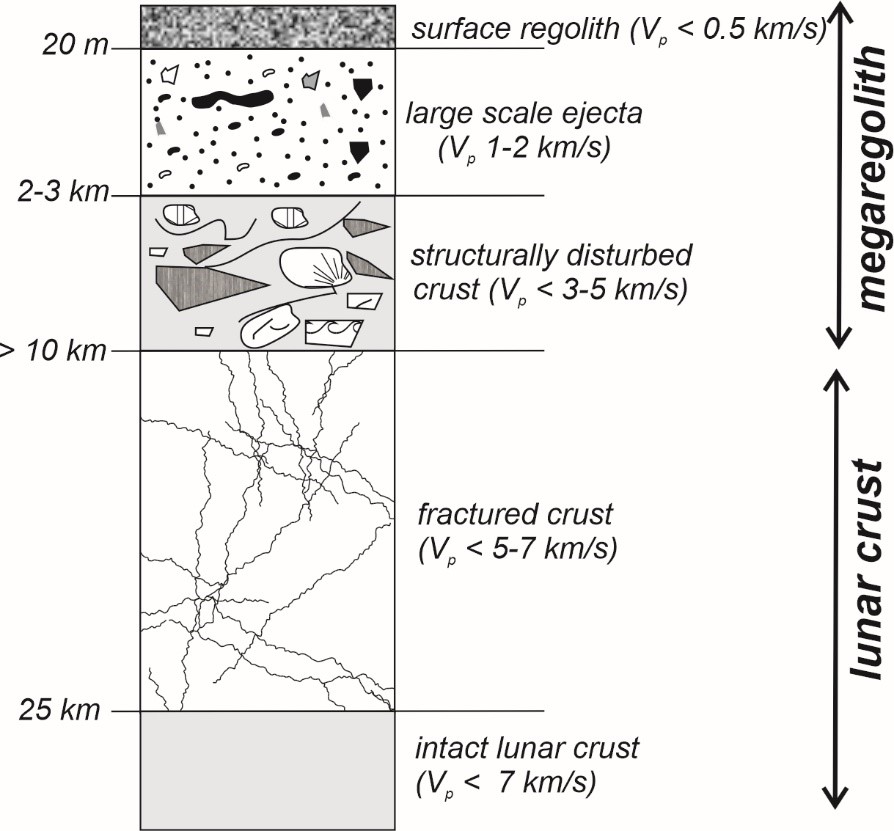}
      \caption{Highly idealized cross-section through the internal structure of the megaregolith and the upper lunar crust (modified after \cite{JAUMANN201215}). The depth scale is highly uncertain because regional variations are expected depending on the degree to which basin-sized impacts have influenced a region. }
      \label{fig:surfacekms}
\end{figure}

At satellite scale, seismic evidence suggests that the cumulative ejecta thickness on the lunar highlands is possibly tens of kilometres thick \cite{JAUMANN201215}. The average depth of mechanically disturbed crust by impacts is not well known, conservative estimates indicate a thickness of the ejecta blankets of at least 2–-3\,km (megaregolith), a structural disturbance to depths of more than 10\,km, and fracturing of the in situ crust down to about 25\,km possibly reaching 60--80\,km depth \cite{gillet_scattering_2017}. (Figure \ref{fig:surfacekms}).
In the upper 1--2\,km of the lunar maria, multiple layers of thick lava with regolith on top have been identified \cite{Oshigami2009}.

The knowledge of the regolith and the \emph{mega regolith} is critical for creating models of wave propagation, simulating the thermal and mechanical ground response to LGWA sensors.  The exact measurements of the overall regolith thickness is challenging because no direct measurements have been done below few meters.  Also the lateral variation of thickness and layering strongly depends to the local geological evolution which shaped the shallow lunar crust.

The thickness and composition of the regolith is one important piece of the puzzle to plan and optimize the deployment scenario for LGWA.

\subsubsection{Ground tilt response}
\label{sec:tilt}
{\it Main contributors:} Marco Olivieri, Daniele Melini, Jan Harms, Giorgio Spada\\

Ground deformation at the LGWA deployment site is a determinant factor for the LGWA platform stability and for its long term operation.  This would require and accurate leveling in the deployment phase, but also later on during the science observations. Ground stabilization could be a long term process consequence of the rheology of the lunar regolith and of thermal and loading effects. In this perspective, the understanding of the regolith properties in PSRs (section \ref{sec:regolith} is crucial to determine the bearing capacity at the landing site \cite{sargeant2020using}. The latter reflects also in soil stability when planning rover's use for siting the instrumentation at distance from the landing point. 
The platform of each LGWA station needs to be leveled with an accuracy of a few tens of microradians to ensure the LIGS functionality at its best (see \ref{sec:payload}). 

In this section we first describe the possible drivers of the tilt angle defined as the variations of the angle between the normal to the surface and the local plumb line. Then, we focus on the tidally-induced tilt providing a theoretical description and estimates of its effects in lunar polar regions. 
As mentioned above, the platform leveling could require post deployment corrections that could be occasional or even continuous, depending on the time variations of the tilt and with the objective of minimizing the data quality degradation during the operation of LGWA. 
The key factors are:
\begin{itemize}
\item {\bf Topography}:  Only a certain amount of ground slope can be tolerated, since it determines how much actuation range the platform leveling system must have \cite{Lognonne2019}. Indeed, deploying the platform in a steep region could result in a significant fraction of the actuation range  being used for initial leveling, shrinking the possibility of further adjustments in response to \textit{e.g.} soil compaction or tidal effects. 
\item {\bf Thermal effects}: PSRs provide a thermally stable environment in which temperature is continuously below 100\,K and, in some PSRs, even continuously below 40\,K. However, even the small residual temperature changes inside a PSR might cause enough thermally induced ground tilt to perturb the GW measurement. These effects must be estimated carefully as part of an evaluation of the possible PSR deployment sites.
\item {\bf Tidal effects}: The Moon is continuously deformed in response to time-varying gravitational attraction due to the Earth and, to a lesser extent, the Sun. The corresponding cyclic ground tilt can be estimated in order to assess its impact on LGWA operation. If the amplitude of the tidally-induced tilt is greater than the leveling tolerance for LGWA operation, periodic adjustment cycles can be scheduled to keep the platform within its operational tilt limits.
\item {\bf Soil properties}: As mentioned above, soil at PSRs is expected to be stable. However, the platform itself will disrupt the local properties of the underlying regolith in reason of its weight, temperature and heat production. This will imply an accommodation time but it could also include fluctuations that reflect in a certain amount of induced seismic noise. Analyses of this effect are important to decide whether a mechanism is required to mount the stations to the ground.
\end{itemize}

Modeling planetary tides is a classical problem in geodynamics, since observations of tidal deformation has provided the first indirect probe of the Earth interior before the advent of global seismology \cite{thomson1863xxvii,love1909yielding}. Tidal deformation of the Moon has been widely studied in the early 1960s in view of possible inferences on the large-scale internal structure on the basis of the geophysical scientific observations planned within the Apollo program \cite{sutton1963tides,harrison1963lunartides} and because of the relevance of tidal deformation in estimating the lunar orbital evolution \cite{groves1960tidal}. Today, accurate models of the tidal potential are available on the basis of ground-based or space-borne selenodetic observations \cite{williams2015moon}, and they are essential for the analysis of spacecraft tracking data.  

The largest contribution to solid lunar tides are due to gravitational attraction of the Earth, with the contribution from the Sun being at least one order of magnitude smaller. Tides raised by the Earth are controlled by the eccentricity and inclination of the lunar orbit, which result in both the amplitude and position of the tidal bulge oscillating monthly by $\sim \pm 5^\circ$. Fourier analysis of the tidal potential \cite{williams2015moon} shows the largest components around the monthly period, with significant contributions also from the bimonthly, trimonthly, yearly and biyearly frequencies. Amplitudes of tidal harmonics vary with position on the Moon surface, with deformation being at the $\sim 0.1$\,m level, peak tilt (relative to the local plumb line) at the $\sim 5 \times 10^{-6}$\,rad level, and peak disturbance to the gravitational acceleration at the $\sim 0.5$\,mGal level. 

Viscoelastic relaxation processes in the Moon interior affect the tidal response by introducing a frequency dependence of the tidal processes and a lag between each tidal harmonic and the corresponding response, which is directly related to dissipation of tidal energy into the Moon mantle \cite[see, \textit{e.g.},][]{williams2015moon}. While the frequency dependent effects are generally small compared to the elastic response, precise observation of the amplitudes and phase lags at different tidal frequencies would provide an indirect probe into the rheology and the structure of the Moon interior \cite{organowski-dumberry-2020,briaud-etal-2023}. 

Deformation of the lunar ground due to temperature changes can introduce excess noise in horizontal displacement measurements. Even in permanently shadowed crater floors, thermo-elastic stresses from the crater rims are present and will lead to long period motion and tilting of the ground. The amplitude of these changes is fundamentally unknown. Mars InSight found a tilt of 10$^{-6}$\,rad and a horizontal acceleration of $5\cdot 10^{-9}$\,m/s$^2$ during a partial Phobos eclipse, which reduced the surface temperature in 1\,m distance by 2\,K over 10s.  \cite{staehler2020} Scaling this to the lunar PSR environment with 200\,K temperature changes in 100-meter distance will produce horizontal signals of $>50$\,pm/s$^2$ at long periods. However, this is expected to be a function of the subsurface rigidity.

\subsubsection{Seismic background noise}
\label{sec:seismicbackground}
{\it Main contributors:} Matteo Di Giovanni, Jan Harms, Hrvoje Tkalcic, Marco Olivieri, Lucia Zaccarelli\\

The SBN might limit the sensitivity of LGWA at frequencies above 0.1\,Hz \cite{Har2022a}, and this is a central part of the studies in preparation of the LGWA technical design and of the mission planning. The SBN is also one of the main targets of investigation with the LGWA pathfinder mission Soundcheck (more details in section \ref{sec:lunarmissions}). Most of our current knowledge of the lunar SBN comes from the Apollo missions as described in the following.

Using the data collected by the Apollo Lunar Surface Experiments Package (ALSEP) between 1969 and 1977 mainly from 4 long-lasting seismic stations, an experimental upper limit to SBN was set at about $10^{-10}\,\rm{m\,Hz^{-1/2}}$ between 0.1\,Hz and 1\,Hz, corresponding to the instrument noise of the Apollo seismometers\cite{Larose2005,CoHa2014c}. A modeled estimate of the SBN was set at about $2\times 10^{-14}\,\rm{m\,Hz^{-1/2}}$ at 1\,Hz \cite{LoEA2009}, which is a full 6 orders of magnitude weaker in amplitude than the terrestrial SBN (depending on frequency). According to ALSEP, there are four distinct categories of natural seismic sources that contribute to the overall SBN: deep moonquakes, shallow moonquakes, thermal moonquakes and meteoroid impacts \cite{nakamura82}.

With more than 7000 events recorded \cite{nakamura82, FROHLICH2009365}, deep moonquakes (DMQ) are the most abundant among all seismic events and occur at depths about halfway between the surface and the center of the Moon, with most of the energy below 1\,Hz and an equivalent body-wave magnitude typically less than 3. Their occurrence was found to be strictly correlated with the tides raised by the Earth and the Sun \cite{nakamura82}. Deep moonquakes were also found to be highly clustered, occurring repeatedly within 300 separate source regions, or nests, 100 of which have been precisely identified \cite{Nak2005, FROHLICH2009365}, each nest having dimensions of 2\,km or less\cite{nakamura78, Nak2005}. As a consequence of occurring repeatedly at nearly the same place, DMQ have almost identical waveforms for a given source region and at a given station \cite{nakamura82, Nak2005}. In practice, this means that the separation of the moonquakes sources is smaller than the typical wavelength at which the waveforms are observed. These nests follow a distribution that supports depths in the 700--1200\,km range and are unevenly distributed in the nearside with only a few of them in the farside \cite{nakamura82, FROHLICH2009365, Nak2005, Garcia2019}. The most active nest, identified as $A_1$, counts more than 400 recorded events in the years 1969--1977 \cite{nakamura78, Nunn2020}. 

The lack of nests on the far side is still a matter of debate whether it is due to an actual absence of seismic activity on the antipodal region, due to the seismic waves propagation mechanism in the interior of the Moon or due to the failure to identify them \cite{Nak2005}. Nevertheless, little progress has been made since then. With time, only 30 nests have been attributed to the far side, but none of them far enough into the far side to gather relevant information about the Moon's interior. This suggests that the area near the antipode could well be aseismic, or the structure of the deeper interior of the Moon is such that no seismic waves will propagate straight through it to get recorded on the near side \cite{Nak2005}.

To date, the exact origin of DMQ remains unclear, and the presence of fluids \cite{saal08} or partial melts \cite{FROHLICH2009365} in the Moon's interior has been hypothesized. Recently \cite{kawamura17}, it has also been argued that tidal stress not only triggers the deep moonquake activity but also acts as a dominant source of excitation, with calculated tidal stresses that are strongest in the 600 -- 1200\,km depth range, which overlaps with the range of estimated deep moonquake hypocentral depths \cite{cheng78, Garcia2019}.

The most frequent contribution to the overall SBN, along with deep moonquakes, are the signals generated by the impacts of meteoroids on the surface of the Moon \cite{nakamura82, Nunn2020,meteorflux}. Between 1969 and 1977, the ALSEP recorded about 1700 impacts. Contrary to moonquakes, which can be observed only on seismic traces, meteoroid impacts have also been detected indirectly using Earth-based observations \cite{suggs2008nasa}, allowing for more precise studies about their rate and spatial distribution to assess the risks associated with working on the surface on the Moon and to interplanetary journeys. If recorded by orbital cameras, the impact locations can serve as Ground Truth locations for calibrating seismic studies of moonquakes and lunar interiors. 

Using Apollo 14 short period seismic data, \cite{dunnebier74} released a mass-dependent meteoroid flux estimate between $1.1\times 10^{-4}\, \rm hr^{-1}\,km^{-2}$ and $1.1\times 10^{-7}\, \rm{1/(hr\,km^2)}$. \cite{dunnebier75} completed the picture using long-period seismic data issuing a rate between $1\times 10^{-7}\, \rm{1/(hr\,km^2)}$ and $1.1\times 10^{-9}\, \rm{1/(hr\,km^2)}$. These figures were also dependent on the estimation of the distance at which the events occurred as inferred from the seismograms and were accompanied by the huge uncertainties in the seismic efficiency of the impacts. In 2014, NASA's Meteoroid Environment Office (MEO), after seven years of Earth-based impact flash observations and the validation of 129 events, released the first estimate of the impact rate which was found to be $7.00\times 10^{-8}\, \rm{1/(hr\,km^2)}$ \cite{suggs}. Between 2018 and 2020 \cite{NELIOTA}, Near Earth Orbit Lunar Impacts and Optical Transients (NELIOTA) detected 79 events with a rate of $2.05\times 10^{-7}\, \rm{1/(hr\,km^2)}$ for sporadic events and $3.9\times 10^{-7}\, \rm{1/(hr\,km^2)}$ for events during meteor showers. The combined rate was found to be $2.3\times 10^{-7}\, \rm{1/(hr\,km^2)}$. Measured kinetic energies of impacts span between $10^6$\,J and $10^9$\,J, corresponding to a mass range from $1\times 10^{-4}$\,kg to 8\,kg \cite{suggs, NELIOTA}. Nevertheless, estimations of the mass and energy of an impactor are subject to a certain degree of uncertainty caused by the assumptions on its velocity and on the efficiency with which impact energy is converted into luminous energy \cite{suggs, NELIOTA}. Among the other significant results from these new impact observations, there is evidence of a difference between $10\% - 20\%$ in the impact rate measured for equatorial and polar regions. Presently, it is not clear whether this depends on the actual distribution of meteoroids in space or on the difficulty of detecting light flashes in the polar regions\cite{NELIOTA}.

Concerning the morphology of the seismic signals generated by impacts, seismograms recorded by ALSEP show two distinct parts, high-frequency, and low-frequency components \cite{dunnebier74}. This work suggested that the high-frequency part could be associated with body wave energy while the low-frequency part contains mostly surface wave energy.

The third category of seismic events recorded on the Moon are shallow moonquakes \cite{nakamura82, Nunn2020} which are rare (only 28 recorded in the ALSEP data) and larger than DMQs, with equivalent body wave magnitude between 3.6 and 5.8. Overall the estimated depth lies between 0\,km and 220\,km \cite{Nunn2020}. Shallow moonquake spectra include high-frequency content of up to 8\,Hz, while DMQs contain less seismic energy above 1\,Hz. Energy for the shallow moonquakes continues up to about 8\,Hz and then rolls off. No correlation between shallow moonquakes and the tides has been observed. Important for the modeling of the lunar GW response is that moonquake observations point to very low attenuation corresponding to Q-factors of a few 1000 (see section \ref{sec:LunarResp}).

The forth class of seismic events is thermally triggered. \cite{DuSu1974} showed that the majority of the many thousands of seismic events recorded on the short-period seismometers were small local moonquakes triggered by diurnal temperature changes. More recently, \cite{DiEA2017} found and categorized 50,000 events recorded by the Lunar Seismic Profiling Experiment at Apollo 17. The events occurred periodically, with a sharp double peak at sunrise and a broad single peak at sunset. The origin of these events might be stress release in the ground, by the lander or payload. It is essential to the functioning of LGWA to avoid the SBN of thermal events, which is why deployment in a PSR with strongly reduced thermal gradients and fluctuations is expected.

It is also critical for the planning of LGWA to continue past efforts on the numerical modeling and simulation of the SBN \cite{LoEA2009}. Simulations of the lunar seismic field in a 3-D mesh are under preparation. They consider the shallow structure, the regolith layer, and the topography of potential LGWA deployment sites. This will give information about the expected noise level at the deployment site but also about the noise correlation between different nodes of the array, similar to what was done for the Einstein Telescope \cite{AnHa2020}. New advances in seismic inter-source correlation are also promising tools for imaging the deep lunar interior and confirming the existence of its deep layers. Through the reciprocity principle, the moonquakes are turned into virtual receivers, while the locations of the seismometers become virtual events \cite{wang2022scanning,wangformation}. Seismic correlations are a key property for the design of the LGWA LIGS array in terms of its diameter and shape, and to predict by how much the impact of the SBN on the GW measurement can be mitigated by optimal filtering of the data  (see section \ref{sec:background}).

\subsubsection{Magnetic fluctuations on the lunar surface}
\label{modeling:magnetic}
{\it Main contributors:} Yoshifumi Futaana \\

The electromagnetic field on the lunar surface is one of the key environmental factors that can be important for LGWA \cite{HaEA2021a}. Magnetic actuators will be used to control the position of the LGWA proof mass (see section \ref{sec:payload}), which introduces a susceptibility to environmental magnetic fields as well. Technological solutions such as magnetic shielding are possible to reduce the magnetic susceptibility of the proof mass, but it is first of all important to assess how much magnetic noise is to be expected in LGWA measurements without mitigation.

Because of the lack of the intrinsic, dynamo magnetic dipole field and extremely low electric conductivity, the Moon is, to the first order, {\it transparent} in terms of the electromagnetic field and its fluctuations. It means that the electromagnetic fields and waves propagate through the body, and thus, the electromagnetic and plasma environment on the lunar surface is expected to be similar to the near-moon space. 

\begin{figure}[ht!]
\centering
\includegraphics[width=0.7\textwidth]{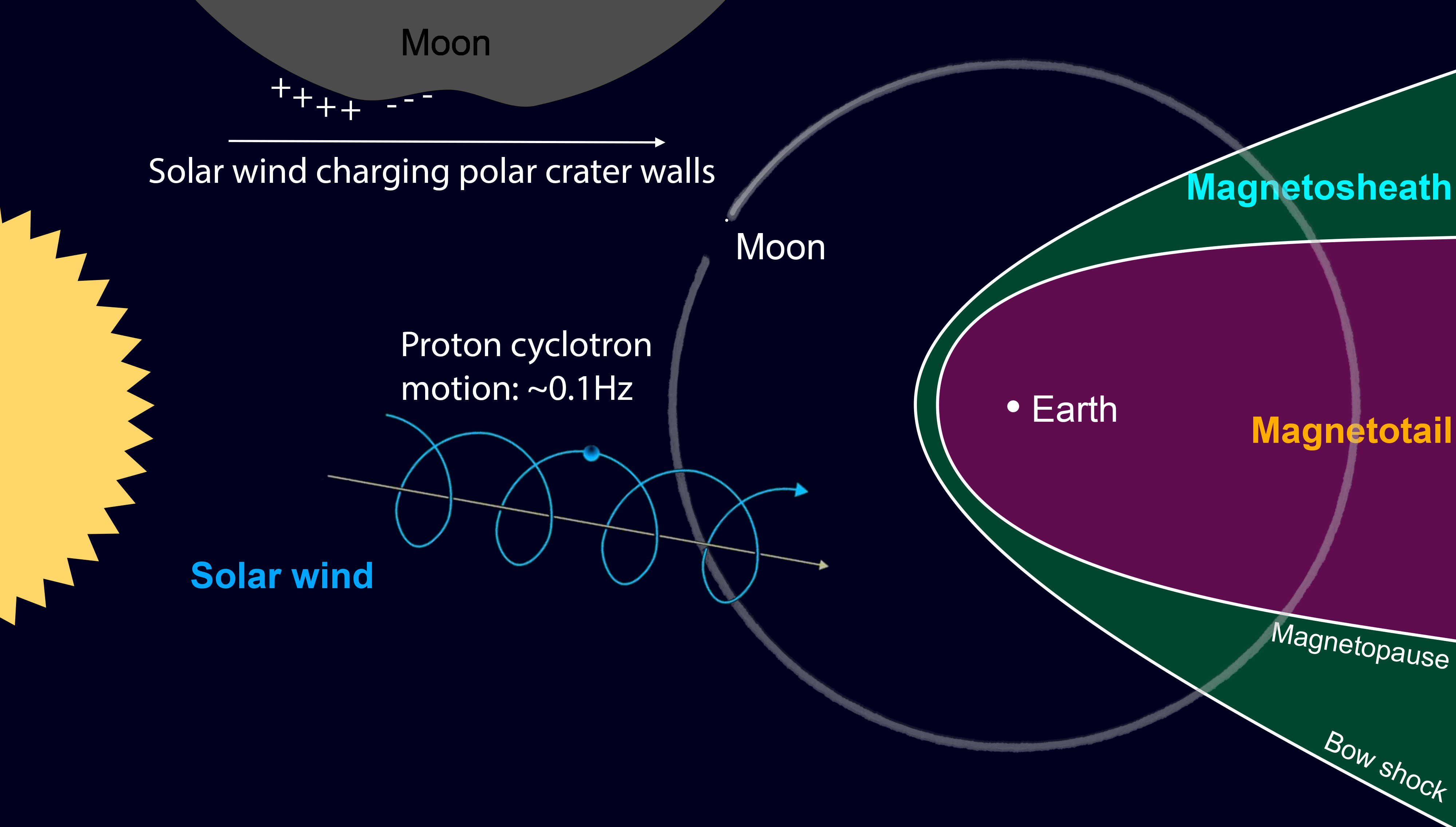}
\caption{Sketch of the plasma and magnetic environment around the lunar orbit.}
\label{fig:magnetic}
\end{figure}

The Moon experiences several domains (different plasma and electromagnetic environments) as it goes around the Earth \cite{dandouras-2023}. The principal domains are shown in figure \ref{fig:magnetic}.
The Moon is in the solar wind for $>60\%$ of the time. The solar wind is a supersonic plasma flow from the Sun, mainly composed of protons and alpha particles. While the electromagnetic environment is relatively stable in the solar wind, the intrinsically unstable nature of plasma excites various electromagnetic waves. Particularly because the proton cyclotron frequency at 1\,AU is approximately $\omega_g=\sqrt{eB/m} \sim 0.1$\,Hz (where $e$ and $m$ are the charge and mass of a proton and $B$ is the magnetic field strength), the deci-Hertz frequency waves (ultralow frequency (ULF) waves) are common near the Moon \cite{harada-2016}. Wave activity is manifested in the foreshock region \cite{greenstadt-1995}, where the reflected solar wind protons by the terrestrial bow shock co-exist with the solar wind plasma. These protons can reach the Moon, producing the ULF wave in the near-Moon environment \cite{salohub-2022}, typically when the Moon is located on the dawn side of the Earth (the lunar phase around 18 to 23 days).

Sometimes, severe solar storms impact the Earth and the Moon, when significant temporal disturbances are present in the Moon's environment.  The disturbance may last several days. In the magnetosheath (the downstream of the bow shock), a disturbed electromagnetic environment is also present.

A specific feature in terms of the lunar surface magnetic field is magnetic anomaly \cite{lin-1998a}. It is a remnant magnetic field of crustal origin distributed patchily but globally. Each anomaly has a spatial scale of 100 to 1000\,km, with a strength of 100s of nano-Tesla (about 1\% of the Earth's dipole field). Despite its small scales (spatially and strength), the crustal magnetic field can stand off the solar wind, preventing the solar wind protons from accessing the lunar surface below the magnetic field (\cite{harnett-2003a}; \cite{wieser-2010a}). This structure is called mini-magnetosphere, and is one of the main scientific topics for the lunar plasma environment (\cite{futaana-2018a}).

The interaction between the crustal magnetic field and the solar wind produces a wealth of electromagnetic disturbances. In particular, when the crustal field is located at the terminator region (applied to the polar regions). A manifestation of the magnetic field has been reported since the Apollo era (sometimes called limb compression) \cite{russell-1975a}. Recent numerical simulations reproduced the limb compression signature when the mini-magnetosphere reflects the solar wind protons back into space (\cite{fatemi-2014a}).

The reflected solar wind protons generate electromagnetic wave activities, relevant for deci-Hertz wave activities due to the resonance with proton cyclotron motion. For example, the reflected ions excite the whistler waves near the crustal magnetic field \cite{tsugawa-2011a}. A global proton reflection from the lunar surface also excites such wave activities \cite{nakagawa-2011a}. In addition, \cite{farrell-1996a}, \cite{nakagawa-2003a} reports the magnetic wave activity around 1\,Hz when the magnetic field is connected to the wake of the Moon.

Almost all the observations are based on orbiters. While the Apollo surface payloads had magnetometers on the lunar surface, there are only a few publications on the fluctuation of the magnetic field on the lunar surface. Recently, an effort to recover the record of the Apollo magnetometer data has been ongoing, and \cite{chi-2013a} reported ion cyclotron waves (with frequency 0.04 -- 0.17\,Hz) detected by an Apollo surface magnetometer when the Moon is inside the terrestrial magnetotail.

At the PSRs near the lunar poles, the local solar wind flows nearly tangential to the surface and interacts with large‐scale topographic features such as mountains and deep large craters. On the leeward side of large obstructions, plasma voids are formed in the solar wind because of the absorption of plasma on the upstream surface of these obstacles. A surface potential is established on these leeward surfaces in order to balance the currents from the electron and ion populations. There are leeward regions where solar wind ions cannot access the surface, leaving an electron‐rich plasma previously identified as an “electron cloud.” A balancing current is required at the surface, and lofted negatively charged dust was proposed as a compensating current source \cite{farrell2010charge}. This charging mechanism was discussed in the context of safety of lunar operation at PSRs and might create a unique electromagnetic environment at the LGWA deployment site.

In summary, very few experiments to characterize the electromagnetic environment on the lunar surface have been conducted. Our knowledge of the electromagnetic environment is based on the measurements by orbiters (typically with altitudes of 10s to 1000s of km) or in the equatorial regions by Apollo magnetometers, e.g., \cite{dyal-1970a}.  The measurements near the south pole are highly required to characterize the electromagnetic fields and their fluctuations near the permanently shadowed region.

\subsection{Soundcheck and Other Relevant Lunar Missions}
\label{sec:lunarmissions}
{\it Main contributors:} Jan Harms, Matteo Di Giovanni, Alessandro Frigeri, Joris van Heijningen \\

\subsubsection*{Soundcheck}
\begin{figure}[ht!]
\centering
\includegraphics[width=0.6\textwidth]{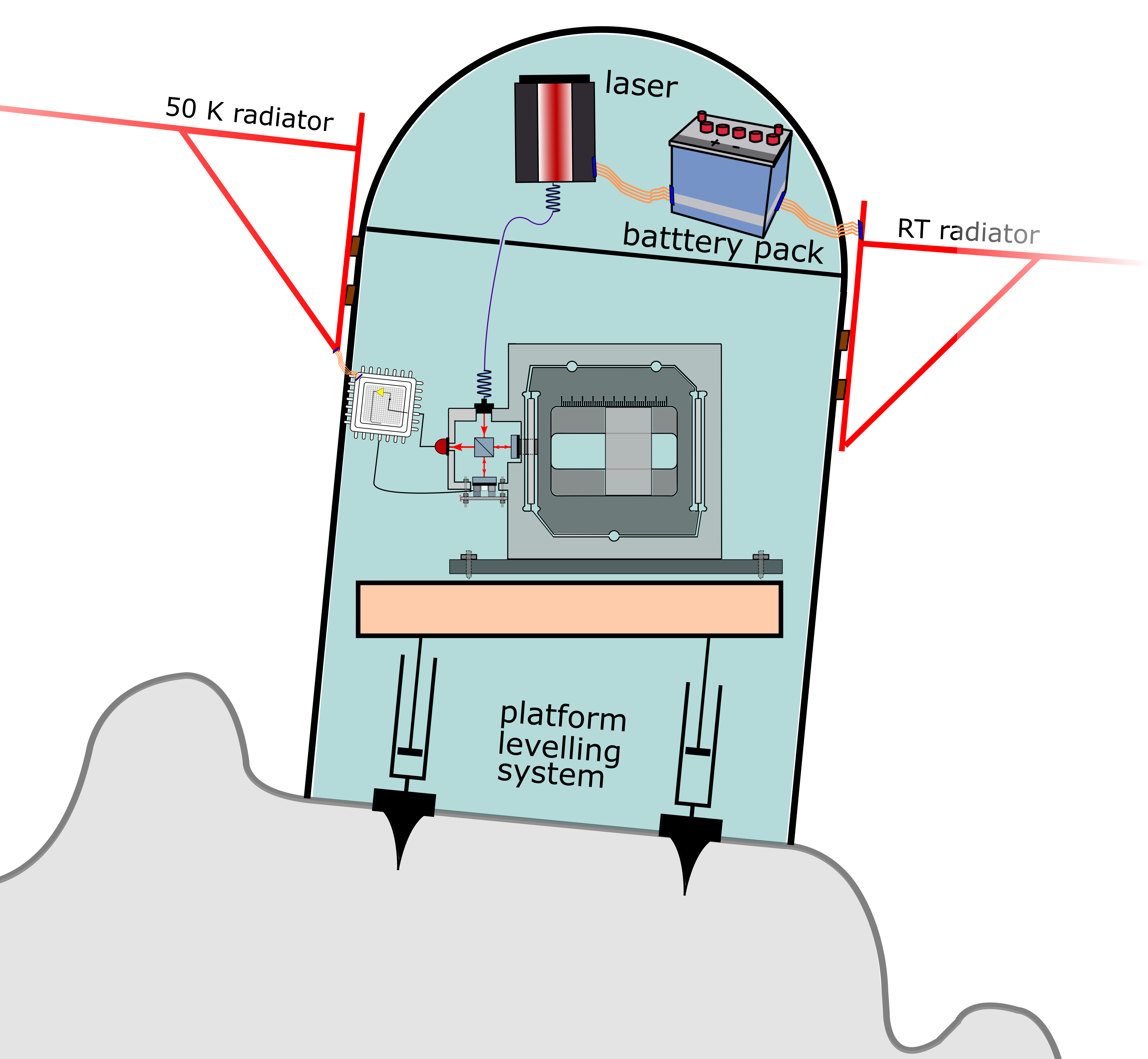}
\caption{Conceptual overview of a Soundcheck seismic station on a tilted surface on the lunar regolith. The sketch can be compared with the LGWA payload concept shown in figure \ref{fig:lgwapay}.}
\label{fig:sndchkpay}
\end{figure}

As shown in the previous sections, the planning of the LGWA mission requires detailed understanding of the lunar geophysical environment inside a PSR, which is not available today. A pathfinder mission is required, which was proposed to ESA in 2022 under the name \emph{Soundcheck}. In addition to carrying out measurements of seismic surface displacement, magnetic fluctuations and temperature, the Soundcheck mission will also be a technology demonstration. An overview of the Soundcheck payload is shown in figure \ref{fig:sndchkpay}.
Soundcheck is expected to reach sub-pm/Hz$^{1/2}$ inertial sensitivity above 0.1\,Hz and probe the SBN with unprecedented precision.

The Soundcheck technology validation focuses on deployment, inertial sensor mechanics and readout, thermal management and platform leveling. The PSRs are as of now unexplored and a deployment strategy in the dark has to be developed. A Watt's linkage inertial sensor for a space application, including a release mechanism, is under development. All parts for (interferometric) position readout and electronics are heritage to be combined into the specific Soundcheck readout. Ambient cryogenic inertial sensing with room temperature laser and battery pack requires an R\&D effort as well. Lastly, platform leveling has heritage in space application\,\cite{FlEA2018}, but not yet in a cryogenic environment.

For Soundcheck a relatively simple approach without force feedback on the proof mass is employed. As the SBN is unknown it is challenging to design an optimal feedback loop and we therefore use multi-fringe readout with sub-pm/$\surd$Hz sensitivity. There are several options, such as homodyne quadrature interferometry\,\citeg{Cooper2018} and deep frequency modulation interferometry\,\citeg{gerberding2015}. While both are theoretically similar in performance\,\cite{EcGe2022}, the homodyne quadrature interferometric readout is less computationally demanding and therefore preferred for Soundcheck given its more stringent power budget.

\begin{figure}[ht!]
\centering
	\includegraphics[width=0.77\textwidth]{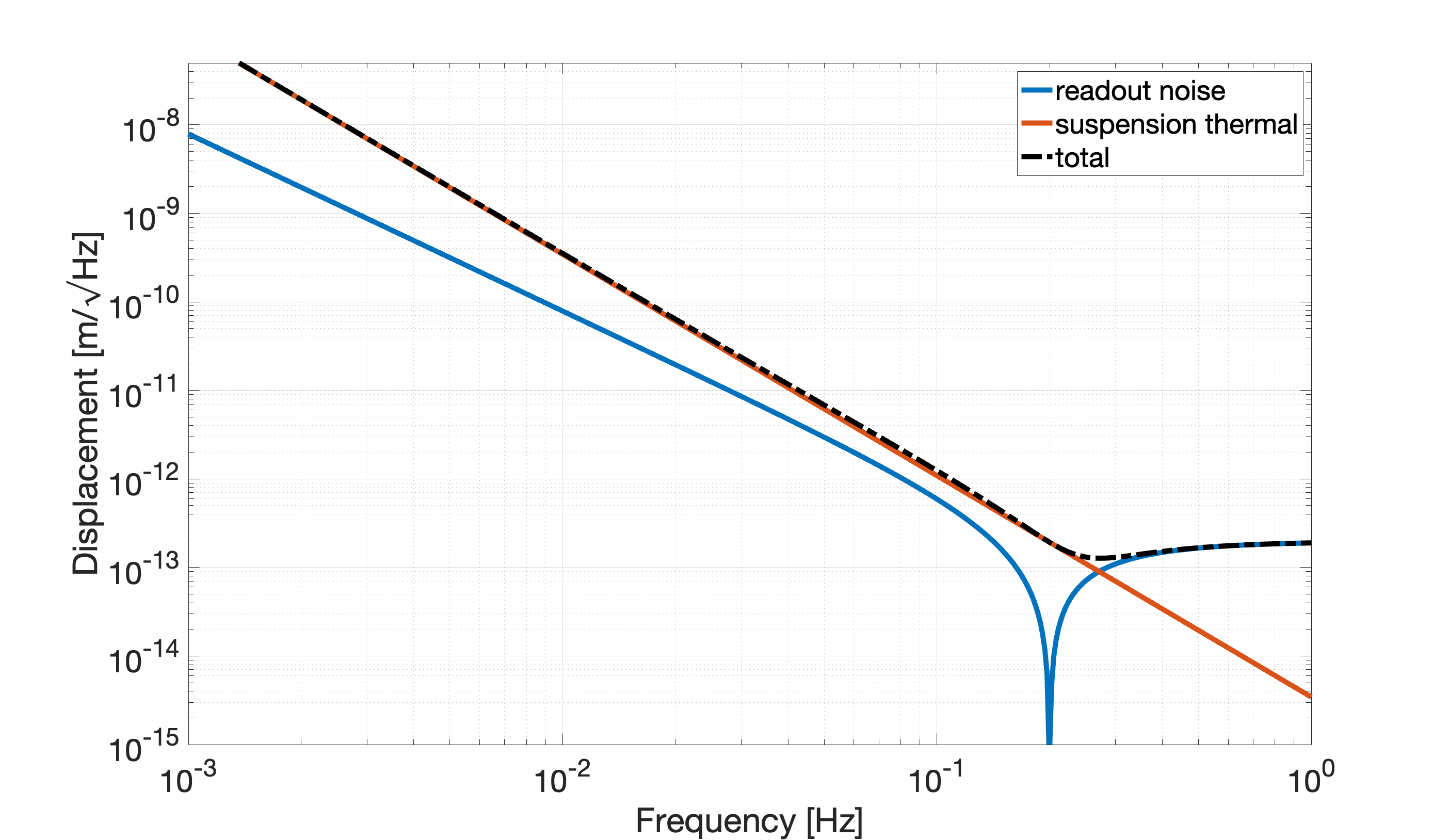}
\caption{Minimum detectable inertial displacement for the Soundcheck using a 3\,kg niobium Watt's linkage with an interferometric readout. More details in the text.}
\label{fig:sndchk}
\end{figure}

\subsubsection*{Other Lunar Missions}
Almost all of our understanding of the lunar seismic field and its sources comes from the Apollo Lunar Surface Experiments Package (ALSEP) starting with Apollo 11's PSEP \cite{Latham1969}. In 2022, seismic data from the ALSEP have been re-organized and archived on the Incorporated Research Institutions for Seismology (IRIS) servers for easy access through the FDSN protocol \cite{nunn2022} at the following DOI:$\rm{10.7914/SN/XA\_1969}$. The catalog of the seismic events recorded by the ALSEP, updated with new detections occurred during recent reviews of the data, can be found in the electronic supplement of \cite{Nunn2020}.

Concerning Earth-based observations, there are currently two active collaborations, namely MEO \cite{suggs} at NASA and NELIOTA \cite{NELIOTA} at ESA, that are monitoring meteoroid impact rates on the Moon. The main goal of these research groups is to understand the spatial distribution of meteoroids near the Earth-Moon system. This can lead not only to a better understanding about the properties of the solar system, but also to predict the small-meteoroid flux that deteriorates space equipment, predict when the next large meteoroid will impact Earth itself and assess the risks for interplanetary journeys. The information about impact rates can also provide useful data about SBN. In March 2023, NELIOTA has also made publicly available an open source software for observatories to detect meteoroid impact flashes on the Moon.

In 2022, ESA approved the Lunar Meteoroid Impact Observer (LUMIO) mission \cite{lumio}. The goal of LUMIO is to monitor meteoroid impacts by use of the same technique of Earth-based observatories, but on a 12U CubeSat structure on a halo orbit at the Earth–Moon L2 point. This is expected to guarantee a higher detection efficiency and reduce the false-alarm rate let alone providing data for the farside, thus complementing Earth-based observations. LUMIO is a consortium of European research institutes led by the Politecnico di Milano (Italy) and, as of mid 2023, is building its working groups.

Of great interest for LGWA is also the Farside Seismic Suite (FSS) \cite{fss}, scheduled to land in 2025 on the far side, near the south pole in the Schrödinger basin. The FSS will deploy a vertical Very Broad Band (VBB) seismometer already developed for the Mars InSight mission and a short-period seismometer \cite{Lognonne2019}. Being the first seismic station on the far side of the Moon, FSS will improve our understanding of the lunar interior, the processes that shaped the lunar crust and the current micrometeoroid impact rate. The sensitivity of the FSS VBB sensor will exceed the sensitivity of the Apollo seismometers by an order of magnitude up to 1\,Hz and therefore provide important new information about the lunar SBN \cite{Branchesi2023}.

The concept of a Lunar Geophysical Network (LGN) was also proposed \cite{LGN}. The LGN is aimed at advancing our knowledge about lunar and planetary science with the deployment of geophysical stations at four locations to enable long-term geophysical measurement with a complete instrument suite providing seismic, geodetic, heat flow, and electromagnetic observations. The details will continue to be optimized throughout the formulation of this mission that has been proposed for launch in 2030.

Important for LGWA are also missions to explore the PSRs such as NASA's Volatiles Investigating Polar Exploration Rover (VIPER). It is a robotic mission designed to explore the South Pole of the Moon in search of ice and other potential resources. The goal of the VIPER mission is to characterize the distribution and physical state of lunar polar water and other volatiles in lunar cold traps and regolith to understand their origin. VIPER is equipped with a number of instruments to help it achieve its goals, including a drill that can penetrate up to 1\,m into the lunar regolith, a mass spectrometer that can identify and measure the abundance of various elements and compounds in the regolith, a radar system that can detect the presence of ice beneath the surface, and cameras that can provide high-resolution images of the lunar terrain. VIPER is scheduled to launch in late 2024 and land at the South Pole of the Moon in early 2025. It will then spend 100 days exploring the lunar surface, collecting data and returning it to Earth. The data collected by VIPER will be used to inform NASA's plans for future human missions to the Moon, including the establishment of a permanent lunar base.

\section{Science Objectives}
\label{sec:science}

\subsection{Lunar science}
\label{sec:lunarscience}
The Moon being the test mass of the LGWA experiment, its elastic properties and the presence of a continuous lunar seismic background are fundamental to the LGWA performance. As explained in section \ref{sec:environment}, these properties of the Moon must be understood very accurately to extract the information from GW signals. As a consequence, LGWA enables studies of the nature and distribution of seismic sources (section \ref{sec:seismicsources}), the Moon's internal structure (section \ref{sec:lunarinternal}), formation history (section \ref{sec:moonform}), and geological processes (section \ref{sec:geologicprocesses}).

\subsubsection{Seismic sources}
\label{sec:seismicsources}
{\it Main contributors:} Taichi Kawamura, Hrvoje Tkal\v{c}i\'{c}, Jan Harms, Sabrina Menina\\

The LGWA will be a unique contribution to a lunar seismic network for the study of moonquakes (see section \ref{sec:lunarmissions} for other proposed lunar seismic stations). Together with its pathfinder mission Soundcheck, LGWA is still the only seismic experiment proposed for deployment inside a PSR, which is expected to enhance the quality of horizontal displacement data; see section \ref{sec:tilt}. Its eight horizontal channels of the four-station array will enable advanced background reduction methods also for moonquake observations (see section \ref{sec:background}). The superior sensitivity of its accelerometer --- sub-picometer precision above 0.1\,Hz (see figure \ref{fig:LGWAsens}) --- will make it possible to  observe weakest lunar seismic events. In the following, we outline our current understanding of lunar seismic sources and a possible role of LGWA for lunar seismology together with other future lunar seismic experiments.

Apollo seismic observation revealed that the Moon is still seismically active today. After about seven years of seismic observation on the Moon, more than 13000 lunar seismic events (known as moonquakes) were identified and cataloged \cite{NaEA1981}. The catalog has been constantly updated with  state-of-the-art methods applied to the data \cite{NAKAMURA2003197,Nunn2020}, and even today, new events are discovered from the data \cite[e.g][]{knapmeyer2015identification}. The Moon has different types of seismic sources, including the endogenic ones. This opened a new question of where and how these seismic events are excited. This is one of the key open questions in lunar seismology and was listed as one of the science goals by Artemis Science Definition Team \cite{ArtemisScienceDefinitionReport}.

When mapping the global scale events distribution for the Moon, the seismicity on the lunar farside is less constrained. Both, the Apollo seismic network and most of the detected moonquake sources, were located on the lunar nearside \cite{nakamura82, Nak2005}. A minimal number of earthquakes on the lunar farside were identified, and it remains uncertain whether this is a result of the Apollo seismic network's limited coverage or if the farside is genuinely aseismic. Given the limited sensitivity of the Apollo seismometers, it is very likely that the network did not have, for the farside, the same detection capability as for the nearside, especially for small magnitude events that dominate the catalogue \cite[e.g.][]{KawamuraCratering2011}. On the other hand, global observation carried out after Apollo revealed that there is a clear dichotomy on the Moon, and there are significant differences in terms of internal structure between the lunar nearside and the farside \cite[e.g.][]{JollifPKT2000, WieczorekGRAIL2013}. These may result in variations of the seismicity rate and spatial distribution, and to elucidate this open question, new observation at different location from Apollo will be essential. The dichotomy on the Moon is one of the key question to understand the evolution of the Moon. Seismicity will inform us about the depth of the dichotomy, which serves as an important constraint on the crust/mantle formation. 

\subsubsection*{Moonquakes}

\begin{figure}[ht!]
    \centering
      \includegraphics[width=1\textwidth]{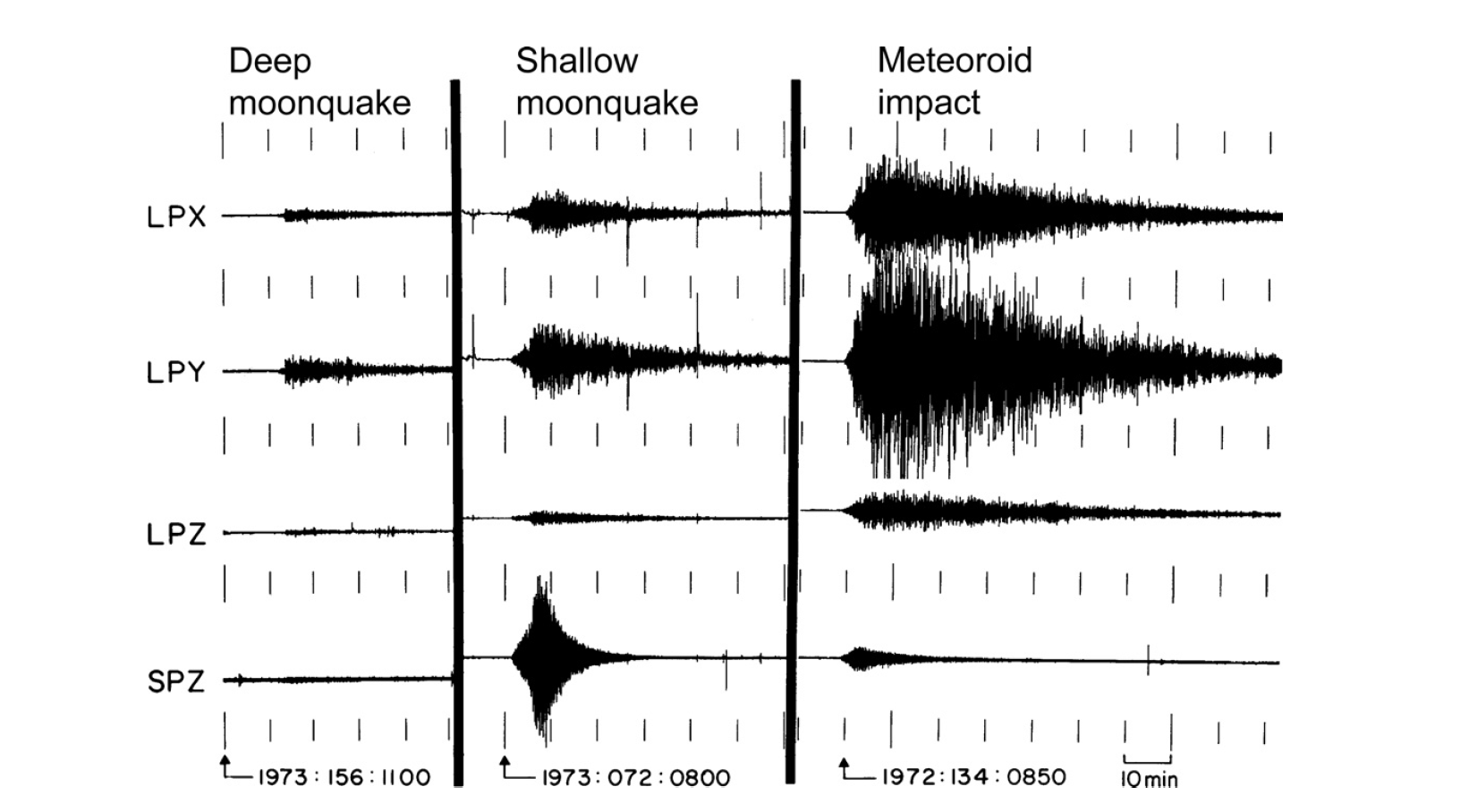}
      \caption{From the left to the right: Deep moonquake, Shallow moonquake and Meteoroid impact lunar seismograms in compressed time scale at the Apollo 16 station. LPX, LPY, and LPZ are the three orthogonal components of a long-period instrument (two horizontal and one vertical), while SPZ is the short-period vertical component (from \cite{nakamura82}).}
      \label{fig:Moon_event}
\end{figure}

The different types of moonquakes are generally classified with their source depth and mechanism (see Figure \ref{fig:Moon_event}). The first type is deep moonquakes, which are most frequently observed seismic event on the Moon, and more than 7000 events were reported \cite{NaEA1981, NAKAMURA2003197, Nunn2020}. Deep moonquake sources are typically located at about 900\,km depth (about midway between the lunar center and the surface) and are known to occur at fixed source regions called nests \cite{nakamura82}. More than 100 nests have been identified and the notable feature is that their waveforms are almost identical for the same set of nest and a station. Thus, deep moonquakes are classified using the cross/auto correlation technique, which also helps us improve signal to noise ratio by stacking. Another interesting feature of deep moonquakes is that their occurrences are correlated with the Earth-Moon tides \cite{nakamura82, FROHLICH2009365, NAKAMURA2003197} resulting from the observation of a monthly periodicity of the occurrence. These events are relatively small events with magnitude of 1--2. There is a longstanding debate whether the tidal stresses are fully responsible of the occurrence of deep moonquakes or they are merely triggered as releases of the accumulated tectonic stresses \cite{cheng78,WeberDMQ2009,kawamura17,TurnerDMQ2022}. The challenge in the discussion is that we have little constraints on the source parameters. Unlike on Earth, obtaining parameters as the source mechanism is challenging on the Moon. This is due to the intense scattering that masks the direct waves from the hypocenter that are necessary for gathering information of the seismic source itself \cite{gillet_scattering_2017}. Generally speaking, going to longer period will result in less scattering but this is not possible for the ALSEP dataset given the limited band width of Apollo seismometers. In addition to this, the small magnitude of deep moonquakes also makes the analyses difficult. 

The second type of moonquakes is shallow moonquakes, which were rarely observed, with only about 30 events identified during the Apollo observation time. Though they are small in number, they are the most energetic events detected and thus have an important impact on the seismic moment release of the Moon as well as for seismic risk assessment for future lunar explorations. The seismic sources of shallow moonquakes are estimated to be at a depth of around 60\,km and shallower than 200\,km \cite{nakamura82, gillet_scattering_2017}. Both studies approached the depth determination by modeling the amplitude variation with distance and coherently concluded that the hypocenter needs to be placed at depth while it is unlikely to be at the Moon's surface. Hence the name of shallow moonquake was given to this type. The source mechanism of shallow moonquakes is not very well understood. Unlike deep moonquakes, no clear correlation with tides nor periodicity were found and the occurrence seems to be almost random \cite{Nakamura1997}. Watters et al. \cite{Watters2019} points out that there is some tendency for shallow moonquakes to occur when the Moon is far from the Earth but this could not explain all the shallow moonquake occurrences. Watters et al. \cite{Watters2019} also points out that some shallow moonquake sources are close to newly found lobate scarps and thus attributing the thrust fault to be responsible for the excitation of shallow moonquakes. However, this does not apply to the whole shallow moonquakes catalogue. This suggests multiple origins for shallow moonquakes with some shallow moonquakes excited by shallow thrust faults visible on the surface, and others, deeper, originating from blind faults not emerging at surface. In both cases, as for deep moonquakes, the main open issue is the source mechanism  of shallow events which remains  poorly constrained. Better understanding the source and nature of shallow moonquakes will shed new light on the dynamics of the Moon and its tectonics, and it will also be informative to evaluate the seismic risk of the Moon.

The third type is meteorite impacts. The Moon is a airless body and is constantly bombarded by meteorites impact with wide range of size from micrometeorites, microscopic dust particles, to kg sized impactors falling on the surface which also generates impact flashes observable even from the Earth. Monitoring such impact events will give us information on current impact rate of the Earth-Moon system. Similar observation would also be possible on Earth but small impactors  are likely missed while filtered by the atmosphere. This broad range of observations complements the study the impact rate for the Earth-Moon system with a further dataset. Impact rate assessment is also interesting to better understand the environmental risk for lunar based activities since even small impactors can create significant risks for future human activities on the Moon. At the same time, it is also important to understand how the impact rate evolved with time and its spatial distribution. Le Feuvre and Wiczorek \cite{LEFEUVRE20111} showed that the impact rate is not uniform on the lunar surface and depends on the latitude and longitude due to relative velocity between the impactor and the revolving Moon. Kawamura et al. \cite{KawamuraCratering2011}  confirmed this hypothesis from seismically detected impacts and pointed out that the larger heterogeneity for the smaller impacts implies that their sources are comets, whereas the sources of large impactors originates from asteroids. Information on the source will be informative to better predict the impact environment and the risk on the Moon both in terms of temporal and spatial variations. Moreover, as pointed out in section \ref{sec:seismicbackground}, impacts are expected to be the prevalent source of background seismic noise for the Moon. So, rate and spatial distribution are key aspects for the prediction and assessment of the expected background noise and its variability over time.

These three main types of moonquakes are all related to key questions in lunar and planetary science. However, the Apollo seismic data has not been sufficient to answer these questions and new observations made in new locations and with state-of-the-art, high performance instruments will be mandatory.

\subsubsection*{Detection of exotic particles}
Among the large number of exogenic events, mostly produced by meteoroid impacts, a few detectable events might be associated with fundamental particles. The interaction of particles with the Moon can be similar to impacts, e.g., in the case of ultra-high-energy cosmic rays or the hypothesised Strange Quark Nuggets (SQNs) \cite{BANERDT2007203,PhysRevD.73.043511}. The ability of some particles to penetrate the ground makes it possible to distinguish them from meteoroid impacts. Other forms of matter, like the hypothesised Ultralight Boson Dark Matter particles (UDMPs), would have a lasting interaction with the Moon. UDMPs are predicted to lie in a mass range between 10$^{-22}$\,eV and 0.1\,eV. They would generate a monochromatic force acting on the Moon leading to vibrations that could be observed by an inertial sensor \cite{Carney_2021,PhysRevD.93.075029}. Finally, the seismic signal of a small primordial black hole with mass of $10^{15}$\,g traversing the Moon was studied \cite{Luo_2012}, and primordial black holes of mass $>10^{19}$\,g would form impact craters with a distinct morphology \cite{Yalinewich_2021}. While all these scenarios are intriguing, it should be emphasized that interactions with the Moon would be very rare and even the existence of any of these particles is speculative.

\subsubsection{Lunar internal structure}
\label{sec:lunarinternal}
{\it Main contributors:} Angela Stallone, Lucia Zaccarelli, Hrvoje Tkal\v{c}i\'{c}\\

Within the Apollo missions,  ALSEP (Apollo Lunar Surface Experiments Package, \cite{BLK1979}) data provided the first direct information on the Moon's internal structure. In particular, we have gained insights on the approximate thickness of its crust, mantle and core, as well as their elastic properties.

Similar to Earth's PREM (Preliminary Earth Reference Model, \cite{DzAn1981}), the one-dimensional model characterizing Earth's average properties in terms of seismic velocities, density, and attenuation as a function of depth, seismic data analysis resulted in a lunar equivalent model known as VPREMOON (Very Preliminary Reference Moon model) \cite{garcia2011very}. This model has been recently updated through a thorough review of seismic data and events \cite{Garcia2019, Nunn2020}. 

In VPREMOON, the mantle and crust velocity structure are defined to match geodetic (lunar mass, polar moment of inertia, Love numbers) and seismological data (direct and secondary P- and S-wave arrival times). The crustal thickness was also determined to align with the results from receiver function analyses, which provide a more direct indication of the mantle-crust boundary below the station \cite{LOGNONNE200327, Vinnik_RF_2001}. VPREMOON integrates knowledge from prior research and physical constraints into a comprehensive framework, serving as the primary reference model for the Moon's interior. The crustal model - P-, S-velocity and density profiles up to a depth of $~30\,$km - is extracted from the seismic model determined by \cite{gagnepain2006seismic}. The lunar 'Moho' discontinuity (the crust–mantle interface) is set at a depth of $40\,$km \cite{chenet2006lateral} to fit geodetic observations. This, in turn, imposes a constraint on the density contrast between crust and mantle, resulting in about $\sim 0.55$. In maintaining a homogeneous lunar mantle mineralogy to match the absence of significant P- and S-velocity variations and using a model of the internal temperature field, the mantle density profile can be constructed  \cite{khan2007joint}.

The average reference model of the Moon describes physical properties, including core radius and density. A core radius of $380 \pm 40\,$km and an average core mass density of $5200\pm 1000 \, \text{kg/m}^3$ were found. This considerable uncertainty arises from limited constraints on the S-wave velocity profile at the base of the mantle and inaccuracies in deep moonquake location data.

\begin{figure}[htp!]
    \centering
    \includegraphics[width=0.9\linewidth]{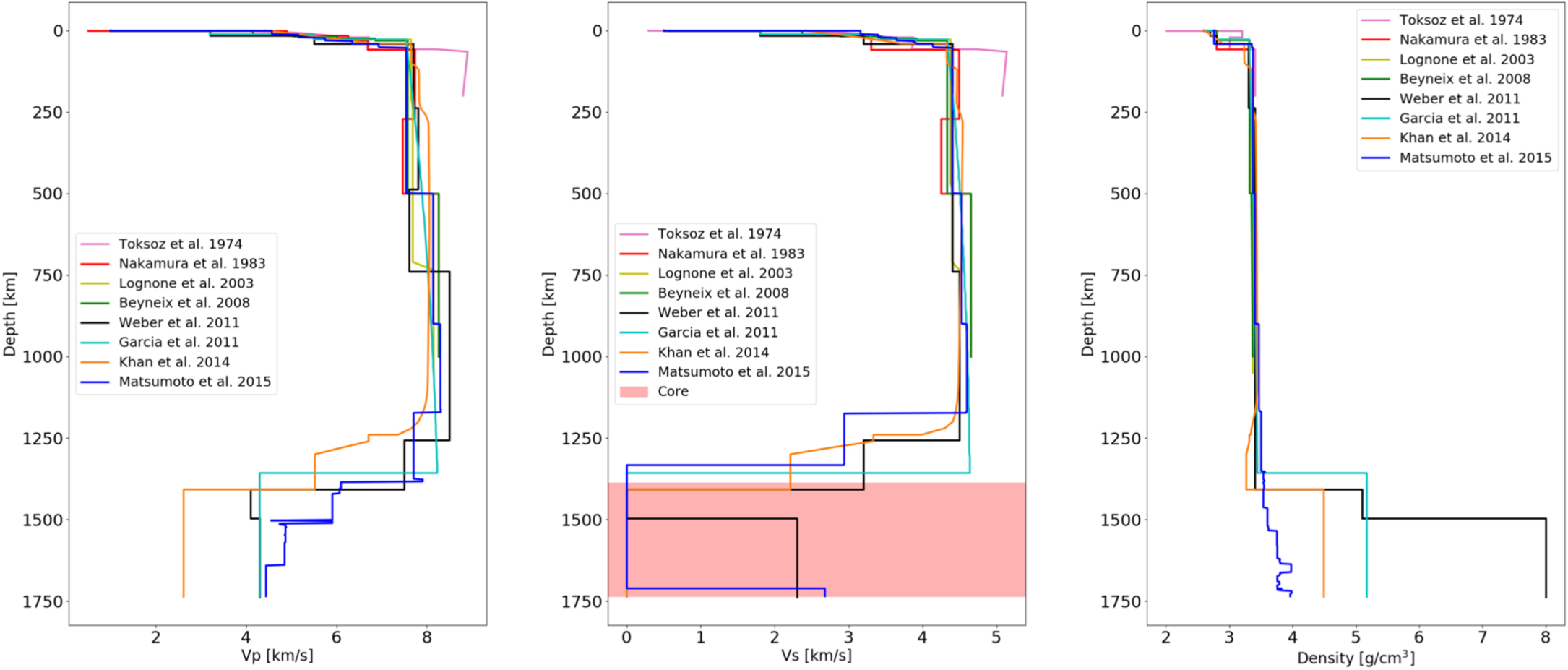}
    \caption{P-wave velocity (left), S-wave velocity (center), and density  (right) profiles from the Moon's surface to the core (identified in red in the middle panel) for several seismic velocity models. Data used for generating these plots are from \citep{Garcia2019}. Image taken from \cite{HaEA2021a}.}
    \label{fig:elastic_profiles}
\end{figure}

Several velocity models for the Moon, derived from the inversion of P- and S-wave arrival times, are available in the literature. They are depicted in Fig. \ref{fig:elastic_profiles}, which also features the VPREMOON model by \cite{garcia2011very}, represented in cyan. These models are overall consistent up to a depth of $\sim 1200\,$km. At larger depths, uncertainties become too high and additional data (e.g. geodetic and electromagnetic sounding data), as well as a priori information, are included to constrain the velocity profiles. Variations in the inverted profiles primarily result from differences in the inversion technique and model parameterization. Beyond depths of $1200\,$km, differences primarily stem from the type of complementary information used. Over the course of different Apollo missions, more than 13,000 seismic events were recorded, although only a subset of them --- those with the best signal to noise ratio and clear P and S identification --- were used to infer the lunar internal structure.

In addition to seismic waves, also normal modes of the Moon can provide information about its internal characteristics. These free oscillations refer to a standing wave that form when a dynamic system, as for the case of planetary bodies, is excited. These oscillations manifest as long-period signals capturing a body response to an excitation. Each normal mode is characterized by a distinct vibrational pattern and a unique frequency. Then, the specific set of normal modes for any planetary body is determined by its size, composition, density distribution, and material properties. Consequently, studying normal modes can provide valuable information on its one-dimensional layered structure along with its physical properties. While normal modes are generally excited by earthquakes on Earth, on the Moon, their primary source are moonquakes, along with contributions from thermal stress and tidal forces. Since normal modes are insensitive to the specific excitation source, they serve as an ideal tool for investigating the Moon's interior. This is particularly advantageous because the precise timing and locations of all meteoroid impacts are not accurately known. \citep{khan2001new} inverted lunar normal modes recorded by the Apollo instruments to gain new insights into the Moon's interior. However, Gagnepain-Beyneix et al \cite{gagnepain2006seismic} raised doubts, demonstrating that the normal modes utilized by \cite{khan2001new} fell below the instrument noise. The high signal-to-noise ratio of the GW detectors and the advanced noise-cancellation techniques pursued for LGWA will enable noise-free normal-mode observations, providing a new valuable source of information for the Moon's interior. Similar advancements have already been demonstrated for Mars (see \cite{bissig2018detectability}).

Similar to Earth, the main source of data to probe the Moon's interior is seismic data. However, the Apollo seismic dataset had limitations due to a narrow observation band, and stations being located only on the Moon's near side. These limitations limit our capability to gain a comprehensive understanding of the Moon's internal structure.

There is currently a renewed interest in lunar exploration within the scientific community, evident in the extensive list of planned lunar missions. Several of these missions include the deployment of seismographs on the lunar surface as part of their objectives, with LGWA and its pathfinder mission Soundcheck being two of them. This development promises to deliver new seismic data of exceptional quality and broader frequency coverage in the coming years. As a result, we are on the cusp of a new era in the exploration of the Moon's interior, with the potential for significant discoveries and insights into its inner structure.

One could argue that the number of seismographs will still be too small, thus limiting the sampling of the Moon's layers and making inferences more challenging than for the Earth. A recent paper by \cite{wangformation}, however, demonstrates the feasibility of utilizing even a single seismograph for probing planetary interiors through global-scale moonquake coda cross-correlations between seismic events. This methodology has already proven successful in the study of Mars interior \cite{wang2022scanning}, and we propose to apply global-scale waveform cross-correlations to lunar seismic data as well, taking advantage of the forthcoming high-quality seismic data. Particularly, the lunar events' codas are expected to be significant due to the Moon's smaller size and relatively small attenuation compared to Earth.  Moreover, the two horizontal components of ground displacements will allow the application of other techniques aimed at the Moon interior's exploration like receiver functions or seismic noise interferometry \cite{GALETTI20121}. In particular, the latter has been already applied to the higher frequency ALSEP data \cite{Larose2005}, making it possible to describe the very shallow layers, but lowering the frequencies will provide an image of the Moon's deeper structure.

\subsubsection{Moon's formation history}
\label{sec:moonform}
{\it Main contributors:} Alessandra Mastrobuono Battisti, Matteo Di Giovanni\\

The Moon's internal structure and its formation history are closely interconnected. The presence of a distinct core suggests that the Moon was formed in a hot, violent event, such as a giant impact. The composition of the mantle and crust aligns with the concept that the Moon was once a molten body that cooled and differentiated. Insights into the mechanism that formed the Moon can be inferred from considerations such as the amount of differentiation, iron content, and the relative size of the core compared to the other layers. The analysis of the characteristics of seismic waves can provide information on the energy released during the formation process and refine the timeline of events that shaped the Moon.

Our understanding of the Moon's internal structure has been greatly enhanced by data collected from the Apollo missions, which returned samples of rocks and soil from the lunar surface. These data, along with seismic observations and studies of the Moon's gravitational field, have provided insights on the Moon's composition and the arrangement of its different layers.

In this context, LGWA might be the first lunar mission to observe distinct lunar normal modes excited by moonquakes, which would provide important data on the deep lunar internal structure. At the same time, site effects on seismic signals provide information about the Moon's shallow structure. The study of the lunar interior with LGWA is outlined in section \ref{sec:lunarinternal}. 

This is important because the Earth-Moon system is unique in several ways, including the Moon's size relative to Earth, its small core, and the high angular momentum of the system. Different theories for the Moon's origin, including capture, fission, co-accretion, and a giant impact have been explored so far \citep{1975Icar...24..504H, Cameron_Ward1976, Wood1986_mo}.

The giant impact theory, first proposed in the mid-1970s \citep{1975Icar...24..504H, Cameron_Ward1976}, is currently the prominent one, due to its ability to explain several key features, as shown by \cite{2001AGUFM.U51A..02C}, \cite{Canup2004a} and \cite{Canup2004b}.  In this scenario, the Moon forms as the result of a relatively low-velocity, oblique impact between the proto-Earth and a planetary embryo of mass similar to Mars, called Theia, occurring during the latest stages of the formation of the Earth, approximately 4.5 billion years ago \citep{Agnor1999, Jacobson2014}. Simulations indicate that, following the impact, a significant portion ($\geq 60\%$) of the material that forms the Moon originates from the impactor mantle. Therefore, the composition of the Moon is expected to mirror that of Theia, which, in principle, should differ from Earth, as observed with other planets in the solar system. However, analysis of lunar samples returned from the Apollo missions revealed an exceptional degree of isotopic similarity between the Earth and the Moon.

\begin{figure}[ht!]
    \centering
      \includegraphics[width=0.7\textwidth]{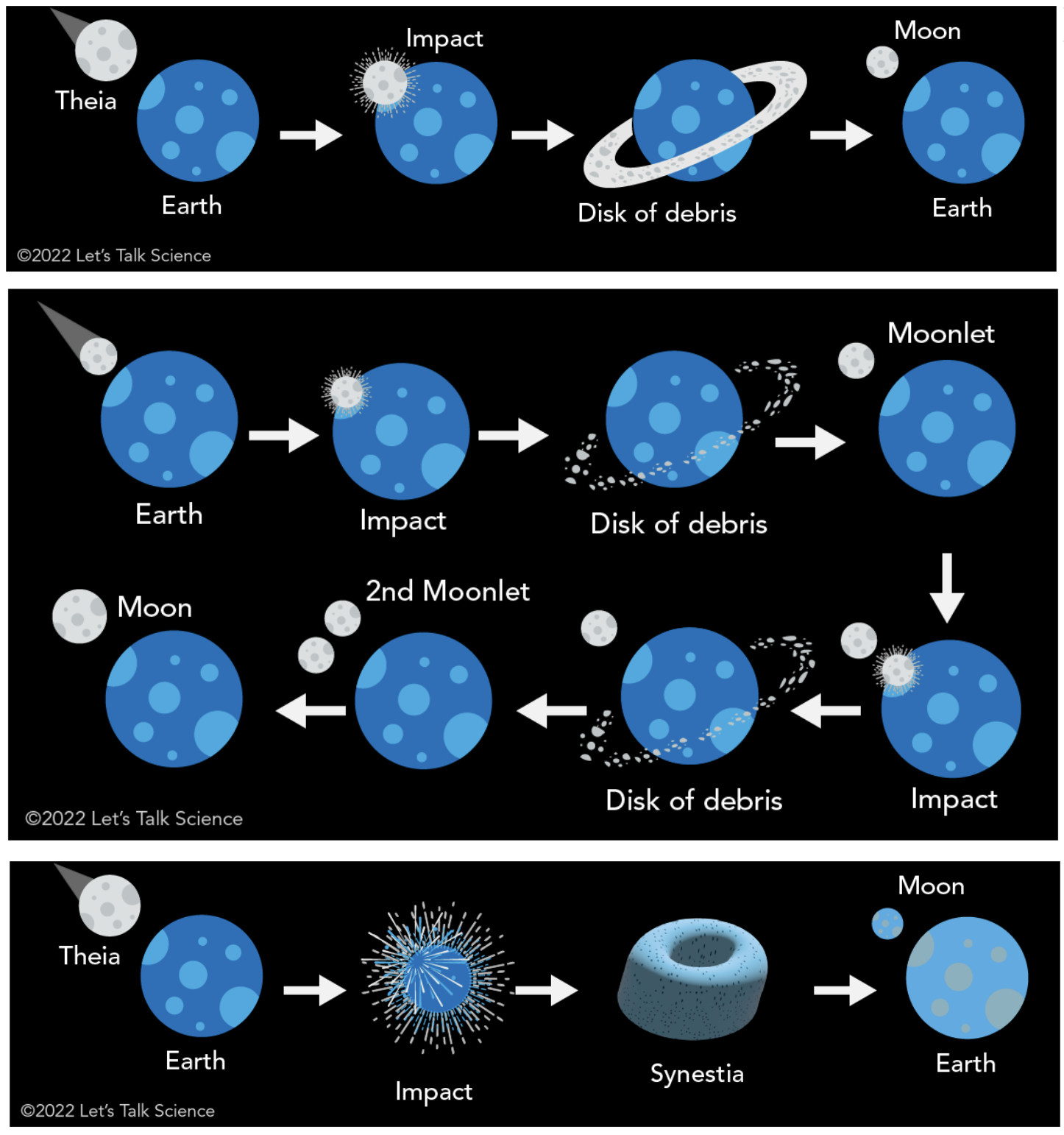}
      \caption{A cartoon picture representing the giant impact scenario (top panel), the multiple impacts scenario (middle panel) and the ``synestia'' model (bottom panel). Image credits are provided on the sketches.}
      \label{fig:moon_formation}
\end{figure}

The oxygen isotope composition, a proxy for overall Moon composition, shows a difference $\Delta^{17}O = 12 \pm 3\,$ppm \citep{Herwartz14} compared to Earth. This difference might be even smaller ($1 \pm 5\,$ppm), as recently found by \cite{Young16} analysing different rock samples. In contrast, Mars has a much larger isotopic difference, $\Delta^{17}O = 321 \pm 13\,$ppm \citep{Franchi99}. The isotopic composition of the lunar and terrestrial mantles is indistinguishable also in terms of other isotopic ratios, including $^{50}Ti/^{47}Ti$ \citep{Zhang12}, and $^{182}W/^{184}W$ \citep{Touboul07}. This unexpected similarity gave rise to what is referred to as the ``isotopic crisis'', challenging the conventional giant impact theory.

One possible solution to this challenge is that Theia and proto-Earth formed at similar distances from the Sun, i.e., from material of similar chemical composition. Various studies have explored and tested this hypothesis by tracking the composition of Earth and Theia in large sets of solar system formation simulations. While \cite{Kaib2015a, Kaib2015b} found this to be a low-probability event, \cite{AMB2015, AMB2017} have demonstrated how late giant impacts frequently occur between a proto-Earth and a last large impactor of approximately identical composition, offering a plausible solution to the giant impact challenge. This finding implies that the resemblance in composition between the Earth and the Moon may naturally arise as a consequence of a late giant impact.

In addition, several variations of the giant impact scenario have been proposed to reconcile the theory with observations. Those include a hit-and-run impact with an even more massive impactor \citep{Reufer12, Timpe23}, an impact between two proto-planets of comparable masses \citep{Canup12}, impacts with a fast-spinning proto-Earth \citep{Cuk12}, multiple impacts \citep{Rufu17} and other more complex scenarios leading to a ``synestia'', that is the vaporization of all material involved in the impact that settles into a torus to then evolve in the Earth-Moon system \cite{Lock18}) (see Fig. \ref{fig:moon_formation}\footnote{The cartoon image of the Moon formation scenarios is from \url{https://letstalkscience.ca/educational-resources/backgrounders/origin-earths-moon}} for a representation of the different models).

All these scenarios aim to achieve a significant mixing fraction between proto-Earth and Theia material, intending to address the challenge associated with the Earth-Moon composition similarity. However, each scenario is influenced by the need for ad hoc assumptions (e.g. the necessity to quickly dissipate a large amount of angular momentum) and the inherent low probability that characterizes them (for instance, the probability of having an impact between two bodies of the same size is extremely low, see e.g. \cite{AMB2017}). Despite numerous proposals, no individual or combined scenarios fully meet all Earth–Moon system constraints. Further studies, encompassing a broader range of parameters and addressing concerns related to angular momentum, isotopic constraints, and disk masses, are essential to comprehensively understand the formation of our satellite. However, different scenarios leave different signatures in the internal structure and composition of the Moon and the Moon's internal structure and its formation history are closely interconnected. 

In all scenarios involving an impact to create the Moon, it is believed that this event generated sufficient heat to form an initially predominantly molten Moon from the materials involved in the impact. As this magma ocean cooled, different minerals crystallized and sank or floated to different depths, depending on their density. This process of fractional crystallization led to the formation of the Moon's distinct layers. As mentioned above, seismic data from LGWA will be instrumental in mapping these layers and understanding their composition, density variations, and structural characteristics. 

Moreover, seismically detected anomalies have indicated the presence of relics of Theia in the Earth's mantle, offering constraints on the giant impact scenario \cite{yuan2023moon}. Seismology of the Moon conducted with LGWA will likely yield similar results, exploring the lunar mantle and potentially providing future insights into its structure and the presence of primordial anomalies related to its formation mechanism.  The detection of unexpected structures or materials will indeed serve as a test for existing theories or will help to refine them, offering a more comprehensive understanding of the Moon's formation and the role of specific impactors.

In conclusion, the formation of the Moon remains an open question, and seismic data from LGWA will be crucial in exploring the internal structure and composition of our satellite, providing important constraints on the various formation scenarios.

\subsubsection{Geologic processes}
\label{sec:geologicprocesses}
{\it Main contributors:} Francesco Mazzarini, Goro Komatsu, Luciano Zuccarello, Alessandro Frigeri \\

The main geological processes on the Moon are impact cratering and regolith gardening \cite{szalay2016impact}, volcanism \cite{head1976volc}, mass-wasting/landslides \cite{scaioni2018slide}, and tectonics. 

The interpretation of composition and morphology of the Moon suggests that effusive and explosive volcanism took place in the past \cite{head1976volc}.  Radio-isotopic dating of first returned lunar samples suggests a volcanism age between 3.1 to 3.8\,Gyr ago \cite{Nyquist1992}, while recent volcanic rocks sampled by Chang'e 5 show younger ages, of about 2\,Ga \cite{Li2021volcanism}.  While there is no evidence for active volcanoes on the surface of the Moon, this does not exclude the presence of magmatic bodies at depth.

Tectonics on the Moon are studied through charaterization of inactive fault systems \cite{nahm2023tect} and by investigating any possible recent activity \cite{Watters2019}. Impact cratering and landslides are directly amenable to seismic observations with LGWA. Landslides are widespread on the Moon and have been generally considered old in the Lunar geologic record \cite{Pike1970,Xiao2013}.  Studies on recent tectonics \cite{Watters2019,Watters2024} are supporting the existence of recent mass wasting processes. 

In the following, points 1 to 3 are about the contribution of the LGWA to the knowledge of the Moon's interior (see also section \ref{sec:lunarinternal}) in terms of the definition of mechanical layering and regolith formation process using the seismic array of LGWA as "seismic antenna". Points 4 to 5 are focused on the definition and analysis of the main tectonic processes in the area of the deployment site as well as on the cratering processes in terms of crater morphology and frequency. 

1) The regolith represents the surface where the seismic array will be deployed in the LGWA experiment. The seismic LGWA array can be used as a seismic antenna that can investigate the seismic “velocity stratigraphy” of the site by analyzing the seismic noise \cite{zuccarello2016,zuccarello2022}. The “velocity stratigraphy” may provide information on the subsurface layering and the geometry of the bedrock-regolith interface. Moreover, the analysis of seismic-acoustic signals (e.g., \cite{minio2023}) of the seismic array in the near side polar region will likely provide new insights about the lunar seismicity (see also section \ref{sec:seismicsources}) as well as its mechanical stratigraphy providing new clues about the issues of the cooling of the pristine Moon's crust \cite{elkinstanton2014}.

2) Seismic noise cross-correlation applied at pairs of stations unveil information on the Earth's crust characteristics and their variations with time \cite{shapiro2005high}. This noise-based methodology enhances the common features hidden in the seismic recordings coming from different stations. We would apply this methodology to implement new protocols and techniques to perform for retrieving information about the process of the continuous flow of micrometeorites that govern the resurfacing of the Moon landscape as well as the production and evolution of regolith.

3) The LGWA array configuration will guarantee us to analyze the kinematic properties of the seismic wavefield, based on the common waveform model (e.g., \cite{zuccarello2022}). In particular, the array methods permit the estimation of the phase velocity and propagation azimuth of the analyzed signals. The array analysis will allow researchers to characterize surface waves in terms of dispersion curves,  phase velocity, and their variability with depth. Through the use of a second array, it will be possible to better constrain the seismic source(s), especially in terms of depth, increasing the knowledge of the Moon’s interior. In particular, the detectable frequency band of the seismic antenna depends on the sensitivity of the sensors and on the maximum inter-distance between the nodes of the antenna \cite{minio2023}.

4) Morphological and structural analysis along with crater counting may be performed at the LGWA site. The regolith production is mostly due to impacts and the amount of regolith produced is somehow linked to the size of the crater (see chapter 2.4.4 Regolith Composition; \cite{zhang2023}). Thus, these studies about the cratering size distribution and craters' morphology may provide new data for modeling the production rate of regolith and a possible estimation of the regolith thickness in the LGWA site and its surroundings. 

5) The geological, morphological, and structural studies of the site of the LGWA will provide information on the deformation style of the area (compression vs. tension/strike-slip). The spatial distribution and the patterns of the structures in the area (for instance, the wrinkle ridges, and grabens) are linked to the mechanical layering of the crust (e.g., \cite{LI2023115361,Watters2019,schultz2000,Schultz2012-rt}). Cooling and tidal despinning of the Moon suggest a tectonic model inferring compressive tectonics in the equatorial zones, strike-slip tectonics in the middle latitude belts, and extensional tectonics at the poles \cite{JAYMELOSH1977221,Hauber2018}; the seismicity detected by the LGWA array may thus be used for better constraining this tectonic model. Additionally, LGWA observation may provide information about the complex relationships between tectonism and volcanism, which are for the moment inferred mostly from the surface geology \cite{Zhang2023b}.

\subsection{Gravitational-wave science}
\label{sec:gwscience}
The LGWA will observe GWs at frequencies from 1\,mHz up to about 1\,Hz, which makes it the sought-for missing link between the space-based detector LISA and future terrestrial detectors like the proposed ET and CE. No other GW detector can observe astrophysical events that involve WDs like tidal disruption events and SN 1a (sections \ref{sec:GWAstro} and \ref{sec:populations_and_formation_channels}). Only LGWA can observe IMBH binaries in the early Universe and understand their role in the formation of today's SMBHs (see section \ref{sec:populations_and_formation_channels}). Only LGWA can provide early warnings weeks before the mergers of solar-mass compact binaries formed by NSs and BHs together with excellent sky localization (section \ref{sec:cosmology}). The observation of DWDs with precise estimation of their luminosity distance and having identified their host galaxies with known redshifts will provide a new and independent approach to measure the Hubble constant (section \ref{sec:cosmology}). Especially LGWA's unique role as partner for multiband observations of GW sources together with terrestrial detectors (or space-based if present) enables a new level of precision in waveform measurements and searches for imprints of a new fundamental physics (section \ref{sec:fundamental}).

\subsubsection{Astrophysical explosions and matter effects on waveforms}
\label{sec:GWAstro}
{\it Main contributors:} Alexey Bobrick, Martina Toscani, David Vartanyan, Tsvi Piran, Jan Harms, Suvodip Mukherjee, Sourav Roy Chowdhury, Aayushi Doshi, Suraj N K, Trisha V, Kiranjyot Gill \\

\changes{The decihertz band harbors signals from a multitude of astrophysical processes not accessible to the currently operating GW observatories. Their detection would transform the study of these objects. However, the GW signatures of many astrophysical objects are either uncertain or unknown, making it harder to estimate their expected rates and the information one might extract from the signals. If no significant improvement of the models is made, it would be harder to detect and analyze the signals. With this caveat in mind, however, thanks to the diversity of the potentially observable sources, some will likely be observed during the mission lifetime, possibly in conjunction with EM observations, this way significantly impacting the respective fields. We also note that the rates of individual detectable astrophysical events are sufficiently low so that no complication from overlapping waveforms is expected. Below, we review the most likely such sources.}

\paragraph{Stellar mergers --} Stellar mergers are a prominent example where GWs may inform us about physical processes. The LGWA has the unique capability to observe WD/WD mergers. \changes{In the Galaxy, these mergers happen at a rate of about $3$ times per century \cite{Maoz2018}. However, specifying} the properties of two white dwarfs, it is currently not yet established with certainty what specific outcome their merger may produce \citep{Dan2011, Seitenzahl2013, Soker2023c}. The complexity arises because the nuclear detonation or deflagration conditions are extremely hard to resolve in simulations \citep{Khokhlov1997, Iwamoto1999, Hillebrandt2013}. They also have a strong sensitivity to the geometry of the merger and the composition profiles, and yet, they play a decisive role in the outcome of the merger \citep{Perets2019, Pakmor2021}. In addition, these processes happen during the fairly complex gravitational interactions of the two stars \citep{Dan2011, Roy2022}. Moreover, following the onset of deflagration or detonation, the exact nucleosynthesis in the white dwarfs is also challenging to model \citep{Seitenzahl2017, Kushnir2020}.

There are multiple classes of astrophysical transients and outcomes known to be produced by double white dwarf mergers \citep{Jha2019}. As an example, it is known that some double white dwarf binaries may detonate fully and lead to bright supernovae such as normal SNe Type Ia, 91bg-like SNe, SNe Type 91T, or Super-Chandrasekhar SNe (\citep{Ruiter2020}, see also Sec.~\ref{sec:SNe}). \changes{They happen at a Galactic rate of approximately $6$ times less frequent than all DWD mergers, i.e., once per about two centuries \citep{Maoz2018}. However, due to their energetic nature, their GW signal is likely to be brighter \citep{Korol24}. Therefore, while it still remains to be studied, LGWA may potentially see such events extragalactically and hence more frequently.} Similarly, some white dwarf binaries may experience a weak detonation and produce Ca-strong SNe, SNe Type Iax, 08ha-like SNe \citep{Ruiter2020}, \changes{with their deci-Hertz band GW signal being uncertain.} In some other binaries, the ignition may be even more gentle, leaving behind remnants such as another, more massive, white dwarf, an sdB star or an RCorBor star \citep{Dan2011}. \changes{The latter has been recently constrained to happen between once per 200 -- 1000 years in the Galaxy\citep{Karambelkar24}.} LGWA observations may be transformative for our understanding of such events. Identifying the connection between the progenitor white dwarf masses and the transient event they produce would dramatically improve our understanding and modelling efforts for these systems. Moreover, deci-Hertz GW observations may be the only way of probing what happened to the binary material in the first seconds following the detonation, in particular shedding light on the fate of the surviving object and the physics of explosion. The main caveat in this regard is the relatively low rate of double white dwarf mergers in the Galaxy and the local universe, \changes{as well as the uncertainties related to their GW signal.}

Less explored avenues involve the physics of mergers of white dwarfs with neutron stars, stellar black holes \citep{Metzger2012,fer13} and intermediate-mass black holes \citep{rosswog09}. \changes{WD-NS mergers, in particular, happen at Galactic rates of about once per few thousand years \citep{bob17,Toonen2018}, but would be observed more frequently if extragalactic events were detectable. Specifically for extragalactic events, the EM signal for the event may allow one to search for the GW counterpart using waveform templates, potentially enhancing GW detectability.} In all these cases, such mergers are expected to produce an optical transient event and are governed by a complex interplay of nuclear evolution, magnetohydrodynamics and radiative transfer, especially in the vicinity of the compact object, which is the dominant feedback source in such events \citep{zen20,bob22,MoranFraile2023}.

\begin{figure}[ht!]
    \centering
    \includegraphics[width=0.975\textwidth]{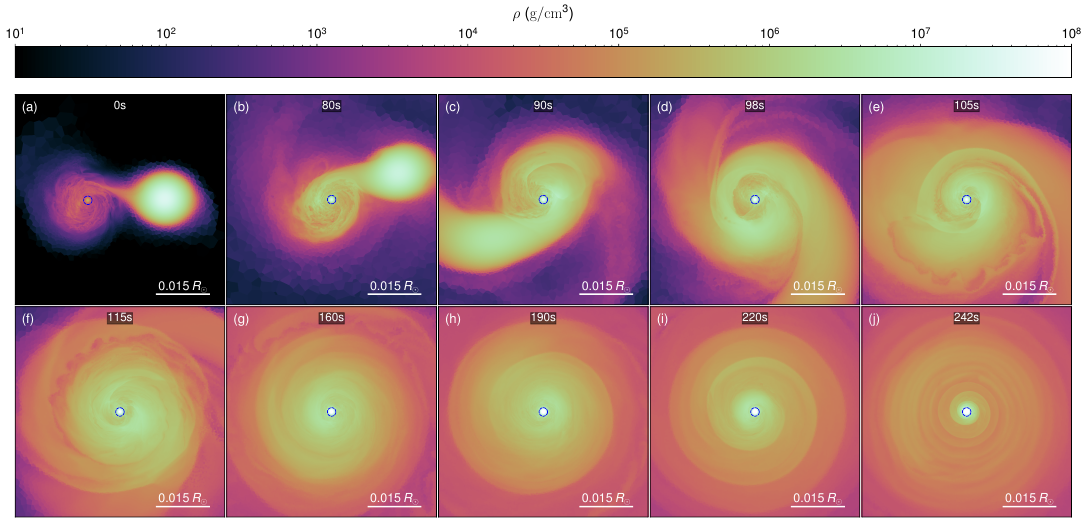}
    \includegraphics[width=0.99\textwidth]{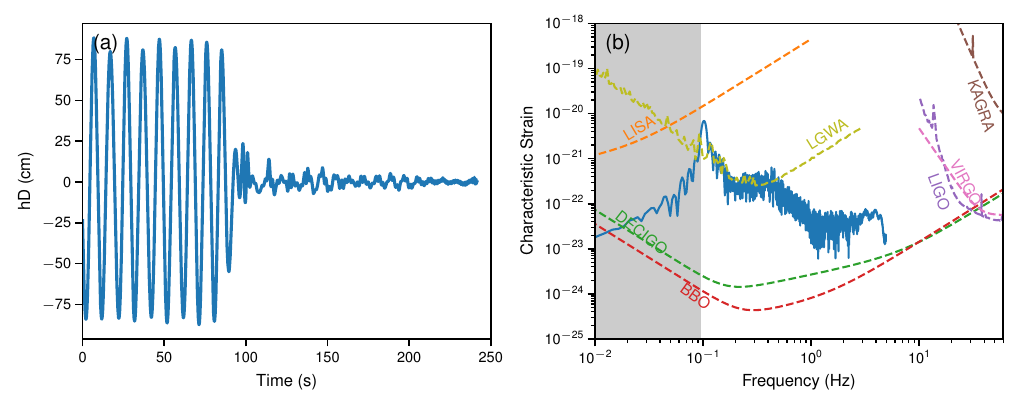}
    \caption{\changes{{\it Top panel:} 3-dimensional magnetohydrodynamic simulation of the merger between a 1.4\,M$\odot$ neutron star and a 1\,M$\odot$ carbon-oxygen white dwarf companion, as presented in \cite{mor24}. Panels (a)–(j) depict density slices in the orbital plane, starting from the beginning of the simulation (panel a) and continuing until t=242\,s (panel j). {\it Bottom panel:} Corresponding gravitational wave signal assuming a distance of 50\,kpc. The left plot shows the strain versus time, while the right plot presents the characteristic strain versus gravitational wave frequency.}}
    \label{fig:NSWDmerger}
\end{figure}

Figure~\ref{fig:NSWDmerger} represents an example of the evolution of GW strain over the last $\sim 250$ seconds prior to the merger of a 1.4 M$_\odot$ NS with a 1 M$_\odot$ carbon-oxygen WD companion, as modeled within the 3-dimensional magnetohydrodynamic framework \cite{mor24}. The simulation covers the last 100 seconds of the inspiral and $\sim$150 seconds after the disruption of the WD. Contrary to the ``chirping" GW amplitude expected for NSNS or BHBH binaries (e.g., \cite{abbott2018gw170817}), this example does not exhibit a chirp prior to the merger. However, small oscillations are noticeable, attributed to the binary's eccentricity in the final stages before the merger (Fig.~\ref{fig:NSWDmerger}, panel a). The GW emission diminishes abruptly shortly after the tidal disruption of the WD. The authors estimate that the merger would not significantly contribute to the signal-to-noise ratio; thus, the inspiral phase will dominate the observed signal \cite{setoIdentifyingDisappearanceWhite2023}.

\paragraph{Gravitational waves from tidal disruption events --}

Tidal Disruption Events \citep{hills75,1982Natur.296..211C,1986ApJS...61..219L,1988Natur.333..523R} occur when a star of mass and radius ($M_*,R_*$), orbiting a MBH of mass $M_{\bullet}$, is torn apart by the MBH tides which overcome the stellar self-gravity. After the disruption, roughly half of the stellar debris is expected to circularize around the MBH and eventually form an accretion disc. These occurrences lead to multimessenger emission, including GWs, electromagnetic radiation, and neutrinos, offering a comprehensive view of the event from the initial disruption to the later stages (see Sec.~\ref{sec:TDE} for further details).

In order for a star to be disrupted by the MBH, it needs to be in a region of the phase-space called \textit{loss cone}. In terms of angular momentum, the loss cone is defined as \citep{1976MNRAS.176..633F}
\begin{align}
    L_{\rm lc}\approx \frac{2GM_{\bullet}R_{\rm t}}{\beta},
\end{align}
where $R_{\rm t}$ is the tidal radius, i.e., the maximum distance from the MBH to have the disruption of the star 
\begin{align}
R_t \approx \left(\frac{M_\bullet}{M_\star}\right)^{1/3}R_\star \approx 2\times10^{-6}{\, \rm pc}\left(\frac{R_*}{R_{\odot}}\right)\left(\frac{M_*}{M_{\odot}}\right)^{-1/3}\left(\frac{M_\bullet}
{10^6\text{M}_{\odot}}\right)^{1/3},
\end{align}
(see \citep{Ryu2020} for relativistic corrections to this formula) and $\beta=R_{\rm t}/R_{\rm p}$ is the penetration factor, that is the ratio between the tidal radius and the stellar pericenter. In reality, the exact extent of the disruption depends on the specific characteristics and internal structure of the star itself \citep[see, e.g.,][]{2013ApJ...767...25G}. In general, we can categorize three distinct cases: when $\beta \gtrsim 1$ the star experiences full disruption; when $\beta \approx 1$ the star undergoes partial disruption, possibly leading to quasi-periodic eruptions (e.g., \citep{2022A&A...661A..68T,2023A&A...670A..93M}; see Sec.\ref{sec:QPEs}, for details); and when $\beta \lesssim 1$ no disruption of the star occurs.

The main channel responsible for TDEs is two-body relaxation. Two-body relaxation occurs when a stellar orbit is altered by the gravitational interaction with another star. In the steep potential of the MBH this interaction is typically weak, meaning the stellar trajectory is usually not significantly affected by this deviation (collisionless system). However, in high-density stellar environments like globular clusters and nuclear stellar clusters, the cumulative effect of numerous collisions can considerably impact the star trajectory (collisional system). As a result, a star may have its pericenter falling within the tidal sphere, increasing the likelihood of a TDE occurrence.

While so far TDEs have been detected mainly through EM emission, with the advent of space-based GW observatories we are on the verge of witnessing the first observations of GW-TDEs \citep{2022MNRAS.510.2025P}. In particular, the most interesting scenario would be the case of a WD revolving around a light massive BHs ($10^{3}-10^{5}$). This system would in fact produce a GW signal, which could be a single burst in case of full disruption of the star or a more continue emission in case of periodic stripping \citep[][]{2010MNRAS.409L..25Z,2020MNRAS.498..507T}. This emission would present peak frequencies around the deci-Hz, that is the frequency window where LGWA is the most sensitive. The detection of such a signal, together with the electromagnetic and neutrinos counterparts of these sources, could be a decisive factor for a full characterization of these systems.

Let us focus on the burst emitted during the star disruption, arising from the time variation of the mass quadrupole of the black hole-star system. Analytical order-of-magnitude estimates for the peak amplitude and frequency of this signal read \citep{2004ApJ...615..855K,2022MNRAS.510..992T}
\begin{align}
    h_{\rm gw} &\approx 2\times 10^{-22}\beta\times \left(\frac{M_{*}}{\text{M}_{\odot}}\right)^{4/3}\times\left(\frac{R_*}{\text{R}_{\odot}}\right)^{-1}\times\left(\frac{M_{\rm bh}}{10^{6}\text{M}_{\odot}}\right)^{2/3}\times\left(\frac{d}{16\,\text{Mpc}}\right)^{-1},\\
    f_{\rm gw} &\approx \beta^{3/2}\times 10^{-4}\text{Hz}\,\times \left(\frac{M_*}{\text{M}_{\odot}}\right)^{1/2}\times \left(\frac{R_{*}}{\text{R}_{\odot}}\right)^{-3/2}.
\end{align}
From these formulas, we see that a typical scenario involving a Sun-like star disrupted by a $10^{6} \,\text{M}_{\odot}$ MBH at a distance of 20$\,\text{Mpc}$ produces a strain around $10^{-22}$ with frequencies ranging from $10^{-4}\sim 10^{-3}\,\text{Hz}$ depending on how much close the star is to the MBH. For WD-TDEs instead, considering as typical parameters describing the star a mass and radius of $M_*=0.5\,\text{M}_{\odot}, R_{*}=0.01\,\text{R}_{\odot}$, we expect a similar strain $h\sim 10^{-22}$ but at higher frequencies $10^{-2}\sim 1\,\text{Hz}$, due to the huge compactness of the WD. Hence, WD-TDEs emit GWs in a region of the frequency interval where LGWA is the most sensitive.

To produce GW-TDE waveforms we need to study this signal numerically. Toscani et al. 2022 \citep{2022MNRAS.510..992T} have started exploring the huge parameter space describing this emission, which is described by nine parameters: the MBH mass and spin, the mass of the star, the penetration factor, the eccentricity and inclination of the stellar orbit, the viewing angle, the orbital orientation parameter and the source distance. In their work, Toscani et al. 2022 employed \textsc{phantom}, a general relativistic smoothed particle hydrodynamics code \citep{2018PASA...35...31P, 2019MNRAS.485..819L} to simulate TDEs and calculate the resulting GWs. They have studied how different parameters affect the signal and started to collect the waveforms in a catalog. However, to be fully prepared for the first observations of TDEs, it is crucial to expand this collection of theoretical waveforms by exploring a wider range of scenarios and parameter combinations.

\paragraph{Interactions in binaries --} Interactions in stably transferring binaries are also physically rich, and involve accretion disc physics, hydrodynamics, possible jet and outflow formation, and still poorly understood mechanisms of mass loss \citep{Tauris2006, Paxton2015, Decin2020, Tauris2023}. With GW observations, it may be possible to constrain the mass transfer rates in these systems, this way at least indirectly constraining the nature of interactions \citep{Sberna2021}. Coupling such observations with optical, X-ray and radio data may be particularly constraining for understanding the interaction physics in such binaries \citep{Breivik2018, Burdge2023}. For the Galactic sources, the timescales in the inner accretion disc match the LGWA band and, therefore, for sufficiently massive accretion discs, such observations may potentially be important \citep{Mineshige2002, Romero2010}. \changes{However, both the GW signal and the rates of detectable mergers need to be studied in detail.}

Mergers involving bigger objects may appear too slow to produce a measurable signal in the LGWA band. However, the compact central cores of geometrically large objects may have much shorter and much more relevant timescales. For example, common envelope merger events of giants and compact stars that initially happen on year-long timescales have been recently shown to produce a measurable GW signal for LISA \citep{Ginat2020, Renzo2021, MoranFraile2023}. Similarly, in mergers of massive main sequence stars, which take place on tens of minutes timescales, the geometrically smaller cores may merge on the factor of tens timescales shorter \citep{Zou2022}. Such mergers are common in the Galaxy \changes{and, for example, Renzo et al. \citep{Renzo2021} estimate that between $0.1$ and $100$ sources are detectable by LISA during its mission lifetime, implying comparable rates for LGWA}. For similar reasons, mergers of white dwarfs with main sequence stars or white dwarfs with the cores of red giants or stars of the asymptotic giant branch may be \changes{visible, although modelling of their GW signatures and LGWA detection rates is still necessary. Generally, we envision that GW signatures of all types of stellar mergers mentioned here need to be studied in detail.}

Tidal interactions in highly eccentric systems may also have their unique GW signature. Such binaries form in hierarchical triple stellar systems \citep{Toonen2016, Toonen2022}. While the effective timescale of the periaston passage may be in tens of minutes timescale, the tidal oscillations induced in the tidally deformed stars may have a factor of several higher frequency harmonics. Potentially detecting such signals will be informative both of the tidal interactions and the triple stellar dynamics for systems containing tidally interacting stars. \changes{However, the GW signatures of these systems also remain unknown so far.}

\paragraph{Gravitational waves from helium flashes --} Another unique avenue is the detection of GW signature from the helium flash at the tip of the red-giant branch. The ignition of helium is expected to be off-center, exhibit non-spherical formation, and occur over timescales of tens to hundreds of seconds \citep{Paxton2011, Kippenhahn2013}. Due to the complexity of the helium flash onset, limited attempts have been made thus far at nucleo-hydrodynamic modeling \citep{2010A&A...520A.114M, 2012ASSP...26...87M}. Given these flashes occur deep within the center, GWs and neutrino observations may be the only way to constrain these events. \changes{While the Galactic rates of these events can be easily estimated as the formation rate of the red giant stars in the Galaxy \citep{Kroupa2002}, thus being approximately once per decade, their GW strain amplitudes are far from being understood, necessitating further research.}

\paragraph{Core-collapse supernovae --} In the last two decades, significant efforts have been made in investigating the GW emission from CCSNe (see \cite{Radice19, Abdikamalov2020, 2021Natur.589...29B, 2023PhRvD.107j3015V, 2023IAUS..362..215M} for recent reviews). Non-spherical explosions are required to produce GWs, which happens during three distinct phases: a) immediately at bounce (through time-dependent rotational flattening and post-shock convection, lasting tens of ms); b) the growing neutrino-driven turbulence (lasting 50--200\,ms); and through c) PNS modal oscillations \cite{Kuroda16, Kuroda2017, Janka16, Andresen17, Oconnor18, Kawahara18, Andresen19, Powell19, Radice19}. These phases, characterised by GW emission in the $\sim$100--1000\,Hz frequency range, are accompanied by an additional low-frequency ($\lesssim$50\,Hz) signal, henceforth known as the GW linear memory \cite{Vartanyan2022, 2024arXiv240513211G} (see also Sec.~\ref{sec:populations_and_formation_channels}) that is generated through aspherical, large-scale matter ejecta motions coupled with anisotropic neutrino emission \cite{epstein1978,burrows1996,emuller,Vartanyan2020,2022PhRvD.106d3020M,Richardson2022}. The neutrino memory signal shows a more secular time-evolution than the matter component, which is due to the cumulative time-integral of the anisotropy-weighted neutrino luminosity \cite{Mukhopadhyay_2021}. Sustained turbulent accretion in more massive progenitors results in higher neutrino luminosities and generally more anisotropic explosions. As with the matter component, the neutrino component is more pronounced for delayed explosions of models with higher compactness \cite{Burrows2020}. 

\begin{figure}[ht!]
    \centering
    \includegraphics[width=0.7\textwidth]{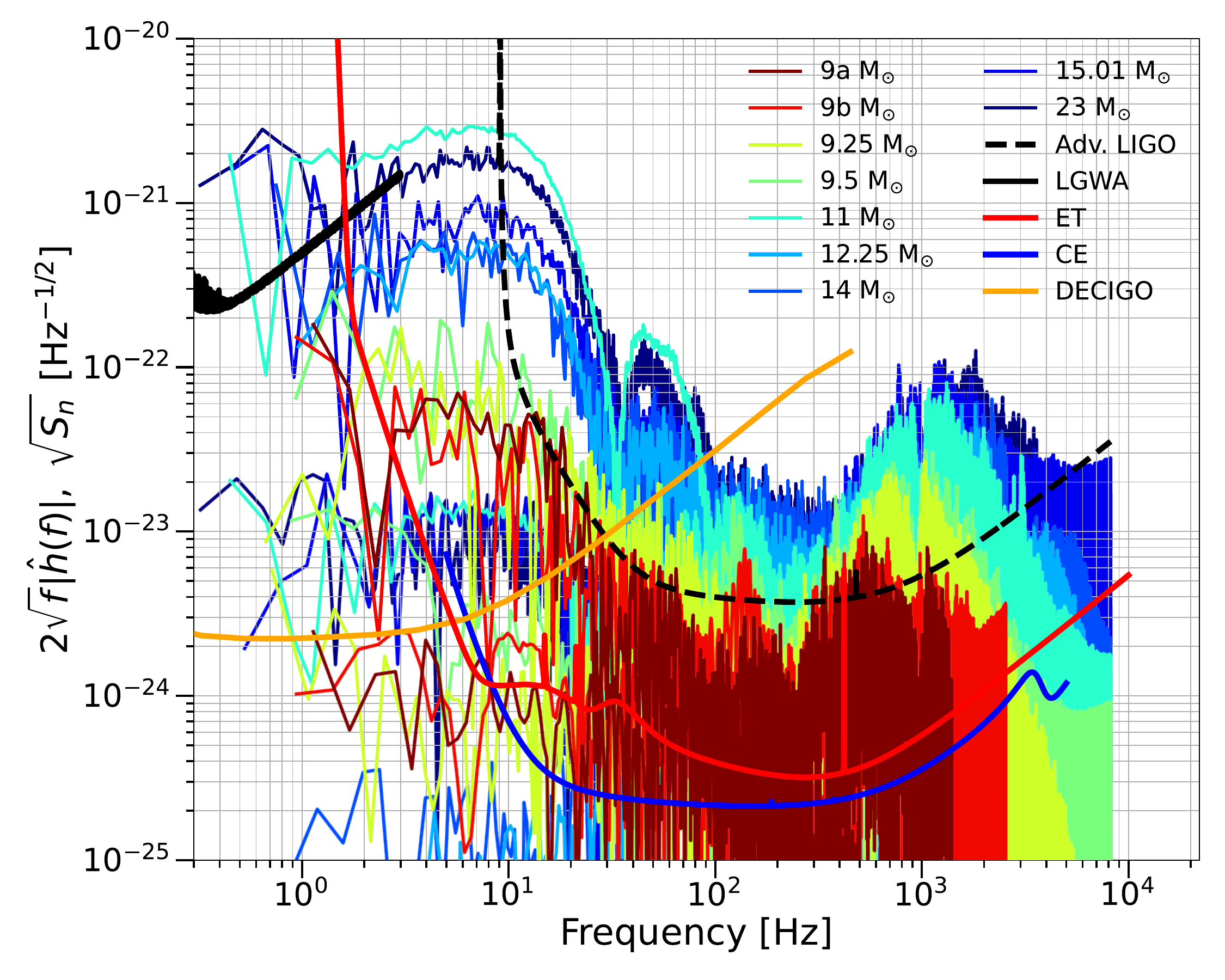}
    \caption{Gravitational-wave spectra $2\sqrt{f}|\tilde h(f)|$ for CCSNe with different stellar masses as a function of time projected against strain noise curves $\sqrt{S_n}$ for various GW detectors (adapted from \cite{2023PhRvD.107j3015V}). The simulations, extending out to $\sim$6 seconds after the bounce, are unable to constrain the spectrogram in the region 0.1--1.0\,Hz. \changes{The sources are assumed to be at a distance of 10\,kpc.}}
    \label{fig:vartanyan2023}
\end{figure}

Large-scale SN ejecta asymmetries (where jet-like structures may emerge) are expected (see e.g. \cite{2001AIPC..565...40F, 2023MNRAS.518.5242U, 2023ApJ...951L..30G, Soker2023,Soker2023b}) and are seen observationally using spectropolarimetry \cite{2005ASPC..342..330L, 2008ARA&A..46..433W, 2016MNRAS.457..288R, 2021NatAs...5..544T, Nagao2023, 2023A&A...675A..83H}, but their GW emissions are associated with frequencies that are beyond the LGWA observation band (100--1000\,Hz) \cite{Mezzacappa2023, Pajkos2023, Dalya2023, Powell2023} (see also Fig.~\ref{fig:vartanyan2023}). 

While recent simulations do not cover the decihertz band (see for instance \cite{2023PhRvD.107j3015V} and Fig.~\ref{fig:vartanyan2023} \changes{for 10\,kpc event signals}) the linear and nonlinear memory effects can lead to detectable GW signals below 1\,Hz \cite{Favata2010, 2022PhRvD.106d3020M}. In this way, even with LGWA, one would be able to study the GW emission generated by the neutrino processes \cite{Sedda2020, Vartanyan2020, 2023PhRvD.107j3015V}. As discussed above, the range $>$50\,Hz primarily arises from matter asymmetries while the range $<$10\,Hz carries information regarding the anisotropy of the neutrino emission. Although these frequency ranges may overlap in reality, they are, in principle, distinguishable. The polarization of the signal provides insights into the asymmetries of the event along with the orientation of the source with respect to the line of sight. Notably, a simultaneous detection of the low-frequency memory signal by LGWA and the high-frequency component by terrestrial detectors would offer an invaluable, unprecedented understanding of CCSNe. The memory signal would provide information about the progenitor star while frequencies above 100 Hz would provide insight into the newly formed PNS. Although the exploration of lower-frequency CCSNe signals and their detectability has only begun in recent years, significant progress has been made in extending simulations to several seconds post-bounce. The near-future goal of extending these simulations to tens of seconds holds promise for producing consistent results at low frequencies.

\paragraph{Gravitational waves from relativistic jets --}

\changes{The acceleration of a relativistic jet is another source of gravitational radiation \cite{Segalis2001,Piran2002,Sago2004}. The resulting GW signal is a memory-type signal, namely, they take the form of a step leaving a permanent deformation of spacetime after they pass. The detailed features of the signal depend on the acceleration (and possible deceleration) of the jet, on its overall duration and orientation relative to the observer \cite{Birnholtz2013,Leiderschneider2021,Piran2022}. The overall amplitude of the GW signal is, as expected on dimensional grounds, of order $G E_{\rm jet}/c^4 d$, where $E_{\rm jet}$ is the energy of the jet and $d$ the distance of the source. The frequency is the smaller of the inverse acceleration time and the jet duration. Natural candidates for jet-GWs are long GRB jets whose typical energy is of order $10^{51}$\,erg and typical frequency of order tens of deciHz. Short GRB jets are somewhat weaker and have higher frequency. As those are more frequent, it might be easier to detect. Most remarkable are hidden jets that take place in some powerful supernovae \cite{Piran2019}. Those jets are choked inside supernovae and are difficult to observe directly. Their unique GW signature might enable us to confirm their existence. The energies of choked supernovae jets are comparable to those of long GRB jets \cite{Piran2019}. As supernovae are much more frequent than GRBs it might be expected to detect these signals first \cite{Piran2022}. In particular, the GWs from a powerful hidden jet in supernovae are likely stronger than the GWs from the collapse event itself. A Jet-GW signal from a galactic supernova will have a strain of  $\gtrsim 10^{-21}$ in the frequency range of LGWA.}

\subsubsection{Populations and formation channels of GW sources}
\label{sec:populations_and_formation_channels}
{\it Main contributors:} Elisa Bortolas, Manuel Arca Sedda, Pau Amaro Seoane, Stefano Benetti, Marica Branchesi, Adam Burrows, Enrico Cappellaro, Roberto Della Ceca, Kiranjyot Gill, Ines Francesca Giudice, Elisabeth-Adelheid Keppler, Chiaki Kobayashi, Valeriya Korol, Francesco Longo, Ferdinando Patat, Sourav Roy Chowdhury, Roberto Serafinelli, Paola Severgnini, Martina Toscani, David Vartanyan\\

This section describes the astrophysical sources that could generate a GW signal accessible to the LGWA. We provide an overview of the main formation channels for each source and other astrophysical aspects crucial for LGWA's source detection and interpretation. We start by detailing the astrophysics of compact stellar binaries, which can be stellar black holes, neutron stars, white dwarfs and any combination thereof. We then describe massive BHs and I/EMRIs: for the former, both bodies in the binary have a mass at least $> 100 M_\odot$; for the latter, the mass of the secondary body is typically $< 10^{-2}$ times lighter than the primary. We then describe further possible LGWA sources as stellar tidal disruption events, asymmetric neutron stars and finally, GWs from type II supernovae via the memory effect. Table~\ref{tab:sourcesLGWA} summarizes the main LGWA sources with their properties. The detection horizons for equal mass binaries and I/EMRIs are shown in Fig.~\ref{fig:horizon-et-lgwa-lisa}.

\begin{table}[t]
    \centering
    \begin{tabular}{c|c|c|c}
       \textbf{Source} & \textbf{Formation channel(s)/cause} & $M_{\rm tot}$ [$M_\odot$] & $q$  \\\hline
       Stellar binaries & Stellar binary evolution, dynamical interactions & 1--10$^2$   & $0.1-1$ \\
       MBH binaries & Galaxy-galaxy mergers  & $10^3-10^6$ &  $0.01-1$  \\
       IMRIs & Two-body relaxation, AGN disks &  $10^2-10^5$ & $10^{-4}-10^{-2}$  \\
       TDEs & Two-body/resonant relaxation & $10^3-10^6$ & $10^{-5}-10^{-3}$   \\
       Asymmetric NSs & Non-axisymmetric rotating NS & $1-4$?   & ---  \\
       \hline
    \end{tabular}
    \caption{Here we list the main GW sources for LGWA with their typical formation channels, together with the total mass, mass ratio accessible to the LGWA for each of them.}
    \label{tab:sourcesLGWA}
\end{table}

\paragraph{Compact-binary mergers of stellar-mass compact objects --}

Compact binary systems, comprising WDs, NSs and BHs populate our local Universe. Approximately $10^6$\,yr before their merger,  these systems emerge as prime targets for space-based GW observatories. Missions like LISA \citep{LISA2017}, TianQin \citep{TianQin}, and Taiji \citep{Taiji}, set to operate at milli-Hz frequencies, are anticipated to be functional in the 2030s. Lunar-based projects, such as LGWA (operating in the deci-Hz regime), although capable of producing exciting science by themselves, can complement space-based missions, especially at the high-frequency end of their sensitivity window, effectively bridging the gap with terrestrial detectors. Furthermore, LGWA can enhance the scientific scope of compact binaries to approximately 1\,Hz, enabling the detection of events like DWD and NS-WD mergers \citep{HaEA2021a}. Notably, some of these mergers are not accessible to the terrestrial detectors nor to the lower-frequency space missions.

\noindent\underline{Stellar black-hole binaries} --
\noindent The detection of GWs by the LIGO-Virgo-Kagra (LVK) collaboration, led by the discovery of GW150914 \citep{AbEA2016a}, has provided proof of BHs,  enriching our knowledge of known merging BHs up to the present population of nearly 100 objects with masses above 30 M$_\odot$. This was well above the limit set by previous observations, supporting the evidence for the existence of binary BHs, and for the formation of new BHs as by-product of a BH mergers. Such a growing catalogue of events \citep{abbott2021gwtc21} allows constraints to be placed on the cosmic population of merging BHs. Among all observed mergers, there are several intriguing sources with properties that clearly deviate from the overall distributions. One of them is GW190521 \citep{2020PhRvL.125j1102A,2021ApJ...907L...9N}, the first GW source to leave behind an IMBH with a  mass of $\sim 150 M_\odot$.

With a primary mass of $\sim 85$ M$_\odot$, GW190521 was also a challenging source to interpret. Stellar evolution models predict a gap in the BH mass spectrum roughly between $40-300$ M$_\odot$ owing to the onset of pair instability and PPISN in their progenitors \citep{barkat67}. Stars $\gtsim 300M_\odot$ may collapse to BHs (see \cite{nom13} for a review). The actual range of masses affected by pair instability and PPISN is highly uncertain, and it depends on the details in stellar evolution models. Also, star collisions and binary interactions can alter the stellar structure such that the star bypasses the pair instability and PPISN phases and instead undergoes direct collapse, possibly leaving a BH remnant with a mass in the gap. BHs in the gap could also originate from previous BH mergers, following the so-called hierarchical merger scenario.

It is important to understand the origin of merging BHs to interpret GW observations. From a theoretical standpoint, the main formation channels for compact binary mergers are via binary stellar evolution \citep{2022MNRAS.516.5737B,2022ApJ...931...17V,2016MNRAS.460.3545D,2019MNRAS.485..889S,2010ApJ...715L.138B,2012ApJ...759...52D,2016A&A...588A..50M,2021ApJ...910...30T,2019MNRAS.482..870E,2022MNRAS.516.2252O,2023MNRAS.524..426I, 2020ApJ...902L..36F,2019MNRAS.490.3740N}, dynamical interactions in star clusters \citep{2019MNRAS.483.1233R,DCEA2019,2020MNRAS.497.1043D, 2019MNRAS.486.5008A,2016ApJ...831..187A,2020PhRvD.102l3016A,2021Symm...13.1678M,2021MNRAS.505..339M,2016PhRvD..93h4029R,2018PhRvL.120o1101R,2017MNRAS.464L..36A,2014MNRAS.441.3703Z,2009MNRAS.395.2127O,2010MNRAS.407.1946D,2010MNRAS.402..371B,2016ApJ...824L..12O,2018MNRAS.473..909B,2021MNRAS.500.3002B,2023arXiv230704807A,2023MNRAS.520.5259A,2021ApJ...920..128A,2018ApJ...866...66M,2020ApJ...898...25T}, and primordial BH interactions \citep{2009PhRvL.102p1101S,2016JCAP...12..031G,2016PhRvL.117f1101S,2017JCAP...09..037R,2017PDU....15..142C,2017PhRvD..96l3523A,2018ApJ...864...61C,sasaki2018pbh,2020PhRvD.101d3015V,2021JCAP...05..003D}. 

Identifying the imprint of different formation channels is not trivial, owing to the many processes, and related uncertainties that can affect the merging BH population (see for example \citep{2020ApJ...894..133A,2021ApJ...910..152Z,2022MNRAS.511.5797M,2023ApJ...953...80B,2023MNRAS.520.5259A}). The cosmic star formation, the metallicity distribution, the star-cluster formation and evolution, the natal BH mass, spin, kicks, and the parameters of BH binaries are only a few of such processes and quantities that can dramatically affect the properties of observed merging BHs \citep{2023MNRAS.520.5259A}. However, only a handful of parameters, like the binary eccentricity, the mass ratio, and the effective spin distribution, might be the key to understanding such population of merging objects. 

Among all, the binary eccentricity is considered to be a quantity that can clearly tell whether a binary merger is formed dynamically or not \citep{2011A&A...527A..70K,2016PhRvD..94f4020N,2017MNRAS.465.4375N,2021ApJ...914...75R}. Several processes, like common envelope dynamics, mass transfer, and GW emission, are expected to dampen binary eccentricities and lead to almost circular orbits ($e<10^{-3}$) even before the binary emits GWs in the frequency range of terrestrial detectors ($>10$ Hz) (see e.g. \citep{2011A&A...527A..70K,2020FrASS...7...38M}). However, SN kicks or common envelope physics may lead to moderately eccentric binaries emitting high-frequency GWs \citep{2021hgwa.bookE..16M,2022PhRvD.106d3014T}. In dynamical environments, instead, binaries are assembled via strong dynamical interactions, which can naturally lead to the formation of highly eccentric binaries \citep{2018ApJ...860....5G,2018PhRvD..98l3005R,2018MNRAS.481.5445S,2018PhRvD..97j3014S,2019ApJ...871...91Z,2021A&A...650A.189A,2023arXiv230307421D,2023arXiv230704807A}, depending on the environment in which they develop. For example, eccentric binaries can form via GW captures, secular processes, and binary--single encounters in quiescent and active galactic nuclei \citep{2018ApJ...860....5G,2018ApJ...856..140H,2022Natur.603..237S,2023Univ....9..138A}, or via three- and four-body interactions in young and globular clusters \citep{2019ApJ...871...91Z,2019MNRAS.483.1233R,2021A&A...650A.189A,2018PhRvD..98l3005R,2023arXiv230704807A}. For dynamical mergers, the environment can also affect the critical value of the binary separation below which GW emission becomes the dominant evolutionary process and the binary decouples from the overall dynamics. This means that an eccentric merging binary forming in a young massive cluster will likely be circularised by the time it reaches the 1--10\,Hz band, but can appear eccentric at lower frequencies observable with LGWA, whilst a binary forming in a massive and dense galactic nucleus may form with a sufficiently small separation to be detectable as eccentric source even with high-frequency detectors (e.g. \citep{2019ApJ...871..178G,2021ApJ...914...75R,2021A&A...650A.189A}). Despite some of the detected GW sources could be eccentric \citep{2020ApJ...903L...5R,2021ApJ...921L..31R,2022NatAs...6..344G,2022ApJ...940..171R,2023arXiv230803822T}, the effects of eccentricity on detected signals are very similar to those induced by binary precession, thus making it hard to place stringent constraints on the eccentricity \citep{2022PhRvD.105b3003F,2023MNRAS.519.5352R}. Advances in data analysis and waveform templates are recently enabling the premises to measure the eccentricity of binaries provided that it exceeds O$(10^{-2})$ \citep{2010PhRvD..81b4007B,2014PhRvD..90d4018D,2017PhRvD..96d4028C,2018PhRvD..98h3028L,2020CQGra..37v5015M,2020PhRvD.101d4049L,2022ApJ...936..172K}. With the possibility to follow the binary over many orbits, lower frequency detectors may be the key to untangle the two effects and robustly determine the binary eccentricity \citep{Sedda2020,2022PhRvD.105b3019L}, helping us to pin down the origin of at least some sub-population of observed sources \citep{2016PhRvD..94f4020N,2017MNRAS.465.4375N}. In these regards, multiband observations with different detectors can help to increase the accuracy of measured parameters. Multiband detections can be sub-threshold in LGWA becoming significant only with the help of the other GW detectors involved in the observation.

Another interesting, possibly distinctive, feature of merging BHs is the distribution of spins. Ideally, a merging BH binary formed from isolated binary stellar evolution is expected to lead to two BHs with aligned spins \citep{2002MNRAS.329..897H}, although some level of misalignment can be induced during the BH formation process, e.g. during the supernova phase \citep{2000ApJ...541..319K,2002MNRAS.329..897H,2020A&A...636A.104B,2021PhRvD.103f3007S,2021ApJ...920..157C,2022ApJ...938...66T}. Conversely, dynamical BH mergers are expected to have spins isotropically distributed in space, owing to the chaotic nature of the interactions that drive their formation \citep{2020FrASS...7...38M}. This suggests that a precise measurement of the BH spins and their mutual orientation would help identifying the binary formation origin. Spins in current GW data analysis are poorly constrained, and generally the spin orientation of merging binaries is inferred from the so-called effective spin parameter $\chi_{\rm eff}$ \citep{2011PhRvL.106x1101A}, which is a projection of the spins onto the binary angular momentum, weighted with the binary component masses. A population of mergers with aligned spins would be characterised by a $\chi_{\rm eff}$ distribution clearly peaked around a positive value, whilst such a peak would be centered on zero for mergers with isotropically distributed spins. 

Finally, identifying the location of the merger would definitely help to pin down the formation scenario. Combining observations performed with different detectors can significantly help reduce the localization area \changes{down to a few arcmin$^2$ (see section \ref{sec222}), possibly providing clues about the merger birth site, at least on a statistical basis and for relatively closeby sources.}

The LGWA can play a crucial role in measuring the aforementioned quantities. Working in the deci-Hz frequency range, depending on its final sensitivity it could follow merging BHs for months and weeks before the merger, ensuring high precision measurements of the masses, the spins, and possibly the eccentricity evolution of the binary (see Sec.~\ref{sec:intrestimates}). These are fundamental for further understanding the evolution of (single) massive stars, and the formation and evolution of massive star binaries.\\

\noindent\underline{White-dwarf and neutron-star binaries} --
\noindent Mergers of WDs and NSs are extremely important for understanding the origin of elements \cite{kob20sr}. NS mergers are proposed for one of the astrophysical sites of the rapid neutron capture process \cite{lat74}, which has been confirmed by the detection of GW and EM transient `kilonova' \cite{LIGOScientific:2017zic}. WDs mergers are thought to become sub-Chandrasekhar-mass Type Ia supernovae \citep{ibe84,web84}, which are responsible for at least some of the Fe production particularly in dwarf galaxies \citep{kob20ia}. Compact stellar systems composed of WDs and NSs are guaranteed GW sources, as some of them have been already discovered with electromagnetic (EM) telescopes (see also Section \ref{sec:DWD-WDNS}). Some of the shortest period and better characterised of the known binaries have also been proposed for testing space-based GW detectors to maximise their scientific output (e.g. \citep{Stroeer2006,KuEA2018,lit18,fin22}). Some of such systems, known in literature as {\it verification binaries} \citep[e.g.,][]{Stroeer2006}, can be detected by LGWA \citep{HaEA2021a}. \changes{They are guaranteed multi-messenger sources offering an} opportunity to study the astrophysics of compact sources using both their GW and EM emission 
(e.g. \citep{mar11,Korol2017,Breivik2018,tau18,lit19}). \changes{They would also be ideal calibration sources for LGWA given their known location and well-modeled GW amplitudes eventually measured precisely by LISA (see section \ref{sec:LunarResp}.}

From a theoretical perspective, compact binaries are expected to be several orders of magnitude more abundant than currently observed by EM telescopes (e.g. \citep{Napiwotzkietal2020,2023A&A...678A.187C}), with binaries composed of two WDs being the most abundant in the Milky Way and its immediate neighbourhood. Their formation involves up to several mass transfer episodes; at least one of these episodes has to be unstable leading to a {\it common envelope}, a phase during which the binary can shrink dramatically \citep{pac76,web84,liv88,deK87,iva15}. Even though this is one of the least understood phases in binary evolution, the fact that we do observe close double compact objects and mergers is a strong indication that something like a common envelope phase happens \citep{2021ApJ...920...86K}. Based on these observations, several theoretical modelling efforts are being made (e.g. \cite{2023LRCA....9....2R,2023ApJ...944...87D}). Thus, the demographics of compact binaries observed via GW radiation will provide opportunities to learn new physics and answer key scientific questions related to the formation and evolutionary processes of close binary systems. Besides the common envelope physics, this includes questions related to the stability and efficiency of mass and angular momentum transfer, tides, and accretion onto compact objects, as well as details of their destruction in supernova explosions \citep[for a review see][]{LISAastroWP,2022LRR....25....1M}.

At frequencies $<0.1\,$mHz, binary population synthesis simulations predict ${\cal O} (10^7)$ double WDs, followed by ${\cal O} (10^5)$ NSWDs in the Milky Way alone \citep[for a review see][]{LISAastroWP}. Studying these binary populations with EM facilities has proven to be technically challenging due to the compact size of these binaries and WD/NS stars themselves, and so GW detectors -- characterised by different selection effects  -- are ideally suited for discovering and characterising them in bulk \citep{Rebassa-Mansergasetal2018,kor22}. Importantly, GW selection effects enable the study of EM-dim double compact objects at unprecedentedly large distances throughout the Milky Way, including its most massive satellites and possibly reaching out as far as the Andromeda galaxy, which otherwise impossible with EM telescopes \citep{kor18,roe20,wil21,kei22}.
\changes{Recent binary population synthesis studies forecast between 2 and 200 double WDs in the Milky Way with orbital periods of less than 200 seconds (GW frequencies greater than $10^{-2}$\,Hz) that LGWA could potentially capture in the inspiral phase \cite{Korol2017,Lamberts2019,2020ApJ...898...71B,Li2020,2023ApJ...945..162T,2024arXiv240520484T}. Observation-based forecasts are even more optimistic, predicting up to 1000 inspiralling double WDs in the same frequency regime in the Milky Way alone \citep{maoz,Maoz2018,kor22}.} \\

\noindent\underline{Supernovae Type Ia and white-dwarf binaries} --
SNe~Ia are believed to originate from WDs which, due to mass transfer or generally, interactions in close binary systems, grow above the Chandrasekhar limit ($M_{Ch}\sim 1.4\, M_\odot$) and are incinerated by a thermonuclear explosion under degenerate matter conditions. This idea dates back to the early '70 \cite{whelan} but the actual nature of the secondary star remains undetermined. Two alternative scenarios have been proposed: the donor is a hydrogen or helium main sequence or a giant star (the single degenerate scenario, or SD), or another WD (the double degenerate scenario, or DD), which would lose orbital momentum via GW radiation and eventually merge. In the case of SD systems, mass transfer onto the accreting WD occurs when the secondary donor star fills its Roche lobe. This sets a lower limit for the separation between the two stars and the orbital period (of the order of 10 minutes for the most compact, non-degenerate donors \cite{Postnov2014}). In this configuration, the GW emission is weak and does not significantly affect the binary evolution. On the contrary, in DD systems the orbital radius naturally shrinks as an effect of angular momentum loss via GW emission \cite{web84}. Because of the small tidal radius of WDs, this allows a minimum orbital period of tens of seconds (see below). Since the intensity of the GW emission increases with increasing frequency, this is also expected to produce a much stronger signal, falling in the 0.01--0.1\,Hz domain, right in the frequency range where LGWA's sensitivity is at its best. This difference in orbital periods creates a clear-cut case, in which GW detection can act as an effective discriminant between the two channels. Using Kepler's third law relating the period $P$ and the semi-major orbital axis $a$, and considering that the GW frequency is given by $f_{GW}=2/P$, it can be readily shown (see for instance \cite{MSV2018}) that
\begin{equation}
\label{eq:kepler}
f_{GW} = \frac{\sqrt{G}}{\pi} (M_1+M_2)^{1/2} \; a^{-3/2}\simeq 0.1 \textrm{Hz} \;\left ( \frac{M_1+M_2}{2M_\odot}\right)^{1/2} \; \left (\frac{a}{0.02R_\odot}\right)^{-3/2},
\end{equation}
where $M_1$ and $M_2$ are the masses of the two stars. Given the total mass constraint ($M_1+M_2\gtrsim~M_{Ch}$), the typical WD radius ($R_{WD}\sim0.01\,R_\odot$; \cite{Arseneau2024}) and the minimum WD separation ($a\sim3R_{WD}$; \cite{Eggleton1983}), Eq.~\eqref{eq:kepler} yields a maximum GW frequency $f_{GW,m}\sim$ 0.1\,Hz when the merger occurs (see also \cite{Dan2011}, Fig.~21 therein). The time it takes for a binary system to merge, under the simplifying assumption of a circular orbit and neglecting possible mass transfer, hydrodynamical and radiative effects was computed by Peters \cite{Peters1964} (see Eqs. (5.9)-(5.10) therein), and can be expressed as:

\begin{equation}
\nonumber
t_m = \frac{5}{256} \frac{c^5}{G^3} \frac{a^4}{M_1 M_2 (M_1+M_2)},
\end{equation}
which, taking into account Equation~\ref{eq:kepler}, can be rewritten as:
\begin{equation}
\label{eq:peters}
t_m(M_c,f_{WG}) = \frac{5}{256} \left(\frac{1}{\pi}\right)^{8/3} \; c^5 \; G^{-5/3} \; M_c^{-5/3}\; f_{GW}^{-8/3}
\simeq 9.3\; \textrm{yr}\; \left( \frac{M_c}{M_\odot}\right)^{-5/3} \left( \frac{f_{GW}}{0.1 \textrm{Hz}}\right)^{-8/3}
\end{equation}
where $M_c=(M_1+M_2)^{-1/5} (M_1 M_2)^{3/5}$ is the chirp mass \cite{MSV2018}.

The SN~Ia rate in the Galaxy is well known from direct observations in the local universe: $r_{\rm Ia}$=(5.4$\pm$1.2)$\times$10$^{-3}$ yr$^{-1}$ \cite{Li2011}. This corresponds to an event every $\sim$200 yr, which is definitely too low to provide a significant chance of detection of a WD binary merger during the lifetime of a GW campaign. 

However, in order to understand the progenitor of SN~Ia, statistical approach can be followed. If the bulk of SNe~Ia comes from DD systems, there needs to be enough such systems with the suitable properties (mass, merging time) to account for the observed SN~Ia rate. Let us assume that $n(M_{\rm c},f_{\rm GW})$ is the probability density function describing the distribution of DD systems. Let us then also indicate with $M_{\rm c,l}$ and $M_{\rm c,h}$ the lower and upper chirp mass limits for having a total mass $M_1+M_2\gtrsim M_{\rm Ch}$. These limits are set considering the maximum and minimum mass of a C-O WD: $\sim 1.0 M_\odot$ \cite{Althaus2021} and  $\sim 0.3 M_\odot$ \cite{Prada2009} ($M_1=0.4 M_\odot$, $M_2=1.0 M_\odot$ and $M_1=M_2=1 M_\odot$, corresponding to an approximate chirp mass range 0.5-0.9 $M_\odot$). Finally, let us indicate with $f_{\rm GW,l}$ the lower limit of the frequency range covered by the GW detector, which corresponds to the maximum merging time $t_{\rm m,h}$ and maximum orbital period $P_h$. With these settings, the total expected number of suitable DD systems found by the GW detector is:
\begin{equation}
\nonumber
\label{eq:nch}
N_*=\int_{M_{\rm c,l}}^{M_{\rm c,h}} \int_{f_{\rm GW,l}}^{f_{\rm GW,m}} n(M_c,f_{\rm GW}) \;dM_c\; df_{\rm GW},
\end{equation}
where, as we have shown above, $f_{GW,m}\sim$0.1\,Hz. Since, by construction, all these systems will have merged (and hence exploded) by $t=t_{m,h}$, the corresponding explosion rate is simply given by 
\begin{equation}
\nonumber
r_{Ia}=\frac{N_*}{t_{m,h}} 
\end{equation}
Since $r_{Ia}$ is well known (see above), the argument can be reversed and one can predict $N_*$. For illustration purposes, using this relation and Eq.~\eqref{eq:peters}, we have compiled Table~\ref{tab:enrico} for some representative values of $P_h$. For the sake of simplicity, we have assumed $M_1=M_2=0.7 M_\odot$ ($M_c\simeq0.6 M_\odot$). We note that the above calculations are a good approximation only for $P\gtrsim100$\,s. For lower periods (50-100\,s, depending on the WD masses) Roche lobe overflow ensues. \changes{Recent 3D hydrodynamical simulations show that for sufficiently massive primaries, the onset of unstable dynamical accretion of helium from the secondary white dwarf, just prior to the binary merger, heats up the helium shell on the primary white dwarf. As a result, a thermonuclear runaway occurs, igniting a helium detonation on the surface of the primary white dwarf and eventually leading to the ignition of the carbon-oxygen core \citep[e.g.,][]{2010ApJ...709L..64G,pakmor,2024A&A...683A..44M}.}
This would lead to a merger timescale that is much shorter than what is predicted by Eq.~\eqref{eq:peters}. 

\begin{table}
\tabcolsep 10mm
\begin{tabular}{cccc}
\hline
$P_{\rm h}$ & $f_{\rm GW,l}$ & $t_{\rm m,h}$  & $N_*$ \\
(s)   & (Hz)      & (yr)   &    \\
\hline
40000 & 5.0$\times$10$^{-5}$ & 1.3$\times$10$^{10}$& (7.1$\pm$1.6)$\times$10$^7$\\
3600  & 5.6$\times$10$^{-4}$ & 2.2$\times$10$^7$ & (1.2$\pm$0.3)$\times$10$^5$ \\
2000  & 10$^{-3}$            & 4.6$\times$10$^6$ & (2.4$\pm$0.6)$\times$10$^4$ \\
200   & 10$^{-2}$            & 9.9$\times$10$^3$            & 52$\pm$12 \\
\hline
\end{tabular}
\caption{\label{tab:enrico} GW frequency, merging time and number of DD systems with the suitable properties required to reproduce the observed Galactic SN~Ia rate for some example values of the maximum orbital period. The largest period ($\sim$11 hours) corresponds to a merging time equal to the Hubble time.}
\end{table}

As shown in Table \ref{tab:enrico}, for $f_{\rm GW,l}$=0.01\,Hz, which corresponds to a maximum merging time $t_{\rm m,h}\sim 10^4$\,yr, $N_*=52\pm 12$. Therefore, to match the observed SN~Ia rate, at any given time the Galaxy needs to host 40-60 double-WDs with $M_1+M_2\gtrsim M_{\rm Ch}$ orbiting with a period of $\lesssim$200\,s. If the number of DD systems found in the frequency range 0.01--0.1\,Hz is not enough to account for the observed rate, either SD contributes in a significant way to the census of progenitor systems, or SN~Ia can originate also from smaller mass systems. These have longer merging times and hence one would need an even higher number of systems in the considered frequency range. On the other hand, they may remain undetected because the intensity of their GW emission is lower than for more massive systems. We notice that whereas the sub-Chandrasekhar scenario was not the favoured option to explain the majority of SN~Ia \cite{mazzali}, in recent years it gained support also in accounting for the observed SN~Ia diversity, and it is considered as one of the possible scenarios \cite{Hillebrandt2013,flors}. The simplified approach outlined above is meant only to illustrate the basic concept. In order to achieve an accurate prediction of the SN rate to be compared with the observed value, one needs also to consider the actual distribution of WD masses, the GW strain at a given frequency and the spatial distribution of sources.

The calculations presented above refer to possible detections within our own Galaxy. However, Fig.~\ref{fig:horizon_dwd} shows a detection horizon for DWD binaries enabling us to detect DWD systems on the verge of merging in nearby galaxies. Even if the actual horizon limit depends on the details of the merging process, which is complex and not fully understood, it is possible to estimate a lower limit on the rate of DWD mergers detectable by LGWA from the local SN~Ia rate, in the assumption that SN~Ia originate from DWD binaries.  

\changes{To derive the rate of SN Ia events in the Local Universe, we use the SN observed by ZTF and ATLAS from 2018 to 2023. These optical surveys continuously cover the sky and their nearby SN candidates are typically spectroscopically classified by programs such as PESSTO\cite{Smartt:2014rpa}. Collecting data from TNS \footnote{www.wis-tns.org} and the Bright Supernova Catalogue \footnote{www.rochesterastronomy.org}, the number of spectroscopically confirmed SN Ia as a function of distance results to be 4 within 10\,Mpc, 11 within 20\,Mpc, 104 within 50\,Mpc and 520 within 100\,Mpc resulting in the SN Ia rates per year  given in Table \ref{tab:snrate} (first row). These numbers has been corrected by a factor of about 1.5 considering the average coverage of 65$\%$ of the sky by the above surveys (second row of the Table).

The numbers thus estimated are conservative. A non-negligible fraction of SNe are not classified and could potentially increase the SN Ia number, and detection incompleteness can start to arise at large distance due to the limited surveys' sensitivity. However, both biases can be considered negligible within 20\,Mpc.}
\begin{table}[ht!]
\tabcolsep 6mm
\begin{tabular}{|c| c c c c|}
\hline
 & \multicolumn{4}{c|}{Maximum distance}\\
 & 10\,Mpc &  20\,Mpc&  50\,Mpc & 100\,Mpc \\
\hline
Observed rate & $0.7\pm0.3$ &$1.8\pm0.5$& $17.3\pm1.7$ &  $86.7\pm3.8$ \\
\hline
Corrected rate & $1\pm0.4$ & $2.7\pm 0.7$& $26\pm2.6$ & $130\pm5.7$ \\
\hline
\end{tabular}
\caption{The table shows the number of SNe~Ia per year. The first row gives the number per year of SN~Ia based on the observations of ZTF and ATLAS. The second row gives the rate corrected by the  effective sky areas of about 65\% surveyed by ZTF and ATLAS.}
\label{tab:snrate}
\end{table}

Considering a merging frequency of 0.1\,Hz for a 1+1 $M_\odot$ DWD system and the LGWA DWD horizon distance that continuously increases with observation time up to 10\,yr, we estimate a total number of about 20 DWD mergers producing SNe Ia possibly detectable by LGWA over the 10 years of mission lifetime. This number does not take into account that some SN Ia are not detected by ZTF/ATLAS due to obstruction of the EM radiation, but such systems would not escape detection in GWs.

These detections enable the interesting possibility of measuring DWD systems in galaxies that in the past hosted a SN Ia. In that case, host galaxy distances derived from the GW signal might be used as an independent probe to calibrate the SN distances for cosmology. It should also be noted that DWD mergers cannot produce a SN Ia when the total mass of the binary lies below the Chandrasekhar limit. This means that many more DWD mergers should be expected to be observable with LGWA than the number inferred from SN Ia observations, and those could be used equally well to calibrate SN distances.

\paragraph{Massive Black Hole Binaries and extreme/intermediate mass ratio inspirals --} 

Massive BHs inhabit the nucleus of many galaxies \citep{2013ARA&A..51..511K,2020ARA&A..58..257G}, including our own Milky Way, and `feedback' from AGN is one of the most important processes that determine the evolution of galaxies. Understanding the formation of such heavy objects remains the subject of open debate \citep{2021NatRP...3..732V}. Still, several channels have been proposed in the literature: objects formed with initially up to $\sim 10^3 M_\odot$ (the so-called \textit{light seeds}) can be the remnants of Population III stars \citep{Madau01,tay14}; otherwise, rarer and heavier objects ($10^4$--$10^5 M_\odot$, referred to as \textit{heavy seeds}) can form through the direct collapse of pristine gas possibly going through an intermediate phase of \textit{supermassive star} that eventually collapses into a MBH \citep{2003ApJ...596...34B,2006MNRAS.370..289B}. Furthermore, MBHs can be generated through runaway collisions of young stars formed into a very dense star cluster \citep{2004Natur.428..724P} or through repeated mergers of smaller stellar black holes again within very dense stellar environments \citep{2015MNRAS.454.3150G, 2002MNRAS.330..232C}. The latter scenario is of particular interest for the LGWA,  as the GW signal produced the repeated inspirals of black holes can be probed by future GW missions \citep{2021MNRAS.500.4628P};  (this aspect is further covered below when addressing EMRI/IMRI sources). After their formation, these objects are expected to grow through gas accretion, accretion of stellar objects and mergers with other MBHs to reach their present-day masses, correlating with the host galaxy masses ($M_{\rm BH}$--$\sigma$ relation, \citep{2013ARA&A..51..511K,2020ARA&A..58..257G}). The GW detection of these sources across a broad range of masses and redshifts would thus provide unprecedented constraints on their still debated origin and growth across cosmic time.

\noindent\underline{Massive Black hole binaries with nearly equal masses} -- 
\noindent MBH binaries can form as a result of galaxy-galaxy mergers, a very common process according to the hierarchical clustering paradigm predicted by the $\Lambda$-CDM cosmological model. Given many galaxies are observed to host quasars since the very early Universe, it is reasonable to expect that many MBH pairs have found themselves in the centre of the same galaxy merger remnant across the history of our Universe. If the separation between the  MBH pair shrinks down to separations $\ll1$ pc, GWs drive their final coalescence \citep{1980Natur.287..307B}. The GWs released in the process can be observed by current and upcoming GW detectors. In particular, LGWA will be mainly sensitive to merging MBHs in the mass range $10^3$--$10^6 M_\odot$. The detections of GWs emitted from these sources can provide robust constraints on the MBH merger rate and thus unprecedented information on the formation and evolution of MBHs over cosmic time. Additionally, combining GW detections and EM observations will provide insights on the origin of the $M_{\rm BH}$--$\sigma$ relation.

At the early stage of the galaxy merger, the two MBHs are initially separated by hundreds of kpc. A series of mechanisms are required to ensure their sinking to the GW driven regime. Initially, the dynamical friction against dark matter, stars and gas induce the orbital decay \citep{1943ApJ....97..255C}, even if large-scale torquing sources such as bars, spirals, massive clumps and other irregularities may randomize the inspiral timescale \cite{2020MNRAS.498.3601B, 2022MNRAS.512.3365B}. If the mass of the two MBHs (and merging galaxies) is not too dissimilar ($q = m_2/m_1 \sim 0.1-1$) the inspiral is likely to be completed within a timescale of the order of a Gyr, and the two MBHs find themselves bound in a pair. At this point, repeated three-body interactions with stars passing close to the binary, possibly combined with the presence of a gaseous circumbinary disk, are expected to sustain the shrinking all the way down to the GW driven regime in a timescale of a fraction of a Gyr \citep{2009ApJ...700.1952H}, although recent numerical studies found that the physics of gaseous dynamics around the MBH can have a non-trivial impact on the binary evolution (see e.g. \citep{2020ApJ...901...25D,2023MNRAS.522.2707S,2020ApJ...889..114M,2020ApJ...900...43T}). Note that, especially when the bound binary evolution is inefficient (e.g. if the properties of the gaseous disc do not induce negative torques on the binary, or the surrounding stellar population is scarce), another MBH can be brought onto the evolving binary by a further galaxy merger, inducing a triple interaction \citep{2018MNRAS.477.3910B}. The dynamical evolution of the triple system often results in the merger of at least two of the three involved MBHs \citep{2018MNRAS.477.3910B}.

Several studies have assessed the merger rates of MBH binaries as a function of the mass of the merging bodies and redshift \citep{2012MNRAS.423.2533B, Bonetti:2018tpf} employing semi-analytical models for the cosmic evolution of galaxies and MBHs. The obtained predictions can vary substantially depending on the selected black hole seed mass (light versus heavy seeds) and physical prescriptions implemented to describe the physics of the inspiral; the intrinsic number of MBH mergers is typically of order a few tens to a few hundreds per year, with light-seed models predicting a larger number of events dominated by mergers of MBHs with $<10^4 M_\odot$, while the heavy seed models tend to feature fewer events, mainly involving MBHs of $10^4-10^6 M_\odot$. The LISA GW detector (which is sensitive to larger masses and higher redshift, if compared with LGWA) will likely have its detection rates (again of tens to hundreds of events per year) dominated by mergers occurring at $z\sim 5$ \citep{Bonetti:2018tpf}; this suggests that the LGWA will be able to detect at least a few (tens) of events, and possibly have its detection rates dominated by lower-$z$ mergers bridging the gap. \changes{Possible observations of MBH mergers with LGWA in the mass range $<10^4$ M$_\odot$ would bring new insights about the rate of low-mass galaxy mergers and the possible collision of massive star clusters (see also Figure \ref{fig:horizon-et-lgwa-lisa}).}\\

\noindent \underline{Extreme and Intermediate Mass-Ratio Inspirals} --
\noindent Binaries that have largely disparate mass ratios are referred to as EMRIs or as IMRIs. In the literature, EMRIs are generally defined as sources with mass ratio $q<10^{-4}$, while IMRIs identify binaries having $q=10^{-4}-10^{-2}$. Given that LGWA would be sensitive to E/IMRIs occurring around MBHs with $\sim 10^2-10^5$, according to Fig~\ref{fig:horizon-et-lgwa-lisa}, many sources involving an MBH and a stellar compact object (WD, NS, or stellar BH) would fall into either of these categories. 

As shown in Sec.~\ref{sec:horizons}, LGWA is suited to detect binary mergers with component masses below $\sim 10^5 M_\odot$, generally referred to as IMBHs. The current debate about the existence and origin of IMBHs can be summarized with a question: are IMBHs the link between stellar BHs and SMBHs, or do they require a different formation channel from SMBHs?  If IMBHs are the precursor of SMBHs, they must form from the first stars. Little is known about their origin as well. If these stars can attain masses above $10^3$ M$_\odot$, the so-called supermassive stars (SMS), they likely undergo direct collapse owing to general relativistic instability (SMSs) (see \citep{woods19} for a review). However, recent numerical simulations predict that these ancient stars form with smaller masses, although this depends on the background ionization and accretion rate (see \citep{Klessen:2023qmc} and reference therein). Alternatively, SMSs can form through runaway collisions in star clusters, provided that its mass overcome the range affected by the PISN mechanism. The IMBH seed formed from the collapse of the SMS can further grow via accretion of stellar material or by merging with other BHs. Note that, however, the amount of matter accreted onto a BH during a star-BH interaction is unknown, and gravitational recoil prompted by asymmetric GW emission during BH mergers can eject the BH remnant from the cluster and hamper the development of further hierarchical mergers. Therefore it is important to obtain observational constraints for a wide range of MBHs, not only for stellar astrophysics but also for galaxy evolution. 
 
 This gives the mission a tantalizing advantage over milliHz detectors for probing lower mass systems such as the nuclei of dwarf galaxies or IMBHs in star clusters, for which observational evidence is severely limited (for recent reviews see \citep{2017IJMPD..2630021M,2020ARA&A..58..257G}). The few observational constraints about these elusive objects come from the detection of IMBHs in dwarf galaxies (see e.g. \citep{2015ApJ...813...82R,2018ApJ...863....1C}), the recent observation of GWs from merging BHs whose final remnants have masses $<200\Ms$, like GW190521 \citep{2020PhRvL.125j1102A},  as well as systems for which a stellar tidal disruption occurred, which temporarily illuminates the MBH's environment \citep{2020ApJ...898L..30M,2022ApJ...937....7S,2022NatAs...6...26R,2020ARA&A..58..257G,2021ApJ...918...46W}. Further observations of IMBH candidates in the $10^3-10^4\Ms$ mass range have been observed in galactic and extragalactic globular clusters, but all such observations are controversial and widely debated \citep{2010ApJ...712L...1I,2010ApJ...719L..60N,2010ApJ...710.1032A,2012ApJ...750L..27S,2013A&A...552A..49L,2013ApJ...769..107L,2010ApJ...710.1063V,2017Natur.542..203K,2017MNRAS.468.2114P,2019ApJ...875....1M,2018ApJ...862...16T,2018NatAs...2..656L,2019ApJ...884L...9A,2022MNRAS.514..806V,2022RAA....22k5007W,2023MNRAS.522.5740V}. 
 As seen in Fig~\ref{fig:horizon-et-lgwa-lisa}, detection of an IMRI merger with a primary of $10^5 M_{\odot}$ is possible, albeit at relatively low redshift ($z\lesssim 10^{-1}$).

The formation of binaries containing IMBHs may occur via several pathways, such as within dense stellar environments like dense young massive, globular, or nuclear clusters, or in the accretion disc of active galactic nuclei (see e.g. \citep{2002ApJ...576..899P,2002MNRAS.330..232C,2009A&A...497..255G,2012MNRAS.425..460M,2015MNRAS.454.3150G,2021MNRAS.507.5132D,2021MNRAS.501.5257R,2023arXiv230704807A}). These events may be a result of 
 \begin{itemize}
\item[(i)] repeated collisions among massive stars occurring in the core of star clusters that build up a very massive star which ultimately collapses to an IMBH;
\item[(ii)] accretion of stellar material onto stellar-size BHs;
\item[(iii)] multiple stellar BH mergers (hierarchical mergers).
\end{itemize} 
In the bright AGN systems for which a radiatively efficient accretion model is invoked to explain the emission, the gas configuration is expected to be dense enough to influence the orbits of surrounding stars and BHs \citep{1991MNRAS.250..505S,2021PhRvD.103j3018P,2020ApJ...898...25T}. Such disks may also become vulnerable to fragmentation in certain conditions, in which case they form stars and subsequent BHs in situ \citep{2007MNRAS.374..515L,2003MNRAS.339..937G,2004ApJ...608..108G,2023MNRAS.521.4522D}. Accretion disks in AGN are considered potential factories for IMBH mergers and light IMRIs given that the gas-rich environment can facilitate the interaction between BHs, IMBHs, and MBHs, in addition to providing them material to accrete and grow. In these environments the formation mechanisms of each of these events are intimately connected, and LGWA will fill in a piece of the puzzle with detections of IMBH interaction. This will be complementary to milliHz space-based detectors, which will be sensitive to more extreme mass ratio events (EMRIs), while potentially missing out on the lower mass binaries which could be the building blocks (see e.g. \citep{2012MNRAS.425..460M,2016ApJ...819L..17B}). 

Recent analysis of the latest LVK detections suggest that at least a fraction of detected BH-BH mergers may occur within dense stellar environments or AGN (e.g., \cite{2020A&A...638A.119G}, although see \cite{2023MNRAS.526.6031V}). The motivation for this is to explain events with particularly high masses, residual evidence of eccentricity \citep{2022Natur.603..237S,2021ApJ...921L..43Z}, spin alignment vs. antialignment \citep{2022ApJ...928L...1L} and events which have associated EM counterparts \citep{2023PhRvD.108l3039M,2023AAS...24132603G}, as variations of each of these may be a result of either dynamically-driven or gas-driven encounters.  The AGN-assisted formation channels may skew towards high mass ratio mergers \citep{2022PhRvD.105f3006L}, strengthening the case for IMRI production in these environments. LGWA will provide valuable information on the earlier stages of these systems, as well new insight into the more massive systems which may be natural by-products of hierarchical and gas-assisted evolutionary pathways. The detection of IMRIs involving IMBHs by LGWA could thus help to shed light on the formation channel of these elusive objects. In particular, the IMBH  formation, and related probability, intrinsically depends on the IMBH formation channels discussed above.

So far, we have discussed EMRIs and ``light IMRIs'' involving a stellar object as a secondary. Another class of IMRIs is often identified, involving an IMBH and a SMBH residing in a galactic nucleus \citep{2001ApJ...562L..19E,2005ApJ...618..426M,2006ApJ...641..319P,2018MNRAS.477.4423A,2019MNRAS.483..152A,2022ApJ...939...97F}. Several formation scenarios can contribute to the formation of these heavy IMRIs, like galaxy mergers, star cluster orbital segregation, and migration in AGN discs. The first channel assumes that an IMBH-SMBH binary can form in consequence of a minor merger. In the second scenario, massive and dense star clusters forming in the inner region of the galaxy nurture the formation of an IMBH while being dragged by dynamical friction toward the galaxy centre. Eventually, the infalling cluster disrupts and its IMBH librates in the galactic nucleus, further spiraling in via dynamical friction as well and finally pairing with the galactic SMBH. In the third scenario, an IMBH forms in an AGN disc, possibly via a combination of processes like migration trap dynamics and repeated mergers (see previous paragraphs) and slowly migrates inward via gaseous dynamical friction, eventually binding to the SMBH. 

To build a comprehensive picture of MBH, IBMH, and BH interactions across the mass spectrum, LGWA presents a promising case to bridge the gap between the SMBH sources detectable in the milliHz regime (with LISA or TianQin) and high frequency sources that will be detected with increasing sensitivity with next-generation terrestrial detectors. In this gap lie the early inspirals of stellar-origin BH binaries, the mergers of IMBHs, and the potential interplay between the two (IMRIs), all of which may form in dense stellar environments and in galactic nuclei. Each scenario has peculiarities in terms of occurrence frequency or IMBH mass spectrum, thus the future detection of GWs emitted by both light and heavy IMRIs will represent a key to unravelling the true nature of IMBHs. 

\paragraph{Tidal disruption events --}
\label{sec:tde_formchan}
As discussed in Sec.~\ref{sec:GWAstro}, the predominant and ubiquitous generation mechanism for TDEs is believed to be two-body relaxation: stars in the vicinity of the central massive black hole tug each other's orbit, so that the trajectory of an unlucky star may eventually attain a very small pericentre and get disrupted (e.g. \citep{2020SSRv..216...35S}). Relying on the so-called \textit{loss-cone theory}, theoretical models predict typical TDE rates between $10^{-6}-10^{-3}$ events per galaxy per year \citep{2004ApJ...600..149W, 2016MNRAS.455..859S}.

The accumulation of TDE observations over the last few years has made it possible to compare those predicted rates with the observed ones and even estimating rates as a function of the black hole and galaxy mass \citep{vanVelzen18, Yao23}, with reasonable agreement with theoretical models (at least at the order-of-magnitude level). A peculiar aspect of the host galaxies of TDEs is the fact that post-starburst systems (which underwent a substantial episode of star formation in the last $\sim 1$ Gyr) appear to be overrepresented by a factor of a few tens \citep{2016ApJ...818L..21F,2021ApJ...908L..20H}.

LGWA is going to predominately observe TDEs occurring about relatively low-mass MBHs ($<10^6 M_\odot$), likely inhabiting dwarf galaxies or globular clusters in the local Universe \citep{2020ARA&A..58..257G}. This poses a series of challenges for the proper modelling of the rates, as the occupation fraction of MBHs in those low-mass systems remains debated \cite{2021MNRAS.507.3246H}, as well as their innermost stellar density profile (which remains virtually unresolvable at the scale of a fraction of the MBH influence radius \cite{2023MNRAS.520.4664H}, where most of TDEs are expected to be generated \cite{2022MNRAS.511.2885B}). Although earlier works have suggested TDE rates would be larger about low mass MBHs \cite{2004ApJ...600..149W, 2016ApJ...820..129B}, this idea has been recently revised and it seems instead that rates about low-mass MBHs drop significantly over time after an initial burst \cite{2022MNRAS.511.2885B,2022MNRAS.514.3270B}. Overall, the event rates about low-mass MBHs in the local universe remain under close investigation, and upcoming surveys as LSST/\textit{Vera Rubin Observatory} as well as ULTRASAT will greatly improve our understanding of this in the next few years, thus allowing to refine the science we expect to obtain from the LGWA.

\paragraph{Gravitational waves from asymmetrical neutron stars --}
Rotating neutron stars emit continuous GWs if they present some sort of non-axisymmetry \citep{Lasky_2015,Riles:2022wwz}. The simplest form of this is an equatorial ellipticity $\varepsilon$, defined as $\varepsilon=(I_{xx}-I_{yy})/I_{zz}$ with $I_{ij}$ being the moment of inertia tensor of the star and the rotation axis being in the $z$ direction. The GW signal is nearly monochromatic at a frequency $f$ equal to twice the rotation frequency of the star, and its amplitude $h_0$ at distance $d$ from the source is
\begin{equation}
\label{eq:h0}
h_0 = \frac{{4\pi^2 G}}{{c^4}} \frac{{\varepsilon I_{zz} f^2}}{{d}}\ .
\end{equation}
For neutron stars observed as pulsars, the frequency and frequency derivatives are measured and the GW amplitude $h_0^{spdwn}$ that could account for the observed kinetic energy loss (i.e for the observed spin-down) can be calculated as
\begin{equation}
\label{eq:h0spdwn}
h_0^{spdwn} = \dfrac{1}{d}\sqrt{\dfrac{5GI}{2c^3}\dfrac{|\dot{f}|}{f}}\ .
\end{equation}
The spin-down amplitude is the maximum GW signal amplitude that one can reasonably expect from a star rotating at frequency $f$ and with a spindown $\dot{f}$, and it is an important benchmark because if the sensitivity of a search does not at least reach this level, it could be argued that such search does not stand a chance of detecting a signal. 

\begin{figure}[ht!]
\centering
\includegraphics[width=0.49\textwidth]{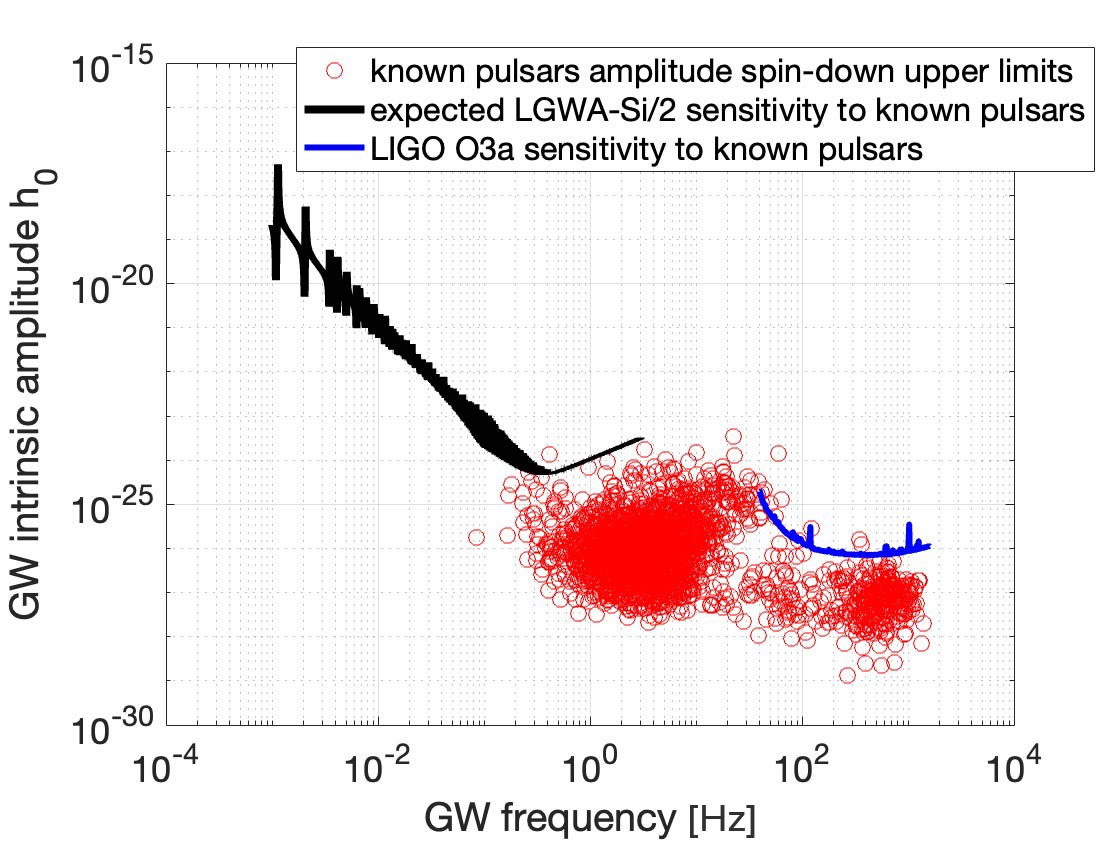}
\includegraphics[width=0.49\textwidth]{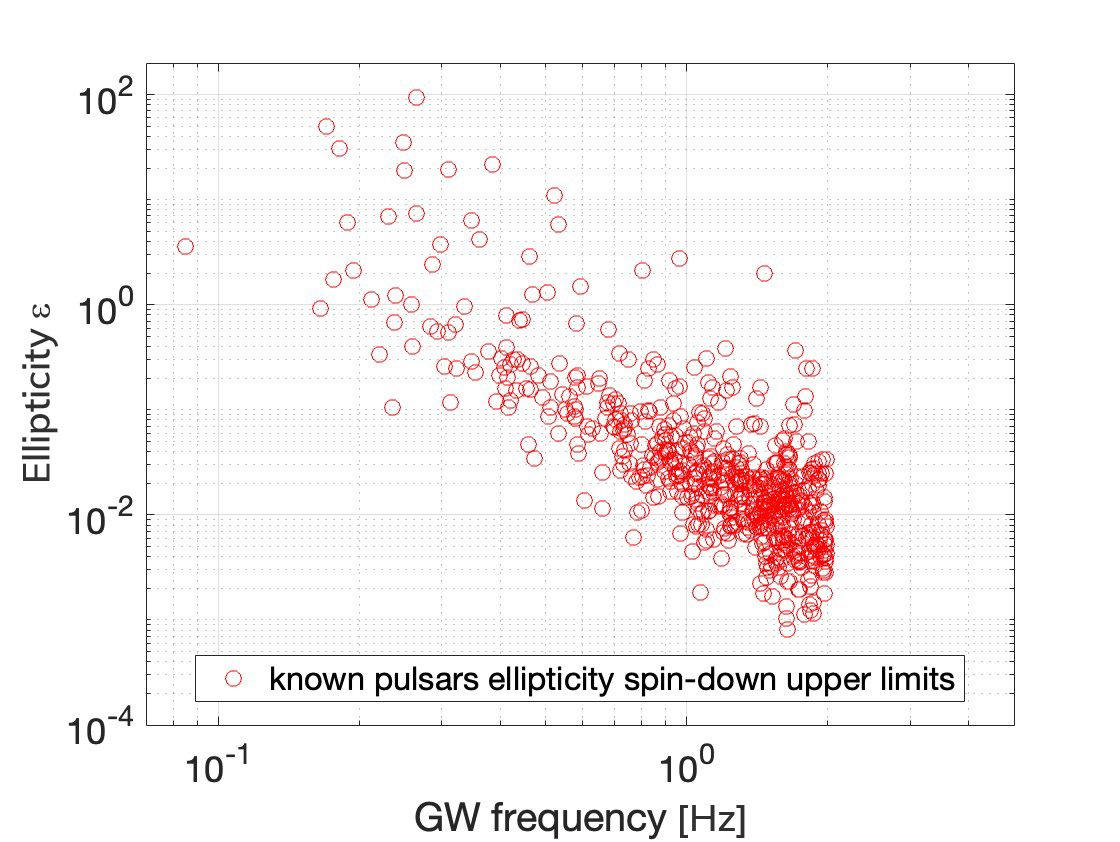}
 \caption{Lines: smallest detectable signal from known pulsars assuming 1 year of observation time. Circles: spin-down upper limits for known pulsars. The left plot shows the intrinsic GW amplitude, whereas the right plot the ellipticity associated with the spin-down limits.} 
 \label{fig:knownPulsarsSensitivity}
\end{figure} 

Searches for emissions by pulsars with positions and frequency evolution known from radio observations are the most sensitive among all GW searches. In order to gauge the performance of LGWA to continuous GWs, this is a good place to start. The left panel of Fig.~\ref{fig:knownPulsarsSensitivity} shows that the bulk of the known pulsars emitting GWs in the LGWA frequency band are emitting below the smallest signal detectable by LGWA. The latter is estimated by assuming a search over 1 year of data with a detector having the power spectral density of the Si configuration divided by 4 (see Sec.~\ref{sec:sensModels}). Furthermore, the spin-down ellipticities of the known pulsars rotating with frequency above 0.05\,Hz (GW frequency 0.1\,Hz) are all greater than $10^{-4}$, as shown in the lower panel of Fig.~\ref{fig:knownPulsarsSensitivity}. Since the maximum ellipticity that a neutron star crust can sustain before breaking is likely not greater than $10^{-5}$ \citep{Johnson-McDaniel:2012wbj,Gittins:2021zpv,Gittins:2020mll,Morales:2022wxs}, emission at the spin-down level for these objects appears somewhat unlikely setting stronger requirements on the sensitivity to observe their GW emissions.

Assuming a factor of $\approx 2$ better sensitivity, corresponding to searching about a decade of data corresponding to the targeted lifetime of LGWA, the number of detectable pulsars -- i.e., with spin-down amplitude larger than the expected sensitivity level -- grows to about 10, but all with very large ellipticities. 

The low frequency range is very challenging for the detection of continuous GWs because, for a given physical scenario, say a given ellipticity value, the GW amplitude scales with a fairly high power of the frequency ($\propto f^2$ for deformations, from 
Eq.~\eqref{eq:h0}, and $\propto f^3$for r-mode emission), making the signal much weaker than at higher frequencies. Whereas signals at low frequencies are emitted by neutron stars with a broader range of magnetic field values than those emitting in the band of terrestrial detectors \citep{Pagliaro:2023bvi}, the prospects for detection would significantly increase with a sensitivity $\approx$ 20 times larger. Relative to the results shown in Fig.~\ref{fig:knownPulsarsSensitivity}, another factor of 3 in sensitivity can be achieved by considering a 10-year observation time with LGWA corresponding to its targeted mission lifetime, but observation of GW emissions from spinning neutron stars remains unlikely.

Massive magnetars are young and highly magnetized neutron stars with an unknown origin. The newly formed massive magnetars spin down mostly through magnetically induced GW emission and magnetic dipole radiation. The evolution equation for a magnetar's angular frequency can be written as
\begin{equation}
   \dot{f} = - \frac{2 \pi^2 B_d^2 R^6 f^3}{3Ic^3} - \frac{512  \pi^4 G \epsilon^2 I f^5}{5c^5}\ .
\end{equation}

The dipole magnetic field of the magnetar at the pole is $B_d$. Different interior magnetic field configurations, such as poloidal-dominated, toroidal-dominated, and twisted-torus configurations, indicate various relationships between the magnetic field and ellipticity \citep{chowdhury:uni21}. Detectors will probe various configured models, and they will be able to explore the parameter space in these models.

\subsubsection{GW cosmology}
\label{sec:cosmology}
{\it Main contributors:} Francesco Iacovelli, Enis Belgacem, Marica Branchesi, Stefano Foffa, Arun Kenath, Michele Maggiore, Michele Mancarella, Suvodip Mukherjee, Niccolò Muttoni, Masroor C. Pookkillath, Alberto Roper Pol, Sourav Roy Chowdhury \\

In the last few decades, cosmology has been greatly enhanced due to a large number of precise observations, among which the EM ones have played  a major role (refer to e.g. \cite{turner2022road} for a recent review). Nevertheless, since the very first GW detection GW150914 \citep{AbEA2016a}, we got access to a new and uncharted observable that can help to shed light on the expansion history of the Universe by providing both information complementary to EM probes (e.g. \cite{Schutz:1986gp,Holz:2005df,MacLeod:2007jd,CuHo2009,Messenger:2011gi,DelPozzo:2011vcw,Taylor:2012db,Chen:2017rfc,Mukherjee:2018ebj, Feeney:2018mkj,LIGOScientific:2018gmd,DES:2019ccw,LIGOScientific:2019zcs,DES:2020nay,Finke:2021aom,LIGOScientific:2021aug,mastrogiovanni2022cosmology,Nair:2018ign,Mukherjee:2019wcg,Mukherjee:2020hyn,Yu:2020vyy,Vijaykumar:2020pzn,Borhanian:2020vyr,Diaz:2021pem,Gray:2021sew,Ghosh:2022muc,Nimonkar:2023pyt}) and completely new potential observables \cite{Finke:2021aom,Mukherjee:2021rtw, Ezquiaga:2021ayr,Mancarella:2021ecn,Karathanasis:2022rtr,Baker:2022eiz,Leyde:2022orh}. Indeed, even if the current analyses of GWs mainly focused on the estimation of the present-day value of the Hubble parameter, $H_0$, eventually a more interesting set of observables will be the ones related to DE or modifications of GR at cosmological scales. At the background level, a non-trivial EoS for DE can give rise to deviations from the standard $\Lambda$CDM model, while in modified gravity theories perturbations can be different. In particular, tensor perturbations, i.e. GWs, can have a different propagation equation with respect to the one predicted by GR, in such a way that both the velocity of GWs can be different from the speed of light (even though the observation of GW170817+GRB170817A \cite{LIGOScientific:2017zic} put very stringent constraints on modified gravity models), and the `friction term' in the equation of motion for the tensor modes can vary, producing a difference between the luminosity distance estimated through GWs and the standard luminosity distance \cite{Belgacem:2017ihm,Belgacem:2018lbp}. Although with GW observations it might be difficult to improve on the accuracy of electromagnetic experiments for the estimation of the DE EoS (see, e.g. \cite{Belgacem:2019tbw}), their complete independence from other measurement systematics makes GWs an extremely valuable new cosmological probe to solve tensions on  cosmological parameters found by the current EM observations. Furthermore, the modified propagation can be accessed only through GW measurements, which are thus fundamental to constrain this kind of potential deviation from GR (that can result in effects even larger than the one associated to a non-trivial DE EoS).\par\medskip 

In general, a GW signal emitted by a coalescing binary system, differently from EM cosmological probes, carries information on the luminosity distance to the source, which, combined with knowledge of its redshift, can be used to constrain the expansion history of the Universe through the $d_L-z$ relation, which in a flat $\Lambda$CDM model reads
\begin{equation}
d_L(z) = (1+z) \dfrac{c}{H_0} \int_0^z \dfrac{{\rm d}\tilde{z}}{\sqrt{\Omega_{m,0}(1+\tilde{z})^3 + \Omega_{r,0}(1+\tilde{z})^4 + \rho_{\rm DE}(\tilde{z})/\rho_{0}}}\,,
\end{equation}
where $\Omega_{m,0} $ and $\Omega_{r,0}$ denote the present-day values of the density fractions for matter and radiation, respectively, while $\rho_{\rm DE}(z)$ is the dark energy density as a function of redshift and $\rho_{0} = 3H_0^2/(8\pi G)$. The crucial aspect in the GW context is then obtaining information on the sources' redshift. In the best-case scenario, the GW event is followed by a detectable EM counterpart which leads to a precise estimate of $z$, as we will discuss in the following. In case a counterpart is not available, GW detections can still be used for cosmological purposes through statistical methods. As first proposed in \cite{Schutz:1986gp}, it is possible to statistically obtain redshift measurements from potential host galaxies falling within the observed GW's localization error volumes exploiting galaxy catalogs. Also, a reconstruction of the intrinsic source-frame mass distribution of the observed mergers, in particular in the presence of sharp features, can lead to a statistical determination of the sources' redshifts, being the observed quantities in the detector-frame redshifted. Several recent studies have successfully tested these methods and their combination, as well as other similar ones, with both real data \cite{LIGOScientific:2018gmd, Finke:2021aom,Gray:2021sew,DES:2019ccw,DES:2020nay,LIGOScientific:2019zcs,LIGOScientific:2021aug, Karathanasis:2022rtr} and simulations \cite{DelPozzo:2011vcw,Chen:2017rfc,Gray:2019ksv,Leandro:2021qlc,Gair:2022zsa,Ezquiaga:2022zkx,Muttoni:2023prw,Mastrogiovanni:2023emh,Gray:2023wgj,Borghi:2023opd}. For the former method to be informative, detections with narrow error volumes, both in the angular and distance components, are extremely valuable. Another independent approach to measure the expansion history of the Universe as well as the GW bias parameter is using the cross-correlation technique, which can measure the spatial clustering of GW sources with the galaxies \cite{Oguri:2016dgk, Nair:2018ign, Mukherjee:2018ebj, Mukherjee:2020hyn, Calore:2020bpd, Mukherjee:2020mha, Libanore:2020fim, Scelfo:2020jyw, Diaz:2021pem, Afroz:2023ndy}. This method can further explore weak gravitational lensing \cite{Mukherjee:2019wcg, Mukherjee:2019wfw, Balaudo:2022znx} and provide tests of GR modifications at cosmological scales in a model-independent way \cite{Mukherjee:2020mha}. Other methods to employ GW detections for cosmological purposes include, e.g., exploiting multiple images of strongly lensed events \cite{Hannuksela:2020xor,Finke:2021znb,Narola:2023viz}; having prior knowledge on the redshift distribution of the observed sources \cite{Ding:2018zrk,Ye:2021klk}; using tidal distortions of NSs combined with EoS knowledge to extract information on the source-frame masses \cite{Messenger:2011gi,Messenger:2013fya,Dhani:2022ulg}; or measuring the angular barion acoustic oscillation scale from a high number of well localized events \cite{Kumar:2022wvh}.

Already from the about 90 detections in the first three observing runs of the current GW interferometers, the potential of GWs for cosmology has been assessed through different techniques. From the observation of GW170817 and its EM counterpart, i.e. a `bright siren' for which the distance is known from the GW signal and redshift from the host galaxy identification, it has been possible to obtain a few percent measurement of $H_0$ \cite{LIGOScientific:2017adf}. The estimate was $H_{0}=70\substack{+12\\-8}~{\rm km\, s}^{-1}\,{\rm Mpc}^{-1}$ with the dominant error contribution being the degeneracy in the GW signal between the source distance and the observing angle. Breaking the degeneracy using precise measurement on the observing angle obtained with very large baseline interferometry (VLBI) high angular resolution imaging of the radio counterpart of GW170817, the Hubble constant measurement improved to $H_{0}=70.3\substack{+5.3\\-5.0}~{\rm km\, s}^{-1}\,{\rm Mpc}^{-1}$ \cite{Hotokezaka:2018dfi} (see also \cite{Mukherjee:2019qmm} where using a different treatment for the peculiar velocity of the host galaxy it has been estimated $H_{0}=68.3\substack{+4.6\\-4.5}~{\rm km\, s}^{-1}\,{\rm Mpc}^{-1}$ and $H_{0}=68.3\substack{+12\\-8}~{\rm km\, s}^{-1}\,{\rm Mpc}^{-1}$ with and without exploiting the VLBI information, respectively).

For GW-only events, the so-called `dark sirens' without an EM counterpart and the host galaxy identification, statistical approaches can be adopted. Despite the low statistics, it has also been possible to get an interesting accuracy on the local expansion rate of the Universe \cite{LIGOScientific:2018gmd,DES:2019ccw,LIGOScientific:2019zcs,DES:2020nay,Finke:2021aom,LIGOScientific:2021aug}, as well as other modified gravity observables \cite{Finke:2021aom,Ezquiaga:2021ayr,Leyde:2022orh,Mancarella:2021ecn}. As an example, from the binary black-hole merger GW170814 in combination with the Dark Energy Survey it has been possible to set the constraint $H_{0}=75\substack{+40\\-32}~{\rm km\, s}^{-1}\,{\rm Mpc}^{-1}$ \cite{DES:2019ccw}. \changes{Using the sources in the GWTC-3 catalog, the LVK Collaboration obtained a constraint of $H_{0}=50\substack{+37\\-30}~{\rm km\, s}^{-1}\,{\rm Mpc}^{-1}$ adopting the statistical method relying on the source-frame mass distribution reconstruction and a \textsc{Power-Law + Peak} profile, while a tighter constraint of $H_{0}=67\substack{+13\\-12}~{\rm km\, s}^{-1}\,{\rm Mpc}^{-1}$ has been obtained from the same sources with the statistical host identification technique~\cite{LIGOScientific:2021aug}.\footnote{Notice that this value strongly depends on the assumptions for the BBH mass distribution.} Combining with the posterior obtained for GW170817 and its associated counterpart, these constraints shrink to $H_{0}=68\substack{+12\\-8}~{\rm km\, s}^{-1}\,{\rm Mpc}^{-1}$ and $H_{0}=68\substack{+8\\-6}~{\rm km\, s}^{-1}\,{\rm Mpc}^{-1}$, respectively. A more recent result obtained in~\cite{Gray:2023wgj} from the combination of these two techniques applied to the GWTC-3 events is $H_{0}=69\substack{+12\\-7}~{\rm km\, s}^{-1}\,{\rm Mpc}^{-1}$.}

In the coming decades, we expect GWs to thrive in the context of cosmology, thanks to the increasing number of detections and a new generation of GW detectors, which on one hand aim to improve the reach and sensitivity of the current facilities, and on the other hand seek to open the door to new frequency bands. Being sensitive to GWs in the dHz band, a frequency region where both resolvable sources and stochastic backgrounds are expected to be observed, LGWA will be a valuable addition to the GW landscape.

In particular, in the bright siren case, we expect EM counterparts from CBCs in the Hz-kHz band to be produced by mergers involving at least one deformable object, such as a neutron star, thus coalescences at $\gtrsim{\cal O}({\rm kHz})$ frequencies, falling outside the LGWA band. However, LGWA is able to detect the early inspiral of such systems and can be crucial to localize the source in the sky with good accuracy, hence improving the chances of counterpart identification (see Sec.~\ref{sec:multiband_gw_obs}). Even though the merit of joint GW+EM detections for cosmology is particularly high, systems with an associated counterpart are expected to be rare, with ${\cal O}(1)$ forecasted BNS multimessenger detections per year with LGWA (see Sec.~\ref{sec:horizons}). \changes{It should be noted that, even though the rate of multimessenger BNS events with LGWA is expected to be considerably lower than what will be achievable with 3G ground-based detectors, with tens to hundreds of expected multimessenger detections per year (see e.g.~\cite{Ronchini:2022gwk,Branchesi:2023mws}), the improved chances of counterpart identification thanks to the good LGWA localization can be valuable and will raise the number of bright siren events. Moreover, multiband observations of this kind of systems with LGWA and 3G detectors could be a powerful tool to study deviations from GR at the level of propagation of GWs through the Universe~\cite{Baker:2022eiz}, with the possible counterpart identification adding even more information as compared to the BBH case.}

In the dark siren case instead, at least for some of the coalescences falling in the band of terrestrial detectors, in particular at low redshifts, LGWA can play a crucial role, significantly shrinking the sky localization area thanks to the long time spent by the sources in the dHz detector band, and thus the huge effective baseline (see Sec.~\ref{sec:multiband_gw_obs}). Combined with the accuracy in particular third generation terrestrial interferometers can have in the reconstruction of the luminosity distance, such detections can be a game changer for dark-siren GW cosmology, reducing the number of potential host galaxies in such small error volumes to only a few, and thus acting effectively almost as bright sirens. Moreover, the detections of MBH and IMBH mergers that occur in the LGWA band could be used for both the aforementioned methodologies, increasing the statistics and thus the constraining power, with the statistical uncertainty scaling as $\propto 1/\sqrt{N_{\rm events}}$. Another interesting opportunity offered by these massive sources is the possible detection of EM radiation preceding or accompanying the gravitational signal from the environment surrounding the compact objects, making them bright siren candidates highly valuable for cosmology (see Sec.~\ref{sec:MBHbinaries}). Finally, as shown in Fig.~\ref{fig:horizon_dwd}, LGWA will be able to detect coalescences of DWD and NSWD binaries up to $\sim\!100~{\rm Mpc}$. Thanks to the long time this kind of sources spend inspiralling in the detector's band, they can potentially be localized with very high accuracy, and provide a powerful tool to estimate the $H_0$ value in the local Universe, both as bright (see Sec.~\ref{sec:DWD-WDNS} for a discussion regarding the EM transients these systems can give rise to) and dark sirens in combination with galaxy catalogs if counterpart detections are not available. The estimation of all the other cosmological parameters impacting the $d_L-z$ relation at higher redshifts coming from the other classes of sources discussed above would greatly benefit from this low redshift measurement.

Another observable and yet elusive source for current detectors is a SGWB of cosmological origin (CGWB). This refers to tensor modes possibly produced by various physical processes happening in the early universe, including, e.g., phase transitions, inflation, cosmic strings or primordial black holes (see e.g. \cite{Maggiore:1999vm,Caprini:2018mtu,Maggiore:2018sht} for reviews). It is, in a sense, the GW equivalent of the Cosmic Microwave Background EM radiation, whose discovery in 1965 has been one of the most relevant in cosmology. The detection of a CGWB would also bring outstanding scientific value, providing direct information about our Universe in its very first instants of evolution that would not be accessible by other means.\footnote{\changes{As studied for example in~\cite{Lehoucq:2023zlt}, an astrophysical GW background generated by stellar origin CBCs might dominate the SGWB also in the decihertz band. Sophisticatd methods will have to be applied to reduce its contribution for CGWB observations \cite{CuHa2006,SmTh2018,ShHa2020}.}} The characteristic frequency $f_0$ of a CGWB at present time is determined by the Hubble rate at the time of generation of the background, redshifted to the present time \cite{Caprini:2018mtu}
\begin{equation}
    f_0 = H_* \frac{a_*}{a_0} \simeq 1.65 \times 10^{-7} {\rm \, Hz} \, \frac{T_*}{{\rm GeV}} \left(\frac{g_*}{100} \right)^{{1}/{6}} \ ,
\end{equation}
where $T_*$ and $g_*$ correspond to the temperature scale and number of relativistic degrees of freedom at the time of generation, and $a_*$ and $a_0$ to the scale factors at the time of generation and today, respectively. In particular, if the source is causal and short, e.g., phase transitions, the resulting CGWB will peak at a frequency $f_p \sim f_0/(R_* H_*)$, where $R_*$ is the characteristic scale of the source, given as a fraction of the Hubble scale, $R_* H_* \leq 1$. Hence, LGWA allows to explore e.g. phase transitions occurring in the early universe at temperature scales up to $T_* \sim {\cal O} (10^6)$~GeV.\footnote{Signals sourced by phase transitions at high temperatures $T_* \gtrsim {\cal O}(10)$~GeV may experience a different redshift, see, e.g., \cite{Kolesova:2023mno}.}

\begin{figure}[t]
    \centering
    \includegraphics[width=.9\textwidth]{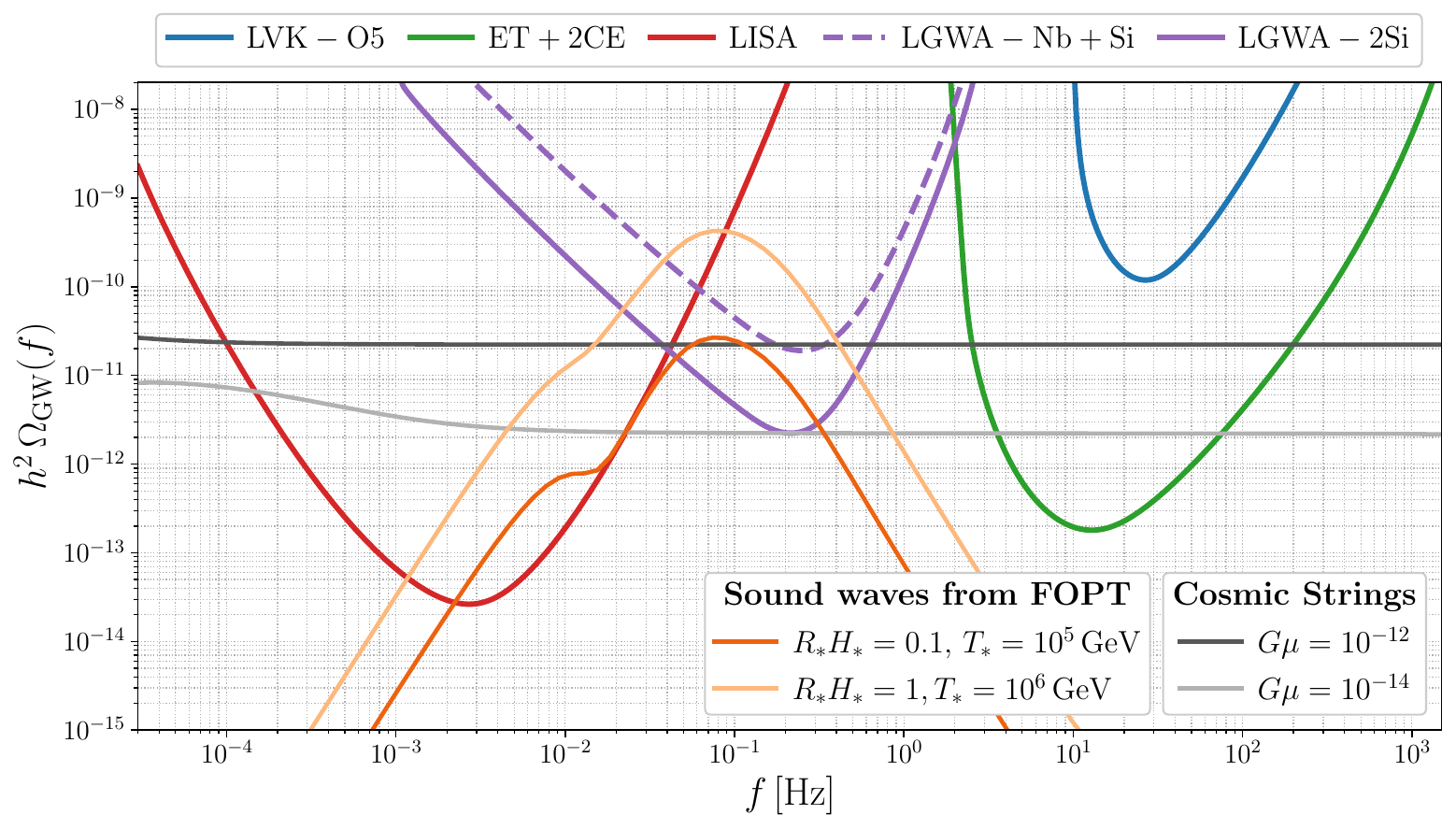}
    \caption{Power-law integrated sensitivity curves [normalised to $h=H_0/(100~{\rm km\,s}^{-1}{\rm Mpc}^{-1})$] for the LIGO-Virgo-KAGRA network with sensitivities representative of O5, ET+2CE network, LISA, and LGWA in two configurations considering two antipodal detectors. For LGWA we used sensitivities for a Nb detector plus an Si one, and two Si detectors, so to have a proxy of the output from correlation with a non-dedicated and a dedicated experiment, respectively. All the curves are obtained considering an observational time of $T_{\rm obs} = 1~{\rm yr}$ and an SNR of 1. We further report examples of CGWBs generated by sound waves produced during a first-order phase transition (with different characteristic scales and temperatures, such that $T_*/(R_* H_*) = 10^6$~GeV, with a phase transition strength parameter $\alpha=0.1$, and wall velocity $\xi_w=0.4$) \cite{RoperPol:2023dzg} and cosmic strings (with different string tension parameters $G\mu$) \cite{Sousa:2020sxs} observable at LGWA.}
    \label{fig:PLSvariousdets}
\end{figure}

In Fig.~\ref{fig:PLSvariousdets} we show examples of possible CGWB generated by two different possible sources: cosmic strings and first-order phase transitions. Cosmic strings are line-like topological defects formed in grand unified theories by symmetry breaking following a phase transition \cite{Vilenkin:2000jqa,Damour:2000wa,Jeannerot:2003qv} that can emit bursts of beamed gravitational radiation, whose incoherent superposition during the evolution of the Universe can give rise to a non-Gaussian SGWB \cite{Damour:2000wa} and can be used to probe particle physics beyond the Standard Model at energy scales not accessible by accelerators. We adopt the template developed in \cite{Sousa:2020sxs,LISACosmologyWorkingGroup:2022jok} with a loop-size parameter $\alpha=10^{-1}$ and different values for the string tension parameter $G\mu$ (related to the energy scale $\eta$ at which the strings are formed by $G\mu\sim10^{-6}[\eta/(10^{16}{\rm GeV})]^2$). In a first-order phase transition (FOPT), GWs can be produced by the collision of the broken-phase bubble walls \cite{Kosowsky:1991ua,Kosowsky:1992vn,Caprini:2007xq,Huber:2008hg,Jinno:2017fby,Cutting:2018tjt}, the production of sound waves \cite{Hindmarsh:2013xza,Hindmarsh:2015qta,Hindmarsh:2016lnk,Hindmarsh:2017gnf,Hindmarsh:2019phv,RoperPol:2023dzg,Sharma:2023mao}, and the development of magnetohydrodynamical turbulence in the primordial plasma \cite{Kamionkowski:1993fg,Kosowsky:2001xp,Caprini:2006jb,Gogoberidze:2007an,Caprini:2009yp,RoperPol:2019wvy,RoperPol:2021xnd,RoperPol:2022iel,Auclair:2022jod}. In Fig.~\ref{fig:PLSvariousdets}, we show the CGWB produced by sound waves, which is expected to produce the dominant signal from a FOPT \cite{Caprini:2015zlo,Caprini:2019egz,LISACosmologyWorkingGroup:2022jok}, based on the results of \cite{RoperPol:2023dzg} for a FOPT strength $\alpha = 0.1$ and a bubble wall velocity $\xi_w = 0.4$.

With various models and sources that predict different trends and amplitudes as a function of frequency, the LGWA sensitivity in the dHz band could be extremely helpful to perform a CGWB detection, filling the gap between terrestrial detectors and LISA. Already as a single detector (in which case the form of the signal has to be known in order to extract it from the noise), LGWA could have an interesting sensitivity to SGWBs, since the minimum detectable GW background amplitude as a single instrument for a given SNR scales as $\propto S_n(f) f^3$ (with $S_n(f)$ being the detector PSD, \cite{Maggiore:2007ulw}): comparing, for example, the sensitivity of a single LIGO instrument with the best forecasted O5 sensitivity at 300~Hz to LGWA--Si at 0.3~Hz we see that $[S_n(300~{\rm Hz}) (300~{\rm Hz})^3]_{\rm LIGO-O5} / [S_n(0.3~{\rm Hz}) (0.3~{\rm Hz})^3]_{\rm LGWA-Si} \approx 5\times10^4$.

Moreover, it could be possible to cross-correlate the output of LGWA with other non-dedicated experiments present on the Moon surface: their sensitivity could, of course, be lower than LGWA, yet the correlation of two or more detectors is much more promising for detectability, and no prior knowledge on the form of the signal is needed. In this context, it is also interesting to study the possibility of having a second antipodal detector consisting of a single seismometer with the same sensitivity of LGWA, which is a more optimistic scenario for SGWB searches. We show in Fig.~\ref{fig:PLSvariousdets} the power-law integrated sensitivity curves (introduced in \cite{Thrane2013}) computed as in App.~A of \cite{Branchesi:2023mws}, for the LIGO-Virgo-KAGRA detector network with sensitivities representative of the O5 run; ET in the reference triangular configuration with 10~km-long arms in a network with two CE detectors, one with 40~km-long arms and one with 20~km-long arms, both located in the U.S.; LISA (with the procedure outlined in \cite{Caprini:2019pxz});  LGWA always considering the reference 4-seismometer configuration with Si sensitivity on one pole and another single detector on the other, both with a lower Nb sensitivity and with the same Si one to simulate a non-dedicated and dedicated experiment, respectively.\footnote{Notice that, since in Fig.~\ref{fig:PLSvariousdets} we report a power-law integrated sensitivity curves for the full LIGO-Virgo-KAGRA network during O5, we do not observe the same $\sim{\cal O}(10^4)$ improvement reported in the text in the case of a single LGWA compared to a single LIGO.}
From Fig.~\ref{fig:PLSvariousdets} we can appreciate how LGWA could, e.g., improve on the LVK results for cosmic strings \cite{LIGOScientific:2021nrg} or access novel sources peaking in its frequency range.

Notice that, for antipodal detectors, in the frequency region covered by LGWA, the so-called overlap reduction function \cite{Fla1993} of the detectors has a nearly constant value of 1. Indeed, using the results in \cite{CoHa2014c, CoHa2014}, it is straightforward to see that, for two antipodal seismometers, the overlap reduction function $\gamma(f)$ at frequency $f$ reduces to $\gamma(f)=15 ~j_2(\Phi(f))/\Phi^2(f)$, where $\Phi(f)={4\pi f R}/{c}$ (with $R$ the Moon radius) and $j_2$ denotes the spherical Bessel function of the second order. Observe that $\Phi(f)$ is the phase difference between the two antipodal seismometers for a GW propagating along the line connecting them. In the frequency range relevant for LGWA (between $1$~mHz and a few Hz) the function $\gamma(f)$ is, to a very good approximation, constant and equal to 1.

\subsubsection{Fundamental physics with GWs}
\label{sec:fundamental}
{\it Main contributors:} Parameswaran Ajith, Chandrachur Chakraborty, Bradley J. Kavanagh, N. V, Krishnendu, Prayush Kumar, Andrea Maselli, Suvodip Mukherjee, R. Prasad, Vaishak Prasad\\ 

\paragraph{Parametrized tests of gravity --} Observations of low-frequency signals emitted by compact binaries evolving in the LGWA band represent golden sources for tests of gravity in the strong field regime and for searches of new fundamental physics \cite{Berti:2015itd,Barack:2018yly}. Among the variety of approaches developed to probe the existence of GR deviations, agnostic tests stand as a powerful and flexible tool, that has been extensively applied to actual observations \cite{LIGOScientific:2016lio,LIGOScientific:2021sio}. 

The ppE approach builds upon the parametrized post-Newtonian formalism \cite{Will:1971wt,Will:1971zzb,Will:1972zz, Nordtvedt:1972zz}, and introduces shifts in both the phase and the amplitude of the waveforms, which can be written in frequency domain as
\begin{equation}
h(f)=h_\textnormal{GR}(f)(1+\alpha u^{\beta})e^{i\delta u^\gamma}\ ,
\end{equation}
where $u\equiv(\pi {\cal M}f)^{2/3}$, ${\cal M}$ is the binary chirp mass, $f$ is the Fourier frequency, and $(\alpha,\beta,\gamma,\delta)$ are the ppE parameters \cite{Yunes:2009ke,Yunes:2013dva}. $\beta$  and $\zeta$ identify the class of modification introduced, while $\alpha$ and $\delta$ control the magnitude of the deviation, and can be constrained by data. Bounds (or measurements) of such parameters can be eventually mapped to fundamental couplings of specific theories of gravity, effectively reducing the parameter space possible in alternative theories of gravity \cite{Nair:2019iur,Perkins:2021mhb,Lyu:2022gdr}.

With agnostic tests being designed to modify the entire pN expansion of the waveform, LGWA would be particularly relevant to constrain the so-called pre-Newtonian (or \textit{negative} pN corrections)  effects, entering the signal before the quadrupolar contribution.  A notable example is given by the $-1$PN dipolar emission, that arises in many extensions of GR which include extra fields coupled to the gravity sector \cite{Barausse:2016eii,Cardoso:2023dwz}. Such negative pN terms are more effective at lower frequencies, i.e. when  systems orbit far from the coalescence \cite{Chamberlain:2017fjl,Gnocchi:2019jzp,Perkins:2020tra}. Indeed, constraints on agnostic parameters would benefit from the thousands of cycles accumulated by stellar mass binaries in the deci-Hz band, allowing LGWA to provide smoking gun signatures on the existence of such new fields. LGWA would also significantly improve measurements of parameters  that work at higher frequencies (\textit{positive} pN corrections), that are more effective in the band of terrestrial detectors. In this case multi-band observations between LGWA and a 3G detectors like ET or CE would strongly reduce correlations among the waveform  parameters, narrowing the bound on the possible GR shift.

The ppE formalism is flexible enough to incorporate a variety of effects deviating from the vanilla vacuum-GR scenario. Indeed, along with changes in the gravity sector, this approach has been recently exploited to describe environmental effects, i.e. changes in the waveform due to the presence of matter, with either a baryonic or a dark component, in the binary surroundings \cite{Barausse:2014tra,Eda:2013gg, Macedo:2013qea,Eda:2014kra,Edwards:2019tzf,Cardoso:2019rou,Hannuksela:2019vip,Yue:2019ndw,Kavanagh:2020cfn, Coogan:2021uqv, Cole:2022fir,Cole:2022ucw}. Environmental signatures typically affect the signal at very pre-Newtonian orders, and hence are best studied by low frequency instruments. In this regard, deci-Hz detectors are expected to provide constraints on non-vacuum spacetimes with exquisite precision \cite{Cardoso:2019rou}.

\paragraph{Testing the nature of black holes and other compact objects --}
Gravitational wave signals emitted by merging binaries provide a new, golden tool, to probe the nature of BHs \cite{Cardoso:2019rvt}, exploiting observations of the inspiral and the post postmerger phases of the coalescence, and studying different but complementary signatures within the GW spectrum. \phantom{a}\\

\noindent\underline{Quasi normal mode oscillations} -- 
\noindent
In GR, the inspiral and subsequent merger of two BHs will result in the formation of a distorted black hole that will settle down to an equilibrium state given by the Kerr metric. The gravitational radiation from the final stages can be described by a superposition of the QNMs of the Kerr spectrum \cite{Vish1970Sch, Chandra1975, Press:1971wr, Teukolsky1973a}. According to the no-hair theorem \cite{NoHair1967, NoHair1971} of GR, the oscillation frequencies and damping times of these QNMs can be characterized solely by the mass and spin of the remnant BH \cite{Kokkotas:1999bd,Ferrari:2007dd,Berti:2009kk}. Thus, detecting multiple modes will provide crucial information about the nature of the remnant, and enable interesting tests of GR~\cite{BHOvertonesGiesler2019, BHSpecBerti2006, BHSpecBrito2018, BHSpecBustillo2021, BHSpecCabero2020, BHSpecBerti2016, BHSpecCapano2020, BHSpecCarullo2019, BHSpecForteza2020, BHSpecOta2020, BHSpecSwetha2020, Bhagwat:2017tkm,Yi:2024elj}. LGWA will be able to measure the QNMs of massive BHs ($M \sim 10^2-10^4 M_\odot$) with high SNRs, thus enabling precision BH spectroscopy. Deviations from the predicted structure can be mapped to specific beyond GR models or exotic compact objects~\cite{Cano:2020cao,Cano:2021myl,Srivastava:2021imr,Wagle:2021tam,Pierini:2021jxd,Pierini:2022eim,Wagle:2023fwl,Carullo:2021dui,Maselli:2023khq,NoHairTestGossan2012,Meidam:2014jpa,Carullo:2018sfu,Maselli:2019mjd}. Additionally, multi-band observations between LGWA and terrestrial detectors will allow us to simultaneously measure the masses and spins of the BHs in the binary as well as that of the remnant BH. This will allow us to do interesting consistency tests between these estimates~\cite{Hughes:2004vw,Ghosh:2016qgn,Ghosh:2017gfp}. \\

\noindent\underline{Spin-induced multipole moments} -- 
\noindent Multipole moments provide yet another set of observables which can be exploited to test the nature of compact objects \cite{Krishnendu:2017shb,Raposo:2018xkf,Bena:2020see,Vaglio:2023lrd,Loutrel:2023boq}. They affect the GW emission, and for BHs in GR their imprint on the signal is  uniquely characterized by the BH mass and spin. However, other compact objects like neutron stars and boson stars, or BHs in alternative theories, will have spin-induced moments that depend on their additional properties (e.g., the equation of state of the object, or additional parameters in the theory). Measurements (or bounds) of these moments can be used to constrain the nature of compact objects or the theory of gravity~\cite{Krishnendu:2017shb}. The multipole structure of the compact object can also be affected by the presence of an external tidal gravitational field (e.g., produced by the binary companion). The corresponding deformations, quantified by the \emph{tidal Love numbers} contain imprints of the internal structure of the compact object~\cite{Flanagan:2007ix,Cardoso:2019rvt,Johnson-Mcdaniel:2018cdu}.  These effects are most accurately calculated for the inspiral part of the GW signal. By measuring the long inspiral signal from compact binaries, LGWA will be able to provide tight constraints on the nature of compact objects. 

\paragraph{Probing dark matter  --} The presence of over-dense `spikes' of particle DM around compact object binaries can lead to detectable effects on GW waveforms~\cite{Eda:2013gg,Eda:2014kra,Macedo:2013qea,Barausse:2014tra}. In particular, dynamical friction, the accretion of the DM particles, and the non-point-like gravitational potential of the spike can alter the rate of inspiral and therefore the GW phase compared to the vacuum case~\cite{Yue:2019ozq,Kavanagh:2020cfn,Becker:2021ivq,Speeney:2022ryg,Cardoso:2022whc,Nichols:2023ufs}.  While the effect is typically small, these effects can accumulate over many orbits, leading to a roughly percent-level change in the number of GW cycles observed while the system is in band~\cite{Kavanagh:2020cfn}. These dense DM spikes may arise from the adiabatic growth of astrophysical BHs at the centres of DM halos~\cite{Gondolo:1999ef,Ullio:2001fb,Sadeghian:2013laa}. Alternatively, if PBHs are produced in the early Universe~\cite{Green:2020jor}, they will naturally be surrounded by large overdensities of particle DM (assuming that the PBHs do not constitute all of the DM themselves)~\cite{Mack:2006gz,Adamek:2019gns,Boudaud:2021irr}. PBHs themselves might significantly contribute to dark matter \cite{bird2023pbh}, and they might have acted as seeds ot today's supermassive black holes \cite{kawasaki2012primseeds}, whose early presence in the Universe --- especially of those revealed by the latest JWST discoveries \cite{larson2023jwstbh} ---  challenges models of their formation.

Feedback effects are expected to disrupt the DM spike if the secondary compact object is too massive~\cite{Kavanagh:2018ggo,Kavanagh:2020cfn}, which means that a sufficiently large mass ratio is required for the spike to survive around the primary BH. However, DM spikes around supermassive BHs are likely to have been disrupted by the motion of stars and gas in their environments~\cite{Ullio:2001fb,Bertone:2005hw}. Intermediate-mass BHs instead are expected to live in more pristine environments~\cite{2020ARA&A..58..257G} where a possible DM spike is more likely to survive until today. This points towards intermediate mass ratio inspirals as promising targets, with typical systems having $m_1 \in [10^3, 10^5]\,M_\odot$ and $m_2 \in [1, 10]\,M_\odot$. Such systems give rise to GWs in the range $f_\mathrm{GW} \sim 0.1-1\,\mathrm{Hz}$ in the last few years leading up to their merger, making deci-Hz observatories such as LGWA ideal probes.

The small DM-induced dephasing in IMRIs should be sufficient to detect the presence of a cold DM spike and to estimate its density profile~\cite{Cardoso:2019rou,Coogan:2021uqv,Cole:2022ucw}. In addition, the dephasing due to a spike of cold DM particle should be distinguishable from other possible sources of dephasing, such as that from an ultra-light boson cloud or from a baryonic accretion disk~\cite{Cole:2022yzw}. The detection of a dense DM spike would also shed light on the ultimate nature of DM, ruling out models which cannot achieve such a high density, such as light fermionic DM or self-annihilating DM~\cite{Hannuksela:2019vip}. Previous estimates of the detectability of this DM dephasing effect have generally focused on LISA, while we expect that for LGWA the enhanced sensitivity at decihertz frequencies should lead to even better prospects. 

Observations of IMRIs by LGWA could also uncover sub-solar mass BHs. Such BHs could be of primordial origin (see, e.g.,~\cite{sasaki2018pbh}) or could be produced by the accretion of particle dark matter into neutron stars or white dwarfs (dark matter transmutation; see, e.g.,~\cite{Goldman:1989nd}). 

\subsection{Multi-messenger observations}
\label{sec:multimessenger}
The LGWA science case is particularly rich in astrophysical and matter phenomena. As a consequence, many of the GW sources are expected to emit detectable EM signals as well. Most of these events are specific to the deci-Hertz band. Some of these are low-probability events that may be extremely rewarding to detect. In this section, we outline the multi-messenger context of LGWA signals. The prospects of EM observations of binaries containing WDs is described in Sec.~\ref{sec:DWD-WDNS}. The detection of SNe together with associated GW signals would be a major breakthrough (Sec.~\ref{sec:SNe}). Rare signals observed in EM and GW might come from TDEs of WDs near the horizon of BHs (Sec.~\ref{sec:TDE}). Another multi-messenger study concerns MBH binaries in AGNs, where  X-ray observations can complement the GW detection (Sec.~\ref{sec:MBHbinaries}). Finally, a new class of EM signals called QPEs are best explained by X-ray emissions from gas in E/IMRI systems (Sec.~\ref{sec:QPEs}). \changes{We refer to section \ref{sec222} for an estimate of LGWA's sky-localization capabilities, which plays a crucial role for most multi-messenger studies.}

\subsubsection{Compact binaries with white dwarfs and neutron stars}
\label{sec:DWD-WDNS}
{\it Main contributors:} Michael Coughlin, Valeriya Korol, Javier Morán-Fraile, Tsvi Piran, Silvia Piranomonte, Jacopo Tissino\\

Compact binaries composed of NSs and WDs with short orbital periods are identified as prime candidates for LGWA (c.f. Sec.~\ref{sec:populations_and_formation_channels}). An increasing number of these binaries are being detected at longer orbital periods just outside the sensitivity band of LGWA, millions of years before their eventual merger \citep{Kupfer2023}. Concurrently, a rise in the number of transient events, which could be a result of the mergers of these same binaries, is being observed. Yet, the link between the progenitors and these transient events is often left ambiguous. LGWA offers a unique opportunity for multi-messenger observations both pre- and post-mergers of DWD and NSWD binaries. This section describes the electromagnetic observations of both NS/WD binaries and transients that may originate and outlines the synergies between LGWA and electromagnetic facilities.

\paragraph{Electromagnetic observations of DWD binaries --}
Recent advances in high-resolution interferometric observations over the past decade have demonstrated the ubiquity of multiple star systems \citep{offner2022origin}. Given that WDs represent the natural end stage of low-mass star evolution ($\sim$95\% of all stars), double white dwarf binary systems (DWDs) should be abundant in nature. Indeed, the Galactic population of DWDs is projected to number in the hundreds of millions \citep{Nelemansetal2001}. However, EM observations of these systems is challenging due to the inherent physical characteristics of WD stars, which are characterised by a mass comparable to that of the Sun but compressed into the size of an Earth-like planet \cite{Rebassa-Mansergasetal2018}. Such a small size makes WDs inherently dim (with absolute magnitude in the range 10--14 mag in the Sloan g-band). While the pronounced gravitational force of these extremely dense stars leads to significant pressure broadening of the observable (spectroscopic) lines, gravitational settling also results in predominantly light elements (mainly Hydrogen and Helium) being most easily observed in their atmospheres. 

The most effective detection strategy is to analyse the compact binary orbital motion via spectroscopy. By identifying radial velocity variations of the order of a few \(\times 100\,\text{km\,s}^{-1}\), it is possible to identify a short period DWD binary, which photometrically resembles a single WD. Over the past two decades, advances in large spectroscopic surveys have enhanced the observation of DWD systems. The Supernova Ia Progenitor surveY \citep{Napiwotzkietal2001} targeted around 1000 bright WDs, confirming several dozen DWDs \citep{Napiwotzkietal2020}. More recently, \cite{Brownetal2020}, aiming at the extremely low-mass WD ($<0.3\,$M$_\odot$), identified an additional hundred systems. However, spectroscopic surveys like these are relatively limited both in brightness and in numbers, motivating the use of time-domain photometric surveys to discover DWDs. 

Time-domain large-scale photometric surveys search for these systems by observing eclipses or light curve modulations. Though less efficient, they may also help to evaluate the efficiency of more common spectroscopic identification techniques, by comparison of estimated rates and systematics. Surveys such as the Asteroid Terrestrial$-$impact Last Alert System (ATLAS) \cite{Tonryetal2018} and the Zwicky Transient Facility (ZTF) \cite{Bellmetal2019}, among others, are beginning to detect WD binaries regularly, with spectroscopy following for confirmation and characterization \citep{BuEA2020}. In addition, DWD binaries can be found in exoplanet-focused time-domain photometric surveys, as exemplified by detections from \textit{Kepler} and \textit{TESS} space missions \citep{Howelletal2014,Mundayetal2023,Rickeretal2014}. In the near future, the \textit{Gaia} survey \citep{Brownetal2021} is poised to offer new insights into the field \cite{Renetal2023}. Additionally, state-of-the-art facilities like the Vera C. Rubin Observatory \cite{Ivezic2019}, along with tools specifically tailored for spectroscopic follow-ups like SOXS \cite{Schipanietal2022}, promise to significantly contribute to the area. They are anticipated to yield hundreds, if not thousands, of new identifications \citep{Korol2017, Li2020}. This wealth of data will facilitate the selection of a subset of short-period, accurately measured systems for dedicated monitoring with GW observatories.

Moving to higher energies, DWD binaries are difficult to detect since they can sustain only moderate accretion, if any, to feed high energy emission. Still, UV and soft X-rays observations for those relatively nearby systems detectable with present-day instruments are useful, for instance, to better constrain the spectral energy distribution of the components of the binary system. In turn, this might allow to better determine the size of the sources and characterize the evolution of the binary. The best example is likely RX\,J0806.3+1527 \cite{Israeletal2002}, modeled as a double degenerate binary with orbital period of 321\,s, first identified analyzing ROSAT data \cite{Israeletal1999}. In this system, mass transfer should come from a Roche lobe filling WD to another more massive WD. This kind of systems are also known as part of the AM\,CVn category, and are of particular interest for GW astronomy since the condition that a companion star fills its Roche lobe translates into a short orbital period, i.e. less than $\sim 2$\,hours. Several tens of these binaries are presently known in this category \citep{Ramsay2018, Kupfer2023}. GW astronomy could also provide a reliable way to measure mass transfer and other system parameters to better constrain their still poorly known formation channel(s) \cite{Wangetal2023}.

Independently of the identification scenario, currently, only a couple hundred DWD systems are known. Given the variability in detection techniques, ranging from instrument resolution to sample definition, it remains challenging to derive statistically significant conclusions about the Galactic DWD population. However, recent estimates indicate that the DWD fraction stands at roughly $\sim6$\,\%, aligning with predictions from population synthesis \citep{Toonenetal2017}. Out of these, merely around $\sim10$\,\% \citep{Brownetal2022} exhibit sub-hour orbital periods. These particular systems are bound to merge within a Hubble time due to the emission of GWs.

\paragraph{Electromagnetic observations of NSWD binaries --}
Being even more challenging to observe than DWD systems, NSWD binaries are also less numerous \cite{Toonen2018, Korol2023}. However, there are a few ways that NSWDs can reveal themselves. For example,  when a NS accretes from a WD companion, these systems are visible as Ultra-Compact X-ray Binaries (UCXBs). During the accretion process, the infalling material emits high-energy X-rays, making them visible in X-ray wavelengths. Additionally, if the NS behaves as a radio pulsar, its radio emission can be picked up as periodic signals, revealing its presence. However, Doppler smearing of radio pulsations occurs when the combined effect of the binary system's motion and the pulsar's inherent spin blurs the received radio signals. This smearing can introduce a bias against the detection of rapidly spinning millisecond radio pulsars in acceleration searches. Recently, Pol et al \cite{Pol2021} have yielded accurate models of the observed binary pulsar population, highlighting a high likelihood of detecting NSWD systems with orbital periods under 15 minutes with Arecibo-like radio telescope surveys. This probability is anticipated to increase with the forthcoming Square Kilometre Array \citep{Weltman_2020}. Moreover, they argued that unequal mass NSWD systems appear more detectable than those with near-equal masses. However, these binaries start interaction at orbital periods between 25 - 15 minutes, which can impede radio detection as discussed above.

\paragraph{Transients that may arise from DWD/NSWD binaries --}
Mergers involving double WD binaries and NSWD binaries can lead to explosive or transient events and the formation of exotic objects that typically do not emerge from the evolutionary path of single stars. Many studies have delved into the mergers of double WD binaries, mostly in the context of potential SN-Ia progenitors, as reviewed by various authors such as \cite{LivioMazzali2018,Soker2018,Wang2018,Toonen2018,Hamers2018}. These binaries are created through common envelope evolution in binary systems, and in the closest double WD binaries, the emission of GWs causes their orbits to shrink and initiates mass transfer between the WDs within a Hubble timescale. The critical factor influencing the final result is whether mass transfer remains stable or diverges, potentially leading to the tidal disruption of the donor star and binary merger \citep{Marsh2004}. Mass transfer's stability depends on the response of the WDs to the process, their masses, and various angular momentum transport mechanisms, such as torques due to accretion discs or tidal bulges \citep{Iben_1998,Piro_2011}. Subsequent evolution after WD mergers can yield diverse outcomes, including He-rich hot subdwarfs, R Coronae Borealis stars, massive carbon/oxygen or oxygen/neon WDs, SN-Ia, accretion-induced collapse to a neutron star or even a black hole. A chart summarizing these potential outcomes of WD mergers can be found in Fig.~3 of \citep{Shen_2015}. The case of SN~Ia holds broader historical significance and particular scientific interest for LGWA, and is therefore discussed in detail in Sec.~\ref{SNIa}.

Mergers involving NSWD and BHWD binaries are also expected to fall within the LGWA sensitivity band and generate electromagnetic transients. However, simulating theoretical predictions about their observational signatures to estimate them numerically presents more challenges. The main challenge lies in dealing with a wide range of scales and complex physical processes inherent in this problem. This is in contrast to mergers involving objects of similar size, such as DWD or NSNS/BH, which are more manageable and have more well-developed observational predictions, as demonstrated in studies like \cite{Dan2014, 2016ARNPS..66...23F, pakmor}. The merger events involving NS or BH with a WD have been extensively studied through various hydrodynamical and nuclear-hydrodynamical simulations. Some notable references in this research include \cite{Metzger2012, fer13, zen19, bob22, mor24}. It is generally anticipated that all NSWD mergers result in unstable mass transfer \citep{bob17}. In such case, the WD is tidally disrupted on dynamical timescales, forming an extended debris disk around the NS. The evolution of this disk is primarily governed by viscosity, but nuclear burning also plays a significant role. Together, disk viscosity and nuclear burning drive outflows throughout the disk \citep{Metzger2012}. Nuclear burning within the disk progresses steadily, potentially with a weak detonation, and it results in the production of small amounts of  $^{56}$Ni, typically at most 10$^{-2}$ M$_\odot$ \citep{zen19}. This limited amount of $^{56}$Ni is expected to lead to optical transients, but they are significantly fainter compared to typical type Ia supernovae, which generate over an order of magnitude more $^{56}$Ni. The transients resulting from NSWD mergers could potentially represent an entirely distinct class of supernovae. These events might be observable primarily in nearby galaxies using large telescopes or, perhaps, with next-generation survey instruments like Rubin-LSST. 

\paragraph{EM-LGWA synergies --}
Binaries with a WD component are not loud emitters of gravitational waves, due to their low masses and long orbital period. Most such binaries will be detected in their inspiral phase integrating the quasi-monotonic signal over years, similarly to what is done for continuous wave searches by current GW detectors. A few of these detected binaries will merge during the LGWA mission lifetime. The presence of eccentricity -- expected for NSWD binaries if the NS forms after the WD \citep{Toonen2018} -- gives rise to higher-frequency harmonics of the emission \cite{PeMa1963, Maggiore:2007ulw}, which boost overall emitted power as well as placing it closer to the most sensitive band for LGWA. As a summary of what is discussed in Sec.~\ref{sec:horizons}, deci-Hz GW sources in our own Milky Way are very likely to be observed regardless of their orientation, while sources outside our galaxy might be seen mostly if they have favorable orientations and masses.

\changes{Multi-messenger detections of WD mergers within our galaxy should not be expected}, even though the Milky Way hosts an estimated \( \mathcal{O}(10^{8}) \) DWD and \( \mathcal{O}(10^{7}) \) NSWD binaries \citep{LISAastroWP}. This is because only a small fraction exists as short-period binaries.anges{Hence, LGWA's capability to see binaries with WDs out to a few 10\,Mpc (see section \ref{sec:horizons}) is crucial for the success of a EM-GW multi-messenger campaign. For reference, the detection rates of DWD merger derived from SN~Ia observations are estimated in section \ref{sec:populations_and_formation_channels}. More details of a DWD and SN~Ia association and a possible coincident detection can be found in section \ref{sec:SNe}.}

Nevertheless, pinpointing even a limited number of ultra-short period DWD/NSWD binaries can yield significant astrophysical insight. Optical follow-ups could be essential in resolving parameter ambiguities that arise from relying solely on GW signals. These parameters include system mass, distance, and orbital period, especially if light-curve modulations are discerned. Collectively, such observations can elucidate various domains, from white dwarf structures and binary stellar evolution to accretion physics, general relativity, and the overarching structure of our galaxy. The association of a gravitational event with an electromagnetic counterpart, such as SN-Ia explosion, would be a groundbreaking discovery, much like for the first multi-messenger observation of binary NS merger GW170817 \citep{GW170817}. 

Concerning NSWD mergers, in the event the NS does not collapse into a BH post-merger, Morán-Fraile et al \cite{mor24} estimate jet velocities similar to those seen in galactic microquasars ($\gtrsim 0.9c$), leading to an event spectrum that initiates with a soft X-ray or UV flash and transitions into lower frequencies, resembling fast blue optical transients (FBOTs, \cite{2014ApJ...794...23D}), potentially associated with NSWD mergers. The merger could generate a prolonged, bright radio signal, with the transient's luminosity potentially exceeding typical kilonovae due to the ejecta's mass and velocity, rather than r-process yields, which are obscured in kilonovae by lanthanides. If the NS transitions into a BH, this could result in ultrarelativistic jet speeds, leading to long or ultra-long GRB-like transients, marking a significant observational shift

\subsubsection{Supernovae}
\label{sec:SNe}
{\it Main contributors:} Ferdinando Patat, Stefano Benetti, Marica Branchesi, Enrico Cappellaro, Kiranjyot Gill, Elisabeth-Adelheid Keppler, Francesco Longo, Sourav Roy Chowdhury, David Vartanyan \\

\paragraph{Thermonuclear Supernovae --}\label{SNIa}
Thermonuclear explosions, better known as Type Ia Supernovae (SN~Ia), play a crucial role in modern cosmology. In addition to providing the first evidence that the expansion of the Universe is accelerated by some sort of dark energy, they recently led to a tension with the value of the Hubble constant derived from the Cosmic Microwave Background (CMB) modelling (see \cite{cappellaro} for a review). Despite their importance, the nature of their progenitors is still unclear and two alternative scenarios have been proposed: the single-degenerate (SD) scenario and the double-degenerate (DD) scenario (see Sec. \ref{sec:populations_and_formation_channels}). The pieces of evidence collected so far in favour of one or the other option are all indirect. As examples of relevant results, we can list: i) the presence \cite{patat} or absence \cite{sand} of signatures of circumstellar gas surrounding the progenitor that favour SD and DD, respectively; ii) the failed search for a surviving companion of historical SN~Ia in our Galaxy, favoring DD \cite{ruiz-lapuente}; iii) the failed search of a precursor star in pre-explosion archival images \cite{graur} favoring DD; iv) the search for candidate DD systems that did not find a sufficient number of short period, massive WDs to account for the observe SN~Ia rates \cite{Napiwotzkietal2020} favoring SD, or otherwise sub-Chandrasekhar explosions; v) the distribution of delay time from star formation to explosion derived from statistics of SN~Ia rate that tends to favor the DD scenario \cite{maoz}, although the strength of this argument is debatable \cite{greggio}.

The bottom line is that, after decades of research, none of the results is sufficiently conclusive and, in some cases, the findings appear to be even in contradiction. Moreover, observations have shown that about 30\% of the discovered SNe Ia largely deviate from the properties of ``normal''  thermonuclear events, such as peak luminosity, light curve morphology and spectral features (for a recent review see \cite{Liu2023}). The rising era of GW astronomy opens a new avenue for obtaining independent and direct insights to the SN~Ia progenitors' nature, which remains one of the burning, open questions in modern astrophysics \cite{Livio2001}.

Arguably, the ultimate proof of the DD-SNIa connection would be the observation of the merging event followed by the detection of its electromagnetic (EM) counterpart. However, while in some theoretical scenarios the SN explosion and hence the appearance of the EM counterpart, follows a few seconds after the merger \cite{pakmor}, others allow for a long delay between the merger and the explosion, extending to $10^4$ yr \cite{shen}. This implies that, even when a DWD merger with the appropriate mass occurs and its GW emission is detected, it may not be associated with a prompt SN~Ia EM event. Hence, missing the long-sought and much-desired smoking gun signature. The comparison between the known rate of SNe~Ia and the rate of GW sources exhibiting properties expected for merging DD systems does not require the DD-SN~Ia association to answer this key question (as described in Sec.~\ref{sec:populations_and_formation_channels}). \changes{As shown in Sec.~\ref{sec:horizons}, LGWA can observe DWD signals out to several tens of Mpc. Using SN~Ia observations to estimate the corresponding DWD merger rate (assuming that all SN~Ia are produced by DWDs), we found in section \ref{sec:populations_and_formation_channels} that about 20 DWDs producing SN~Ia would be observed in 10 years with LGWA. This would be enough detections to make a meaningful comparison of rates and infer a possible DD progenitor scenario.}

Interestingly enough, the simultaneous detection of a ``peculiar'' SN~Ia in both GW and EM domains is expected for thermonuclear events arising from WD-WD systems originating from a CO-WD and a He-WD. These objects are anticipated to explode via the detonation of a thin He-layer piled-up on the CO-WD through a stable mass transfer that is driven by GW emission with maximum expected frequency of several mHz (see \cite{Wong2023} for details). Multi-messenger astronomy could help determining the contribution of double-detonating WDs to the observed SNe Ia rate and, more broadly, allow one to infer the chirp masses for some of the transient, hence constraining the SN progenitor. 

\paragraph{Core-Collapse Supernovae --}
Following the gravitational collapse of a massive star ($M\geq8M_\odot$), the inner core rebounds and collides with the infalling stellar material producing a shock wave (see \cite{Janka1012} for a general review). The shock wave tends to stall within the infalling material. Extensive theoretical work done in recent years has demonstrated that it is the neutrino heating coming from the accreting proto-NS (coupled to the neutrino-driven turbulent convection between the shock cavity and the stellar surface) that re-powers the shock and leads to subsequent explosion (\cite{Glas2019, Burrows2020, Vartanyan2022} and references therein). The shock break-out, i.e., the moment when the blast wave reaches the stellar surface, occurs minutes to days post collapse, and shortly after starts emitting electromagnetic radiation, appearing as a SN. The star is destroyed leaving behind a compact remnant.

It is clear that CCSNe are natural-born candidates for multi-messenger detection, as they are sources of EM, GW radiation and neutrinos. The CC SN1987A indeed marked the birth of neutrino astronomy \cite{Bionta1987, Hirata1987}. The temporal coincidence of the EM and neutrino signal produced by SN1987A has clearly shown the association between the neutrino emission and the (supposed) core-collapse, hence providing direct support to the underlying theory. 

Although more frequent than SNe~Ia, CCSNe are rare events. Their estimated rate for the Milky Way is 1.63$\pm$0.46 (100\,yr)$^{-1}$, which corresponds to an average recurrence time of 61$_{-14}^{+24}$\,yr \cite{Rozwadowska2021}. \changes{As described in section \ref{sec:GWAstro}, key for the detectability of the GW signal from CCSNe is the memory effect, which produces relatively large amplitudes below 10\,Hz and into the decihertz band.} In this way, the LGWA coupled with other future, planned GW detectors will open the possibility of observing these garden-variety events using all three messengers: EM, GW, neutrinos. Clearly, the simultaneous detection of all three would be the next watershed detection in the field of GWs.

\subsubsection{Tidal Disruption Events}
\label{sec:TDE}
{\it Main contributors:} Martina Toscani, Francesca Onori, Elisa Bortolas, Marica Branchesi, Roberto Della Ceca, Giovanni Miniutti\\

Tidal disruption events (see  Sec.~\ref{sec:GWAstro}) are especially interesting transient phenomena for several reasons. First of all, they are exquisite EM sources, as they produce very bright flares (luminosities of order of $10^{43-44}$ erg s$^{-1}$) in different bands of the EM spectrum, extending from X-rays to radio wavelengths (for recent reviews see \citep{2020SSRv..216...81A,vanVelzen20,2021SSRv..217...18S} and references therein). Thanks to this powerful emission, they have been an unparalleled tool to unveil the presence of otherwise quiescent MBHs in the cores of galaxies. In addition to this, they also emit astrophysical neutrinos during the later stages, when they could trigger the formation of jets (for details see \citep{2021NatAs...5..436H}). Finally, they are expected to emit GWs in the low frequency regime $f_{\rm gw}\sim 10^{-4}-10^{-1}\,\text{Hz}$ \citep{2022MNRAS.510..992T}.

While the majority of EM observations of these events suggest that TDEs typically occur with MBH in the mass range of $\sim 10^{6}-10^{8}\,\text{M}_{\odot}$, a particularly fascinating prospect lies in the detection of a TDE involving a WD around a MBH in the intermediate-mass range, $\sim 10^{3}-10^{5}\,\text{M}_{\odot}$. These occurrences have proven challenging to detect thus far, despite some suggestive clues  \citep{2011ApJ...743..134K, Shcherbakov:13aa, Peng:19aa}. Observing such a WD-TDE would be a smoking gun of the existence of these elusive IMBHs \citep{2020MNRAS.498..507T}, which have remained difficult to detect until now. It is worth noting that these WD-TDEs represent the best TDE sources of GWs in the LGWA frequency band. 

Another interesting scenario involving tidal disruptions is the case of repeating partial TDEs, where a star, likely in an evolved state, follows a highly eccentric bound orbit and loses portions of its envelope during each pericenter passage due to the gravitational forces of the central MBH. This results in relatively short-lived, quasi-periodic electromagnetic flares, driven by accretion and/or shocks. Partial TDEs are considered to be the prevailing explanation for a new class of detected EM nuclear transients, the so-called repeating nuclear transients (RNTs), a small number of which has been observed in the optical (e.g. ASASSN-14ko \cite{2021ApJ...910..125P,2022ApJ...926..142P}) or X-ray bands (e.g. eRASSt~J045650.3-203750 \cite{2023A&A...669A..75L} and Swift~J023017.0+283603 \citep{2023arXiv230902500E,2023arXiv230903011G}).

\indent Similarly to the case of X-ray quasi-periodic eruptions discussed in Sec.~\ref{sec:QPEs}, RNTs are likely to represent the EM counterpart of extreme mass-ratio binary systems, opening up a possible window for multi-messenger synergies. If RNTs are indeed due to periodic mass transfer events at pericenter from the envelope of evolved donor stars, they are likely to evolve into MBH-WD (or WD-like) extreme mass-ratio binary systems, once the whole envelope has been stripped off.

\paragraph{Electromagnetic emission from TDEs --}
TDEs were anticipated to exhibit exceptional brightness in X-rays  \cite{hills75,1988Natur.333..523R}. Indeed, the first candidates \cite{komossa99, greiner00} were revealed as luminous X-ray sources in the {\it ROSAT} all sky survey archive and subsequently other X-ray TDEs were identified thanks to dedicated searches or serendipitous discoveries by using {\it CHANDRA}, {\it XMM-Newton} and {\it Swift} satellites \cite{esquej07,esquej08, saxton12, gezari12, komossa15}. However, over the past decade, wide-field optical transient surveys such as ASASSN, ATLAS, Pan-STARRS, iPTF, and ZTF have played a crucial role in making the optical band the primary channel for discovering TDEs. Consequently, the sample of TDEs has rapidly grown from few candidates to tens of confirmed TDEs, revealing a heterogeneous population of transients and enlightening a number of well established key observational properties both in optical and in X-ray bands \cite{vanVelzen20,2021SSRv..217...18S}.

\indent Flares produced by WD-TDEs are instead relatively rare with respect to that emitted by TDEs involving main sequence stars (MS-TDEs) and, although it has been currently reached a TDE discovery rate of $\sim$10 events per year, no confirmed WD-TDEs have been reported so far, and only a few possible candidates. However, recent theoretical studies and simulations have been able to accurately predict their expected EM observational signatures \cite[][]{maguire20}. In the following, we briefly review the main properties of their EM emission distinguishing between MS-TDEs and WD-TDEs.\\

\noindent\underline{MS-TDEs} -- MS-TDEs manifest themselves as large amplitude blue flares (g-r$<$0) in the core of quiescent galaxies (with an overabundance of TDE detection in post-starburst/E+A host galaxies), characterized by a roughly constant optical colours. The light-curves usually rise to the peak in approximately $<$30 days and show a power-law decline broadly consistent with the $\sim$t$^{-5/3}$ law \citep{1989IAUS..136..543P} on time-scales of months to years. They are characterized by hot and near-constant black body temperatures ($T_{\rm BB} \sim$10$^{4}$ K) with a lack of cooling in the post peak phase \cite{vanvelzen21}. X-rays detection of optically selected TDEs are rare, with only few exceptions, also including some events showing delayed X-ray flares (e.g. ASASSN-14li, ASASSN-15oi, AT\,2019dsg, AT\,2019qiz, AT\,2019azh and AT\,2017gge \cite{holoien16a, gezari17,cannizzaro21,nicholl20,liu22, Onori:17gge}, respectively). Such an observational dicotomy has been explained as the result of the intrinsic physical properties of the EM emitting region coupled with viewing angle effects (see the TDE unification model proposed in \cite{dai18}). When detected, the X-ray emission of MS-TDEs is typically in the soft energy band (0.1-2.5 keV), with peak luminosities between 10$^{42}$ and 10$^{44}$ erg s$^{-1}$ and extremely soft spectra near the peak, well modeled by power-law with spectral indices $\Gamma_x$=3-5 (or, alternatively by blackbody model with temperatures $kT_{\rm bb}$=0.04-0.12 keV). Some events have shown a spectral hardening with time. The overall lightcurve decline follows quite well the $\sim$t$^{-5/3}$ power-law, but some events have shown also a fast variability on timescales of minutes to hours (see the reviews of \cite{gezari21, saxton20}). Notably exceptions to this picture are the so-called relativistic TDEs (AT2022cmc \cite{2023NatAs...7...88P}, Sw J1644+57 \cite{2011Sci...333..203B}, Sw J2058.4+0516 \cite{2012ApJ...753...77C} and Sw J1112.2-8238 \cite{2015MNRAS.452.4297B}), extremely luminous, hard X-ray events (L$_{\rm X}\sim$10$^{47-48}$ erg s$^{-1}$ at peak) in which the detection of a non-thermal component has been explained with the launch of face-on relativistic jets.

\begin{figure}[ht!]
    \centering
    \includegraphics[width=0.7\textwidth]{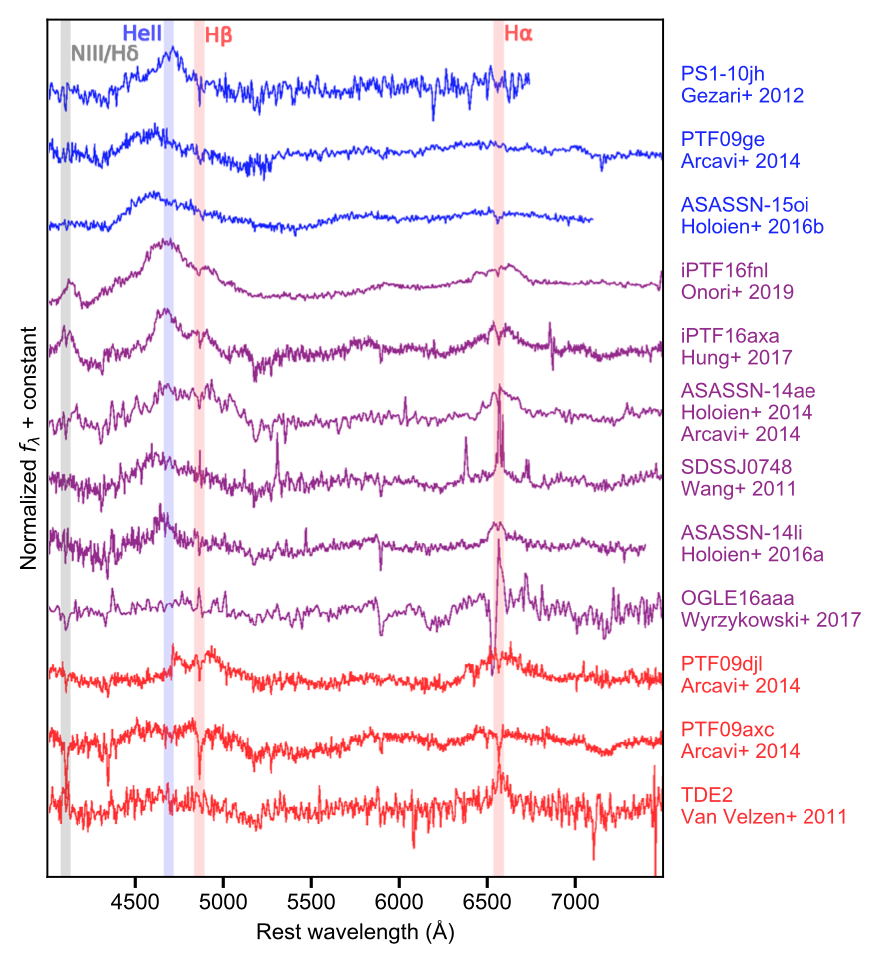}
    \caption{Sequence of continuum subtracted TDE optical spectra. Very broad H and/or HeII emission lines are observed with different relative ratios. The TDE sub-classes are highlighted by different colors: TDE He (in blue), TDE H-He (in purple), TDE H (in red). Those showing additional NIII features are referred N-rich/Bowen TDEs. Figure from \cite{vanVelzen20}.}
    \label{fig:TDEspec}
\end{figure}

Optical spectroscopy is required to unambiguously identify a TDE. The early spectra are dominated by a strong blue continuum with superimposed very broad (FWHM$\sim$10$^{4}$ km/s) H and/or HeII $\lambda$4686 emission lines, with different strengths and relative ratios \cite{arcavi14}. The line profiles are of pure emission with the FHWM decreasing with time. However, line asymmetries, double peaked line profiles and outflows components have been reported for some events. Recently, the notable discovery of TDEs exhibiting broad Bowen fluorescence emission lines such as O III 3760 and N III 4100,4640 (iPTF15af \cite{Blagorodnova:15af}, iPTF16fnl \cite{Onori:16fnl}, AT2018dyb \cite{Leloudas:18dyb}) have led to the mapping of the TDE spectral diversity into four main sub-classes, the He-rich,  H+He-rich, the H-rich and N-rich TDEs \cite{vanVelzen20}, and represented a strong evidence for the presence of an obscured and reprocessed EUV/X-ray emission. In Figure \ref{fig:TDEspec} a compilation of continuum subtracted TDE optical spectra is shown.

\noindent\underline{WD-TDEs} -- The WD-TDEs EM emission differs from that observed in MS-TDEs since the WD disruption is expected to induce a super-Eddington accretion phase, which in turn results in the lunch of relativistic jets. Moreover, the MBH gravitational influence on a compact object such as a WD may produce an additional source of energy in case of deep encounters ($\beta >$3, with $\beta$ defined as the ratio of the tidal radius above the stellar pericenter). In particular, the strong tidal compression experienced by the WD at the first pericenter passage, could results in a thermonuclear runaway reaction before the disruption and can trigger an optical transient powered by the radioactive decay of $^{56}$Ni. The results is a highly asymmetric SNIa-like event \cite{maguire20}. Given the involvement of IMBHs in this process, also the host environment is different with respect the MS-TDEs, with WD-TDE expected to be detected either in the nuclei of dwarf galaxies, or in the outskirt regions of galaxies, or even in globular cluster.

The observational properties of this class of transient are highly viewing angle dependents. For systems observed face-on, a luminous X-ray jet precedes an optical afterglow, which is followed by the SNIa-like transient. After the thermonuclear transient starts to fade, a radio afterglow arise. In case of off-axis observers, instead, the early time emission is dominated by a soft X-ray component coming from a newly formed accretion disk, the SNIa-like transient outshines the optical afterglow and it is in turn followed by the radio afterglow \cite{macleod16}. When the high energy emission is detected, it is extremely luminous (with typical peak jet luminosities of $\sim$10$^{47-50}$ erg s$^{-1}$), fast rising (with timescales of $\sim$10$^{2-4}$s) and long lasting (with a super-Eddington accretion phase lasting months to years) \cite{macleod16}. 

As introduced before, in the case of WD-TDEs there are two sources behind the UV/optical emission: the thermonuclear reaction and the ionization of the stellar debris. 

The transient resulting from the thermonuclear runaway reaction is expected to mimic the optical emission of SNIa but with lower luminosities (M$_{B}$ between -16 and -18 mag, L$_{\rm bol} \sim$10$^{40-42}$ at peak), bluer colors (g-r$<0$) and a faster time evolution (rise to the peak in $\sim$5-10 days). In particular, the early UV/optical spectra are expected to be mainly featureless and dominated by a hot blue continuum, with the characteristic spectral features developing after $\sim$15 days from the light-curve peak \cite{maguire20}. At this phase the spectra broadly resemble those of SNe, with broad absorption features and P-cygni profiles in correspondence of Si II, Ca II and Ti II (the lines intensities and widths have a viewing angle dependence). Key distinctive features are the presence of Doppler shift in the spectral lines time series (up to $\pm$12,000 km/s), which is the results of the WD-TDE unbound ejecta orbital motion, and the the detection of an X-ray emission \cite{macleod16}.
 
The ionization of the stellar debris tail will produce broad emission lines such as C IV $\lambda$1550; C III$\lambda$977; [OIII]$\lambda \lambda$4959,5007 and [O III]$\lambda$4636, but are expected to appear in the UV/optical spectra at very late times ($>$100 days), when the WD-TDE is in the sub-Eddington accretion phase \cite{rosswog09}. Their strength should decline with time following the $\sim$t$^{-5/3}$ law, with the [OIII]$\lambda$5007 having the slowest decline and thus to be still visible in the optical spectra taken a decade after the WD-TDE occurrence.

Although no confirmed WD-TDE has been reported so far, some candidates have been proposed such as the class of relativistic TDEs \cite{2011ApJ...743..134K}, fast X-rays transients \cite{jonker13, irwin16} and Ca-rich transients \cite{2015MNRAS.450.4198S}. However, their nature is still highly debated.\newline

\begin{figure}[ht!]
    \centering
    \includegraphics[width=0.7\textwidth]{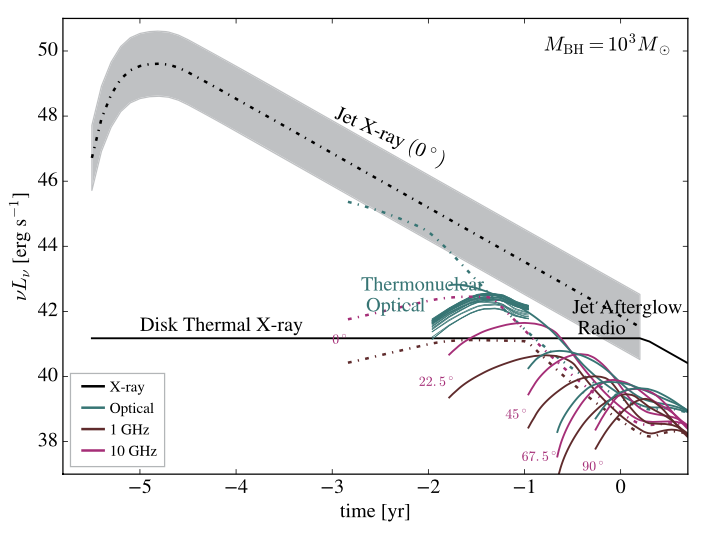}
    \caption{EM observational features expected in the case of a WD-TDE involving a M$_{\rm WD}$=0.6 WD and a $\text{M}_{\odot}$ and a $M_\bullet$ = 10$^{3}$ $\text{M}_{\odot}$ IMBH. Times are from the WD periapse passage. Dot-dashed lines indicate the signal seen by an observer placed along the jet axis, while solid lines refer to an off-axis observer. Figure from \cite{maguire20}.}
    \label{fig:WDTDE_EM}
\end{figure}
\indent Additional ways to identify candidate TDEs are the detection of reverberation signals caused by the interaction of the TDE EM prompt emission with a host environment enriched in dust and gas 
(see the review of \cite{vanvelzen21a}). In particular, the \textit{WISE} \citep{Wise2010} and the more recent \textit{NEOWISE} \citep{Neowise2012} projects turn out to have a crucial role in finding the mid-infrared echoes resulting from the re-radiation of the TDE emission by dust \cite{jiang21}. Furthermore, the remarkable detection of transient high ionization coronal emission lines in the late-time optical and IR spectra of the TDE AT\,2017gge \cite{Onori:17gge} represented the first observational evidence of the connection between the occurrence of a TDE and the presence of these emission lines in inactive galaxies. This finding strongly supporting the theory that extreme coronal line emitter (ECLE) galaxies may have indeed hosted a TDE in the past \cite{komossa08, yang13}.

\paragraph{Neutrino emission from TDEs --}
The idea of TDEs as neutrinos sources was first proposed after the discovery of the peculiar class of jetted-TDEs \cite[][]{wang11, daifang17, lunardiniwinter17}. In particular, the neutrinos production was considered in the framework of the jet models as the result of different state in the accretion disk evolving from the debris circularization of stellar debris to super- and sub-Eddington to radiatively inefficient accretion flows and/or of the interaction of the TDE ejecta with the hosting circumnuclear environment \cite[][]{2019ApJ...886..114H, fang20}. 

Indeed, ZTF EM follow-up searches of IceCube alerts identified TDE AT2019dsg \cite{Stein:19dsg} as a possible optical counterpart of a very high energy neutrino ($\sim$0.2 PeV, IceCube event IC191001A), which represents the first multi-messenger observation of a TDE. Later, other two TDE candidates have been associated with two high-energy neutrino events: AT2019fdr (IceCube event IC200530A \cite{Reush:19fdr}) and AT2019aalc (IceCube event IC191119A \cite{vanVelzen:19aalc}). Interestingly, all the three neutrino-emitting TDEs discovered so far show some common observational features, such as the delayed neutrino signal ($\sim$100 days after the BB peak), a delayed strong dust echoes in the IR and the X-ray detection. This can be ascribed to the physical properties of the system after the star disruption and raise important question on the mechanisms behind the neutrino production itself \cite[][]{winterlunardini23}. \changes{However, it should be noted that recent reanalyses of IceCube data suggest that the coincidences between TDE and neutrino observations are consistent with background \cite{IceCubeTDE2023}.}
 
\paragraph{Multimessenger perspectives and observational strategy --}
GW-detected TDEs will mark the moment of disruption of the star, otherwise undetectable. Such a detection could serve as crucial information for telescopes to know where to search for subsequent EM/neutrino counterparts, which may arise within minutes to days after the disruption phase (depending on the exact onset of the accretion process).

Multimessenger detections of TDEs open the possibility to perform exciting scientific analyses. For example, measuring the time delay between the GW signal and subsequent EM flares would allow us to discriminate among different type of TDE EM mechanisms, to date still debated \citep{2021SSRv..217...16B}. Moreover, the combined observation of these signals will allow us to constrain cosmological parameters, e.g., the Hubble constant \citep{2023MNRAS.523.3863T,2023arXiv230108407W}.

Given these exciting perspectives, it is important to outline a strategy for the multimessenger detections of these sources. First, the GW identification from a TDE can allow us to localise an area in the sky where to look for the expected EM counterpart. In particular, in the case of MS-TDEs, their identification as EM counterparts could come within weeks-months from the GW alert, considering the delay between the disruption of the star (where the main GW burst is produced) and the beginning of the accretion. In this case, there is plenty of time to start promptly a multi-wavelength follow-up not only to identify the transient as a TDE candidate, but also to characterize it (i.e., photometric analysis to derive BB properties, a spectral sequence to monitor the development of broad lines and their profile and a proper X-ray follow-up). In particular, for MS-TDEs we have observational evidence of their overabundance in post-starburst/E+A galaxies. Thus, blue nuclear flares from such galaxies placed in the GW localization area would be the first objects to look for.

Given the short time delay between GW production and EM flares in the case of WD-TDEs, EM emission could actually be used as an alert for GW counterparts. Indeed, detected electromagnetic emission could be used as valuable information to analyse the interferometer's data stream and to look for GW signals compatible with the properties of the  observed system.

Moreover, we recall that GW emission form TDEs is also possible during the circularization phase, that is as the same time as the production of EM radiation. Yet this GW emission is roughly 2 orders of magnitude weaker than the one emitted during the disruption phase \citep{2022MNRAS.510..992T}.

It is worth noting that some late-time identifications of TDEs as EM counterpart of GW signals may come from the detection of mid-infrared echoes and/or transient high-ionization coronal emission lines in quiescent galaxies inside the GW localization area, making late-time searches important.

\subsubsection{Intermediate and Massive Black Hole Binaries}
\label{sec:MBHbinaries}
{\it Main contributors:} Paola Severgnini, Cristian Vignali, Elisa Bortolas, Valentina Braito, Roberto Della Ceca, Alessia Franchini, Giovanni Miniutti, Deeshani Mitra, Alessandra De Rosa, Roberto Serafinelli, Jacopo Tissino \\

\paragraph{Astrophysical relevance --}
In the current multi-messenger era, the search for and the characterization of intermediate-mass (IMBH, 100 M$_{\odot}<M_{\rm BH}< 10^5 M_{\odot}$) and massive black hole (MBH, $M_{\rm BH}\geq 10^5 M_{\odot}$) binaries are among the hottest and most challenging topics of modern observational astrophysics. Theory predicts that super-Eddington accretion onto stellar BHs (e.g., \cite{Kocsis11}) and a combination of stellar collisions, star-BH interactions, and BH-BH mergers (e.g., \cite{2017MNRAS.467.4180S,2022MNRAS.512..884R,2022ApJ...929L..22R,2023arXiv230704805A,ArcaSedda23,2023arXiv230704807A}) are the main channels for the formation of IMBHs, that will then become the seeds for MBHs through further accretion and mergers (see Sec.~\ref{sec:populations_and_formation_channels} and \cite{Madau01, Silk17, Natarajan21}). Since most galaxies harbor a central MBH, the current $\Lambda$CDM cosmological paradigm predicts a large population of MBH binaries as a natural outcome of galaxy mergers. These systems, which are amongst the loudest emitters of GWs, will eventually coalesce by forming a more massive BH. Thus, detecting and tracing the evolution of IMBH and MBH binary systems across cosmic time would be the key to understanding the hierarchical structure formation, the growth and demographics of BHs, and the accretion and feedback that regulate the MBH/galaxy interplay.

Despite their relevance, both IMBH and MBH binaries remain observationally very elusive.  Shortly, upcoming facilities at various electromagnetic bands are expected to play a pivotal role in building up catalogues of EM-detected MBH/IMBH candidates for GW follow-up (see {\it Electromagnetic emission}). GWs will open the way to unambiguously recognize merging binaries, especially at low masses; the concurrent (multimessenger) detection of GWs, electromagnetic radiation, and possibly neutrino emission from the same source will substantially enhance the scientific return of GW observations \cite{LISAastroWP}. Assessing the potential for multimessenger astronomy in the context of LGWA requires a significant step forward in our theoretical understanding of the accretion mechanism in binary systems at different stages of their evolution, besides a more exhaustive  comprehension of the localization capabilities once their GW signal is detected.

\paragraph{Numerical simulations --}
A necessary condition for an electromagnetic emission to be produced by the accreting binary is that the host galaxies are sufficiently rich in gas, and gas can efficiently reach the close vicinity of the MBHs. The first condition is likely to be  true for LGWA sources, which are most probably embedded in dwarf galaxies at moderate redshifts (see {\it Electromagnetic emission}).

\begin{figure}[ht!]
    \centering
    \includegraphics[width=0.5\textwidth]{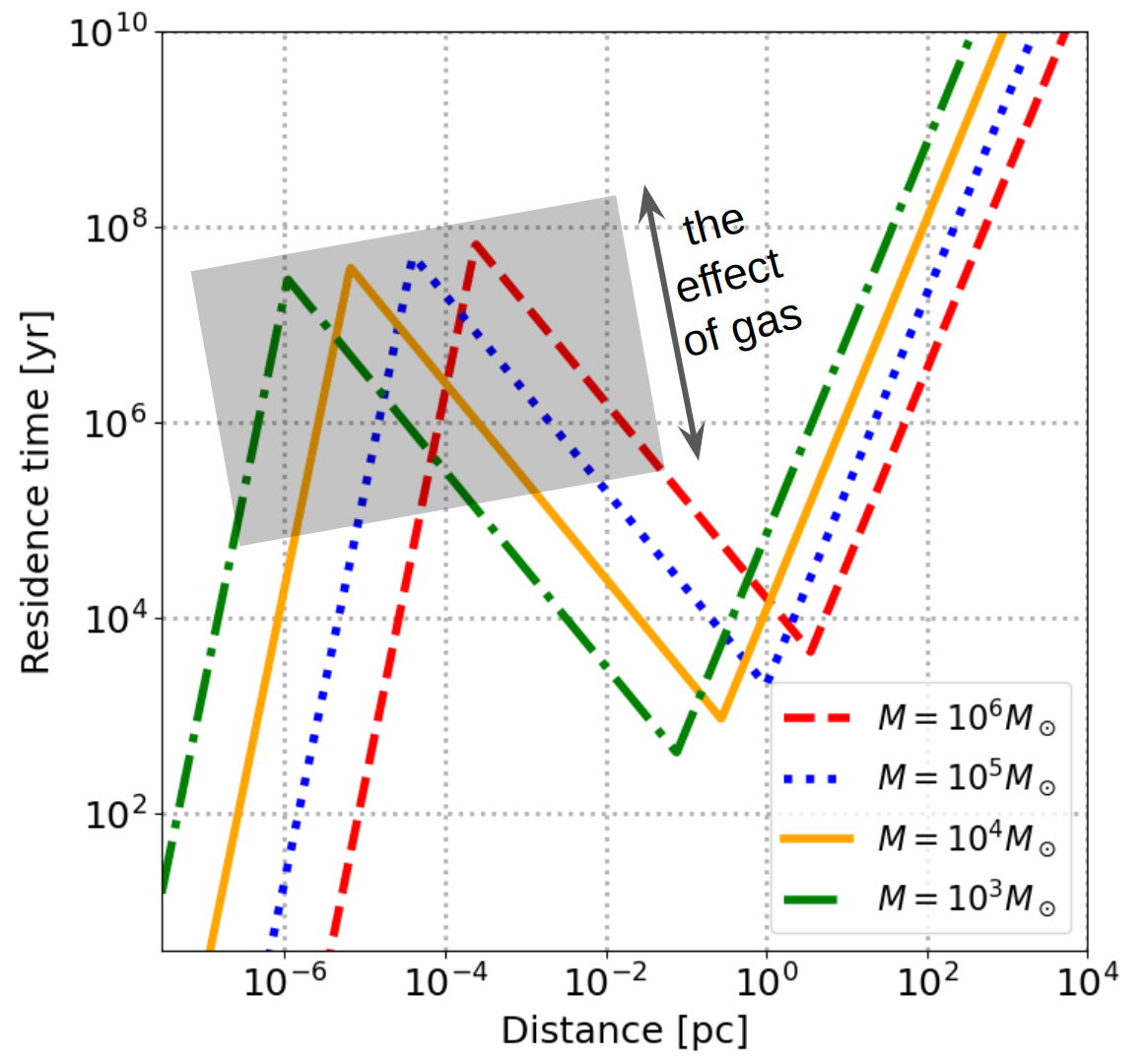}
    \caption{
    \textit{Residence time}, i.e. the characteristic time a binary spends in each logarithmic bin of separation, as a function of the separation itself; different lines refer to different MBH masses. The residence time is computed as the separation $r$ of the binary divided by its time derivative, $r/(dr/dt)$. In the transition between stellar hardening and GWs, the presence of a circumbinary disc can shorten or perhaps enhance the residence time; the related uncertainty is highlighted in grey. Credit: Elisa Bortolas.}
\label{fig:MBHB_residence_time}
\end{figure}

The probability of electromagnetically spotting an accreting MBH along its orbital decay is also related to the typical timescale the binary spends at each given separation. Keeping in mind the evolutionary path of MBH binaries and the definition of the various ``phases" (see  Sec.~\ref{sec:populations_and_formation_channels} and the summary provided below), Fig.~\ref{fig:MBHB_residence_time} qualitatively sketches how much time the binary spends at each distance \citep{1980Natur.287..307B}. The right-most trend at large scales is to be referred to dynamical friction, the central slope is the stellar hardening, and the left-most slope change is to be attributed to the onset of the GW phase. When the two MBHs are still unbound and separated by several kpc, each of them behaves independently of the other. This largest-scale inspiral is likely to take the longest among the pairing phases, so it is intrinsically more likely to spot MBHs at such distances. Furthermore, large projected separations are  easier to spatially resolve for imaging campaigns. Several works have investigated the accretion of MBH pairs in the large-scale, dynamical friction-driven inspiral. In particular, simulations of merging galaxies have shown that the perturbation induced by the merger itself (which renders the two galaxies distorted) can efficiently torque the gas and promote its loss of angular momentum, so that it may reach the vicinity of either MBH and get efficiently accreted
\citep{2015MNRAS.447.2123C, 2017MNRAS.465.2643C,2023hxga.book..126C}, resulting in variations in the mass ratio of the two MBHs. Accretion feedback released by an inspiralling MBH can evacuate gas from the MBH vicinity, resulting in the so-called \textit{negative dynamical friction}, which would further lengthen the inspiral itself \citep{2021ApJ...916..110L} and perhaps enhance our chance to detect binaries in this stage. Further perturbations to the decay timescale that can impact the residence time can be induced by the presence of bars, spirals, star-forming clumps and, in general, morphological distortions that are likely to appear in merging gas-rich systems \citep{2013ApJ...777L..14F, 2017MNRAS.464.2952T,2020MNRAS.498.3601B,2022MNRAS.512.3365B}.

As the pairing proceeds below separations of a few pc, the binary gets bound in a Keplerian pair. The transition between the dynamical friction and stellar hardening phase is so prompt -- unless the binary mass ratio is very small, $\lesssim 10^{-2}$ \citep{2017ApJ...840...31D} -- that it is very unlikely to observe the system at this stage (Fig.~\ref{fig:MBHB_residence_time}). The presence of gas in the vicinity of the binary can give rise to electromagnetic signals that would reveal the presence of the binary itself. The gas would settle into a circumbinary disc with an evacuated cavity, carved by the binary, with a size roughly twice the semi-major axis. Under some conditions (namely, if the disc temperature is high enough or if the disc viscosity is large), a significant amount of gas can leak from the cavity's inner edge forming structures called ``mini-discs" around the binary components. The interaction of the binary with the surrounding gas has recently received significant attention as the delicate balance between the positive gravitational torque (adding angular momentum to the binary) provided by the mini-discs and the negative torque of the circumbinary disc (removing angular momentum from the binary) eventually determines whether the binary inspirals or outspirals \citep{2020ApJ...901...25D,2021ApJ...914L..21D,Munoz2019}. Recent numerical works have shown this balance to depend on the disc aspect ratio (i.e. on its temperature) \citep{2020ApJ...900...43T,Heath2020}. In more recent simulations at higher resolution, \cite{2022ApJ...929L..13F} and \cite{2023MNRAS.522.1569F} have shown that the sign of the total torque acting on the binary also depends on the viscosity of the disc, a parameter that is challenging to measure or constrain, as well as whether the simulation incorporates a dynamically evolving binary system.

It is worth noting that, neglecting the effect of the gas, the binary would spend most of its time at the transition between the stellar hardening and GW emission stage, at or below mpc separations (Fig.~\ref{fig:MBHB_residence_time}). As both GW emission and stellar hardening are the least efficient in the transition, this is exactly where gas-driven torques can be the most impactful and affect the orbit, as well as where binaries are more likely to be observed, based on these timescale arguments \citep{2021ApJ...918L..15B}.

Although so far we only mentioned gas accretion, the two MBHs can also reveal themselves as they disrupt a star passing in their vicinity, giving rise to a TDE (Sec.~\ref{sec:TDE}). TDEs can in principle occur at virtually any stage of the inspiral. An enhancement in the TDE rates is likely to occur when the MBHs are still unbound, at the galaxy merger stage \citep{2019MNRAS.488L..29P}, and at the binding stage \citep{2019MNRAS.484.2851L, 2019ApJ...883..132L}. Furthermore, the Kozai-Lidov mechanism may enhance TDE rates occurring about the secondary MBH even in the hardening stage \citep{2023arXiv230605510M, 2023arXiv230605472M}, provided that the secondary MBH is small enough; since the MBH mass inferred from the TDE light curve would be much smaller (because corresponding to the smaller BH) than the one estimated through scaling relations (since the potential of the galactic nucleus is dominated by the larger BH), this could suggest the presence of a binary system. As our monitoring capabilities in both optical/UV and X-rays improve, we might expect to be able to select a few binary candidates by detecting quasi-periodic variability of the TDE-induced accretion flow due to the orbital motion of the secondary object about the primary. TDEs can also be produced around the newborn MBH as it receives a GW-induced kick at its formation, even if the associated TDE rates are expected to be comparable or smaller than those to be expected around single MBHs \citep{2009ApJ...699.1690M, 2017ApJ...834..195L}. It is important to note that TDEs are not likely to aid the search for a counterpart of an MBH merger, as the intrinsic TDE rates are too low (at the very best, one every $\sim$100-1000 years, Sec.~\ref{sec:TDE}) to allow for prompt identification of the counterpart after the GW event has been observed; yet, they can be fundamental allies for the broader study of the MBH (binaries) population.

\paragraph{Electromagnetic emission} -- As shown in Fig.~\ref{fig:horizon-et-lgwa-lisa}, LGWA will be able to detect pairs of BHs in the last stage of their inspiral and in the merger phases with masses up to about a few $10^6$ M$_{\odot}$. The LGWA will confirm the existence of 10$^{3-4}$ M$_{\odot}$ BHs up to $z>10$ and detect binaries containing MBHs of 10$^{5}$ M$_{\odot}$ up to $z\sim\ $5--6, while for higher mass (10$^6$ M$_{\odot}$) LGWA will be limited to $z<0.5$. The electromagnetic characterization of MBH/IMBH binaries, both before and after GW detections, is of utmost importance. This characterization not only enhances the sky localization accuracy of GW detectors and provides better constraints on the physical parameters of the binary system, but is also essential for studying the environments in which these mergers occur \cite{Arzoumanian20, Songsheng21}. Unfortunately, observations lag behind theoretical predictions and models. Currently and in the coming years, it will be possible to provide samples of reliable electromagnetic MBH/IMBH candidates, which can be definitively confirmed as such only by GW detectors.

\noindent\underline{Where to look for electromagnetic candidates for LGWA emitters ?} -- The tight correlations observed between the host-galaxy properties and the BH mass indicate that associated EM signals produced from the more massive BHs detected by LGWA, i.e., upwards of 10$^{5}$ -- 10$^6$ M$_{\odot}$, should be mainly searched for in dwarf galaxies. Even more challenging is the case of lower-mass BHs (IMBHs). Indeed, besides some indications from GWs \cite{2020ApJ...900L..13A, 2020PhRvL.125j1102A}, any conclusive electromagnetic evidence for the existence of even a single IMBH is still missing. Among the best candidates for accreting IMBHs, we can include ultra-luminous X-ray sources (ULXs; e.g., \cite{Webb12,Cseh15}). ULXs are defined as point-like, off-nuclear (i.e., sources showing offsets from the dynamical center of their hosts) X-ray sources in nearby galaxies; they are characterized by X-ray luminosity above the Eddington limit for a stellar BH (L$_X$$>$10$^{39}$~erg~s$^{-1}$). In particular, the still-limited (in number) high-mass end of ULX distribution could be powered by IMBHs. ULXs are largely associated with star-forming regions in spiral and irregular galaxies and possibly blue compact dwarf galaxies; however, cases where they trace the old stellar population in elliptical galaxies are also present (see \cite{Kaaret2017}). Since low-mass BHs in shallow potentials (as, e.g., those of dwarf galaxies) are likely to never settle at the center of their galaxies \cite{Bellovary19}, we may expect that the location of IMBHs is consistent with the offset position of ULXs. Another astrophysical environment in which low-mass BHs may form is that of globular clusters \cite{ArcaSedda23}. As dense stellar clusters evolve, the compact remnants sink to the center of the cluster and segregate in mass. \changes{Although old globular clusters are not known for their large gas content, dynamical interactions in the densest regions may convey material (possibly originated by gas lost by stars during their evolutionary phases). If the gas available in such systems is sufficient to fuel efficient accretion, electromagnetic signatures are likely to be expected.} Wandering IMBHs in the halo of galaxies are also expected as a result of BH-BH encounters in globular clusters that can eject IMBHs. These objects can also represent one of the consequences of the galaxy merging process if seed BHs were common in the first galaxies that merged to form today's massive galaxies.

\noindent\underline{What are the electromagnetic signatures expected for MBHs/IMBHs binaries?} -- Due to the compactness of binary systems, directly imaging the two BHs requires angular resolutions that are beyond the capabilities of most of the current facilities. The only exceptions are very high-resolution interferometric (VLBI) observations, which allow us to identify double radio cores, if present, down to about parsec/tenths of parsec of relative separation in the local Universe. Observations of individual AGN in dwarf galaxies revealed that they generally show radio jets (e.g. \cite{Wrobel06,Nyland12,Nyland17,Reines20,Molina21,Mezcua18,Mezcua19}). The jet powers are very high ($\sim$10$^{42-44}$~erg~s$^{-1}$), with efficiencies $>$10\%, similar to those of more massive galaxies \cite{Davis22}. However, these sources are rare, and dedicated VLBI observations performed on, e.g., radio sources with X–shaped morphology (originally thought to be an expected signature of a binary BH; see \cite{2012RAA....12..127G} and references therein) found very few reliable systems so far \cite{Rodriguez06,Bansal17,Kharb17}, all with masses larger than $\sim$10$^7$ M$_{\odot}$. The forthcoming/proposed radio facilities, such as the Square Kilometer Array (SKA, \cite{Braun2019}) and the Next-Generation Very Large Array (ngVLA, \cite{Burke-Spolaor2018}), thanks to their high-resolution, sensitivity, and large-scale survey capabilities are expected to significantly increase the fraction of binary AGN detected at parsec separation in the local Universe \cite{Paragi15}. By resolving the sub-microJy population, these facilities will detect the radio emission of IMBHs in the Milky Way globular clusters and will unveil about 60\% of wandering BH population in the halo of our Galaxy (if accretion is triggered, for instance, by the encounter with a molecular cloud, \cite{Seepaul22}). 

At closer separations and/or higher redshifts, only indirect methods can be adopted, namely spectroscopic identification and time variability. At sub-pc distances from the accreting BH, the gas, photoionized by the central engine and rapidly moving in its potential well, produces broad emission lines observable in the UV and optical bands. In the presence of a binary system, variable asymmetric/shifted/double-peaked broad emission lines are expected \cite{Loeb10,Popovic12,Popovic21} owing to the orbital motion of the two BHs. Detailed studies were also performed to explore the potential contribution to the broad line emission from both the circumbinary disc and gaseous streams flowing towards the BH in the presence of a binary MBH \cite{Montuori11,Montuori12}. Although the presence of asymmetric/shifted/double-peaked broad emission lines is not ubiquitously associated with a binary (i.e., they can also be produced by the chance superposition of two distinct galaxies and the presence of outflows), these electromagnetic signatures have been used to spectroscopically select several samples of SMBH binary candidates \cite{Tsalmantza11,Eracleous12,Decarli13,Liu15,Runnoe15,Runnoe17,Ju13,Shen13,Wang17,Guo19}. This method works well for unobscured systems with relative separations above the physical distance of the broad emission line regions from the central engine. This translates into the capability of identifying local AGN binary candidates with masses higher than $\sim$10$^{6-7}$ M$_{\odot}$, i.e., with relative separations larger than about 0.01~pc. Current spectroscopic surveys like the Sloan Digital Sky Survey (SDSS; \cite{York2000}) are strongly biased against lower-mass AGN and IMBHs; in these plausibly fainter accreting systems, if the broad emission lines are produced, they are difficult to detect and, crucially, proper modeling does not exist yet. 

Furthermore, in the presence of a BH binary, periodic modulations are expected in the UV/optical, X-ray, and radio continuum emission, with periods comparable to the binary period or semi-period; we note, however, that also in this case alternative explanations are possible, like jet precession and warped accretion disc of a single MBH. In the presence of binaries, flux modulations can be produced by (a) the periodic feeding of the mini-disc bound to each BH due to the torque exerted onto the circumbinary disc (\cite{Bowen18,dAscoli18}, and references therein), (b) the Doppler boosting associated with the emission produced in the two mini-discs (e.g., \cite{DOrazio15}), (c) the approaching of two relativistic jets, (d) the effect of a secondary BH crossing the accretion disc of the primary BH. In this regard, the most studied example is the blazar OJ287 (0.1 pc separation at z=0.3056), which was unveiled via a complex modeling of its optical and radio variability and post-Newtonian dynamics (e.g., \cite{Valtonen08,Valtonen13,Britzen18,Dey19}). Besides OJ287, many other promising binary MBH candidates have been discovered through light-curve variations. They have a typical mass higher than 10$^7$ M$_{\odot}$ and separations even down to milli-pc; their putative periods range from several months to a few years, depending on the mass (e.g., \cite{Boroson09,Ju13,Shen13,Liu15,Graham15,Charisi16,Sandrinelli16,Sandrinelli17,Runnoe17,Wang17,Severgnini18,Serafinelli20,ONeill22}). Many of these candidates have been questioned; distinguishing real periodicities from stochastic processes acting in a single AGN is still a challenging process (e.g. \cite{Vaughan16,Zhu20}). 

In principle, flux modulations could allow us to select lower mass ($<$10$^7$ M$_\odot$) binary candidates with relative separations much closer than 10$^{-3}$~pc, periods lower than one day, and coalescence time of hours/years (depending on the mass), i.e., MBH/IMBH binaries producing GW detectable by LGWA. However, from the electromagnetic point of view, detecting these systems is still demanding. Due to the decreasing residence time (e.g., \cite{Xin21} and see Fig.~\ref{fig:MBHB_residence_time}), the number of BH binaries detectable through electromagnetic signatures is expected to decrease going to very low relative separations. This, combined with the expected short periods, implies that, to select significant samples of MBH binaries through photometric periodicity for LGWA, large sky coverage time-domain surveys capable of visiting the same source within a very short period (lower than hours/days) are needed. In addition, high sensitivities are required, as in each observation a signal-to-noise ratio high enough to significantly constrain the presence of periodicity even for lower-mass active BHs is required. This will be possible in the next few years in the optical/IR bands thanks to Euclid \cite{Laureijs2011}, the Vera C. Rubin Observatory \cite{Ivezic2019}, and the Nancy Grace Roman Space Telescope \cite{Spergel2015} (at least for compact MBHs with masses down to $\sim$10$^5$ M$_{\odot}$, \cite{Xin21}), and in the radio band thanks to the next-generation radio facilities, SKA and ngVLA.

Another electromagnetic signature that can be used to trace the relative motion of the two MBHs down to very close separations is the presence of variable and double-peaked iron emission lines in X-rays; in this case, the centroid energies of the line are Doppler-shifted as a consequence of both the orbital motion and the inclination of the two BHs \cite{Sesana12,McKernan15,Severgnini18}). These iron lines are produced either in the two mini-discs (each BH is expected to be surrounded by its own mini-disc) or in the inner edge of the circumbinary disc. Given the sensitivity of the current X-ray facilities, the exposure time needed to characterize and trace the variations of the iron emission lines surpasses the very short timescales over which these variations are expected to occur. These are indeed of the order of a few kiloseconds once the masses of the systems that will be mostly detected by LGWA are considered. Another issue that may hamper the detection of variable double-peaked iron emission lines is the limited spectral resolution of the current X-ray detectors, which are unable to ascertain whether a profile is double-peaked or ascribed to a single broad emission line. This calls for dedicated next-generation X-ray telescopes with large collecting area and a few eV spectral resolution (e.g., Athena, \cite{Nandra2013}). 
Still remaining in the field of future X-ray facilities, X-ray detections of binary systems will also take advantage of the increased sensitivity and superb spatial resolution of the envisaged missions AXIS \cite{Mushotzky2018} and Lynx \cite{Lynx2018}, due to their sharp PSF, of the order of one arcsec, kept almost constant over a large portion of their large field of views ($\sim$150--4000~arcmin$^2$), which will allow the detection of the X-ray counterparts of in-spiral and coalescing BHs of $\sim$10$^{5-7}$ M$_{\odot}$ up to z$\sim$2. 

It is worth reminding that most of the electromagnetic signatures discussed above do not have a unique interpretation in terms of a binary system. However, their joint detection with GWs would remove all possible degeneracies and help constraining the physical and orbital parameters of the pair (see Sec.~\ref{sec:populations_and_formation_channels}).

\subsubsection{Extreme/Intermediate Mass-ratio Inspirals}
\label{sec:QPEs}
{\it Main contributors:} Riccardo Arcodia, Elisa Bortolas, Roberto Della Ceca, Margherita Giustini, Giovanni Miniutti, Paola Severgnini, Cristian Vignali \\

Binary systems with extreme or intermediate mass ratio are scientifically very interesting, relatively long-lived GW sources. Depending on the mass ratio ($q=M_2/M_1$), they are generally referred to as EMRIs (with $q\leq 10^{-4}$) or IMRIs (with $q \simeq 10^{-4}-10^{-2}$). In the presence of gas, these systems (collectively called EMRI hereafter) are also potential sources of electromagnetic radiation, and could therefore represent an important avenue for multi-messenger observations. Here, we discuss in some detail a newly discovered X-ray variability phenomenon that is currently best explained in the context of EMRIs; the so-called X-ray QPEs.

QPEs are soft X-ray flares that repeat quasi-periodically, with observed duration within 0.5--few hours and recurrence every $\sim2.5-20\,$hours \cite{miniutti19,giustini20,arcodia21,arcodia2024:qpes34}. They were so far observed at the centre of galaxies up to $z\sim0.05$ with stellar masses $\approx 10^{9-10} M_{\odot}$ \cite{arcodia21,arcodia2024:qpes34}, harboring low-mass SMBHs with $M_{\rm BH}\approx10^{5-7} M_{\odot}$, as inferred from the stellar velocity dispersion \cite{Wevers+2022:qpehosts,arcodia2024:qpes34}. These low SMBH mass estimates are supported by the presence of thermal-like soft X-ray emission in the quiescent time between QPEs, likely the Wien tail of a radiatively-efficient accretion disk (with peak temperature $kT \sim40-70\,$eV, \cite{miniutti19,giustini20,arcodia21,arcodia2024:qpes34}). QPEs are characterised by a sharp flux increase of more than one order of magnitude in the soft X-rays over this quiescence emission, reaching a soft X-ray luminosity of $\sim10^{42}-10^{43}$\,erg\,s$^{-1}$ at their peak \cite{miniutti19,giustini20,arcodia21,arcodia2024:qpes34}. The spectrum during the QPEs is hotter when brighter with an asymmetric energy-dependence, namely the spectrum on the rise is hotter than that of the decay at the same luminosity level \cite{arcodia22,2023A&A...670A..93M,arcodia2024:qpes34}. Strikingly, this spectral evolution is seen consistently across sources with largely different timing behavior in terms of regularity in the bursts recurrence and amplitude \cite{miniutti19,giustini20,arcodia21,arcodia22,2023A&A...670A..93M,arcodia2024:qpes34}, suggesting a common physical mechanism. So far, six sources with recurrent X-ray eruptions were found to follow these observational properties \cite{miniutti19,giustini20,arcodia21,arcodia2024:qpes34}, with two further candidates showing a similar energy dependence but only $0.5-1.5$ flares \cite{Chakraborty+2021:qpecand,Quintin+2023:qpecand}. A harder component, more akin to the hot corona in active galactic nuclei, is either absent or compatible with background thus unconstrained \cite{miniutti19,giustini20,arcodia21,arcodia2024:qpes34}. So far, no simultaneous variability was observed in the optical, UV, IR and radio wavebands \cite{miniutti19,giustini20,arcodia21,arcodia2024:qpes34}. However, the multi-wavelength photometry currently available is likely contaminated or dominated by either the host galaxy emission or that of the quiescence accretion disk, and therefore can not be unambiguously associated with the X-ray QPEs.

The origin of QPEs is currently still actively debated, although most models identify QPEs as a high-mass ratio binary \cite{Sukova+2021:qpemodel,Xian+2021:qpemodel,Zhao+2022:qpemodel,Wang+2022:qpemodel,Metzger+2022:qpemodel,King+2022:qpemodel,Krolik+2022:qpemodel,Linial+2023:qpemodel,Lu+2023:qpemodel,franchini23,Tagawa+2023:qpemodel,Linial+2023:qpemodel2}. Some of the latest proposed that QPEs are the outcome of the interaction between the smaller orbiter and the accretion disk around the primary massive black hole \cite{Xian+2021:qpemodel,Sukova+2021:qpemodel,Lu+2023:qpemodel,franchini23,Tagawa+2023:qpemodel,Linial+2023:qpemodel2}. For instance, in \cite{franchini23} the QPE signal (and the departure from exact periodicity) can be explained considering a geometrically thick and low mass ($\sim1M_\odot$) disk undergoing Lense-Thirring precession around a spinning central massive black hole. The disk is regularly pierced by an inspiralling stellar black hole \cite{franchini23} or star \cite{Xian+2021:qpemodel,Linial+2023:qpemodel2} that is misaligned with respect to the plane of the disc. Each impact between the orbiter and the disk results in an  initially optically thick gas cloud expelled from the disk plane, which adiabatically expands giving rise to the quasi-regular X-ray eruptions \cite{franchini23,Linial+2023:qpemodel2}. Quite interestingly, some models have proposed that the compact accretion disk around the primary massive black hole may be the fed via a precursor TDE \cite{Xian+2021:qpemodel,Wang+2022:qpemodel,franchini23,Tagawa+2023:qpemodel,Linial+2023:qpemodel2}. This connection between QPEs and previous multi-wavelength behavior akin to that of TDEs is supported by observations of the discovery QPE source GSN 069, such as its UV spectrum \cite{Sheng+2021:tdegsn} and long-term X-ray evolution \cite{Shu+2018:decay,2023A&A...670A..93M}, and by the two additional candidate QPEs \cite{Chakraborty+2021:qpecand,Quintin+2023:qpecand}. If confirmed, this interpretation of the QPE signal would imply that QPEs are the counterparts of the still-to-be-detected GWs from EMRIs. The study of those X-ray sources would be fundamental to anticipate, complement and enhance the study of EMRIs through their gravitational wave emission. For completeness, it is worth mentioning that alternative models have been proposed for QPEs, suggesting their origin to be related to disk instabilities \citep{Raj+2021:qpemodel,Sniegowska+2023:qpemodel,Kaur+2023:qpemodel,Pan+2023:qpemodel}. 

So far the signal observed from QPEs is restricted to the soft X-ray band ($E< 2$ keV) and the abundance rates of these transients are low ($\approx0.6\times10^{-6}\,$Mpc$^{-3}$; \cite{arcodia2024:rates}). Therefore, for the purpose of QPE discovery, a soft X-ray telescope with a large effective area and a large field of view is desirable. The WFI onboard Athena may discover $\approx2$ QPE source per 15-square-degree survey \citep{arcodia2024:rates}. WFI will also be able to study spectral and timing properties of QPEs deeper in redshift, up to $z\sim 0.5$ for the most common QPE sources peaking at $\sim10^{42}$\,erg\,s$^{-1}$ and up to $z\sim 1.3$ for the much rarer ones peaking at $\sim10^{43}$\,erg\,s$^{-1}$. The large effective area and high spectral resolution of the X-IFU onboard Athena will also allow to assess the presence of ionized outflowing/inflowing gas before, during, and after QPEs.

\acknowledgments

This article has benefited from the program \emph{Lunar Gravitational-Wave Detection} (ICTS / LGWD2023/04) organised by the International Centre for Theoretical Sciences (ICTS), Bengaluru, India. 

AS acknowledges the financial support provided under the European Union's H2020 ERC Consolidator Grant ``Binary Massive Black Hole Astrophysics'' (B Massive, Grant Agreement: 818691). GM is supported by grant No.~PID2020-115325GB-C31 funded by MCIN/AEI/10.13039/50110001103. MManc is supported by ERC Starting Grant No.~945155--GWmining, Cariplo Foundation Grant No.~2021-0555, MUR PRIN Grant No.~2022-Z9X4XS, and the ICSC National Research Centre funded by NextGenerationEU. The work of MManc received support from the French government under the France 2030 investment plan, as part of the Initiative d’Excellence d'Aix-Marseille Universit\'e – A*MIDEX AMX-22-CEI-02. TP's work is supported by an advanced ERC grant MultiJets. The work of MMagg is supported by the Swiss National Science Foundation (SNSF) grants 200020$\_$191957  and CRSII5$\_$213497; SF, FI and NM are supported by the SNSF grant 200020$\_$191957; EB is supported by the SNSF grant CRSII5$\_$213497. ARP is supported by the Swiss National Science Foundation (SNSF Ambizione grant no.~182044). RA was supported by NASA through the NASA Hubble Fellowship grant \#HST-HF2-51499.001-A awarded by the Space Telescope Science Institute, which is operated by the Association of Universities for Research in Astronomy, Incorporated, under NASA contract NAS5-26555. MWC acknowledges support from the National Science Foundation (NSF) with grant numbers PHY-2010970 and PHY-2117997. AM acknowledges financial support from MUR PRIN Grant No. 2022-Z9X4XS, funded by the European Union - Next Generation EU. CC acknowledges the support of P Sreekumar and the LGWA discussion meeting hosted by the International Centre for Theoretical Sciences (ICTS), Tata Institute of Fundamental Research, India, in April 2023. MAS acknowledges funding from the European Union’s Horizon 2020 research and innovation programme under the Marie Skłodowska-Curie grant agreement No.~101025436 (project GRACE-BH, PI: Manuel Arca Sedda). MAS acknowledge financial support from the MERAC Foundation. OG is funded by the Deutsche Forschungsgemeinschaft (DFG, German Research Foundation) under Germany’s Excellence Strategy – EXC 2121 ``Quantum Universe" – 390833306. SCS is funded by the Deutsches Zentrum f\"ur Luft- und Raumfahrt (DLR) with funding from the Bundesministerium f\"ur Wirtschaft und Klimaschutz under project reference 50OQ2302. RS acknowledges support from the agreement ASI-INAF eXTP Fase B - 2020-3-HH.1-2021 and the INAF-PRIN grant “A systematic Study of the largest reservoir of baryons and metals in the Universe: the circumgalactic medium of galaxies” (No. 1.05.01.85.10). FO acknowledges support from MIUR, PRIN 2020 (grant 2020KB33TP) ``Multimessenger astronomy in the Einstein Telescope Era (METE)''. PS, RDC, RS, CV, ADR, VB acknowledge a financial contribution from the Bando Ricerca Fondamentale INAF 2022 Large Grant, ‘Dual and binary supermassive black holes in the multi-messenger era: from galaxy mergers to gravitational waves’. MG is supported by the ``Programa de Atracci\'on de Talento'' of the Comunidad de Madrid, grant number 2022-5A/TIC-24235. The work of SM is a part of the $\langle \texttt{data|theory}\rangle$ \texttt{Universe-Lab} which is supported by the TIFR and the Department of Atomic Energy, Government of India.  MZ is funded by the Fonds National de la Recherche Scientifique (FNRS) under projet de recherche STELLAR (T.0022.22). AB acknowledges support for this project from the European Union's Horizon 2020 research and innovation program under grant agreement No 865932-ERC-SNeX. PA, NVK, PK, RP and VP were supported by the Department of Atomic Energy, Government of India, under Project No. RTI4001. AMB acknowledges funding from the European Union’s Horizon 2020 research and innovation programme under the Marie Sk\l{}odowska-Curie grant agreement No 895174. DV acknowledges support from the NASA Hubble Fellowship Program grant HST-HF2-51520. M.T. acknowledges support from MUR through the FARE 2020 scheme, project 'B Massive-LISA', code: R203NMYZJ5.

\paragraph{Note added.} The main coordinators of the white paper are J Harms, A Frigeri, A Maselli,  M Olivieri, R Serafinelli, and P Severgnini. Important section coordination was done by P Ajith, M Arca Sedda, R Arcodia, A Bobrick, E Bortolas, MW Coughlin, J van Heijningen, F Iacovelli, V Korol, M Mancarella, F Patat, M Toscani, and C Vignali.

\bibliographystyle{JHEP}
\bibliography{references1.bib,references2.bib,references3.bib}

\end{document}